\documentclass[12pt]{report}

\usepackage[a4paper,left=35mm,right=25mm,top=40mm,bottom=35mm]{geometry}
\usepackage{inputs/masthesis}
\usepackage{numprint}
\usepackage{verbatim}
\usepackage{amsmath}
\usepackage{graphicx}
\usepackage{tabularx}
\usepackage{subfigure}
\usepackage{float}
\usepackage{longtable}
\usepackage{url} 
\usepackage{color}
\usepackage{times,natbib}
\usepackage{amsmath,amsfonts}
\usepackage{fixltx2e}
\usepackage[labelfont=bf]{caption}
\usepackage{mathtools}
\usepackage{placeins}
\usepackage{multirow}
\newcommand{\hl}[1]{#1}
\usepackage{titlesec}
\titleformat{\chapter}{\normalfont\centering\fontsize{16pt}{1em}\bfseries}{}{0em}{}
\titleformat{\section}{\normalfont\fontsize{14pt}{1em}\bfseries}{\thesection.}{0.5em}{}
\titleformat{\subsection}{\normalfont\fontsize{12pt}{1em}\bfseries}{\thesubsection.}{1em}{}
\titlespacing*{\chapter}{0pt}{0pt}{40pt}
 \usepackage{nomencl} 
\makenomenclature
\begin{document}

  \phdtitle{Bayesian Hierarchical Modelling for Inferring Genetic Interactions in Yeast}
           {Jonathan Heydari}          
           {fig/shield}                 
           {February, $2014$}                

  \thispagestyle{empty}
  \cleardoublepage

 \begin{abstract}
Identifying genetic interactions for a given microorganism, such as yeast, is difficult. 
Quantitative Fitness Analysis (QFA) is a high-throughput experimental and computational methodology for quantifying the fitness of microbial cultures.  
QFA can be used to compare between fitness observations for different genotypes and thereby infer genetic interaction strengths. 
Current ``naive'' frequentist statistical approaches used in QFA do not model between-genotype variation or difference in genotype variation under different conditions.
In this thesis, a Bayesian approach is introduced to evaluate hierarchical models that better reflect the structure or design of QFA experiments.
First, a two-stage approach is presented: a hierarchical logistic model is fitted to microbial culture growth curves and then a hierarchical interaction model is fitted to fitness summaries inferred for each genotype.  
Next, a one-stage Bayesian approach is presented: a joint hierarchical model which simultaneously models fitness and genetic interaction, thereby avoiding passing information between models via a univariate fitness summary.
The new hierarchical approaches are then compared using a dataset examining the effect of telomere defects on yeast.
By better describing the experimental structure, new evidence is found for genes and complexes which interact with the telomere cap.
Various extensions of these models, including models for data transformation, batch effects and intrinsically stochastic growth models are also considered.
\\
\\

\end{abstract}
\thispagestyle{empty}
\cleardoublepage

\begin{acknowledgements}
First and foremost I would like to thank both Prof Darren Wilkinson and Prof David Lydall for their support and encouragement during the preparation of this thesis.
Thanks also go to Dr Conor Lawless for his invaluable support and advice.
Further, thanks to the staff and students from both the School of Mathematics and Statistics and the Institute of Cellular and Molecular Biosciences.

Special thanks go to my family and friends for the support and motivation they have provided me throughout my studies.
In particular, I would like to express my love and gratitude to my partner Christina for her encouragement and patience.

Finally, I would like to acknowledge the financial support provided by the Biotechnology and Biological Sciences Research Council and the Medical Research Council.
\end{acknowledgements}
\thispagestyle{empty}
\cleardoublepage
  
  \pagenumbering{roman}                 
\setcounter{secnumdepth}{5}
\setcounter{tocdepth}{5}
  \tableofcontents

  \listoffigures

  \listoftables
 \printnomenclature

  \clearpage                            
  \thispagestyle{empty}                 
  \cleardoublepage                      

  \pagenumbering{arabic}                
\nomenclature{BUGS}{Bayesian inference Using Gibbs Sampling}
\nomenclature{\emph{S. cerevisiae}}{\emph{Saccharomyces cerevisiae}}
\nomenclature{\emph{H. sapiens}}{\emph{Homo sapiens}}
\nomenclature{SLGM+N}{Stochastic logistic growth model with Normal measurement error}
\nomenclature{SLGM+L}{Stochastic logistic growth model with Log-normal measurement error}
\nomenclature{JAGS}{Just Another Gibbs Sampler}
\nomenclature{ORF}{Open reading frame}
  \nomenclature{\emph{orf}$\Delta$}{Open reading frame deletion}
   \nomenclature{QFA}{Quantitative Fitness Analysis}
\nomenclature{IOD}{Integrated optical density}
  \nomenclature{MDP}{Maximum doubling potential}
  \nomenclature{MDR}{Maximum doubling rate}
  \nomenclature{SHM}{Separate hierarchical model}
  \nomenclature{IHM}{Interaction hierarchical model}
 \nomenclature{JHM}{Joint hierarchical model}
   \nomenclature{ACF}{Auto-correlation function}
 \nomenclature{\emph{CDC13}}{A wild type gene}
  \nomenclature{Cdc13}{A wild type protein}
  \nomenclature{\emph{cdc13$\Delta$}}{A null allele (or gene deletion) of \emph{CDC13}}
 \nomenclature{\emph{cdc13-1}}{A point mutation of the wild type gene \emph{CDC13}}
   \nomenclature{E-MAP}{Epistatic Miniarray Profiling}
      \nomenclature{SGA}{Synthetic Genetic Array}
						       \nomenclature{GO}{Gene ontology}
									 \nomenclature{SGD}{Saccharomyces Genome Database}
	 \nomenclature{DAVID}{Database for Annotation, Visualization and Integrated Discovery}				
							\nomenclature{DNA}{Deoxyribonucleic acid}
	\nomenclature{ssDNA}{Single-stranded deoxyribonucleic acid}
	\nomenclature{dsDNA}{Double-stranded deoxyribonucleic acid}
	\nomenclature{DDR}{Deoxyribonucleic acid damage response}
	\nomenclature{DSB}{Double-strand break}
		\nomenclature{RNA}{Ribonucleic acid}
	\nomenclature{DSB}{Double-strand break}
\nomenclature{MCMC}{Markov chain Monte Carlo}
			       \nomenclature{ODE}{Ordinary differential equation}
       \nomenclature{SDE}{Stochastic differential equation}
								 \nomenclature{SLGM}{Stochastic logistic growth model }
											 \nomenclature{RRTR}{\citet{roman} logistic growth diffusion process}
			       \nomenclature{LNA}{Linear noise approximation}
			       \nomenclature{LNAM}{Linear noise approximation of the stochastic logistic growth model with multiplicative intrinsic noise}
			       \nomenclature{LNAA}{Linear noise approximation of the stochastic logistic growth model with additive intrinsic noise}
				       \nomenclature{FP}{False positive}
			       \nomenclature{FN}{False negative}
										
\titleformat{\chapter}{\centering\fontsize{16pt}{1em}\normalfont\bfseries}{}{0em}{Chapter \thechapter.\hspace{0.5em}}
  \begin{chapter}{\label{cha:introduction}Introduction}


High-throughput screening of microbial culture fitnesses is a powerful tool in biology that can be used to learn about the interaction between genes and proteins in living cells.  
Fitness, the ability of organisms to survive and reproduce in a specific environment, is of fundamental importance to every living organism.  
Measuring components of fitness (such as population growth rate) in microbial cultures is a way to directly assess and rank the health of such populations.
{Genome-wide Quantitative Fitness Analysis}~(QFA) is a robot-assisted high-throughput {laboratory} workflow, combining systematic {genetic} techniques {to generate arrays of genetically distinct microbial cultures with} quantification and modelling of growth curves to estimate fitnesses \citep{jove, QFA1}.
An important reason for carrying out QFA is to \hl{compare the fitnesses of cultures with distinct genotypes in order to} quantify epistasis (genetic interaction).

In \citet{QFA1}, a frequentist statistical approach is used to model and make inference for significantly interacting genes in a QFA screen comparison.
Other large-scale quantitative genetic interaction screening approaches exist, such as Epistatic Miniarray Profiling (E-MAP) \citep{emap} and Synthetic Genetic Array (SGA)\sloppy \citep{sgaboone}\sloppy, but we expect QFA to provide higher quality fitness estimates by using a culture inoculation technique which results in a wider range of cell densities during culture growth and by capturing complete growth curves instead of using single time point assays. 
\hl{QFA and alternative genetic interaction screening approaches mentioned above use frequentist statistical methods that cannot account for all sources of experimental variation or estimate evidence of genetic interaction simultaneously and do not partition variation into population, genotype and repeat levels.}
\hl{Further, the frequentist statistical approaches used in the methods above cannot account for relevant prior} information.

The first aim of this thesis is to develop new Bayesian models that will better determine genes which significantly interact than the current frequentist approach. 
Accounting for more sources of variation than the frequentist approach, Bayesian QFA will be able to find genetic interactions within QFA with less error and increased confidence. 
The new Bayesian QFA will be used to help locate genes that are related to telomere activity in suppressor/enhancer analysis as well as other high throughput experiments such as drug screening. 

Analysis of high throughput genetic screen data involves modelling both the experimental structure and its sources of variation. 
Many underlying sources of variation within the data can be identified in the experimental design. 
Without fully modelling variation within the experiment, a model may not be able to identify the more subtle interactions.
With a Bayesian approach \citep{Bayth} there is more flexibility of model choice, allowing model structure to reflect experimental structure or design.
Currently there is no standard frequentist approach which can deal with inference for a hierarchical model that simultaneously models logistic growth parameters and probability of genetic interaction. 
\hl{Using Bayesian hierarchical modelling \citep{GelmanMultilevel}, this study looks to extract as much information as possible from valuable QFA data sets}.
The Bayesian hierarchical approach also allows the borrowing of strength across subjects, helping identify
significantly interacting open reading frame deletions ($\emph{orf}\Delta$s) which otherwise may have been given low significance and overlooked. 

Prior distributions are used to incorporate the existing information known about the possible values for parameters.
Bayesian analysis can allow the use of Boolean indicators to describe the evidence that each $\emph{orf}\Delta$ interacts with the query mutation in terms of probability. 
During the model fitting procedure, we find that $\emph{orf}\Delta$ fitnesses have a long-tailed distribution around their population mean due to unusually fit, dead or missing \emph{orf}$\Delta$s.
In these instances, the scaled $t$ distribution is used to describe these features.

Following the approach for determining epistasis from the comparison of two QFA screens presented by \citet{QFA1}, the present study develops a two-stage approach to this problem: 
$i)$~the separate hierarchical model (SHM) is fitted to cell density measurements to estimate fitness, then $ii)$~fitness estimates are input to the interaction hierarchical model (IHM).
Next, a unified approach, referred to as the joint hierarchical model (JHM), is developed. 
The JHM models mutant strain fitnesses and genetic interactions simultaneously, without having to pass information between two different models.
\hl{The JHM can also allow two important, distinct, microbial fitness phenotypes (population growth rate and carrying capacity) to provide evidence for genetic interaction} simultaneously.

Applying the new Bayesian approaches to QFA screen data, the present study is able to identify new genes and complexes that interact with genetic mutation \emph{cdc13-1} in yeast. 
\emph{cdc13-1} is a genetic mutation which results in dysfunctional telomere maintenance. 
Telomeres are repetitive regions of deoxyribonucleic acid (DNA) at the end of linear chromosomes. They have been of great interest in recent years as they have been shown to have a role in ageing and cancer \citep{telo_sen}.
\\
\\
Current approaches \citep{QFA1} fit  a deterministic logistic growth model to yeast QFA data. 
For logistic growth data sets where stochastic fluctuations are observed, the deterministic model fails to account for the intrinsic noise.
To better describe observed yeast QFA data, a stochastic model can be used.
Stochastic models simultaneously describe dynamics and noise or heterogeneity in real systems \citep{calibayes}.  For example, stochastic models are increasingly recognised as necessary tools for understanding the behaviour of complex biological systems \citep{wilkinson2012stochastic,wilkinson_nature} and are also used to capture uncertainty in financial market behaviour \citep{stochfinance,stochfinance2}. Many such models are written as continuous stochastic differential equations (SDEs) which often do not have analytical solutions and are slow to evaluate numerically compared to their deterministic counterparts.  Simulation speed is often a particularly critical issue when inferring model parameter values by comparing simulated output with observed data \citep{woodtrees}.  

For SDE models where no explicit expression for the transition density is available, it is possible to infer parameter values by simulating a latent process using a data augmentation approach \citep{darren2005}. However, this method is computationally intensive and not practical for all applications. When fast inference for SDEs is important, for example real-time analysis as part of decision support systems or big data inference problems where simultaneous model fits are made to many thousands of datasets (e.g. \cite{heydari}), an alternative approach is needed \citep{heydari_sde}.

The second aim of this thesis is to present a fast approach for stochastic modelling of processes with intractable transition densities and apply this approach to a SDE describing logistic population growth for the first time.
One such approach is demonstrated: developing an analytically tractable approximation to the original SDE, by making linear noise approximations (LNAs) \citep{kurtz1,kurtz2,van}.
The present study introduces two new first order LNAs of a stochastic logistic growth model (SLGM) \citep{capo_slgm}, one with multiplicative and one with additive intrinsic noise, which are labelled LNAM and LNAA respectively.
The LNA reduces a SDE to a linear SDE with additive noise, which can be solved to give an explicit expression for the transition density. 

The Bayesian approach can be applied in a natural way to carry out parameter inference for state space models with tractable transition densities \citep{dynamicmodels}.  
A state space model describes the probabilistic dependence between an observation process variable $X_t$ and state process $S_t$.
The transition density is used to describe the state process $S_t$ and a measurement error structure is chosen to describe the relationship between $X_t$ and $S_t$. 
Transition densities are derived for the LNA approximate models and measurement noise is chosen to be either multiplicative or additive in order to construct a linear Gaussian structure and allow fast inference through the use of a Kalman filter. 
The Kalman filter \citep{kalmanoriginal} is typically used to infer the hidden state process of interest $S_t$ and is an optimal estimator, minimising the mean square error of estimated parameters.
The main assumptions of the Kalman filter are that the underlying system is a linear dynamical system and that all noise is Gaussian (or that the mean and standard deviation of the noise is known).
Here the Kalman filter is used to reduce computational time in a parameter inference algorithm by recursively computing the marginal likelihood \citep{dynamicmodels}.

It is shown that both of the new diffusion equation models have more realistic growth characteristics at the saturation stage when compared to a related model by \citet{roman} (an approximate model approach which is labeled RRTR) and it is shown that a zero-order LNA of the logistic growth SDE with multiplicative intrinsic noise is equivalent to the RRTR.

This study compares the utility of each of the approximate models during parameter inference by comparing simulations with both synthetic and real datasets.
After inference it is shown that the fast approximate methods give similar posterior distributions to the slow arbitrarily exact models. 
Of the approximate models considered, the RRTR model is shown to be the worst at recovering true parameters of logistic growth data.

The LNA models are an improvement over the RRTR and so should be used for better parameter inference of logistic growth data, as they are just as fast but more accurate.
The stochastic modelling approach presented in this study, a LNA followed by a Kalman filter recursion for marginal likelihood computation, is applicable to a range of population growth models or stochastic processes, where fast inference is of importance.  
The approach presented in this study enables stochastic modelling for a big data genome-wide analysis, where previously a deterministic model, unable to capture the information within the stochasticity of a process, is assumed due to the constraints in computational time associated with large volumes of data.
The problems of big data \citep{big} are relatively new and part of an expanding field of research that involves large and complex collections of data sets, typically with large components of noise.

\section{\label{int:QFA}Quantitative Fitness Analysis}
Genome-wide Quantitative Fitness Analysis (QFA) is a robot-assisted high-throughput \hl{laboratory} workflow, combining systematic \hl{genetic} techniques \hl{to generate arrays of genetically distinct microbial cultures with} quantification and modelling of growth curves to estimate fitnesses \citep{jove, QFA1}.  
A QFA screen can be used to compare the fitnesses of cultures with distinct genotypes in order to quantify genetic interaction.

\hl{Genetic interaction strengths are typically estimated by comparing fitnesses in two QFA screens: a control screen and a query screen.
QFA output includes fitness estimates for all \hl{microbial} cultures in an arrayed library including replicate cultures.  For example, such a library could be a systematic collection of all non-essential, single gene deletion strains in the \hl{model eukaryote} \emph{Saccharomyces cerevisiae} (\emph{S. cerevisiae}, brewer's yeast).}
All strains within a query screen differ from their control screen counterparts by a \hl{common} condition such as \hl{a background} gene mutation, drug treatment, temperature \hl{or other treatment}.
\hl{To identify strains that show interaction with the query condition, corresponding fitness responses for each strain in the library under the query and control conditions can be compared.}

An example of the procedure to create mutant strains to test for genetic interaction using QFA screens is as follows.  
First a suitable query mutation is chosen, which is relevant to an area of biology of particular interest \hl{(e.g. \emph{cdc13-1} for its relevance to telomere capping processes)}.  
Next, a library of strains is chosen, within which to search for strains interacting with the query mutation (e.g. a genome-wide library of independent strains with individual, non-essential genes deleted: $\emph{orf}\Delta$s).  
Finally, an appropriate, neutral control background mutation is chosen \hl{(e.g.} \emph{ura3}$\Delta$) to allow the separation of the effect of background condition from that of the library strains.  
In most cases, control and query mutations are crossed with the chosen library using \hl{Synthetic Genetic Array} (SGA) technology \citep{sgaboone}.  
Independent replicate cultures are inoculated and grown across several plates for each strain under each condition to capture biological \hl{and technical} heterogeneity.  
\hl{Cultures are grown simultaneously and time course images captured by photography.
Robotic assistance is required for both culture inoculation and image capture during genome-wide screens which can include approximately 5,000 independent genotypes.}

Raw QFA data (photographs) are converted into cell density estimates using the image analysis software Colonyzer \citep{Colonyzer}.
Observed changes in cell density over time are converted to fitness estimates for both the control and query strain by fitting logistic growth curves to data.
Genetic interactions are identified by finding mutants in the query screen whose fitnesses deviate significantly from predictions given by a theoretical model of genetic independence.

\citet{QFA1} describe using QFA to infer genetic interactions with telomere-specific query mutations. They use least squares methods to fit logistic growth curves to culture time courses, then generate a univariate fitness estimate for each time course.
They use a linear model predicting query strain fitness given control strain fitness, consistent with \hl{Fisher's} multiplicative model of genetic independence, to test for genetic interaction between the query mutation and each \emph{orf}$\Delta$. 
Deviation from the predicted linear relationship between the query and control fitnesses is evidence for genetic interaction between $\emph{orf}\Delta$ and the query mutation.  
The significance of observed interactions is assigned using a simple frequentist linear modelling approach. 
One of the major limitations of the statistical model used in \cite{QFA1} is that it assumes each $\emph{orf}\Delta$ fitness has the same variance. 
It is expected that explicit modelling of heterogeneity will allow \hl{more robust identification of interactions, particularly where variability for a particular strain is unusually high (e.g. due to experimental or technical difficulties).}

\subsection{\label{int:quantifying_fit}Quantifying fitness}  
\hl{Observing changes in cell number in a microbial culture is the most direct way to  estimate culture growth rate, an important component of microbial culture fitness.
Direct counting of cell number on a high-throughput scale is not practical and so cell density estimates are made instead from culture photographs taken during QFA.
Estimates of the integrated optical density (IOD) generated by the image analysis tool Colonyzer \citep{Colonyzer} are used to capture cell density dynamics in independent cultures during QFA.}
Density estimates, scaled to normalise for camera resolution, are gathered for each culture and a dynamic model of population growth, the logistic model $\dot{x}=rx(1 - x/K)$ \citep{Verhulst1847} (see Section~\ref{int:logistic_gro}), is fit to the data. 
Example photographic images of two yeast colonies inoculated by QFA, growing over time, along with corresponding quantitative measures of growth can be seen in Figure~\ref{fig:spot2}.

\begin{figure}[h!]
  \centering
\includegraphics[width=13cm]{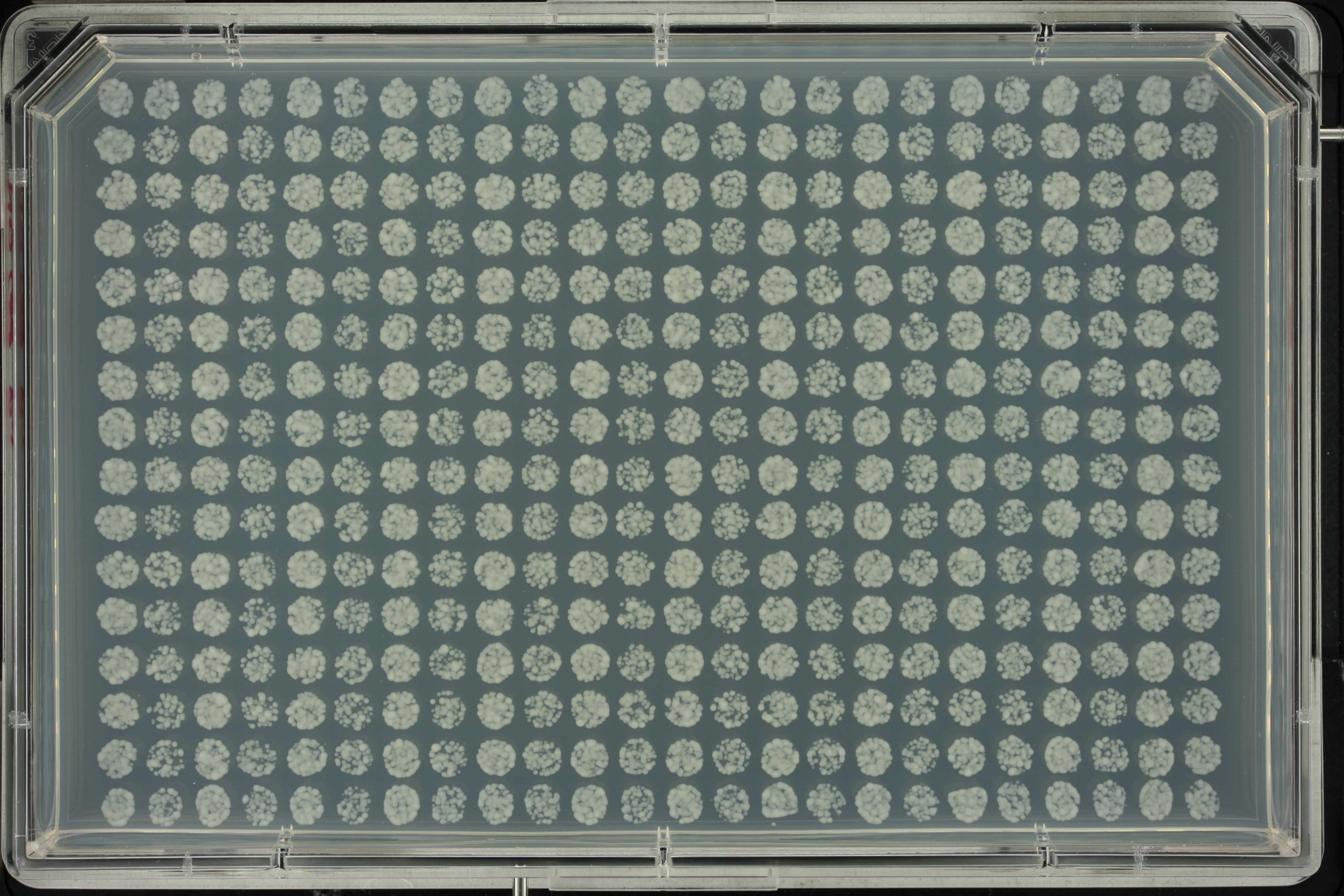}
\caption[Example 384-spot plate image from a yeast quantitative fitness analysis screen]{
Example 384-spot plate image from a yeast quantitative fitness analysis screen, taken approximately 3 days after inoculation.
Yeast cultures are spotted and grown in regular arrays on solid agar plates. 
}
\label{fig:3daysplate}
\end{figure}
\begin{figure}[h!]
  \centering
\includegraphics[width=13cm]{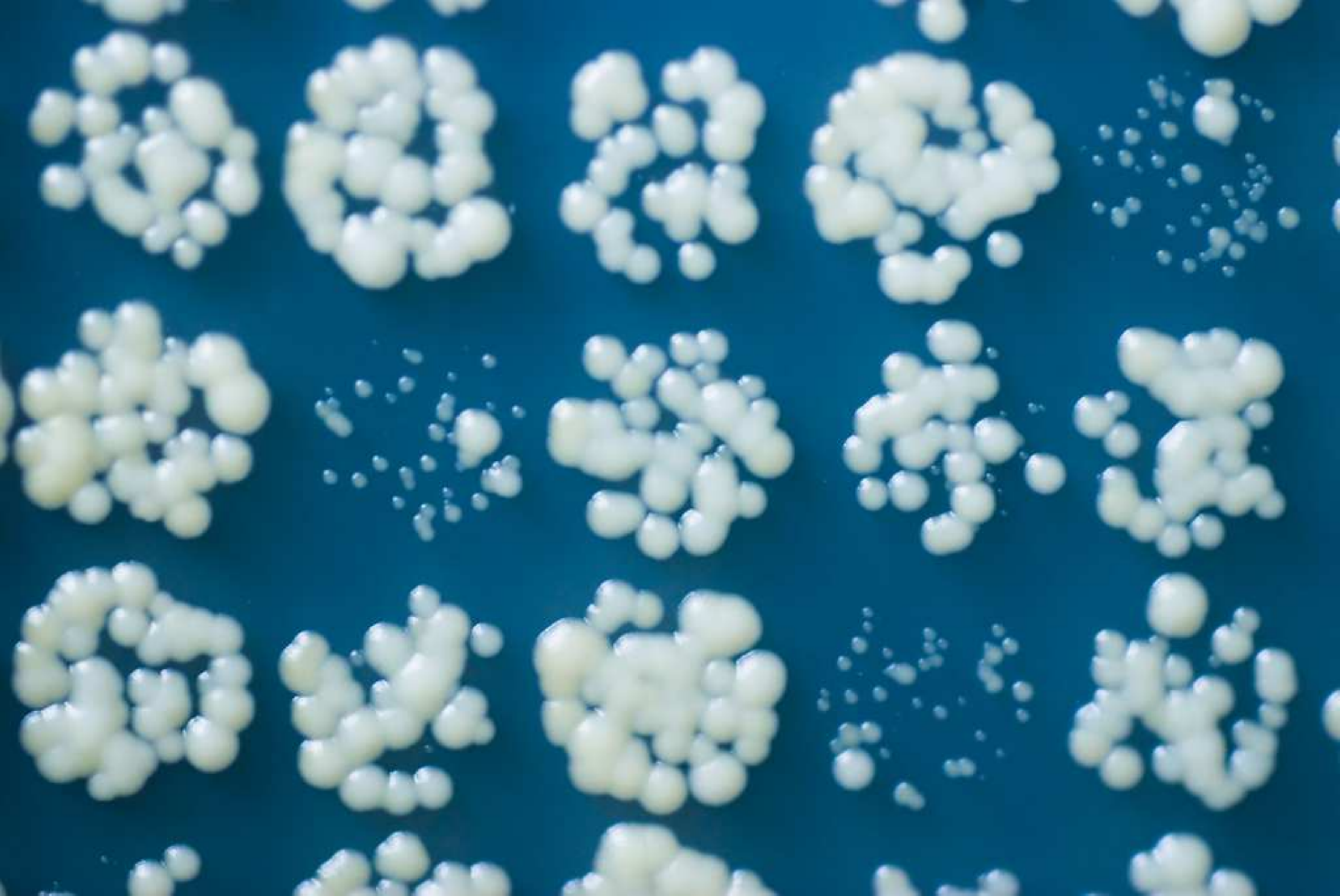}
\caption[Cropped image of 15 out of 384 spotted yeast cultures from a 384-spot plate]{
Cropped image of 15 out of 384 spotted yeast cultures from a 384-spot plate, taken from a quantitative fitness analysis screen. Image taken approximately 3 days after inoculation.
Yeast cultures are spotted and grown in regular arrays on solid agar plates.
}
\label{fig:zoomplate}
\end{figure}
For a QFA screen, cultures are typically grown on 384-spot plates over time, where a process called \emph{spotting} is used to inoculate microbial cultures on the plates.
The spotting process involves a stage where microbial cultures are first diluted and then the diluted culture is spotted to the plate.
Section~\ref{lit:synthetic_gen_arr} describes the spotting process and alternatives in further detail.
An example 384-spot plate of yeast cultures is given in Figure~\ref{fig:3daysplate}. 
Yeast cultures in Figure~\ref{fig:3daysplate} are all alive and have similar culture size. 
A cropped image of 15 yeast cultures from a 384-spot plate is given in Figure~\ref{fig:zoomplate}.
Yeast cultures in Figure~\ref{fig:zoomplate} have different culture sizes, the smaller cultures have had slow growth relative to the larger cultures.
An example of the raw time series data is given in the \hl{Appendix}, Figure~\ref{app:QFA_set_sam}.
Further detail on the QFA workflow and alternative 384-spot plate images can be found at \citep{jove} and \url{http://research.ncl.ac.uk/qfa/}.

After logistic growth model fitting, estimated logistic growth parameters sets can then be used to determine the fitness of a culture. If required, a univariate fitness definition can be chosen to summarise a set of logistic growth parameters (see Section~\ref{int:fitness_def}).
\begin{figure}[h!]
  \centering
\includegraphics[width=14cm]{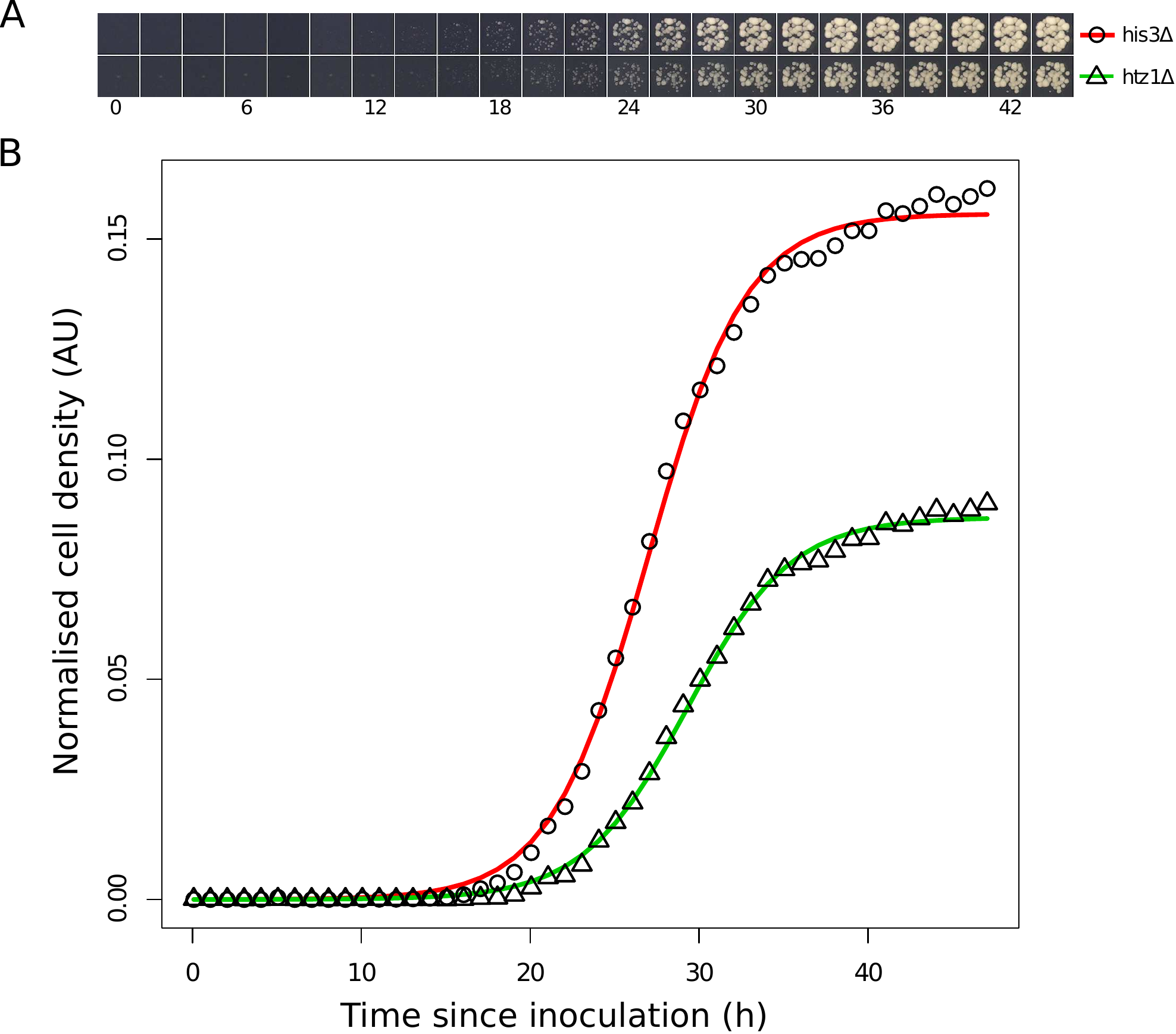}
\caption[Observed yeast data and fitted logistic growth curves]{
\hl{
A) Timelapse images for two genetically modified \emph{S. cerevisiae} cultures with different genotypes (indicated) corresponding to the time series measurements plotted in panel B.
B) Time course cell density estimates derived from analysis of the timelapse images in panel A together with (least squares) fitted logistic growth curves.
}
}
\label{fig:spot2}
\end{figure}

\subsection{\label{int:logistic_gro}The logistic growth model}  
The logistic model of population growth, an ordinary differential equation (ODE) describing the self-limiting growth of a population of size $x(t)$ at time $t$, was developed by \cite{Verhulst1847},
\begin{align}
\label{eq_det}
\frac{dx(t)}{dt}&=rx(t)\left(1-\frac{x(t)}{K}\right).
\end{align}
{The ODE has the following analytic solution:
		\begin{equation}
  x(t;\theta) = \frac{K P e^{rt}}{K + P \left( e^{rt} - 1\right)},
  \label{eq:logistic}
 	\end{equation}
	where $P=x(0)$ and  $\theta=(K,r,P)$.}
The model describes a population growing from an initial size $P$ (culture inoculum density) with an intrinsic growth rate $r$, undergoing approximately exponential growth which slows as the availability of some critical resource (e.g. nutrients or space) becomes limiting \citep{theoryoflogisticgro}.
Ultimately, population density saturates at the carrying capacity (maximum achievable population density) $K$, once the critical resource is exhausted.
Appendix~\ref{app:solving_log_gro} shows how to derive the solution of (\ref{eq_det}), given in (\ref{eq:logistic}).
An example of two different logistic growth trajectories are given by the solid lines in Figure~\ref{fig:spot2}B.
Where further flexibility is required, generalized forms of the logistic growth process \citep{analysisoflogistic,logisticrevisited} may be used instead (see Section~\ref{lit:generalised_log}).

\subsection{\label{int:fitness_def}Fitness definitions} 
Culture fitness is an important phenotype, indicating the health of a culture.
Several distinct quantitative fitness measures based on fitted logistic model parameters (\ref{eq:logistic}) can be constructed.  
\cite{QFA1} present three univariate measures suitable for QFA: Maximum Doubling Rate $(MDR)$ and Maximum Doubling Potential $(MDP)$ detailed in (\ref{eq:MDRMDP}), and their product $MDR\times MDP$, where
  \begin{equation}
  \label{eq:MDRMDP}
  MDR=\frac{r}{log\left(2\frac{K-P}{K-2P}\right)}\:\text{ and }\:MDP=\frac{log\left(\frac{K}{P}\right)}{log(2)}.
  \end{equation}	
MDR is reciprocal of minimum doubling time $T$ which a cell population takes to reach $2x(0)$, assuming the exponential phase begins at $t=0$:
	\begin{equation*}
\frac{x(t)}{x(0)}=2.
	\end{equation*}
	We now rearrange to give the following expression for MDR:
	\begin{equation*}
MDR=\frac{1}{T}=\frac{r}{\log(\frac{2(K-P)}{K-2P})}.
	\end{equation*}
MDP is the number of times population size doubles before reaching saturation, assuming geometric progression: 
	\begin{equation*}
{x(0)}\times 2^{MDP}=K.
	\end{equation*}
	Rearrange to give the following:
	\begin{equation*}
MDP=\frac{\log(\frac{K}{P})}{\log 2}.
	\end{equation*}

$MDR$ captures the rate at which microbes divide when experiencing minimal intercellular competition or nutrient stress.  A strain's growth rate largely dictates its ability to outcompete any neighbouring strains.
$MDP$ captures the number of divisions the culture is observed to undergo before saturation.  A strain which can divide a few more times than its neighbours in a specific environment also has a competitive advantage.

The choice of a single overall fitness score depends on the aspects of microbial physiology most relevant to the biological question at hand.
Typically the fitness definition $MDR\times MDP$ is used in QFA to account for both attributes simultaneously.
Other fitness definitions available include cell count, expected generation number and their approximations \citep{expectgennum}.

\section{\label{int:epistasis}Epistasis}  
Epistasis is the phenomenon where the effects of one gene are modified by those of one or several other genes \citep{epis4}.
Besides the multiplicative model, there are other definitions for epistasis such as additive, minimum and log \citep{epis2}. 
Minimum is a suboptimal approach which may allow ``masking'' of interactions \citep{epis2}.
For a typical yeast QFA screen comparison, \citet{QFA1} assumes a multiplicative interaction model (\ref{eq:epistasis}), but when dealing with measurements on a log scale, it is effectively assuming \hl{an additive} interaction model \citep{epis3}.
\hl{This highlights the point that multiplicative and additive models are equivalent if fitness data are scaled appropriately \citep{cordell2002epistasis}.}

\subsection{\label{int:defining_epi}Defining epistasis}  
\hl{As presented in \cite{QFA1}, this study assumes Fisher's multiplicative model of genetic independence (\ref{eq:epistasis}) \citep{cordell2002epistasis,epis1}, to represent the expected relationship between control strain fitness phenotypes and those of equivalent query strains in the absence of genetic interaction. 
In this study, we interpret genotypes for which the query strain fitness deviates significantly from this model of genetic independence as interacting significantly with the query mutation.
Square bracket notation is used to represent a quantitative fitness measure. 
For example $[wt]$ and $[query]$ represent wild-type and query mutation fitnesses respectively.
``Wild-type'' strictly refers to the genotype that is prevalent among individuals in a natural (or wild) population.
However, during laboratory cultivation of microbes it is more usual to introduce extra gene mutations to an ancestral lineage that is well established within the scientific community.
Working with established lineages allows direct comparison with results from the literature without the confounding effect of sampling genotypes from natural populations, which are considerably more heterogeneous.
Thus in context of this thesis, ``wild-type'' will refer to the reference strain, before additional mutations are introduced.
$\emph{orf}\Delta$ represents an arbitrary single gene deletion strain (i.e. a mutant from the control strain library). 
$query:\emph{orf}\Delta$ represents an arbitrary single gene deletion from the query strain library (e.g. crossed with the query mutation).}
Fisher's multiplicative model of genetic independence is as follows:
\begin{eqnarray}
\label{eq:epistasis}
[query:\emph{orf}\Delta]\times [wt] &=& [query]\times [\emph{orf}\Delta]\\
\Rightarrow [query:\emph{orf}\Delta] &=& \frac{[query]}{[wt]}\times [\emph{orf}\Delta]. \label{eq:linear}
\end{eqnarray}

In (\ref{eq:linear}),~$\frac{[query]}{[wt]}$ is a constant for a given pair of QFA screens, meaning that if this model holds, there should be a linear dependence between $[query:\emph{orf}\Delta]$ and $[\emph{orf}\Delta]$ for all deletions $\emph{orf}\Delta$. 
During genome-wide screens of thousands of independent $\emph{orf}\Delta$s, it can be assumed that the majority of gene mutations in the library do not interact with the chosen query mutations.  
Therefore, even if the query or wild-type fitnesses are not available to us, the slope of this linear model can still be estimated by fitting it to all available fitness observations, before testing for strains which deviate significantly from the linear model. 
Any extra background condition, such as a gene mutation common to both the control and query strains (e.g. triple instead of double deletion strains for the query and control data sets), may change the interpretation or definition of the type of genetic interaction but the same linear relationship is applicable.

\subsection{\cite{QFA1} Quantitative Fitness Analysis screen comparison\label{int:QFAqfa}}  
\cite{QFA1} present QFA where the logistic growth model (\ref{eq:logistic}) is fit to experimental data by least squares to give parameter estimates $(\hat{K},\hat{r})$ for each culture time course (each $\emph{orf}\Delta$ replicate).  
Inoculum density $P$ is assumed known and the same across all $\emph{orf}\Delta$s and their repeats.
\hl{After inoculating approximately 100 cells per culture, during the first several cell divisions there are so few cells that culture cell densities remain well below the detection threshold of cameras used for image capture and so, without sharing information across all $\emph{orf}\Delta$ repeats, $P$ cannot be estimated directly.}
It is therefore necessary to fix $P$ to the same value for both screens, using an average estimate of $P$ from preliminary least squares logistic growth model fits.
Fitting the model to each $\emph{orf}\Delta$ repeat separately means there is no sharing of information within an $\emph{orf}\Delta$ or between $\emph{orf}\Delta$s when determining $\hat{K}$ and $\hat{r}$. 
By developing a hierarchical model to share information across $\emph{orf}\Delta$ repeats for each $\emph{orf}\Delta$ and between $\emph{orf}\Delta$s, estimates for every set of logistic growth curve parameters $(K,r)$ can be improved and therefore for every strain fitness.

Quantitative fitness scores ($F_{cm}$) for each culture were defined (\ref{eq:F}) (see (\ref{eq:MDRMDP}) for definitions of $MDR$ and $MDP$), where
  \begin{equation}
  \label{eq:F}
  F_{cm} = MDR_{cm}\times MDP_{cm}.
  \end{equation} 
The index $c$ identifies the condition for a given $\emph{orf}\Delta$: $c=0$ for the control strain and $c=1$ for the query strain. 
$m$~identifies an $\emph{orf}\Delta$ replicate.
Scaled fitness measures $\tilde{F}_{cm}$ are calculated for both the control and query screen such that the mean across all $\emph{orf}\Delta$s for a given screen is equal to 1.
After scaling, any evidence that $\tilde{F}_{0m}$ and $\tilde{F}_{1m}$ are significantly different will be evidence of genetic interaction.
  
The following linear model was fit to the control and query strain scaled fitness measure pairs $\tilde{F}_{cm}$ for each unique $\emph{orf}\Delta$ in the gene deletion library:
\begin{align}
  \label{eq:lm}
	\begin{split}
	\tilde{F}_{cm} &= \mu+\gamma_{c}+\varepsilon_{cm}, \text{ where $\gamma_{0}=0$}\\
	\varepsilon_{cm} &\sim \operatorname{N}(0,\sigma^{2}), \text{ where $ \varepsilon_{cm}$ is i.i.d.}
	\end{split}
\end{align}
In (\ref{eq:lm}), $\gamma_{1}$ represents the \hl{estimated strength of genetic interaction} between the control and query strain.
If the scaled fitnesses for the control and query strain are equivalent for a particular $\emph{orf}\Delta$ such that they are both estimated by some $\mu$, i.e. no evidence of genetic interaction, we would expect $\gamma_{c}=0$.
The model was fit by maximum likelihood, using the R function ``lmList'' \citep{nlme} with variation assumed to be the same for all strains in a given screen and the same for both control and query screens.
So, for every gene deletion from the library an estimate of $\gamma_{1}$ was generated together with a p-value for whether it was significantly different from zero.
False discovery rate (FDR) corrected q-values were then calculated to determine levels of significance for each $\emph{orf}\Delta$. 
\citet{QFA1} use the Benjamini-Hochberg test \citep{ben_hoc} for FDR correction.
This test is commonly used in genomic analyses as although it assumes independence of test statistics, even if positive correlation exists between tests, the result is that FDR estimates are slightly conservative.
Finally a list of $\emph{orf}\Delta$ names, ranked by $\gamma$ magnitudes, was output and $\emph{orf}\Delta$s with q-values below a significance cut-off of 0.05 classed as showing significant levels of genetic interaction with the query mutation.

\subsection{\label{int:fit_pro}Fitness plots}
Fitness plots are used to show which $\emph{orf}\Delta$s show evidence of genetic interaction from a QFA screen comparison. 
Figure~\ref{fig:old_first} shows an example fitness plot taken from \citep{QFA1}.
Fitness plots are typically mean $\emph{orf}\Delta$ fitnesses for control strains against the corresponding query strains. 
$\emph{orf}\Delta$s with significant evidence of interaction are highlighted in the plot as red and green for suppressors and enhancers respectively.
$\emph{orf}\Delta$s without significant evidence of interaction are in grey.
Solid and dashed grey lines are for a simple linear model fit (corresponding to a model of genetic independence) and \hl{the line of equal fitness} respectively. 

\begin{figure}[h!]
  \centering
\includegraphics[width=14cm]{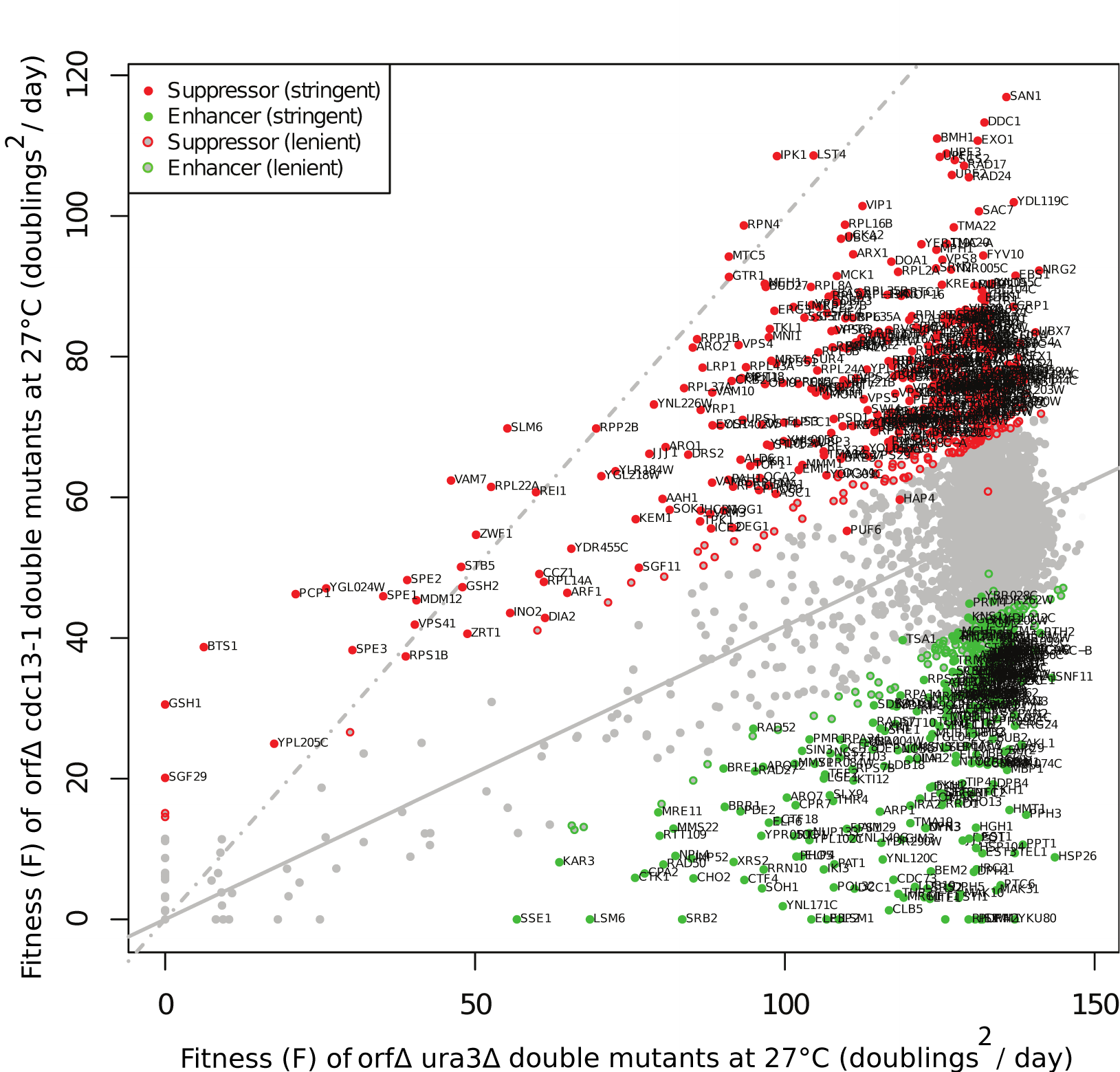}
\caption[Fitness plot taken from \citet{QFA1}]{\label{fig:old_first} Fitness plot taken from \citet{QFA1}.
A yeast genome knock out collection was crossed to the \emph{cdc13-1} mutation, or as a control to the \emph{ura3$\Delta$} mutation. 8 replicate crosses were performed for the query and control strains.
$\emph{orf}\Delta$s with significant evidence of interaction are highlighted in red and green for suppressors and enhancers respectively.
$\emph{orf}\Delta$s without significant evidence of interaction are in grey and have no \emph{orf} name label.
Lenient and stringent classification of significant interaction is based on p-values $<0.05$ and FDR corrected p-values (q-values) $<0.05$ respectively.
For a further description on fitness plots, see Section~\ref{int:fit_pro}.
}
\end{figure}
\clearpage

\section{\label{int:stochastic_logistic_gro}The stochastic logistic growth model}  
To account for uncertainty about processes affecting population growth which are not explicitly described by the deterministic logistic model, we can include a term describing intrinsic noise and consider an SDE version of the model.  Here we extend the ODE in (\ref{eq_det}) by adding a term representing multiplicative intrinsic noise (\ref{eq_det_sde_2}) to give a model which we refer to as the stochastic logistic growth model (SLGM), which was first introduced by \citet{capo_slgm},  
\begin{align}
\label{eq_det_sde_2}
dX_t&=rX_t\left(1-\frac{X_t}{K}\right)dt+\sigma X_t dW_t, 
\end{align}
where $X_{t_0}=P$ and is independent of Wiener process $W_t$, $t\geq t_0$.
The Wiener process (or standard Brownian motion) is a continuous-time stochastic process, see Section~\ref{lit:sde}.
The Kolmogorov forward equation has not been solved for (\ref{eq_det_sde_2}) (or for any similar formulation of a logistic SDE) and so no explicit expression for the transition density is available.
\cite{roman} introduce a diffusion process approximating the SLGM with a transition density that can be derived explicitly (see Section~\ref{sec:roman}).  

Alternative stochastic logistic growth models to (\ref{eq_det_sde_2}) are available.
\citet{allen} derives the stochastic logistic growth models given in (\ref{eq_allen1})~and~(\ref{eq_allen2}) from Markov jump processes \citep{allen,wilkinson2012stochastic}.
Firstly,
\begin{align}
\label{eq_allen1}
dX_t&=rX_t\left(1-\frac{X_t}{K}\right)dt+\sqrt{rX_t}dW_t, 
\end{align}
where $X_{t_0}=P$ and is independent of $W_t$, $t\geq t_0$. Secondly,
\begin{align}
\label{eq_allen2}
dX_t&=rX_t\left(1-\frac{X_t}{K}\right)dt+\sqrt{rX_t\left(1+\frac{X_t}{K}\right)}dW_t, 
\end{align}
where $X_{t_0}=P$ and is independent of $W_t$, $t\geq t_0$.

Note that (\ref{eq_det_sde_2})~(\ref{eq_allen1})~and~(\ref{eq_allen2}) are not equivalent to each other.
(\ref{eq_allen1})~and~(\ref{eq_allen2}) are able to describe the discreteness of the Markov jump processes that they approximate (or demographic noise).
Demographic noise becomes less significant for large population sizes, therefore (\ref{eq_allen1})~and~(\ref{eq_allen2}) describe more deterministic growth curves when population size is large (i.e. large carrying capacity $K$).
Equation~\ref{eq_det_sde_2} introduces an additional parameter $\sigma$, unlike (\ref{eq_allen1})~and~(\ref{eq_allen2}).
The additional parameter in (\ref{eq_det_sde_2}) allows us to tune the amount of noise in the system that is not directly associated with the noise due to the discreteness of the process (demographic noise).
The additional parameter also gives (\ref{eq_det_sde_2}) further flexibility for modelling intrinsic noise than (\ref{eq_allen1})~and~(\ref{eq_allen2}).
As the diffusion terms of (\ref{eq_allen1})~and~(\ref{eq_allen2}) are functions of the logistic growth parameters, for large populations (\ref{eq_allen1})~and~(\ref{eq_allen2}) can confound intrinsic noise with estimates of logistic growth parameters $r$ and $K$. 
For the above reasons, the SLGM in (\ref{eq_det_sde_2}) is the most appropriate model for estimating logistic growth parameters of large populations, as intrinsic noise does not tend to zero with larger population sizes, unlike (\ref{eq_allen1})~and~(\ref{eq_allen2}).

\section{Outline of thesis}
A brief outline of thesis is as follows.
Chapter~\ref{cha:background} gives background to the biological and statistical methods used throughout the thesis. Yeast biology related to the QFA data sets analysed in this study is given as well as an introduction to Bayesian inference. 

In Chapter~\ref{cha:modelling_den_int} the SHM and IHM models for the new two-stage Bayesian QFA approach are presented.
Next, the JHM for the new one-stage Bayesian QFA approach is presented. 
The chapter is concluded by introducing a two-stage frequentist QFA approach using a random effects model.

In Chapter~\ref{cha:case_stu} the new Bayesian approaches are applied to a previously analysed QFA data set for identifying genes interacting with a telomere defect in yeast. 
The chapter is concluded with an analysis of further QFA data sets with the JHM and two extensions of the JHM; included for further investigation and research. 

Chapter~\ref{cha:stochastic_app} begins by introducing an existing logistic growth diffusion equation by \citet{roman}.
Two new diffusion equations for carrying out fast, Bayesian parameter estimation for stochastic logistic growth data are then presented.
The chapter is concluded by comparing inference between the approximate models considered and with arbitrarily exact approaches.

Finally, Chapter~\ref{cha:conclusion} presents conclusions on the relative merits of the newly developed Bayesian approaches and stochastic logistic growth models. The chapter is concluded by discussing the broader implications of the results of the studies presented and scope for further research.
\end{chapter}

  \begin{chapter}{\label{cha:background}Background}
\section{\label{lit:yeast_bio}Yeast biology}  
\emph{Saccharomyces cerevisiae} is a species of budding yeast widely used to study genetics. \emph{S. cerevisiae} was the first eukaryotic genome that was completely sequenced \citep{yeast6000}. Yeast is ideal for high throughout experimentations as it is easy to use and arrayed libraries of genetically modified yeast strains are readily available or obtainable for experiments \citep{yeast}. There are many different observable traits available with \emph{S. Cerevisiae}, such as size, opacity and density. There are about 6000 genes in the \emph{S. Cerevisiae} genome of which 5,800 of these are believed to be true functional genes \citep{sgd}.

Yeasts are ideal for genome-wide analysis of gene function as genetic modification of yeast cells is relatively straightforward and yeast cultures grow quickly. 
Epistasis identified within a species of yeast may exist in the analogous
genes within the human genome \citep{yeastorg}. Therefore, finding genes involved in
epistasis within yeast is of great interest outside the particular experimental species in question.

\subsection{\label{lit:telomere}Telomeres}  
Telomeres are the ends of linear chromosomes and found in most eukaryotic organisms \citep{telo}. Telomeres permit cell division and some researchers claim that telomere-induced replicative senescence is an important component of human ageing \citep{endrep}. They cap (or seal) the chromosome end to ensure genetic stability and are believed to prevent cancer \citep{telo_sen}.   
  \begin{figure}[h!]
  \centering
\includegraphics[width=13cm]{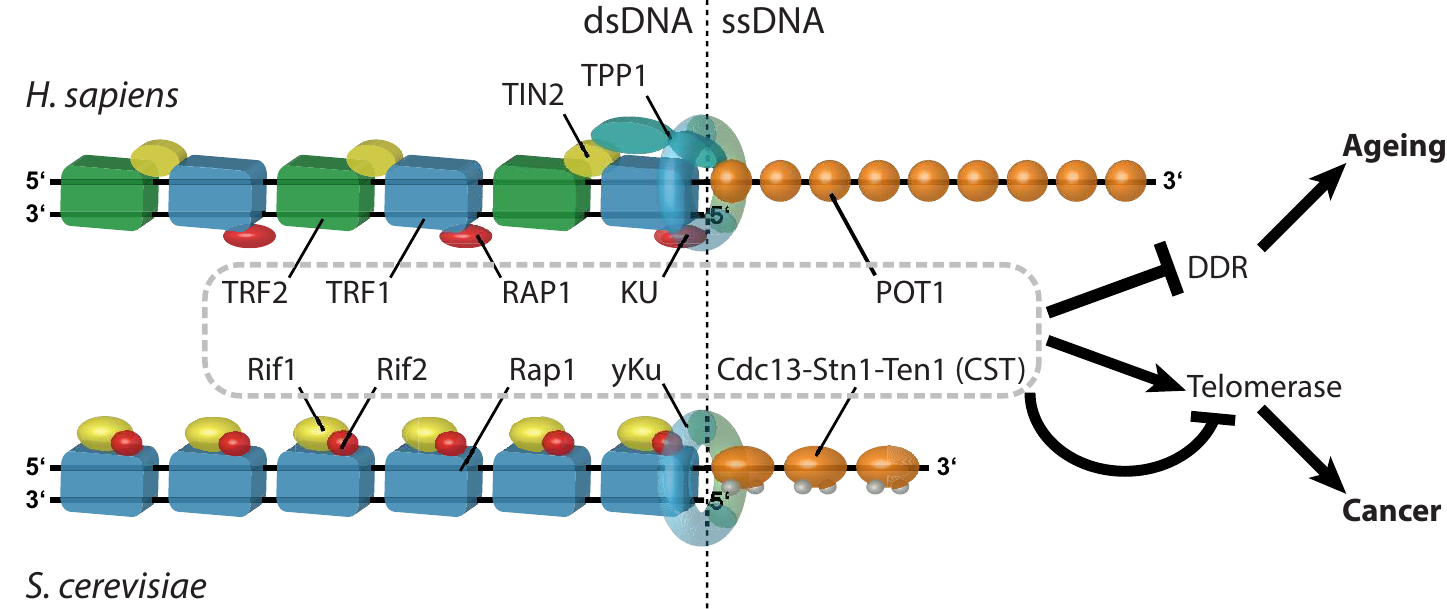}
\caption[Telomere at a chromosome end]{Telomere at a chromosome end (diagram and legend taken from \citet{james}). The telomere cap is evolutionarily conserved. Telomeres are nucleoprotein caps present at the ends of most eukaryotic chromosomes, consisting of double-stranded DNA (dsDNA) with a single-stranded DNA (ssDNA) overhang, bound by dsDNA- and ssDNA-binding proteins. Collectively, the telomere binding proteins ``cap'' the telomere and serve to regulate telomerase activity and inhibit the DNA damage response (DDR). In budding yeast, the telomeric dsDNA is bound by Rap1, which recruits the accessory factors Rif1 and Rif2. In humans, the telomeric dsDNA is bound by TRF1 and TRF2 (held together by TIN2) and TRF2 recruits RAP1 to telomeres. In budding yeast, Cdc13 binds the telomeric ssDNA and recruits Stn1 and Ten1 to form the CST (Cdc13-Stn1-Ten1) complex, while in humans, the telomeric ssDNA is bound by POT1. In human beings, POT1 and TRF1-TRF2-TIN2 are linked together by TPP1, which may permit the adoption of higher-order structures. In both budding yeast and humans, the Ku complex, a DDR component that binds to both telomeres and Double-strand breaks (DSBs), also binds and plays a protective role.}
\label{fig:lit:telomere}
\end{figure}

In Figure~\ref{fig:lit:telomere}, a \emph{S. cerevisiae} chromosome is shown with the telomere single-stranded DNA (ssDNA) at the end, where DNA binding proteins such as Cdc13 are bound.
Figure~\ref{fig:lit:telomere} also shows how telomere maintenance compares between a Homo sapiens (\emph{H. sapiens}) and \emph{S. cerevisiae} chromosome.
\\
Telomere length decreases with each division of a cell until telomere length is very short and the cell enters senescence \citep{hay}, losing the ability to divide. 
Some cancerous cells up-regulate the enzyme called telomerase which can prevent shortening of telomeres or elongate them, potentially allowing cancerous cells to live indefinitely \citep{immortal}. 
\\
It is believed that telomeres are partly responsible for ageing; without the enzyme telomerase, a fixed limit to the number of times the cell can divide is set by the telomere shortening mechanism because of the end replication problem \citep{ageing}.

\subsection{\label{lit:end_rep}The end replication problem}
In eukaryote cell replication, shown in Figure~\ref{fig:lit:drprob}, new strands of DNA are in the $5^\prime$ to $3^\prime$ direction (red arrows), the leading strand is therefore completed in one section whereas the lagging strand must be formed via backstitching with smaller sections known as Okazaki fragments \citep{endrep}. 
Figure \ref{fig:lit:drprob} shows how the lagging strand is left with a $3^\prime$ overhang, with the removal of the terminal primer at the end and how the leading strand is left with a blunt end \citep{endrep2}. Telomerase fixes this problem by extending the $3^\prime$ end to maintain telomere length \citep{ageing}. Without telomerase, the leading strand is shortened \citep{short_strand} and telomere capping proteins such as Cdc-13 in yeast binds to the ssDNA that remains.
Most eukaryotic cells have telomerase activated and may maintain DNA replication indefinitely. Not all mammalian cells have telomerase activated and it is believed this problem then leads to the shortening of their telomeres and ultimately senescence. 

\begin{figure}[h]
  \centering
\includegraphics[width=7cm]{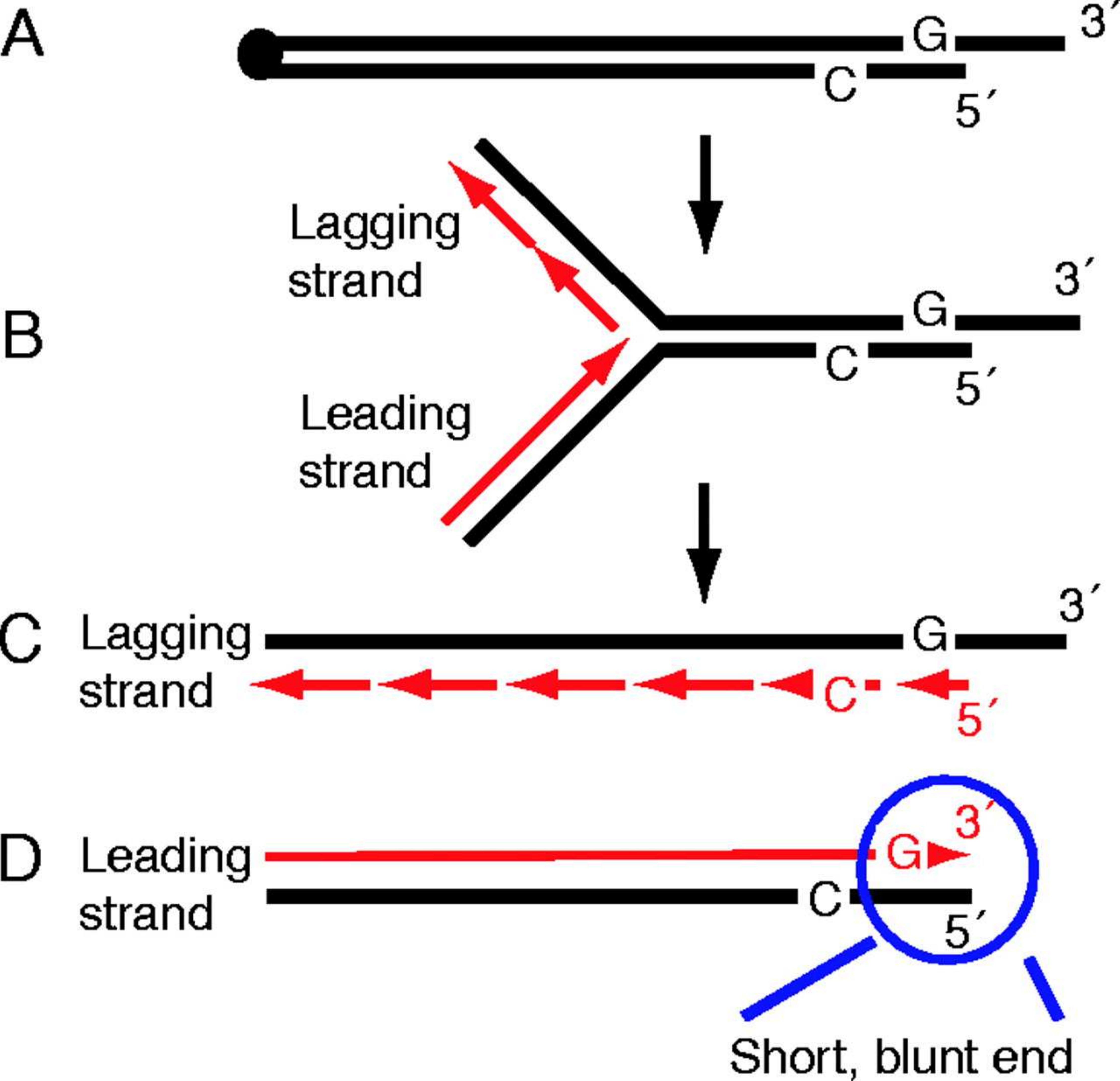}
\caption[The end replication problem]{The end replication problem (diagram and legend taken from \citet{endrep}).
(A) Telomeres in all organisms contain a short $3^\prime$ overhang on the G rich strand. (B) A replication fork moving towards the end of the chromosome. (C) The newly replicated, lagging C strand, will generate a natural $3^\prime$ overhang when the ribonucleic acid (RNA) primer is removed from the final Okazaki fragment, or if the lagging strand replication machinery cannot reach the end of the chromosome. In the absence of nuclease activity the unreplicated $3^\prime$ strand will be the same length as it was prior to replication. (D) The newly replicated leading G strand will be the same length as the parental $5^\prime$ C strand, and blunt ended if the replication fork reaches the end of the chromosome. Therefore the newly replicated $3^\prime$ G strand will be shorter than the parental $3^\prime$ strand and unable to act as a substrate for telomerase because it does not contain a $3^\prime$ overhang. If the leading strand replication fork does not reach the end of the chromosome a $5^\prime$ rather than $3^\prime$ overhang would be generated, but this would not be a suitable substrate for telomerase.}
\label{fig:lit:drprob}
\end{figure}

\subsection{\label{lit:cdc13-1}\emph{CDC13} and \emph{cdc13-1}}  
\emph{CDC13} is an essential telomere-capping  gene  in \emph{S. cerevisiae} \citep{essential}.
The protein Cdc13, encoded by \emph{CDC13}, binds to telomeric DNA (see Figure~\ref{fig:lit:telomere}), forming a nucleoprotein structure \citep{cdc13_summary}. Cdc13 regulates telomere capping and is part of the CST complex with Stn1 and Ten1 \citep{cdc_cst}. This provides protection from degradation by exonucleases such as Exo1. 
\mbox{\emph{cdc13-1}} is a temperature-sensitive allele of the \emph{CDC13} gene that has temperature sensitivity above $26\,^{\circ}\mathrm{C}$, where the capping ability of the protein is reduced \citep{cdc131}.
By inducing the temperature sensitivity of \emph{Cdc13-1}, telomere maintenance is disrupted.
A lot of research activity for telomere integrity focuses on the CST complex and often \emph{cdc13} mutations are considered, like \emph{cdc13-1} and \emph{cdc13-5} \citep[see, for example,][]{cdc,MRX}. 

\subsection{\label{lit:ura3}\emph{URA3}} 
\emph{URA3} is a gene that encodes orotidine 5-phosphate decarboxylase \citep{URA3_again}. 
\emph{URA3} is used as a genetic marker for DNA transformations, allowing both positive and negative selection depending on the choice of media \citep{URA3}.

In \citet{QFA1} \emph{ura3}$\Delta$ is used as a control mutation because it is neutral under the experimental conditions.
For a QFA comparison, constructing a query mutation such as \emph{cdc13-1} typically involves adding selection markers to the genome.
To ensure that the same selection markers are found in both the query and control strains, and that the control and query screens can be carried out in comparable environments, a neutral mutation such as \emph{ura3}$\Delta$ can be introduced to the control strain.
\emph{URA3} encodes an enzyme called ODCase.
Deleting \emph{URA3} causes a loss of ODCase, which leads to a reduction in cell growth unless uracil is added to the media \citep{ura3end}.
\citet{QFA1} include uracil in their media so that \emph{ura3}$\Delta$ is effectively a neutral deletion, approximating wild-type fitness.
As a control deletion, \emph{URA3} is not expected to interact with the query mutation, the library of \emph{orf}$\Delta$s in the control and query screen or any experimental condition of interest such as temperature.

\subsection{\label{lit:synthetic_gen_arr}High-throughput methodology for Quantitative Fitness Analysis}
To collect enough data to perform QFA \citep{QFA1}, a methodology such as high-throughput screening is required \citep{high,HTP}. 
High-throughput screening is most notably used in the field of biology for genome wide suppressor/enhancer screening and drug discovery.
The automation of experimental procedures through robotics, software, sensors and controls allows a researcher to carry out large scale experimentation quickly and more consistently.

Hundreds of microbial strains with various gene deletions need to be systematically created, cultured and then have measurable traits quantified.
The repeatability of microbial culture growth is ideal to give sufficient sample sizes for identifying both variation and significance in high throughput experimentation \citep{microbes_HTP}. 

The quality of the quantitative data is critical for identifying significantly interacting genes.
To measure the phenotypes of different mutant strains of a micro-organism such as yeast \citep{yeast}, a process called \emph{spotting} is used.
This process is different to a typical SGA experiment where \emph{pinning} would be used (see, for example, \cite{sgaboone}).
Pinning is a quicker but less quantitative process where the microbial strains are typically directly pinned to a 1536 plate and allowed to grow until image analysis starts.
Spotting on the other hand has a stage where the cultures are diluted and then the dilute culture is spotted in 384 format to give a more accurate reading in image analysis.
This in turn gives rise to much more accurate time series data for modelling.

Figure~\ref{fig:int:spot} illustrates the spotting process.
An image opacity measure is typically used as a proxy for the density of microbial colonies. 
Time lapse photographs are taken of the 384-spot plates after incubation, using high resolution digital cameras, to measure growth. 
A software package such as Colonyzer \citep{Colonyzer} can then be used to determine a quantitative measure of fitness from the photographs taken of the cultures grown on the plates.
To ensure a consistent method to capture images of microbial colonies, all cameras should be of the same make and model.

\begin{figure}[h!]
  \centering
\includegraphics[width=14cm]{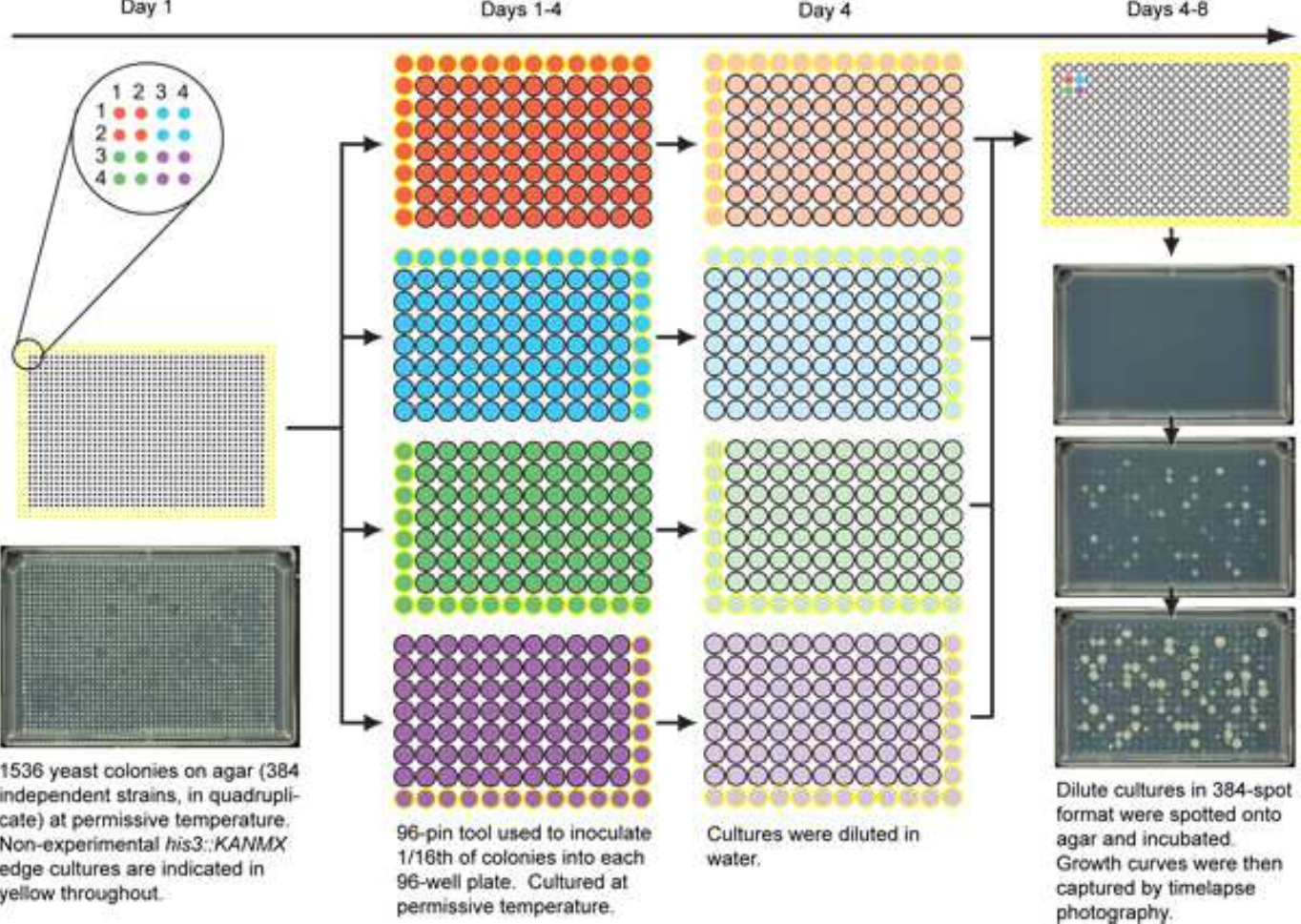}
\caption[The spotting procedure]{The spotting procedure for robotic inoculation of yeast strains in 384-spot format (diagram and legend taken from \citet{jove}).
This procedure begins with 1536 independent cultures per plate (left).
In this typical example, colonies at positions 1,1; 1,2; 2,1 and 2,2 (colored red) are four replicates of the same genotype. 
\emph{his3::KANMX} cultures in yellow, growing on the edge of the plate, have a growth advantage due to lack of competition and are therefore not examined by Quantitative Fitness Analysis.
One of these replicates (e.g. 1,1) is inoculated into liquid growth media in 96-well plates using a 96-pin tool which inoculates 96 out of 1536 colonies each time.
In order to inoculate one replicate for each of 384 gene deletions, four different ``quadrants'' (indicated as red, blue, green and purple) are inoculated into four different 96-well plates containing growth media.
After growth to saturation (e.g. 3 days at 20 °C), cultures are diluted in water, then the four quadrants from one repeat are spotted in 384-format onto a solid agar plate (right) in the same pattern as the original Synthetic Genetic Array plate (as indicated by color).
The process can be repeated to test other replicates: 1,2; 2,1 and 2,2. Example time-lapse images on the right were captured 0.5, 2 and 3.5 days after inoculation.
}
\label{fig:int:spot}
\end{figure}

\section{\label{lit:comparing_res}Comparing lists of genes}
Upon completing a QFA screen comparison, a list of genes ordered by genetic interaction strength can be obtained.
Lists of ordered genes can be used to compare two different statistical approaches for a QFA screen comparison.

A comparison of two lists can be carried out through standard statistical similarity measures such as the Jaccard Index or Spearman's rank correlation coefficient.
Observing only the subset of genes showing significant evidence of genetic interaction, two lists of genes can be compared using the Jaccard Index \citep{Jaccard}, see Section~\ref{lit:jaccard_ind}. 
The Jaccard index does not account for the ordering of genes and is dependent on the number of interactions identified when the cut-off of genes showing significant evidence of interaction is chosen or influenced by the experimenter.
Due to these undesirable properties of the Jaccard index, this method is not appropriate for an unbiased comparison of statistical methods.  
The Spearman's rank correlation coefficient \citep{spearman_book} is able to account for the ordering of genes and is able to account for the whole list of genes available, see Section~\ref{lit:spearmans_cor}.

Gene ontology (GO) term enrichment can be used to suggest which list of genetic interactions has the most biological relevance \citep{GOterm2}.
There are many other alternative approaches available for the comparison of two gene lists \citep{compare_genes,compare_genes2}.

Using both Spearman's correlation coefficient and GO term enrichment analysis of gene lists allows for both an unbiased statistical and biological comparison of two lists of ordered genes.

\subsection{\label{lit:jaccard_ind}Jaccard index}
For two sample sets, the Jaccard index \citep{Jaccard_origin,Jaccard} gives a measure of similarity.
Where $A$ and $B$ are two sample sets of interest, the Jaccard Index is as follows:
\begin{equation*}
J(A,B) = {{|A \cap B|}\over{|A \cup B|}}.
\end{equation*}
The value of J(A,B) can range from 0 to 1, with a larger number for more similarity.

\subsection{\label{lit:spearmans_cor}Spearman's rank correlation coefficient}
The Spearman's rank correlation coefficient \citep{spearman,spearman_book} allows comparison of two variables $X_i$ and $Y_i$, both of sample size $n$. 
First, $X_i$ and $Y_i$ are both converted into ranks $x_i$ and $y_i$. 
Where there are rank ties or duplicate values, the rank equal to the average of their positions is assigned. 
The Spearman's rank correlation coefficient is as follows:
\begin{equation*}
\rho = \frac{\sum_i(x_i-\bar{x})(y_i-\bar{y})}{\sqrt{\sum_i (x_i-\bar{x})^2 \sum_i(y_i-\bar{y})^2}}.
\end{equation*}
The value of $\rho$ can range from -1 to 1. As the relationship between two variables becomes closer to being described by a monotonic function, the larger in magnitude $\rho$ will be.

\subsection{\label{lit:GO_term}Gene ontology term enrichment analysis}
Gene ontology (GO) term enrichment analysis can give insight to the biological functions of a list of genes \citep{GOterm2}.
A list of GO terms can be acquired from a list of genes. For yeast the Saccharomyces Genome Database (SGD) \citep{sgd} can be used to find GO term associations for each gene in the genome.
A statistical analysis is carried out to determine which GO terms are most prevalent in a list of genes.
The experimenter can then look at GO terms of interest, find out which genes they correspond to and how many are identified in the list.

An unbiased Gene Ontology (GO) term enrichment analyses on a list of genes can be carried out using the software R \citep{rprog} and the bioconductoR package GOstats \citep{GOstats}. 
There are many other software packages and online services available to carry out a GO term enrichment such as the Database for Annotation, Visualization and Integrated Discovery (DAVID)  \citep{DAVID,DAVID2} or the Gene Ontology Enrichment Analysis and Visualization tool (GOrilla) \citep{Gorilla,Gorilla2}. 

A GO term clustering analysis is a statistical approach that can be used to follow up a GO term analysis. 
Information on the relation of GO terms is used in a clustering analysis to find functionally related groups of GO terms.
The bioinformatics tool DAVID \citep{DAVID,DAVID2} can be used to carry out GO term clustering (\url{david.abcc.ncifcrf.gov/}). 

\section{\label{lit:bayesian_inf}Bayesian inference}
A classical (or frequentist) statistical approach typically assumes model unknown parameters are constants and uses the likelihood function to make inference.
An alternative methodology is a Bayesian approach \citep{Bayth,BayPriors}, named after Thomas Bayes \citep{bayes1763}. 
In a Bayesian setting, a parametric model similar to the frequentist approach can be assumed but model parameters are treated as random variables.
This feature allows any \emph{prior} knowledge for a given parameter to be incorporated into inference by building a \emph{prior} distribution to describe the information available.
We are interested in the \emph{posterior} distribution, that is the probability of the parameters given the evidence.
Moreover, where $D$ is the observed data, $\theta$ is the set of parameters of interest, we are interested in calculating the \emph{posterior density} $\pi(\theta|D)$.
\emph{A priori} knowledge of $\theta$ is described by $\pi(\theta)$ and the likelihood of data by $L(D|\theta)$. Using Bayes theorem we obtain the following:
\begin{flalign*}
&& \pi(\theta|D)&\propto\pi(\theta)L(D|\theta)&\\
\text{or} && Posterior&\propto Prior\times likelihood.
\end{flalign*}

\subsection{\label{lit:markov_cha_mon_car}Markov chain Monte Carlo}
In Bayesian inference we are typically interested in sampling from the posterior distribution or one of its marginals, but often this is difficult. 
Markov Chain Monte Carlo (MCMC) methods are used for sampling from probability distributions \citep{MCMC,MCMC2}. The Monte Carlo name describes the repeated random sampling used to compute results.
A Markov chain can be constructed with an equilibrium distribution that is the \emph{posterior} distribution of interest.

A Markov chain $\{X_{n},n\in\mathbb{N}^0\}$ is a stochastic process which satisfies the Markov property (or ``memoryless'' property):
for $A\subseteq{S}$, where $S$ is the continuous state space s.t. $X_n\in{S}$,
\begin{equation*}
P(X_{n+1}\in{A}|X_n=x,X_{n-1}=x_{n-1},...,X_0=x_{0})=P(X_{n+1}\in{A}|X_n=x),
\end{equation*}
$\forall x,x_{n-1},...,x_0\in{S}$.
The equilibrium distribution $\pi(x)$ is a limiting distribution of a Markov chain with the following two properties.
First, there must exist a distribution $\pi(x)$ which is stationary. 
This condition is guaranteed when the Markov chain satisfies detailed balance:
\begin{equation*}
\pi(x)p(x,y)=\pi(y)p(y,x),\qquad{\forall}x,y,\end{equation*}
where $p(x,y)$ is the transition density kernel of the chain.
Secondly, the stationary distribution $\pi(x)$ must be unique.
This is guaranteed by the ergodicity of the Markov process; see \citet{MCMC} for a definition and sufficient conditions. 

\subsection{\label{lit:metropolis-has_alg}Metropolis-Hastings algorithm}
The Metropolis-Hastings algorithm \citep{met,hastings} is a MCMC method for obtaining a random sample from a probability distribution of interest (or stationary distribution) \citep{met_hast}.
With the following procedure a sample from the stationary distribution of the Markov chain can be obtained:\\
 \\
1) Initialise counter $i = 0$ and initialize $X_0=x_0$\\
\\
2) From the current position $X_i=x$, generate a candidate value $y^*$ from a proposal density $q(x,y)$.\\
\\
3) Calculate a probability of acceptance $\alpha(x,y^*)$, where
\begin{equation*}
\alpha(x,y)=
\begin{cases}
\min\left\{1,\frac{\pi(y)q(y,x)}{\pi(x)q(x,y)}\right\} & \text{if } \pi(x)q(x,y)>0\\
1 & \text{otherwise.}\\
\end{cases}
\end{equation*}
\\
4) Accept the candidate value with probability $\alpha(x,y^*)$ and set $X_{i+1}=y^*$, otherwise reject and set $X_{i+1}=x$.\\
\\
5) Store $X_{i+1}$ and iterate $i=i+1$.\\
\\
6) Repeat steps 2-5 until the sample size required is obtained.
\\
\\
The choice of proposal density is important in determining how many iterations are needed to converge to a stationary distribution.
There are many choices of proposal distribution \citep{MCMC}, the simplest case is the symmetric chain.
The symmetric chain involves choosing a proposal where $q(x,y)=q(y,x)$, such that step two simplifies to give the following:
\begin{equation*}
\alpha(x,y)=
\begin{cases}
\min\left\{1,\frac{\pi(y)}{\pi(x)}\right\} & \text{if } \pi(x)>0\\
1 & \text{otherwise.}\\
\end{cases}
\end{equation*}
More general cases are random walk chains and independence chains.

For a random walk chain, the proposed value at stage $i$ is given by $y^*=x_i+w_i$, where $w_i$ are i.i.d. random variables. 
The distribution for $w_i$ must therefore be chosen, and is typically Normal or Student's $t$ distribution centred at zero. 
If the distribution for $w_i$ is symmetric, the random walk is a special case of symmetric chains.

For an independence chain, the proposed transition is formed independently of the previous position of the chain, thus $q(x,y)=f(y)$ for some density $f(.)$:
\begin{equation*}
\alpha(x,y)=
\begin{cases}
\min\left\{1,\frac{\pi(y)f(x)}{\pi(x)f(y)}\right\} & \text{if } \pi(x)f(y)>0\\
1 & \text{otherwise.}\\
\end{cases}
\end{equation*}

Parameters within our proposal distribution are known as tuning parameters. They are typically used to adjust the probability of acceptance or improve mixing and must be chosen through some automatic procedure or manually, see Section~\ref{lit:convergence_iss}.

\subsection{\label{lit:gibbs_sam}Gibbs sampling}
The Gibbs sampler \citep{Gibbs_origin,gibbs2} is a MCMC algorithm for obtaining a random sample from a multivariate probability distribution of interest $\pi(\theta)$, where $\theta=(\theta^1,\theta^2,...,\theta^d)$. 
Consider that the full conditional distributions $\pi(\theta_i|\theta_{1},...,\theta_{i-1},\theta_{i+1},...,\theta_{d})$, $i=1,...,d$ are available. 
Where it is simpler to sample from conditional distribution than to marginalize by integrating over a joint distribution, the Gibbs sampler is applicable. 
The following procedure sequentially samples from the full conditional distribution for each parameter, resulting in the probability distribution of interest.
The algorithm is as follows:\\ \\
1) Initialise counter $i=1$ and parameters $\theta_{(0)}=(\theta_{(0)}^1,\theta_{(0)}^2,...,\theta_{(0)}^d)$.\\ \\
2) Simulate $\theta_{(i)}^1\text{ from }\theta_{(i)}^1\sim \pi(\theta^1|\theta_{(i-1)}^2,...,\theta_{(i-1)}^d)$.\\ \\
3) Simulate $\theta_{(i)}^2\text{ from }\theta_{(i)}^2\sim \pi(\theta^2|\theta_{(i)}^1,\theta_{(i-1)}^3,...,\theta_{(i-1)}^d)$.\\ \\
4) $...$\\ \\
5) Simulate $\theta_{(i)}^d \text{ from }\theta_{(i)}^d\sim \pi(\theta^d|\theta_{(i)}^1,...,\theta_{(i)}^{d-1})$.\\ \\
6) Store $\theta_{(i)}=(\theta_{(i)}^1,\theta_{(i)}^2,...,\theta_{(i)}^d)$ and iterate $i=i+1$.\\ \\
7) Repeat steps 2-6 until the sample size required is obtained.\\
\\
To ensure the full conditional distributions for each parameter in a Bayesian model are known and easy to handle, conjugacy can be used.
Conjugacy is where the prior is of the same family as the posterior.
Conjugacy can be induced by the choice of prior, for example if it is known that a likelihood is Normal with known variance, a Normal prior over the mean will ensure that the posterior is also a Normal distribution. 

\subsection{\label{lit:convergence_iss}Convergence issues}
To accept output from MCMC algorithms, all chains are required to have reached convergence \citep{MCMC,converge}.
Convergence is a requirement to gain unbiased samples of a posterior distribution.
Visual and statistical tests can be used to determine if  chains have converged, see Section~\ref{lit:convergence_dia}.

Other issues that we must consider for MCMC sampling algorithms are choice of tuning parameters, burn-in period, sample size and thinning, if required.
Tuning parameters require a good choice of proposal distribution, preferably with high acceptance rates and good mixing.
There are many schemes available for the choice of tuning parameters \citep{tuning}. Typically tuning parameters are determined during a burn-in period.
The burn-in period is a number of iterations which an algorithm must be run for in order to converge to equilibrium.
Sample size depends on how many iterations from the posterior are required for both inference and testing convergence.
Thinning involves discarding output for iterations of a MCMC algorithm, in order to give less dependent realizations from the posterior distribution.

Extending the length of the burn-in period, sample size and thinning leads to increased computational time. 
With large data sets and models with a large number of parameters, computation time can become a problem.
With a Bayesian modelling approach, computational time associated with MCMC can be much longer than a much simpler least squares approach. 
This problem is exacerbated when coupled with poor mixing and is likely to lead the experimenter to simplify their modelling procedure, consequently sacrificing the quality of inference, in order to complete their analysis within a shorter time frame.

\subsection{\label{lit:convergence_dia}Convergence diagnostics}
To determine whether chains are true samples from their target distributions, tests for lack of convergence or mixing problems \citep{MCMC,converge} must be carried out.
Typically multiple tests are used to give confidence that the output has convergence.
There are many convergence diagnostics for testing chains for convergence, for example the Heidelberg-Welch \citep{Heidelberger} and Raftery-Lewis \citep{Raftery} tests.
For many convergence diagnostics, summary statistics such as p-values can be used to decide whether convergence has been reached.
Visual inspection of diagnostic plots can also be used to determine if convergence has been reached. 
Trace plots are used to check if samples from the posterior distribution are within a fixed region of plausible values and not exploring the whole range. 
ACF (auto-correlation function) plots are used to determine serial correlation between sample values of the posterior distribution in order to check for the independence of observations. 
Density plots are used to check whether a sample posterior distribution is restricted by the choice of prior distribution and determine whether choice of prior is appropriate.
Running multiple instances of our MCMC algorithm and comparing chains can also help us decide whether our chains have converged.

\subsection{\label{lit:computer_lan}Computer programming}
To ensure results and inference are reproducible, it is useful to create a computer package so that an analysis can be made in the future without all the code required being re-written.
Using freely available software such as the statistical program R \citep{rprog}, scripts and commands can be built and shared for easy implementation of code. 

Where fast inference is of importance, the choice of programming language is an important consideration.
The software package R can also be used as an interface for running code in the C programming language. 
Statistical code written in the C programming language is typically much faster than using standard R functions or code written in many other programming languages \citep{ccode}.

\section{\label{lit:hierarchical_mod}Hierarchical modelling}  
Hierarchical modelling is used to  is used to describe the structure of a problem where we believe some population level distribution exists, describing a set of unobserved parameters \citep{BayPriors}.
Examples include pupils nested within classes, children nested within families and patients nested within hospitals.
With the pupil-class relationship (2 level-hierarchy), for a given class there may be a number of pupils.
We may believe that by being in the same class, pupils will perform similarly in an exam as they are taught by the same teacher.
Further, we may have a pupil-class-school relationship (3 level-hierarchy).
For a given school, multiple classes exist and in each class there is a number of pupils.
We may believe that being within the same school, classes would perform similarly in an exam as they share the same head teacher or school principal.

Hierarchical modelling is used to describe a parent/child relationship \citep{GelmanMultilevel}. Repeating the parent/child relationship allows multiple levels to be described.
Where a hierarchical structure is known to exist, describing this experimental structure avoids confounding of effects with other sources of variation.

There are many different hierarchical models available, depending on what the experimenter is most interested in \citep{mixedeffects,BayHi}. 
Sharing of information can be built into hierarchical models by the sharing parameters. 
Allowing parameters to vary at more than one level allows an individual child (subject) effect to be examined.
A typical frequentist hierarchical model is built with random effects and has limited distributional assumptions available, whereas a Bayesian hierarchical model is flexible to describe various distributions \citep{Gelmanprior}, see Section~\ref{lit:distributional_ass}.

Plate diagrams allow hierarchical models to be represented graphically \citep{DAGbook,oldDAGbook}. 
Nodes (circles) are used to describe parameters and plates (rectangles) to describe repeating nodes.
The use of multiple plates allows nesting to be described. 

\subsection{\label{lit:distributional_ass}Distributional assumptions}  
The flexibility of the Bayesian paradigm allows for models to be built that are otherwise not practical in the frequentist paradigm. 
More appropriate assumptions can therefore be made to better describe experimental structure and variation in a Bayesian setting \citep{BayPriors}.
For example, inference for a hierarchical \emph{t}-distribution or hierarchical variable section model in a frequentist context is difficult in practise without using MCMC methods that are a more natural fit with Bayesian approaches.

The use of prior distributions allows information from the experimenter and experimental constraints to be incorporated, for instance if a parameter is known to be strictly positive then a positive distribution can be used to enforce this. 
Truncation can be used to reduce searching posterior areas with extremely low probability.

\subsection{\label{lit:indicator var}Indicator variables}  
Indicator variables are used in variable selection models to describe binary variables \citep{indicator}.
A Bernoulli distributed indicator variable can take the value $0$ or $1$ to indicate the absence or presence of an effect and can be used to describe binary outcomes such as gender. 
 
\subsection{\label{lit:t_dist}The three parameter \emph{t}-distribution }  
The Student's \emph{t}-distribution has one parameter, namely the degrees of freedom parameter $\nu$ which controls the kurtosis of the distribution \citep{tdist1}.
The Student's \emph{t}-distribution is as follows:
\begin{equation} \label{eq:t_dist}
t_1(x;\nu)=\frac{\Gamma \left(\frac{\nu+1}{2} \right)} {\sqrt{\nu\pi}\,\Gamma \left(\frac{\nu}{2} \right)} \left(1+\frac{x^2}{\nu} \right)^{-\frac{\nu+1}{2}}, x\in\mathbb{R},\nu\in\mathbb{R}^+.
\end{equation}
The $\nu$ scale parameter has the effect of increasing the heaviness of the distribution's tails. 
Adding an additional location parameter $\mu$ and scale parameter $\sigma$ allows further flexibility with the shape of the distribution \citep{tdist}.
The $\sigma$ scale parameter does not correspond to a standard deviation but does control the overall scale of the distribution.
The three parameter \emph{t}-distribution (or scaled \emph{t}-distribution) is then as follows:
\begin{equation*}
t_3(x;\mu,\nu,\sigma)=\frac{1}{\sigma} t_1\left(\frac{(x - \mu)}{\sigma}; \nu\right), x\in\mathbb{R},\nu\in\mathbb{R}^+,
\end{equation*}
where $t_1$ is given in (\ref{eq:t_dist}).

\section{\label{lit:generalised_log}Generalisations of the logistic growth model}
Where more flexibility than the logistic growth model is required, the logistic growth model (\ref{eq_det}) can be extended by adding parameters \citep{analysisoflogistic,theoryoflogisticgro}.
A common extension of the logistic growth model is Richards' growth model  \citep{GenLog,logisticrevisited}, which adds a single parameter for changing the shape of growth.
A more general case to both the logistic and Richards' growth model is the generalised logistic growth model.
Similarly to the logistic growth model (\ref{eq_det}) and its stochastic counterpart (\ref{eq_det_sde_2}), these more general equations can be extended to diffusion equations if required.

\subsection{\label{lit:richards_gro}Richards' growth model}
Richards' Growth model \citep{GenLog} adds an extra parameter $\beta$ to the logistic growth equation (\ref{eq_det}). The parameter $\beta$ affects where maximum growth occurs and consequently the relative growth rate \citep{analysisoflogistic}.
Richards' Growth model is as follows:
\begin{align}\label{eq:richards}
\frac{dx_t}{dt}&=rx_t\left[1-\left(\frac{x_t}{K}\right)^\beta\right].
\end{align}
The ODE has the following analytic solution:
\begin{align*}
x_t&=\frac{K}{(1+Qe^{-r\beta t})^{\frac{1}{\beta}}},\\
\text{where }Q&=\left[\left(\frac{K}{P}\right)^\beta-1\right]e^{\beta t_o},
\end{align*}
$({\alpha},{\beta})$ are positive real numbers and $t\geq{t_0}$.
When $\beta=1$, Richards' growth model is equivalent to the logistic growth equation.

\subsection{\label{lit:generalised_log}Generalised logistic growth model}
The generalised logistic growth model adds extra parameters $({\alpha},{\beta},{\gamma})$ to the logistic growth equation (\ref{eq_det}). The extra parameters $({\alpha},{\beta},{\gamma})$ affect where maximum growth occurs, the relative growth rate \citep{analysisoflogistic} and give a greater selection of curve shapes than the Richards' growth model (\ref{eq:richards}).
The generalised logistic growth model is as follows:
\begin{align}
\frac{dx_t}{dt}&=rx_t^{\alpha}\left[1-\left(\frac{x_t}{K}\right)^\beta\right]^\gamma,
\end{align}
where $({\alpha},{\beta},{\gamma})$ are positive real numbers and $t\geq{t_0}$.
The generalised logistic growth model cannot in general be integrated to give an analytical solution for $x_t$.
When ${\alpha}=1$, ${\beta}=1$ and $ {\gamma}=1$, the generalised logistic growth model is equivalent to the logistic growth equation.

\section{\label{lit:state_spa_mod}State space models}
A state space model describes the probabilistic dependence between a measurement process $Y_t$ and a state process $X_t$ \citep{dynamicmodels,statespace2}.
The most basic case of a state space model is as follows:
 \begin{align}\label{eq:state_space}
\begin{split}
				 \left(X_t|X_{t-1}=x_{t-1}\right)&\sim f(t,x_{t-1}),\\
		     \left(Y_t|X_{t}=x_t\right)&\sim g(t,x_t),
				\end{split}
				    \end{align}
where $f$ and $g$ are known.
A state space model with a linear Gaussian structure has the advantage of allowing us to carry out more efficient MCMC by integrating out latent states with a Kalman filter, instead of imputing all states.
The probabilistic representation and the ability to incorporate prior information makes Bayesian inference an appropriate choice for parameter estimation of a state space model. 

State space representation provides a general framework for analysing stochastic dynamical systems observed through a stochastic process. 
A state space model allows us to include both an internal state variable and an output variable in our model. 
The state-space representation of a stochastic process with measurement error can be given by (\ref{eq:state_space}) where $f$ is the transition density of the process and $g$ is the assumed measurement error. 
Inference methods are also readily available to carry out estimation of state space models.

\subsection{\label{lit:sde}Stochastic differential equations}
An ordinary differential equation (ODE) can be used to model a system of interest.
For systems with inherent stochastic nature we require a stochastic model.
A stochastic differential equation (SDE) is a differential equation where one or more terms include a stochastic process \citep{wilkinson2012stochastic,sdebook}.
An SDE differs from an ODE by the addition of a diffusion term, typically a  Weiner process, used to describe the intrinsic noise of a given process.
A Wiener process (or standard Brownian motion) is a continuous-time stochastic process.
A Wiener process $W(t)$, $t\geq{0}$, has the following three properties \cite{wiener}:\\
1) $W(0)=0$.\\
2) The function $t\rightarrow W(t)$ is almost surely everywhere continuous.\\
3) $W(t)$ has independent increments with $W(t)-W(s)\sim \operatorname{N}(0, t-s)$, for $0 \leq s < t$.\\
\\
Intrinsic noise from a Weiner process perpetuates the system dynamics of a differential equation.
The intrinsic noise is able to propagate though the process, unlike measurement noise.
Instead of inappropriately modelling intrinsic noise by measurement noise, an SDE allows our process to model both system and measurement noise separately.

The simplest case of a stochastic differential equation is of the form:
\begin{equation*}
dX(t)=\mu dt+\sigma dW(t),
\end{equation*}
where $W$ denotes a Wiener process. 
Parameters $\mu$ and $\sigma$ may depend on time and correspond to the drift and diffusion coefficients respectively.
The transition density of a stochastic process describes the movement from one state to the next and can be found from the solution of the process.

\subsection{\label{lit:em}The Euler-Maruyama method}
The Euler-Maruyama method provides an approximate numerical solution of a SDE \sloppy\citep{embook}.\sloppy
For a stochastic process of the form:
\begin{equation*}
dX_t=f(X_t)dt+g(X_t)dW_t,
\end{equation*}
where functions $f$ and $g$ are given and $W_t$ is a Wiener process.
Given an initial condition $X_0=x_0$ we can build an Euler-Maruyama approximation of $X$ over an interval $[0,T]$.
The Markov chain $Y$ defined below is an Euler-Maruyama approximation to the true solution of $X$.
First we set the initial condition $Y_0=x_0$.
Next, the interval $[0,T]$ is partitioned into $N$ equal subintervals of width $\Delta{t}>0$.
The Euler-Maruyama approximation is then recursively defined for $1\leq{i}\leq{N}$ as follows:
\begin{equation*}
Y_{i+1}=Y_{i}+f(Y_i)\Delta{t}+g(Y_i)\Delta{W_i},
\end{equation*}
where $\Delta{W_i}={W_{t_{i+1}}}-{W_{t_i}}\sim\operatorname{N}(0,\Delta{t})$.
The Euler-Maruyama approximation $Y$ will become a better approximation to the true process $X$ as we increase the size of $N$.

\subsection{\label{lit:kalman_fil}Kalman filter}
The Kalman filter \citep{kalmanoriginal,kalman} is a recursive algorithm that can be used to estimate the state of a dynamic system from a series of incomplete and noisy measurements. 
The main assumptions of the Kalman filter are that the underlying system is a linear dynamical system and that the noise has known first and second moments.  Gaussian noise satisfies the second assumption, for example.

Inference for a state space model (\ref{eq:state_space}) (see Section~\ref{lit:state_spa_mod}), where both $f$ and $g$ are Gaussian, can be carried out using a Kalman filter.
If all noise is zero-mean, uncorrelated and white, then the Kalman filter represents an optimal linear filter \citep{kalman_optimal}, even if the noise is not Gaussian.
An application of the Kalman filter is given in Section~\ref{app:kalman_fil} of the Appendix.

The Kalman filter algorithm is derived as follows: 
$X_{{t_{i}}}$ and  $Y_{t_{i}}$ are the state and measurement processes respectively.
$w_t$ and $u_t$ are the state and measurement error respectively, where $w_t$ and $u_t$ are IID, $E[w_t]=0$, $E[u_t]=0$, $E[w_t{w_t}^T]=W_t$ and $E[u_t{u_t}^T]=U_t$. 
The Kalman filter can be extended where $w_t$ and $u_t$ are not zero mean.
The unobserved latent process is driven by:
\begin{equation*}
X_{{t_{i}}}|X_{{t_{i-1}}}\sim\operatorname{N}(G_{{t_{i}}}X_{{t_{i-1}}},W_{t_{i}})
\end{equation*}
and the measurement error distribution, relating the latent variable to the observed is given by
\begin{equation*}
Y_{t_{i}}|X_{{t_{i}}}\sim\operatorname{N}(F^T_{t_i}X_{{t_{i}}},U_{t_i}),
\end{equation*}
where matrices $F_{t_i}$, $G_{t_i}$, $U_{t_i}$ and $W_{t_i}$ are all given. 
Now, suppose that:
\begin{equation*}
X_{t_{i-1}}|Y_{1:{t_{i-1}}}\sim \operatorname{N}(m_{t_{i-1}},C_{t_{i-1}}).
\end{equation*}
Incrementing time with $X_{t_i}=G_{t_i}X_{t_{i-1}}+w_{t_{i-1}}$ and condition on $Y_{1:{t_{i-1}}}$ to give:
\begin{align*}
X_{t_{i}}|Y_{1:{t_{i-1}}}&=G_{t_i}X_{t_{i-1}}|Y_{1:{t_{i-1}}}+w_{t_i}|Y_{1:{t_{i-1}}}\\
&=G_{t_i}X_{t_{i-1}}|Y_{1:{t_{i-1}}}+w_{t_{i-1}},
\end{align*}
as $w_{t_i}$ is independent of $Y_{1:{t_{i-1}}}$.
We can then show the following using standard multivariate theory:
\begin{equation*}
X_{t_{i}}|Y_{1:{t_{i-1}}}\sim \operatorname{N}(a_{t_{i}},R_{t_{i}}).
\end{equation*}
where $a_{t_{i}}=G_{{t_{i}}}m_{{t_{i-1}}}$ and $R_{t_{i}}=G_{{t_{i}}}C_{{t_{i-1}}}G_{t_{i}}^T+W_{t_{i}}$.
As $Y_{t_i}=F_{t_i}^TX_{t_{i}}+u_{t_i}$, and condition on $Y_{1:{t_{i-1}}}$ to give:
\begin{align*}
Y_{t_i}|Y_{1:{t_{i-1}}}&=F_{t_i}^TX_{t_{i}}|Y_{1:{t_{i-1}}}+u_{t_i}|Y_{1:{t_{i-1}}}\\
&=F_{t_i}^TX_{t_{i}}|Y_{1:{t_{i-1}}}+u_{t_i},
\end{align*}
as $u_{t_i}$ is independent of $Y_{1:{t_{i-1}}}$.
We can then show the following using standard multivariate theory:
\begin{equation*}
Y_{1:{t_{i}}}|Y_{1:{t_{i-1}}}\sim\operatorname{N}(F^T_{t_i}{a_{t_i}},F^T_t{R_{t_i}}F_t+U_{t_i})
\end{equation*}
$Y_{1:{t_{i}}}|Y_{1:{t_{i-1}}}$ and $X_{t_{i}}|Y_{1:{t_{i-1}}}$ are therefore jointly Gaussian with the following mean and covariance:\\
\begin{equation*}
\begin{pmatrix}
	X_{{t_{i}}} \\
  Y_{1:{{t_{i}}}}
 \end{pmatrix}
\sim
MVN\left(
\begin{pmatrix}
	a_{t_i} \\
  Y_{t_i}
 \end{pmatrix},
\begin{pmatrix}
	R_{t_i} & {R_{t_i}}F_t\\
  F^T_t{R_{t_i}} & F^T_t{R_{t_i}}F_t+U_{t_i}
 \end{pmatrix}
\right),
\end{equation*}
Finally, the following multivariate theorem is used:
\begin{align*}
\text{if }\begin{pmatrix}
	Y_1 \\
  Y_2
 \end{pmatrix}
&\sim
MVN\left(
\begin{pmatrix}
	\mu_1 \\
  \mu_2
 \end{pmatrix},
\begin{pmatrix}
	\Sigma_{11} & \Sigma_{12}\\
  \Sigma_{21} & \Sigma_{22}
 \end{pmatrix}
\right),\\
\text{then }Y_1| Y_2=y_2
&\sim
MVN\left(
\mu_1+\Sigma_{12}\Sigma^{-1}_{22}(y_2-\mu_2),\Sigma_{11}-\Sigma_{12}\Sigma^{-1}_{22}\Sigma_{21}
\right),
\end{align*}
to obtain the following:
\begin{align}
\label{eq:recursive}
\begin{split}
X_{{t_{i}}}|Y_{1:{{t_{i}}}}&\sim \operatorname{N}(m_{t_{i}},C_{t_{i}}),\\
\text{where } m_{t_{i}}&=a_{t_{i}}+R_{t_{i}}F(F^{T}R_{{t_{i}}}F+U)^{-1}[Y_{t_{i}}-F^{T}a_{t_{i}}]\\
 \text{and }C_{t_{i}}&=R_{t_{i}}-R_{t_{i}}F(F^TR_{t_{i}}F+U)^{-1}F^{T}R_{t_{i}}.
\end{split}
\end{align}
Parameters $m_0$ and $C_0$ must be initialised first, then using the equations in (\ref{eq:recursive}), $m_{t_i}$ and $C_{t_i}$ can be recursively estimated.

Typically, the Kalman filter is used to make inference for a hidden state process, but it can be used to reduce computational time in algorithms for inferring process hyper-parameters by recursively computing the marginal likelihood $\pi(y_{t_{1:N}})$ \citep{dynamicmodels}, where 
\begin{equation*}
\pi(y_{t_{1:N}})=\prod^N_{i=1}\pi(y_{t_{i}}|y_{t_{1:(i-1)}})
\end{equation*}
and $\pi(y_{t_{i}}|y_{t_{1:(i-1)}})=\int_{X}\pi(y_{t_{i}},x_{t_{i}}|y_{t_{1:(i-1)}})dx_{t_{i}}=\int_{X}\pi(y_{t_{i}}|x_{t_{i}})\pi(x_{t_{i}}|y_{t_{1:(i-1)}})dx_{t_{i}}$ gives a tractable Gaussian integral.
The procedure for computing the marginal likelihood $\pi(y_{t_{1:N}})$ using the Kalman filter algorithm is as follows:\\
\\
1) Initialise with prior knowledge for $X_{0}$ and set $i=1$.\\ 
\\
2) Prediction step from $X_{t_{i-1}}|Y_{1:{t_i-1}}$ to $X_{t_i}|Y_{1:{t_{i-1}}}$ (giving $\pi (x_{t_i}|y_{1:t_{i-1}})$).\\
\\
3) Calculate and store $\pi (y_{t_i}|y_{1:t_{i-1}})$.\\
\\
4) Update step to give $X_{t_i}|Y_{1:{t_i}}$, then iterate $i=i+1$.\\
\\
5) Repeat steps 2-4 (and compute $\pi (y_{t_{1:N}})$ ).\\

\subsection{\label{lit:LNA}Linear noise approximation}
The linear noise approximation (LNA) \citep{kurtz1,kurtz2,van} reduces a non-linear SDE to a linear SDE with additive noise, which can be solved \citep{LNA,komorowski}.
The LNA assumes the solution of a diffusion process $Y_t$ can be written as ${Y_t = v_t+Z_t}$ (a deterministic part $v_t$ and stochastic part $Z_t$), where $Z_t$ remains small for all $t\in\mathbb{R}_{\geq 0}$.
The LNA is useful when a tractable solution to a SDE cannot be found. 
Typically the LNA is used to reduce an SDE to a Ornstein-Uhlenbeck process which can be solved explicitly.
Ornstein-Uhlenbeck processes are Gaussian, time discretising the resulting LNA will therefore give us a linear Gaussian state space model with an analytically tractable transition density available.
The LNA can be viewed as a first order Taylor expansion of an approximating SDE about a deterministic solution (higher order approximations are possible \citep{gardiner2010stochastic}).
We can also view the LNA as an approximation of the chemical Langevin equation \citep{wallace}.
Applications of the LNA to non-linear SDEs are given in Section~\ref{sec:LNAM}~and~\ref{sec:LNAA}.
\end{chapter}                     
\begin{chapter}{\label{cha:modelling_den_int}Modelling genetic interaction}

\section{\label{modelling:intro}Introduction}
In this chapter, alternative modelling approaches are developed to better model a QFA screen comparison than the current frequentist \citet{QFA1} approach.
Section~\ref{modelling:Bay_hie_mod_inf} presents the modelling assumptions for the development of a Bayesian approach.
Two Bayesian approaches are then presented in Sections~\ref{cha:two_stage}~and~\ref{cha:one_stage}, incorporating some model assumptions that are not convenient in a frequentist setting.
So that our Bayesian models can be compared with a frequentist hierarchical modelling approach, a random effects model is then presented in Section~\ref{two:REM}. 

The models in this chapter are compared using previously analysed \emph{S. cerevisiae} QFA screen data in the next chapter.
Historic \emph{S. cerevisiae} QFA screen datasets are used to shape the model assumptions adopted in the following sections.

\section{\label{modelling:Bay_hie_mod_inf}Bayesian hierarchical model inference}
As an alternative to the maximum likelihood approach presented by \cite{QFA1}, we present a Bayesian, hierarchical methodology where \emph{a priori} uncertainty about each parameter value is described by probability distributions \citep{Bayth} and information about parameter distributions is shared across $\emph{orf}\Delta$s and conditions.
Plausible frequentist estimates from across 10 different historic QFA data sets, \hl{including a wide range of different background mutations and treatments} were used to \hl{quantify} \emph{a priori} uncertainty in model parameters. 

Prior distributions describe our beliefs about parameter values. These should be diffuse enough to capture all plausible values (to capture the full range of observations in the datasets) while being restrictive enough to rule out implausible values (to ensure efficient inference). 
Inappropriate choice of priors can result in chains drifting during mixing and becoming stuck in implausible regions.  
Although using conjugate priors would allow faster inference, we find that the conjugate priors available for variance parameters \citep{Gelmanprior} are either too restrictive at low variance (Inverse-gamma), not restrictive enough at low variance (half-t family of prior distributions) or are non-informative or largely discard the prior information available (Uniform).  
Our choice for the priors of precision parameters is the non-conjugate Log-normal as we find the distribution is only restrictive at extremely high and low variances.

We use three types of distribution to model parameter uncertainty: Log-normal, Normal and scaled t-distribution with three degrees of freedom.  We use the Log-normal distribution to describe parameters which are required to be non-negative (e.g. parameters describing precisions, or repeat-level fitnesses) or parameter distributions which are found by visual inspection to be asymmetric.  We use the Normal distribution to describe parameters which are symmetrically distributed (e.g. some prior distributions and the measurement error model) and we use the $t$-distribution to describe parameters whose uncertainty distribution is long-tailed (i.e. where using the Normal distribution would result in excessive shrinkage towards the mean). 
A Normal distribution was considered for describing the variation in  $\emph{orf}\Delta$s but was found to be inappropriate, failing to assign density at the extreme high and low fitnesses.
For example, after \hl{visual inspection of} frequentist $\emph{orf}\Delta$ level means about their population mean, we found there to be many unusually fit, dead or missing $\emph{orf}\Delta$ and concluded that $\emph{orf}\Delta$ fitnesses would be well modelled by the t-distribution.

Instead of manually fixing the inoculum density parameter $P$ as in \cite{QFA1} our Bayesian hierarchical models deal with the scarcity of information about the early part of culture growth curves by estimating a single $P$ across all $\emph{orf}\Delta$s \hl{(and conditions in some of our models).}
Our new approach learns about $P$ from the data and gives us a posterior distribution to describe our uncertainty about its value.    
 
\hl{The new, hierarchical structure implemented in our models \citep{BayHi} reflects the structure of QFA experiments.}
Information is shared efficiently among groups of parameters such as between repeat level parameters for a single mutant strain.
An example of the type of Bayesian hierarchical modelling which we use to model genetic interaction can be seen in \cite{hierarchical1}, where hierarchical models are used to account for group effects.

In \cite{epis1} the signal of genetic interaction is chosen to be ``strictly ON or OFF" when modelling gene activity. 
\hl{We include this concept in our interaction models by using a Bernoulli distributed indicator variable \citep{indicator} to describe whether there is evidence of an $\emph{orf}\Delta$ interacting with the query mutation; the more evidence of interaction, the closer posterior expectations will be to one.} 

Failing to account for all sources of variation within the experimental structure, such as the difference in variation between the control and query fitnesses, may lead to inaccurate conclusions. 
By incorporating more information into the model with prior distributions and a more flexible modelling approach, we will increase statistical power. 
With an improved analysis it may then be possible for a similar number of genetic interactions to be identified with a smaller sample size, saving on the significant experimental costs associated with QFA.

Inference is carried out using Markov Chain Monte Carlo (MCMC) methods. The algorithm used is a Metropolis-within-Gibbs sampler where each full-conditional is sampled in turn either directly or using a simple Normal random walk Metropolis step.
Due to the large number of model parameters and large quantity of data from high-throughput QFA experiments, the algorithms used \hl{for carrying out inference} often have poor mixing and give highly auto-correlated samples, requiring thinning.
Posterior means are used to obtain point estimates where required.

For the new Bayesian approaches (described in Section~\ref{cha:two_stage} and \ref{cha:one_stage}), model fitting is carried out using the techniques discussed above, implemented in C for computational speed, and is freely available in the R package ``qfaBayes'' at \sloppy\url{https://r-forge.r-project.org/projects/qfa}.\sloppy

\section{\label{cha:two_stage}Two-stage Bayesian hierarchical approach}
\hl{In the following sections, a two-stage Bayesian, hierarchical modelling approach (see Section~\ref{two:SHM}~and~\ref{two:IHM}) is presented.
The following two-stage Bayesian approach generates $\emph{orf}\Delta$ fitness distributions and infers genetic interaction probabilities separately.}
For a QFA screen comparison, first the separate hierarchical model (SHM) given in Section~\ref{two:SHM}, is fit to each screen separately and a set of logistic growth parameter estimates obtained for each time-course.
Secondly, each set of logistic growth parameter estimates is converted into a univariate fitness summary and input to the interaction hierarchical model (IHM) given in Section~\ref{two:IHM}, to determine which genes show evidence of genetic interaction.

\subsection{\label{two:SHM}Separate hierarchical model}
The separate hierarchical model (SHM), presented in Table~\ref{tab:SHM}, models the growth of multiple yeast cultures using the logistic function described in (\ref{eq:logistic}). 
In this first \hl{hierarchical model, the logistic model} is fit to the query and control strains separately.

In order to measure the variation between $\emph{orf}\Delta$s, parameters ($K^p$,$\sigma^{K}_{o}$) and ($r^p$,$\sigma^{r}_{o}$) are included at the population level of the hierarchy. 
Within-$\emph{orf}\Delta$ variation is modelled by each set of $\emph{orf}\Delta$ level parameters ($K^{o}_{l}$,$\tau^{K}_{l}$) and ($r^{o}_{l}$,$\tau^{r}_{l}$). 
Learning about these higher level parameters allows information to be shared across parameters lower in the hierarchy.
A three-level hierarchical model is applied to $(K,K^{o}_{l},K_{lm})$ and $(r,r^{o}_{l},r_{lm})$, sharing information on the repeat level and the $\emph{orf}\Delta$ level.
Note that $\emph{orf}\Delta$ level parameters ${K^{o}_{l}}$ and ${r^{o}_{l}}$ are on the log scale ($e^{K^{o}_{l}}$ and $e^{r^{o}_{l}}$ are on the scale of the observed data).

Assuming a \hl{Normal} error structure, random \hl{measurement} error is modelled by the $\nu_l$ parameters (one for each $\emph{orf}\Delta$).
Information on random error is shared across all $\emph{orf}\Delta$s by drawing \hl{$\log \nu_l$} from a normal distribution parameterised by ($\nu_p$,$\sigma^{\nu}$).
A two-level hierarchical structure is also used for both the $\tau_{l}^{K}$ and $\tau_{l}^{r}$ parameters. 

Modelling logistic model parameter distributions on the log scale ensures that parameter values remain strictly positive (a realistic biological constraint).  Truncating distributions allows us to implement further, realistic constraints on the data. Truncating $\log r_{lm}$ values greater than 3.5 corresponds to disallowing biologically unrealistic culture doubling times \hl{faster than about 30 minutes} and truncating of repeat level parameters $\log K_{lm}$ above 0 ensures that no carrying capacity estimate is greater than the maximum observable cell density, which is 1 after scaling.

$\emph{orf}\Delta$ level parameters $e^{K_{o}^{l}}$ and $e^{r_{o}^{l}}$ are on the same scale as the observed data.  Realistic biological constraints (positive logistic model parameters) are enforced at the repeat level, however both $e^{K_{o}^{l}}$ and $e^{r_{o}^{l}}$, which are assumed to have scaled~$t$-distributions, are truncated below zero to keep exponentiated parameters strictly positive.
Most $\emph{orf}\Delta$ level logistic growth parameters are distributed in a bell shape around some mean value, it is the unusually fit, dead or missing $\emph{orf}\Delta$s within a typical QFA screen that require the use of a long tailed distribution such as the scaled~$t$-distribution with 3 degrees of freedom.
The non-standard choice of a truncated scaled~$t$-distribution with 3 degrees of freedom ensures that the extreme high and low values have probability assigned to them regardless of the population level location and scale parameters for a given QFA screen.
 
For example, after \hl{visual inspection of} frequentist $\emph{orf}\Delta$ level means about their population mean, we found there to be many unusually fit, dead or missing $\emph{orf}\Delta$ and concluded that $\emph{orf}\Delta$ fitnesses would be well modelled by the t-distribution.

\begin{table}
  \caption[Description of the separate hierarchical model]{Description of the separate hierarchical model (SHM).
Dependent variable $y_{lmn}$ (scaled cell density measurements) and independent variable $t_{lmn}$ (time since inoculation) are data input to the SHM. 
$x(t)$ is the solution to the logistic model ODE given in (\ref{eq:logistic}).
$l$~indicates a particular $\emph{orf}\Delta$ from the gene deletion library, $m$ indicates a repeat for a given $\emph{orf}\Delta$ and $n$ indicates the time point for a given $\emph{orf}\Delta$ repeat.\label{tab:SHM}}
\begin{align*}
l&=1,2,...,L &&\;\;\;\; \text{$\emph{orf}\Delta$ level}\\
m&=1,...,M_{l} &&\;\;\;\; \text{Repeat level}\\
n&=1,2,...,N_{lm} &&\;\;\;\; \text{Time point level}\\
\intertext{Time point level}
y_{lmn} &\sim \operatorname{N}(\hat{y}_{lmn},({ \nu_{l}  })^{-1} ) \quad 
&\hat{y}_{lmn} &= x(t_{lmn};K_{lm} ,r_{lm},P)\\
\intertext{Repeat level}
\log~K_{lm} &\sim \operatorname{N}(K_{l}^o, ({ \tau_{l}^{K} })^{-1} )I_{(-\infty,0]} &\log~\tau_{l}^K &\sim \operatorname{N}(\tau^{K,p}, ({\sigma^{\tau,K}})^{-1} )I_{[0,\infty)}\\
\log~r_{lm} &\sim \operatorname{N}(r_{l}^o, ({ \tau_{l}^{r} })^{-1} )I_{(-\infty,3.5]} &\log~\tau_{l}^r &\sim \operatorname{N}(\tau^{r,p}, ({\sigma^{\tau,r}})^{-1} )
\end{align*}
$\emph{orf}\Delta$ level
\begin{align*}
e^{K_{l}^o} &\sim t(K^p, ({ \sigma^{K,o} })^{-1},3)I_{[0,\infty)}
&\log~\sigma^{K,o} &\sim \operatorname{N}(\eta^{K,o}, (\psi^{K,o})^{-1} )\\
e^{r_{l}^o} &\sim t(r^p, ({ \sigma^{r,o} })^{-1},3 )I_{[0,\infty)}
&\log~\sigma^{r,o} &\sim \operatorname{N}(\eta^{r,o}, (\psi^{r,o})^{-1} )\\
\log~\nu_{l} &\sim \operatorname{N}(\nu^p, ({ {\sigma}^{\nu} })^{-1} )
&\log~\sigma^{\nu} &\sim \operatorname{N}(\eta^{\nu}, (\psi^{\nu})^{-1} )\\
\intertext{Population level}
\log~K^p &\sim \operatorname{N}(K^\mu, ({\eta^{K,p}})^{-1} )
&\log~r^p &\sim \operatorname{N}(r^\mu, ({\eta^{r,p}})^{-1} )\\
\log~P &\sim \operatorname{N}(P^\mu, ({\eta^{P}})^{-1} )
&\nu^p &\sim \operatorname{N}(\nu^\mu, (\eta^{\nu,p})^{-1} )\\
\tau^{K,p} &\sim \operatorname{N}(\tau^{K,\mu}, ({\eta^{\tau,K,p}})^{-1} )
&\log~\sigma^{\tau,K} &\sim \operatorname{N}(\eta^{\tau,K}, (\psi^{\tau,K})^{-1} )\\
\tau^{r,p} &\sim \operatorname{N}(\tau^{r,\mu}, ({\eta^{\tau,r,p}})^{-1} )
&\log~\sigma^{\tau,r} &\sim \operatorname{N}(\eta^{\tau,r}, (\psi^{\tau,r})^{-1} )
\end{align*}
\end{table}

Identifiability problems can arise for parameters $K_{lm}$ and $r_{lm}$ when observed cell densities are low and unchanging (consistent with growth curves for cultures which are very sick, dead or missing). In these cases, either $K_{lm}$ or $r_{lm}$ can take values near zero, allowing the other parameter to take any value without significantly affecting the model fit.
In the \citet{QFA1} approach identification problems are handled in an automated post-processing stage: for cultures with low K estimates (classified as dead), $r$ is automatically set to zero.
Without correcting for identification problems in our Bayesian models, misleading information from implausible values will be shared across our models.
Computing time wasted on such identifiability problems is reduced by truncating repeat level parameters $r_{lm}$, preventing the MCMC algorithms from becoming stuck in extremely low probability regions when $K_{lm}$ takes near zero values.
Similarly, $\log \tau^{K}_{l}$ parameters are truncated below 0 to overcome identifiability problems between parameters $K_{lm}$ and $r_{lm}$ when $r_{lm}$ takes near zero values.

The SHM in Table~\ref{tab:SHM} is fit to both the query and control strains separately.
Means are taken \hl{to summarise} logistic growth parameter posterior distributions for each $\emph{orf}\Delta$ repeat. 
Summaries $(\hat{K}_{lm},\hat{r}_{lm},\hat{P})$ for each $\emph{orf}\Delta$ repeat are converted to univariate fitnesses $F_{clm}$, where $c$~identifies the condition (query or control), with any given fitness measure e.g. $MDR\times MDP$ (see (\ref{eq:MDRMDP}) and \cite{QFA1}).
A problem of the two-stage approach is that we must choose a fitness definition most relevant to the experiment.
We choose the same definition used in \citet{QFA1}, $MDR{\times}MDP$, for the comparison of our methods.
An alternative choice of fitness definition could be used given sufficient biological justification.
Section~\ref{int:fitness_def} gives the derivations of $MDR$ and $MDP$.
The product of $MDR{\times}MDP$ is used as it accounts for the attributes of two definitions simultaneously.

The flow of information within the model and how each parameter is related to the data can be seen \hl{from} the \hl{plate diagram} in Figure~\ref{fig:SHMDAG} \citep{DAGbook}.

\begin{figure}[t]
\centering
\makebox{\includegraphics[width=12cm]{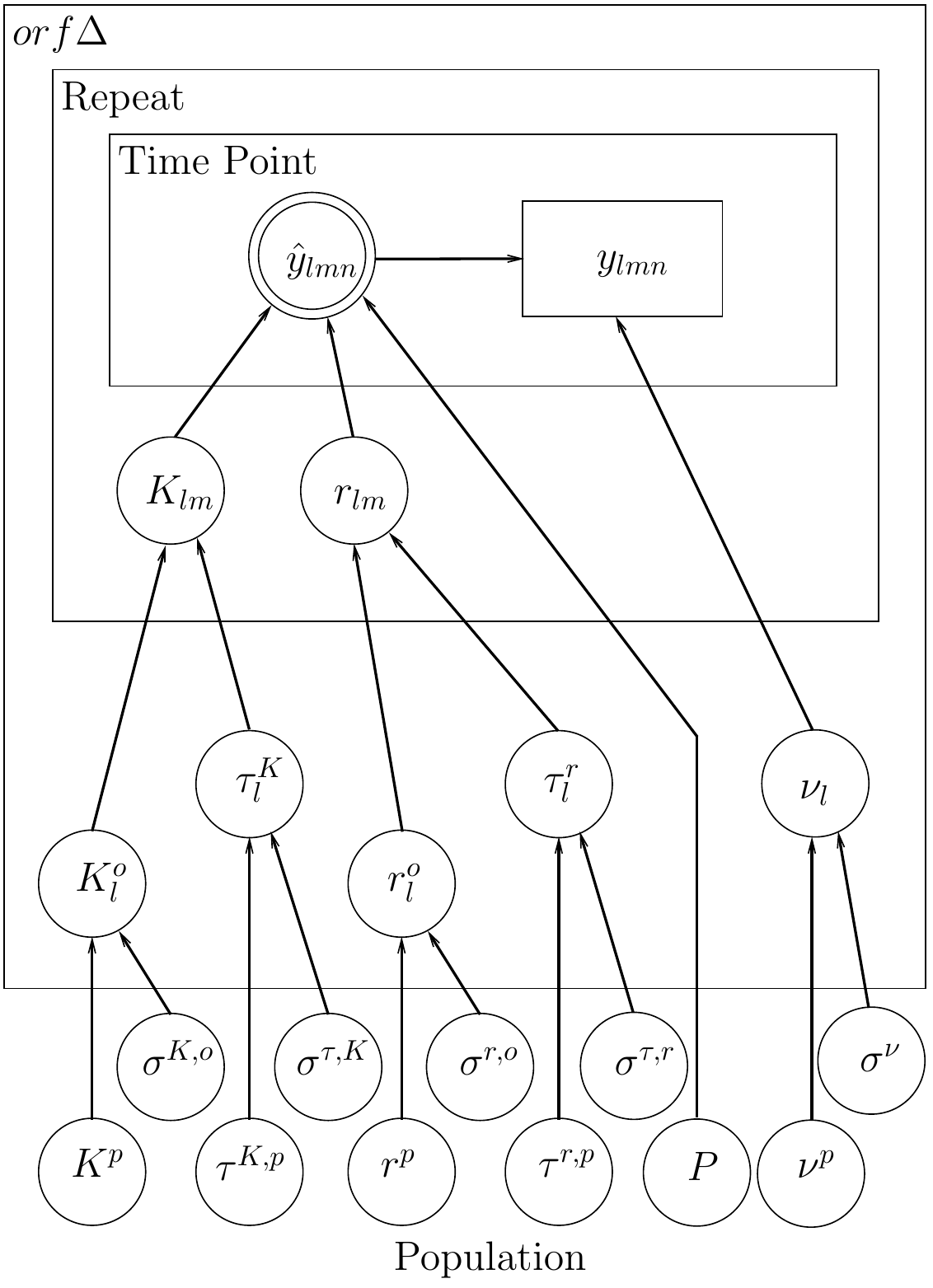}}
\caption[Plate diagram for the separate hierarchical model]{\hl{Plate diagram} for the separate hierarchical model, described in Section~\ref{two:SHM}. This figure shows the four levels of hierarchy in the SHM model, population, $\emph{orf}\Delta$ ($l$), repeat ($m$) and time point ($n$).
Prior hyperparameters for the population parameters are omitted.
A circular node represents a parameter in the model. 
An arrow from a source node to a target node indicates that the source node parameter is a \hl{prior hyperparameter} for the target node parameter. 
Each rectangular box corresponds to a level of the hierarchy. 
Nodes within multiple boxes are nested and their parameters are indexed by corresponding levels of the hierarchy. The node consisting of two concentric circles corresponds to the models fitted values. 
The rectangular node represents the observed \hl{data}.}
\label{fig:SHMDAG}
\end{figure}
\FloatBarrier
\subsection{\label{two:IHM}Interaction hierarchical model}
\hl{After the SHM fit, the IHM, presented in Table~\ref{tab:IHM}, can then be used to model estimated fitness scores $F_{clm}$ and determine, for each $\emph{orf}\Delta$, whether there is evidence for interaction.}

\hl{Fitnesses are passed to the IHM where query screen fitnesses are compared with control screen fitnesses, assuming genetic independence.
Deviations from predicted fitnesses are evidence for genetic interaction.}
The flow of information within the IHM and how each parameter is related to the data can be seen \hl{from} the \hl{plate diagram} in Figure~\ref{fig:IHMDAG}.
\begin{table}
  \caption[Description of the interaction hierarchical model]{Description of the interaction hierarchical model (IHM).
$F_{clm}$ are the observed fitness scores, where $c$~identifies the condition for a given $\emph{orf}\Delta$, $l$~identifies a particular $\emph{orf}\Delta$ from the gene deletion library and $m$ identifies a repeat for a given $\emph{orf}\Delta$.\label{tab:IHM}}
\begin{align*}
c&=0,1 &&\;\;\;\; \text{Condition level}\\
l&=1,...,L_{c} &&\;\;\;\; \text{$\emph{orf}\Delta$ level}\\
m&=1,...,M_{cl} &&\;\;\;\; \text{Repeat level}\\
\intertext{Repeat level}
F_{clm} &\sim \operatorname{N}(\hat{F}_{cl},(\nu_{cl}  )^{-1}) &\hat{F}_{cl} &= e^{\alpha_{c}+Z_{l}+\delta_{l}\gamma_{cl}}\\
\intertext{$\emph{orf}\Delta$ level}
e^{Z_{l}} &\sim t(Z^{p},{({\sigma^{Z}})}^{-1},3)I_{[0,\infty)} &\log~\sigma^{Z} &\sim \operatorname{N}(\eta^{Z},\psi^{Z})\\
\log~\nu_{cl} &\sim \operatorname{N}(\nu^p,{ ({{\sigma}^{\nu}} )}^{-1} ) &\log~\sigma^{\nu} &\sim \operatorname{N}(\eta^{\nu}, \psi^{\nu} )\\
\delta_{l} &\sim Bern(p) &\\
e^{\gamma_{cl}}&=\begin{cases}
1  & \text{if } c=0;\\
t(1,{({\sigma^{\gamma}})}^{-1},3)I_{[0,\infty)} & \text{if } c=1.
\end{cases}
&\log~\sigma^{\gamma}&\sim
\operatorname{N}(\eta^{\gamma},{(\psi^{\gamma})}^{-1})
\\
\intertext{Condition level}
\alpha_{c}&=\begin{cases}
0  & \text{if } c=0;\\
\operatorname{N}(\alpha^\mu,\eta^{\alpha}) & \text{if } c=1.
\end{cases}
\\
\intertext{Population level}
\log~Z^{p} &\sim N(Z^{\mu},{(\eta^{Z,p})}^{-1}) 
&\nu^p &\sim \operatorname{N}(\nu^{\mu}, (\eta^{\nu,p})^{-1} )
\end{align*}
\end{table}

The interaction model accounts for between $\emph{orf}\Delta$ variation with the set of parameters ($Z^{p}$,$\sigma_{Z}$) and within $\emph{orf}\Delta$ variation by the set of parameters ($Z_{l}$,$\nu_{l}$).
A linear \hl{relationship} between the control and query $\emph{orf}\Delta$ level parameters is \hl{specified} with a scale parameter $\alpha_{1}$. 
\hl{Any deviation from this relationship (genetic interaction) is accounted for by the term $\delta_{l}\gamma_{1,l}$.}
$\delta_{l}$ is a binary indicator of genetic interaction for \emph{orf}$\Delta$ $l$.
A scaling parameter $\alpha_{1}$ allows any effects due to differences in the control and query data sets to be scaled out, such as differences in \hl{genetic background}, incubator temperature or inoculum density.

The linear relationship between the control and query fitness scores, consistent with the multiplicative model of genetic independence, described in (\ref{eq:linear}), is implemented in the IHM as: $\hat{F} = e^{\alpha_{c}+Z_{l}+\delta_{l}\gamma_{cl}}=e^{\alpha_{c}}e^{Z_{l}+\delta_{l}\gamma_{cl}}$.
\hl{Strains whose fitnesses lie} along the linear relationship \hl{defined by} the scalar $\alpha_{1}$ \hl{show no evidence for interaction with the query condition}.  \hl{On the other hand, deviation from the linear relationship, represented} by the posterior mean of $\delta_{l}\gamma_{1,l}$ \hl{is} evidence for genetic interaction.
The larger the posterior mean for $\delta_{l}$ is the \hl{higher} the probability or evidence there is for interaction, \hl{while $\gamma_{1,l}$ is a measure of the strength of interaction}. 
Where the query condition has a negative effect (i.e. \hl{decreases fitness on average, compared to the control condition}), query fitnesses which are above and below the linear relationship are suppressors and enhancers \hl{of the fitness defect associated with the query condition} respectively.
A list of gene names are ordered by $\delta_{l}\gamma_{cl}$ posterior means and those $\emph{orf}\Delta$s with $\hat{\delta}_{l}>0.5$ will be classified and labelled as showing ``significant'' evidence of interaction.

\hl{The Bernoulli probability parameter $p$ is our prior estimate for the probability of a given $\emph{orf}\Delta$ showing evidence of genetic interaction. For a typical yeast QFA screen, $p$ is set to 0.05 as the experimenter's belief before the experiment is carried out is that $5\%$ of our $\emph{orf}\Delta$s exhibit genetic interactions.}
Observational noise is quantified by $\nu_{cl}$. 
The $\nu_{cl}$ parameter accounts for difference in variation between condition i.e. the query and control data sets and for difference in variation between $\emph{orf}\Delta$s. 

\begin{figure}[t]
\centering
\makebox{\includegraphics[width=14cm]{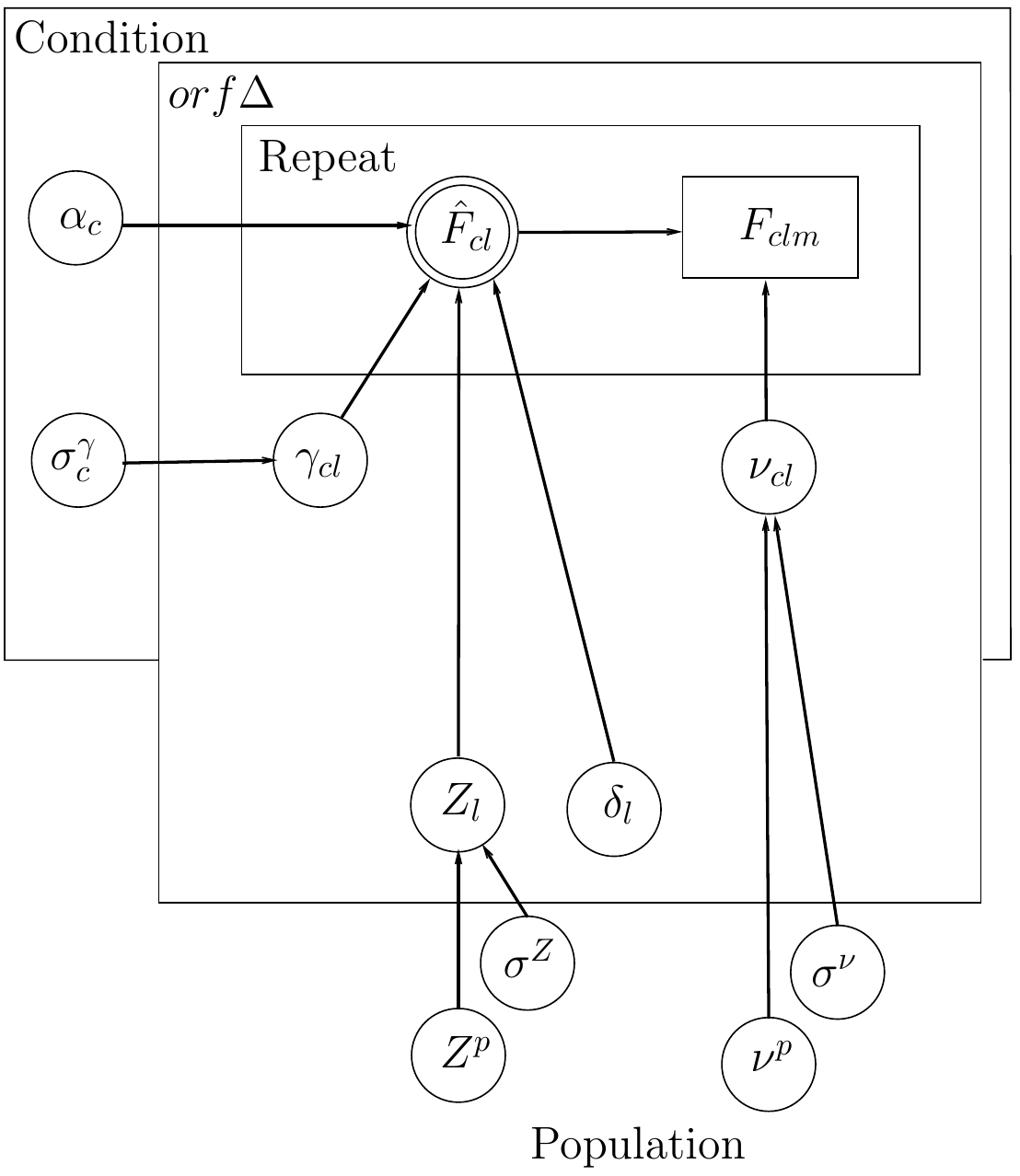}}
\caption[Plate diagram for the interaction hierarchical model]{Plate diagram for the interaction hierarchical model, described in Section~\ref{two:IHM}. 
This figure shows the four levels of hierarchy in the IHM model: population, $\emph{orf}\Delta$ ($l$), condition ($c$) and repeat ($m$).
Prior hyperparameters for population parameters are omitted.
Plate diagram notation as in Figure~\ref{fig:SHMDAG}.
}
\label{fig:IHMDAG}
\end{figure}
\FloatBarrier
\section{\label{cha:one_stage}One-stage Bayesian hierarchical approach}
Following from Section~\ref{cha:two_stage}, a one-stage approach for inferring fitness and genetic interaction probabilities separately is presented.  
All of the SHM and IHM modelling assumptions described in Section~\ref{cha:two_stage}, such as distributional choices and hierarchical structure are inherited by the one stage approach known as the joint hierarchical model (JHM). 

    \subsection{\label{joi:JHM}Joint hierarchical model}
The JHM given in Table~\ref{tab:JHM} is an alternative, fully Bayesian \hl{version of} the two-stage approach described in Section~\ref{two:SHM}~and~\ref{two:IHM}.
\hl{The JHM} incorporates the key modelling ideas from both the SHM and the IHM \hl{with the considerable advantage that we can learn about logistic growth model, fitness and genetic interaction parameters simultaneously, thereby avoiding having to choose a fitness measure or point estimates for passing information between models}. 
\hl{The JHM is} an extension of the SHM with \hl{the presence or absence of genetic interaction} being described by a Bernoulli indicator and an additional level of error to account for variation due to the query condition.  
Genetic interaction is modelled in terms of the two logistic growth parameters $K$ and $r$ simultaneously.
Similar to the interaction model in Section~\ref{two:IHM} in Chapter~\ref{cha:two_stage}, linear relationships between control and query carrying capacity and growth rate (instead of fitness score) are assumed:  $(e^{\alpha_{c}+K^{o}_{l}+\delta_{l}\gamma_{cl}},e^{\beta_{c}+r^{o}_{l}+\delta_{l}\omega_{cl}})$.

\begin{table}
  \caption[Description of the joint hierarchical model]{Description of the joint hierarchical model (JHM).
The dependent variable $y_{clmn}$ (scaled cell density measurements) and independent variable $t_{clmn}$ (time since inoculation) are input to the JHM.
$c$~identifies the condition for a given $\emph{orf}\Delta$, $l$~identifies a particular $\emph{orf}\Delta$ from the gene deletion library, $m$ identifies a repeat for a given $\emph{orf}\Delta$ and $n$ identifies the time point for a given condition and $\emph{orf}\Delta$ repeat.\label{tab:JHM}}
\begin{align*}
c&=0,1  &&\; \text{Condition level}\\
l&=1,...,L_{c}   &&\; \text{$\emph{orf}\Delta$ level}\\
m&=1,...,M_{cl}       &&\; \text{Repeat level}\\
n&=1,...,N_{clm}      &&\; \text{Time point level}\\
\shortintertext{Time point level}
y_{clmn} &\sim \operatorname{N}(\hat{y}_{clmn},({\nu_{cl}})^{-1} )\;
&\hat{y}_{clmn} &= x(t_{clmn};{ K_{clm} } ,{ r_{clm} } , { P })\\
\shortintertext{Repeat level}
\log~K_{clm} &\sim \operatorname{N}(\alpha_{c}+K_{l}^o+\delta_{l}\gamma_{cl},({ \tau_{cl}^K })^{-1})I_{(-\infty,0]}
\; & \log~\tau_{cl}^K &\sim \operatorname{N}(\tau^{K,p}_{c}, ({\sigma^{\tau,K}_{c}})^{-1} )I_{[0,\infty)}\\
\log~r_{clm} &\sim \operatorname{N}(\beta_{c}+r_{l}^o+\delta_{l}\omega_{cl},({ \tau_{cl}^r })^{-1})I_{(-\infty,3.5]}
\; &\log~\tau_{cl}^r &\sim \operatorname{N}(\tau^{r,p}_{c}, ({\sigma^{\tau,r}_{c}})^{-1} )\\
\shortintertext{$\emph{orf}\Delta$ level}
e^{K_{l}^o} &\sim t(K^p, ({ \sigma^{K,o} })^{-1},3 )I_{[0,\infty)}\qquad &\log~\sigma^{K,o} &\sim \operatorname{N}(\eta^{K,o}, (\psi^{K,o})^{-1} )\\
e^{r_{l}^o} &\sim t(r^p, ({ \sigma^{r,o} })^{-1},3 )I_{[0,\infty)}\qquad &\log~\sigma^{r,o} &\sim \operatorname{N}(\eta^{r,o}, (\psi^{r,o})^{-1} )\\
\log~\nu_{cl} &\sim \operatorname{N}(\nu^p,({ \sigma^{\nu} })^{-1})\qquad& \log~\sigma^{\nu} &\sim \operatorname{N}(\eta^{\nu}, (\psi^{\nu})^{-1} )\\
\delta_{l} &\sim Bern(p)\\
e^{\gamma_{cl}}&=\begin{cases}
1  & \text{if } c=0;\\
t(1,{({\sigma^{\gamma}})}^{-1},3)I_{[0,\infty)} & \text{if } c=1.
\end{cases}
\qquad
&\log~\sigma^{\gamma}&\sim
\operatorname{N}(\eta^{\gamma},\psi^{\gamma})  
\\
e^{\omega_{cl}}&=\begin{cases}
1  & \text{if } c=0;\\
t(1,({{\sigma^{\omega}})}^{-1},3)I_{[0,\infty)} & \text{if } c=1.
\end{cases}
\qquad
&\log~\sigma^{\omega}&\sim
\operatorname{N}(\eta^{\omega},\psi^{\omega})
\\
\shortintertext{Condition level}
\alpha_{c}&=\begin{cases}
0  & \text{if } c=0;\\
\operatorname{N}(\alpha^{\mu},\eta^{\alpha})  & \text{if } c=1.
\end{cases}
&\beta_{c}&=\begin{cases}
0  & \text{if } c=0;\\
\operatorname{N}(\beta^{\mu},\eta^{\beta}) & \text{if } c=1.
\end{cases}
\\
\tau^{K,p}_{c} &\sim \operatorname{N}(\tau^{K,\mu}, ({\eta^{\tau,K,p}})^{-1} )
& \log~\sigma^{\tau,K}_{c} &\sim \operatorname{N}(\eta^{\tau,K}, (\psi^{\tau,K})^{-1} )\\
\tau^{r,p}_{c} &\sim \operatorname{N}(\tau^{r,\mu}, ({\eta^{\tau,r,p}})^{-1} )
& \log~\sigma^{\tau,r}_{c} &\sim \operatorname{N}(\eta^{\tau,r}, (\psi^{\tau,r})^{-1} )\\
\shortintertext{Population level}
\log~K^p &\sim \operatorname{N}(K^\mu, ({\eta^{K,p}})^{-1} )
& \log~r^p &\sim \operatorname{N}(r^\mu, ({\eta^{r,p}})^{-1} )\\
\nu^p &\sim \operatorname{N}(\nu^\mu, ({\eta}^{\nu,p})^{-1} )
& \log~P &\sim \operatorname{N}(P^\mu, ({\eta^{P}})^{-1} )\\
\end{align*}
\end{table}
\hl{By fitting a single JHM, we need only calculate posterior means, check model diagnostics and thin posteriors once. However, the CPU time taken to reach convergence for any given data set is roughly twice that of the two-stage approach for a genome-wide QFA.}

The flow of information within the model and how each parameter is related to the data can be seen \hl{from} the \hl{plate diagram} in Figure~\ref{fig:joi:JHMDAG}.



\begin{figure}
\centering
\makebox{\includegraphics[width=14cm]{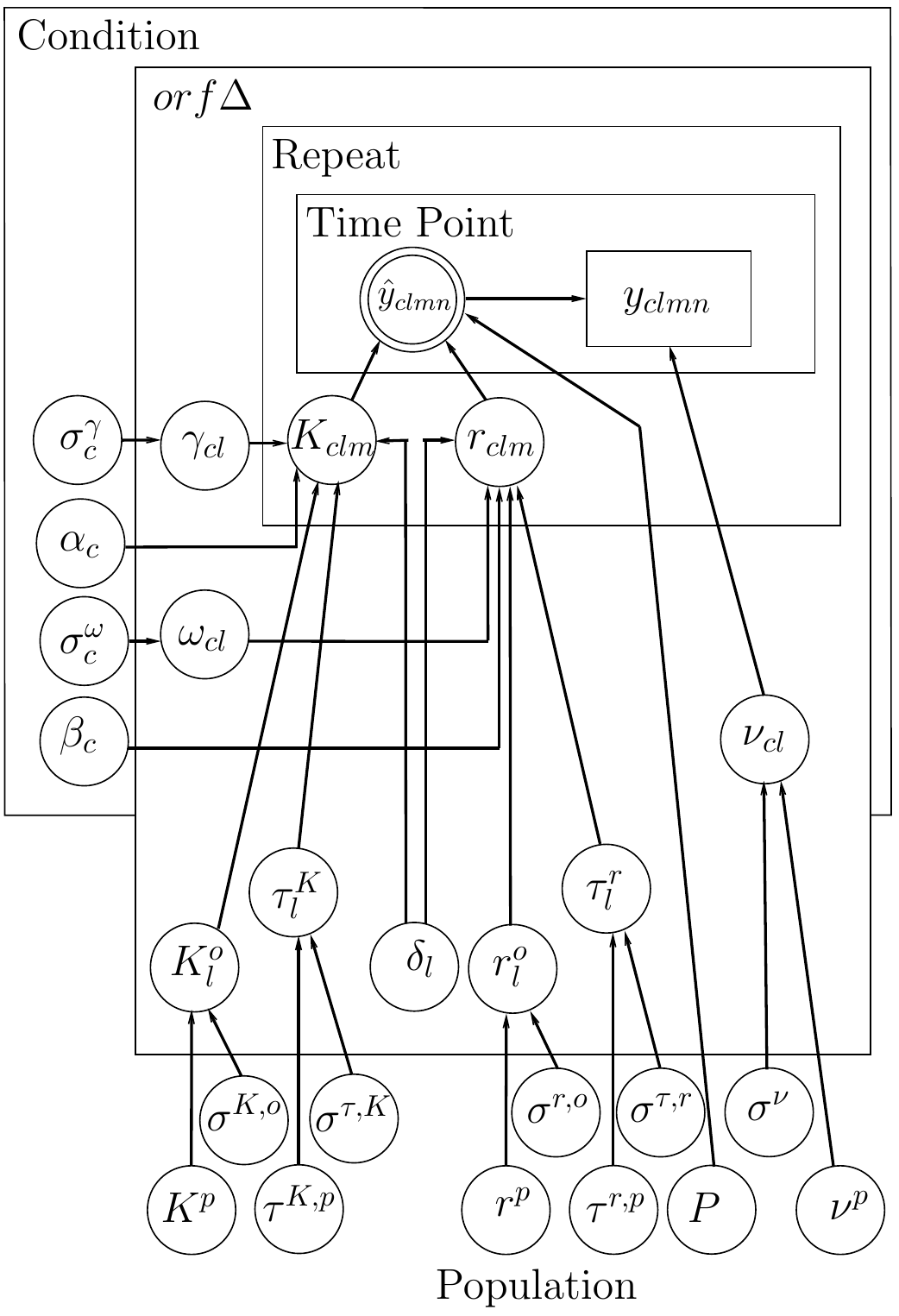}}
\caption[Plate diagram for the joint hierarchical model]{Plate diagram for the joint hierarchical model, described in Section~\ref{joi:JHM}. 
This figure shows the five levels of hierarchy in the JHM model, population, $\emph{orf}\Delta$ ($l$), condition ($c$), repeat ($m$) and time point ($n$). 
Prior hyperparameters for the population parameters are omitted.
Plate diagram notation is given in Figure~\ref{fig:SHMDAG}.
}
\label{fig:joi:JHMDAG}
\end{figure}
\FloatBarrier
\section{\label{two:REM}Random effects model}
\hl{To improve on the \cite{QFA1} modelling approach whilst remaining within the frequentist paradigm, by accounting for the hierarchical structure of the data, a random effects model \citep{mixedeffects,nlme} can be used. 
The random effects model (REM) given in Table~\ref{tab:REM} is used to model estimated fitness scores $F_{clm}$ from (\ref{eq:F}) and estimate evidence of interaction for each $\emph{orf}\Delta$ simultaneously with a single model fit.
Introducing a random effect $Z_l$ allows us to account for between subject variation by estimating a single ${\sigma_Z}^2$.
Unlike the \cite{QFA1} approach, observed values ${F}_{clm}$ are not scaled and instead a parameter to model a condition effect $\mu_c$ is introduced.}

$\gamma_{cl}$ represents the estimated strength of genetic interaction between an $\emph{orf}\Delta$ and its query mutation counterpart. 
For a multiplicative model of epistasis, an additive model is used to describe the log transformed data $f_{clm}=\log(F_{clm}+1)$, where ${F}_{clm}$ are the observed fitnesses.
We use the Benjamini-Hochberg test to correct for multiple testing in order to make a fair comparison with the \citep{QFA1} approach.

Inference for a frequentist random effects model can be carried out most simply with the R package ``lme4'' \citep{lme4}.
For the R code to fit the REM see Section~\ref{app:remcode} of the Appendix.
In the frequentist paradigm some parameters cannot be modelled as random effects since computational difficulties associated with large matrix computations arise with multiple random effects and very large data sets. 
Similarly, a more appropriate model with a log-link function in order to model repeat level variation with a normal distribution cannot be fit, due to computational difficulties that arise with non-linear model maximum likelihood algorithms and large data sets.
Such computational difficulties cause algorithms for parameter estimation to fail to converge.
\begin{table}
  \caption[Description of the random effects model]{Description of the random effects model (REM). 
$c$~identifies the condition for a given $\emph{orf}\Delta$, $l$~identifies a particular $\emph{orf}\Delta$ from the gene deletion library and $m$ identifies a repeat for a given $\emph{orf}\Delta$.\label{tab:REM}}
\begin{align*}
f_{clm}&= \mu_c+Z_l+\gamma_{cl}+\varepsilon_{clm}\\
\mu_{c}&=\begin{cases}
\mu+\alpha  & \text{if } c=0;\\
\mu & \text{if } c=1.
\end{cases}\qquad
&\gamma_{cl}&=\begin{cases}
0  & \text{if } c=0;\\
\gamma_{l} & \text{if } c=1.
\end{cases}\\
Z_l&\sim \mathcal{N}(0,{\sigma_Z}^2)
&\varepsilon_{clm} &\sim \mathcal{N}(0,\sigma^2)
\end{align*}
\end{table}

\end{chapter}

\begin{chapter}{\label{cha:case_stu}Case Studies}

\section{\label{case:intro}Introduction}
In this chapter, the new Bayesian models developed in Chapter~\ref{cha:modelling_den_int} are applied to previously analysed QFA screen data.
The one-stage and two-stage Bayesian approaches are compared with the two-stage \citet{QFA1} and random effects model (REM) approaches for a QFA screen comparison designed to inform the experimenter about telomere biology in \emph{S. cerevisiae}. 

After comparing the approaches developed, the one-stage Bayesian joint hierarchical model (JHM) is found to best model a QFA screen comparison.
The JHM is then applied to further examples of \emph{S. cerevisiae} QFA screen data to demonstrate the JHM's ability to model different experiments.
Two extensions of the JHM are then considered, to account for a batch effect and a transformation effect within a QFA screen comparison.
Fitness plots for the further case studies and extensions of the JHM are included for further investigation and research.

The new one-stage Bayesian QFA will be used at first to help identify genes that are related to telomere activity, but the analysis is general enough to be applicable to any high-throughput study of arrayed microbial cultures (including experiments such as drug screening).

\section{\label{sec:ura3_cdc13-1_27_27}\emph{cdc13-1}~$\boldsymbol{{27}^{\circ}}$C~vs~\emph{ura3}$\Delta$~$\boldsymbol{{27}^{\circ}}$C suppressor/enhancer data set}

The following analysis is for a QFA experiment comparing query \mbox{\emph{cdc13-1}} \emph strains with control \emph{ura3$\Delta$} strains at~${27}^{\circ}$C, previously analysed by \citet{QFA1}, to identify genes that show evidence of genetic interaction with the query mutation \mbox{\emph{cdc13-1}}. 
The ability of the Cdc13 protein produced by \mbox{\emph{cdc13-1}} strains to cap telomeres is reduced at temperatures above $26\,^{\circ}\mathrm{C}$ \citep{cdc131}, inducing a fitness defect. 

The experimental data used are freely available at \sloppy\url{http://research.ncl.ac.uk/colonyzer/Addi​nallQFA/}.\sloppy 
\cite{QFA1} present a list of interaction strengths and p-values for significance of interaction, together with a fitness plot for this experiment.  We will compare lists of genes classified as interacting with \mbox{\emph{cdc13-1}} by the non-hierarchical frequentist approach presented by \cite{QFA1} and the hierarchical REM with those classified as interacting by our hierarchical Bayesian approaches. 

4,294 non-essential $\emph{orf}\Delta$s were selected from the yeast deletion collection and used to build the corresponding double deletion query and control strains. 
Independent replicate culture growth curves (time course observations of cell density) were captured for each query and control strain. 
The median and range for the number of replicates per $\emph{orf}\Delta$ is 8 and $[8,144]$ respectively.
There are 66 $\emph{orf}\Delta$ strains that have greater than 8 replicates (for both the control and query screen).
More replicates have been tested for this subset of $\emph{orf}\Delta$s as a quality control measure to check if 8 replicates are sufficient to generate a stable fitness summary for each $\emph{orf}\Delta$. $\emph{orf}\Delta$s with high replicate number include a small number of mutations whose phenotypes are well understood in a telomere-defective background, together with some controls and a range of mutations randomly selected from the deletion library.
Including genotypes with well characterised phenotypes allows us to leverage expert, domain-specific knowledge to assess the quality of experimental results.
The modelling approaches considered can accommodate different numbers of replicates for each $\emph{orf}\Delta$, therefore we don't expect systematic bias from the number of repeats.
The range for the number of time points for growth curves captured in the control experiment is $[7,22]$ and $[9,15]$ in the query experiment.  
Raw \mbox{\emph{cdc13-1}}~${27}^{\circ}$C time series data is given in Figure~\ref{app:appendix_label}, for example.

As in the \cite{QFA1} analysis, a list of 159 genes are stripped from our final list of genes for biological and experimental reasons. 
Prior hyper-parameters for the models used throughout this chapter are provided in Table~\ref{tab:SHM_priors}.
\hl{Although our priors are informed by frequentist estimates of historical QFA data sets, we ensure our priors are sufficiently diffuse that all plausible parameter values are well represented and that any given QFA data set can be fit appropriately.}

\hl{The Heidelberg-Welch \citep{Heidelberger}\sloppy and Raftery-Lewis \sloppy\citep{Raftery}\sloppy convergence diagnostics are used to determine whether convergence has been reached for all parameters.
Posterior and prior densities are compared by eye to ensure that sample posterior distributions are not restricted by the choice of prior distribution.
ACF (auto-correlation) plot diagnostics are checked visually to ensure that serial correlation between sample values of the posterior distribution is low, ensuring that the effective sample size is similar to the actual sample size.} 

To assess how well the logistic growth model \hl{describes cell density observations} we generate plots of raw data with fitted curves overlaid. 
Figures~\ref{fig:diagABC}A, \ref{fig:diagABC}B and \ref{fig:diagABC}C show time series data for three different mutant strain repeats at~${27}^{\circ}$C\hl{, together with fitted logistic curves}.
We can see that each $\emph{orf}\Delta$ curve fit well represents the repeat level estimates as each $\emph{orf}\Delta$ level (red) curve lies in the region where most repeat level (black) curves are found.
Sharing information between $\emph{orf}\Delta$s will also affect each $\emph{orf}\Delta$ curve fit, increasing the probability of the $\emph{orf}\Delta$ level parameters being closer to the population parameters.
Comparing Figures~\ref{fig:diagABC}A, \ref{fig:diagABC}B and \ref{fig:diagABC}C shows that the separate hierarchical model (SHM) \hl{captures heterogeneity at} both the repeat and $\emph{orf}\Delta$ levels.

Figure~\ref{fig:diagABC}D demonstrates the hierarchy of information about \hl{the logistic model parameter $K$} generated by the \hl{SHM} for the $\emph{rad50}\Delta$ control mutant strain (variation decreases going from population level down to repeat level). 
Figure~\ref{fig:diagABC}D \hl{also shows that the posterior distribution for $K$ is much more peaked than the prior, demonstrating that we have learned about the distribution of both the population and $\emph{orf}\Delta$ parameters.}
Learning more about the repeat level parameters \hl{reduces} the variance of our $\emph{orf}\Delta$ level estimates.
The posterior for the first time-course repeat \hl{ $K_{clm}$ } parameter shows exactly how much uncertainty there is for this particular repeat in terms of carrying capacity $K$.

\begin{figure}[h!]
  \centering
\includegraphics[width=14cm]{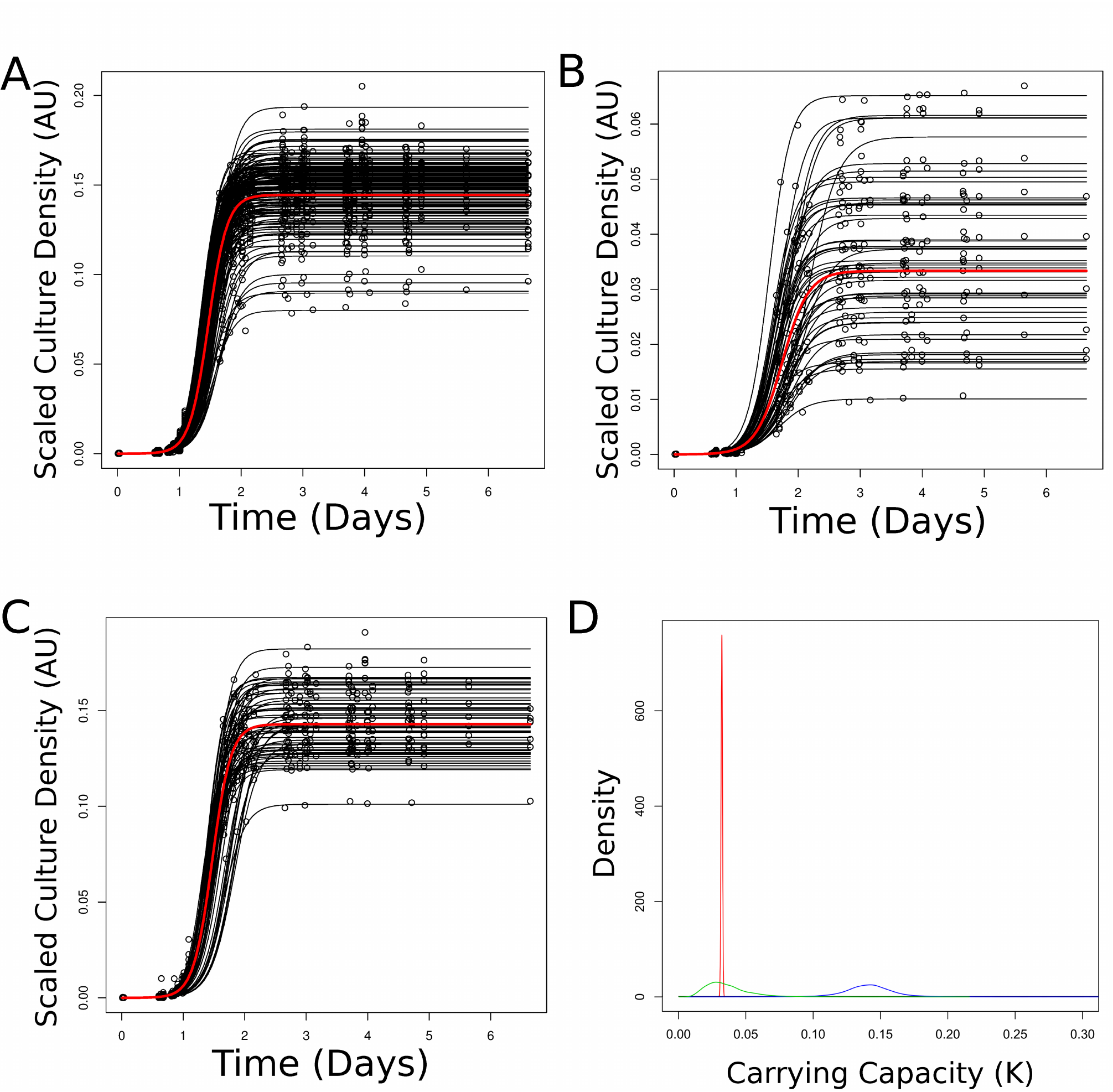}
 \caption[Separate hierarchical model logistic growth curve fits]{
Separate hierarchical model (SHM) logistic growth curve fits.
Data for $\emph{orf}\Delta$ repeats have been plotted in A, B and C, with SHM fitted curves overlaid in black for repeat level parameters and red for the $\emph{orf}\Delta$ level parameter fit. 
A) SHM scatter plot for 144 \emph{his3}$\Delta$ \emph{ura3$\Delta$} repeats at~${27}^{\circ}$C. 
B) SHM scatter plot for 48 \emph{rad50}$\Delta$ \emph{ura3$\Delta$} repeats at~${27}^{\circ}$C. 
C) SHM scatter plot for 56 \emph{exo1}$\Delta$ \emph{ura3$\Delta$} repeats at~${27}^{\circ}$C. 
D) SHM density plot of posterior predictive distributions for \emph{rad50}$\Delta$ \emph{ura3$\Delta$} 
carrying capacity $K$ hierarchy. 
The prior distribution for $K^p$ is in black.
The posterior predictive for $e^{K^o_l}$ is in blue and for $K_{clm}$ in green.
The posterior distribution of the first time-course repeat $K_{clm}$ parameter is in red.
\hl{Parameters $K^p$, $e^{K^o_l}$ and $K_{clm}$ are on the same scale as the observed data.}
}

\label{fig:diagABC}
\end{figure}
\FloatBarrier

\subsection{\label{sub:two_sta_fre_app}Frequentist approach}


Figure~\ref{fig:old}A is a $MDR\times MDP$ fitness plot from \cite{QFA1} \hl{where growth curves and evidence for genetic interaction are modelled using} the non-hierarchical frequentist methodology discussed in Section~\ref{int:QFAqfa}.
Figure~\ref{fig:REM}B is a $MDR\times MDP$ fitness plot for the frequentist hierarchical approach REM, described in Table~\ref{tab:REM}, applied to the logistic growth parameter estimates used in \cite{QFA1}.
The number of genes identified as interacting with \emph{cdc13-1} by \cite{QFA1} and by the REM are 715 and 315 respectively (Table~\ref{tab:sup_enh}).
The REM has highlighted many strains which have low fitness. In order to fit a linear model to the fitness data and interpret results in terms of the multiplicative model we apply a log transformation to the fitnesses, thereby affecting the distribution of $\emph{orf}\Delta$ level variation.

The REM accounts for between subject variation and allows for the estimation of a query mutation and $\emph{orf}\Delta$ effect to be made simultaneously, unlike the model presented by \cite{QFA1}.
Due to the limitations of the frequentist \hl{hierarchical} modelling framework, the REM model assumes equal variances for all \emph{orf}$\Delta$s and incorrectly describes \emph{orf}$\Delta$ level variation as Log-normal, assumptions that are not necessary in our new Bayesian approaches.

\begin{figure}[h!]
  \centering
\includegraphics[width=14cm]{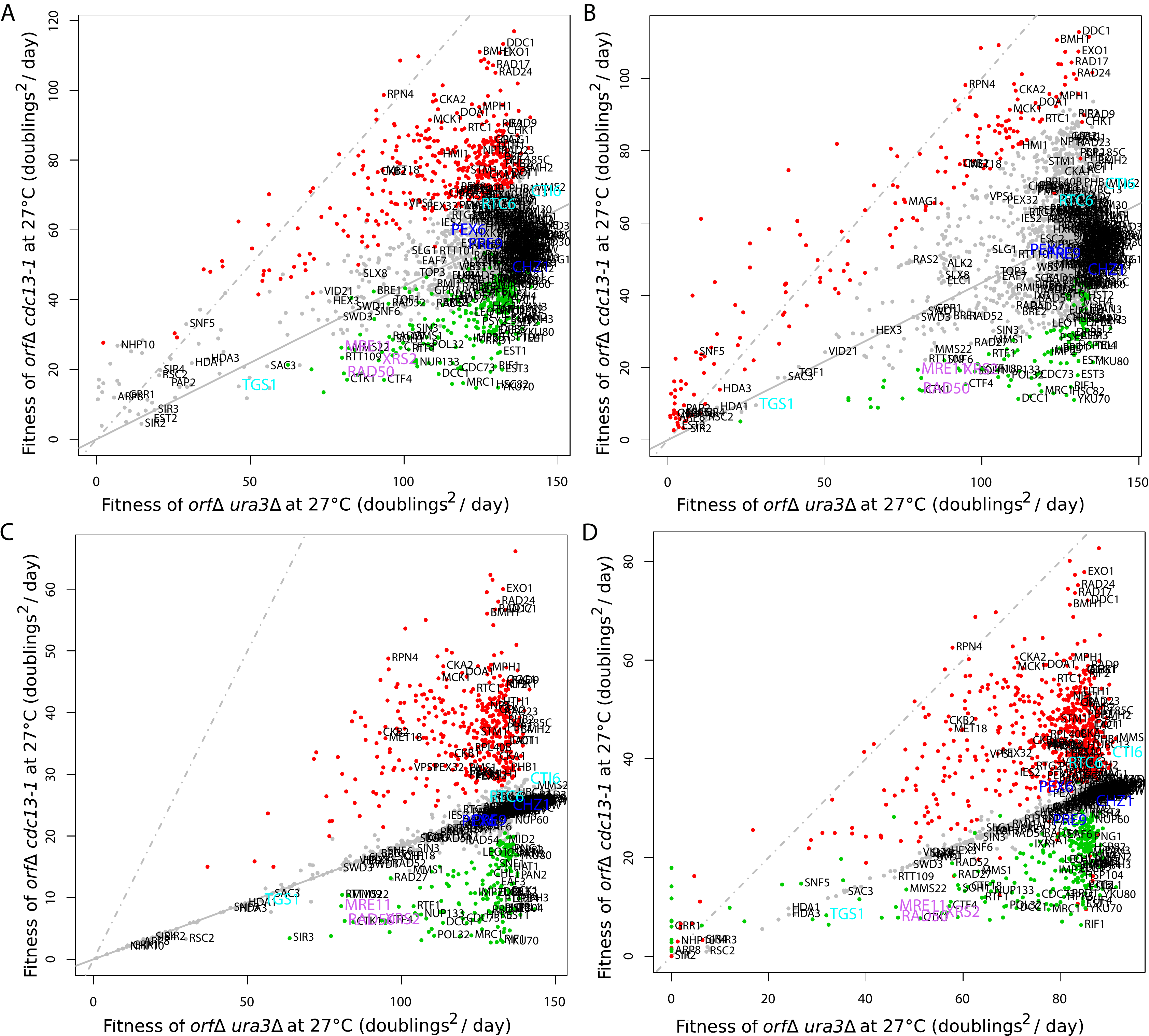}
\caption[Fitness plots with \emph{orf}$\Delta$ posterior mean fitnesses]{Fitness plots with \emph{orf}$\Delta$ posterior mean fitnesses.
Mean $\emph{orf}\Delta$ level fitness are plotted for the control strains against the corresponding query strains. 
$\emph{orf}\Delta$s with significant evidence of interaction are highlighted in red and green for suppressors and enhancers respectively.
A) Non-Bayesian, non-hierarchical fitness plot, based on Table~S6 from \cite{QFA1} $(F=MDR\times MDP)$.
B) Non-Bayesian, hierarchical fitness plot, \hl{from fitting the REM to data} in Table~S6 from \cite{QFA1} $(F=MDR\times MDP)$.
C) IHM fitness plot with $\emph{orf}\Delta$ posterior mean fitness. 
$\emph{orf}\Delta$s with significant evidence of interaction are highlighted on the plot as red and green for suppressors and enhancers respectively $(F=MDR\times MDP)$. 
D) JHM fitness plot with $\emph{orf}\Delta$ posterior mean fitnesses.
$\emph{orf}\Delta$ strains for the JHM plot are classified as being a suppressor or enhancer based on analysis of growth parameter $r$, meaning occasionally strains can be more fit in the query experiment in terms of $MDR\times MDP$ but be classified as enhancers (green).
For panels A and B significant interactors are classified as those with FDR corrected p-values $<0.05$.
For panels C and D significant interactors have posterior probability $\Delta>0.5$.
To compare fitness plots, labelled genes are those belonging to the following GO terms in Table~\ref{tab:sup_enh}: ``telomere maintenance'', ``ageing'', ``response to DNA damage stimulus'' or ``peroxisomal organization'', as well as the genes identified as interactions only in $K$ with the JHM (see Figure~\ref{fig:JHM_only}) (blue), genes interacting only in $r$ with the JHM (cyan) and the MRX complex genes (pink).
Solid and dashed grey fitted lines are for the 1-1 line and linear model fits respectively.
Alternative fitness plots with each of the GO terms highlighted are given in Section~\ref{app:GO_fit} of the Appendix.
\vspace{0.2in}
}
\label{fig:old}
\label{fig:REM} 
\label{fig:IHM}
\label{fig:JHM}
\end{figure}
\clearpage
\begin{table}
\caption[Number of genes interacting with \emph{cdc13-1} at ${27}^{\circ}$C]{\label{tab:sup_enh}Number of genes interacting with \emph{cdc13-1} at ${27}^{\circ}$C identified using each of four approaches: Add \citep{QFA1}, REM, IHM and JHM.
 Number of genes annotated with four example GO terms (telomere maintenance, ageing, response to DNA damage stimulus and peroxisome organisation) are also listed.
For the \citet{QFA1} and REM approach, significant interactors are classified as those with FDR corrected p-values (q-values) $<0.05$.
The label ``half data'' denotes analyses where only half of the available experimental observations are used.
The JHM uses a $MDR\times MDP$ summary after model fitting to classify suppressors and enhancers, comparable with the other three approaches.
The full lists of GO terms for each approach considered are given in a spreadsheet document, freely available online at \sloppy\url{http://research.ncl.ac.uk/qfa/HeydariQFABayes/}.\sloppy
}
\centering
\includegraphics[width=14cm]{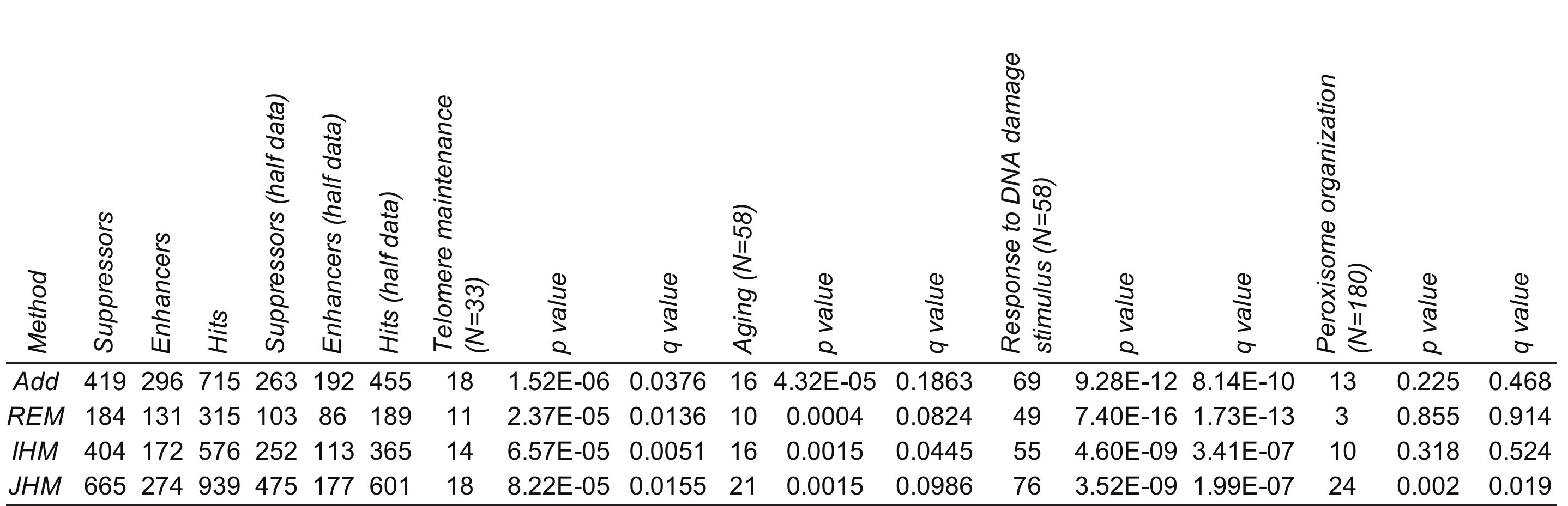}
\end{table}
\FloatBarrier

\subsection{\label{sub:two_sta_bay_app}Two stage Bayesian approach}
Figure \ref{fig:IHM}C is an interaction hierarchical model (IHM) fitness plot with $\emph{orf}\Delta$ level fitness measures generated using the new Bayesian two-stage methodology with fitness in terms of $MDR\times MDP$. 
576 genes are identified by the IHM as genetic interactions (Table~\ref{tab:sup_enh}).
Logistic parameter posterior means are used to generate fitness measures. 
For a gene $(l)$ from the gene deletion library, $(e^{Z_{l}})$ is the fitness for the control and $(e^{\alpha_{1}+Z_{l}+\delta_{l}\gamma_{c,l}})$ for the query in the IHM.
For a gene $(l)$ in the query screen, with no evidence of genetic interaction i.e. $\delta_{l}=0$, fitness will be a linear transformation from the control counterpart $(e^{\alpha_{1}+Z_{l}})$.
Similar to Figures~\ref{fig:old}A~and~\ref{fig:REM}B, Figure~\ref{fig:IHM}C shows how the majority of control strains are more fit than their query strain counterparts, with a mean fitted line lying below the line of equal fitness. 
Comparing the fitted lines in Figures~\ref{fig:old}A~and~\ref{fig:REM}B with Figure~\ref{fig:IHM}C, the IHM shows the largest deviation between the fitted line and the line of equal fitness, is largely due to the difference in $P$ estimated with the SHM for the control and query data sets being scaled out by the parameter $\alpha_{1}$.
If we fix $P$ in our Bayesian models, similar to the frequentist approach, genetic interactions identified are largely the same, but we then have the problem of choosing $P$. We recommend estimating $P$ simultaneously with the other model parameters because if the choice of $P$ is not close to the true value, growth rate $r$ estimates must compensate and don't give accurate estimates for time courses with low carrying capacity $K$.

It can be seen that many of the interacting $\emph{orf}\Delta$s have large deviations from the genetic independence line. 
This is because of the indicator variable in the model, used to describe genetic interaction. 
When there is enough evidence for interaction the Bernoulli variable is set to 1\hl{, otherwise it is} set to 0. 
It is interesting to note that non-significant $\emph{orf}\Delta$s, marked by grey points, lie amongst some of the significant strains. 
Many \hl{such points} have high variance and therefore we are \hl{less confident that these interact with the query mutation}.
This \hl{feature} of our new approach \hl{is an improvement over that} presented in \cite{QFA1}, which always shows evidence for an epistatic effect when mean distance from the genetic independence line is large, regardless of strain fitness variability.
An extract from the list of top interactions identified by the IHM is included in Table~\ref{app:IHM_interactions}.
\FloatBarrier

\subsection{\label{sub:one_sta_app}One stage Bayesian approach}
Figure~\ref{fig:JHM}D is a JHM $MDR\times MDP$ fitness plot using the new, \hl{unified} Bayesian methodology. 
The $MDR\times MDP$ fitness plot given in Figure~\ref{fig:JHM}D is for visualisation and comparison with the $MDR\times MDP$ fitness plots of the other approaches considered: the JHM does not make use of a fitness measure.
939 genes are identified by the JHM as genetic interactions (Table~\ref{tab:sup_enh}).
Posterior means of model parameters are used to obtain the following fitness measures. 
With the JHM we can obtain an \emph{orf}$\Delta$ level estimate of the carrying capacity and growth rate $(K,r)$ for a gene ($l$).
For a gene ($l$) from the gene deletion library, carrying capacity and growth rate $(e^{K^{o}_{l}},e^{r^{o}_{l}})$ are used to evaluate the fitness for the control and $(e^{\alpha_{1}+K^{o}_{l}+\delta_{l}\gamma_{c,l}},e^{\beta_{1}+r^{o}_{l}+\delta_{l}\omega_{c,l}})$ for the query.
For a gene $(l)$ in the query screen, with no evidence of genetic interaction i.e. $\delta_{l}=0$, carrying capacity and growth rate will be linear transformations from the control counterpart $(e^{\alpha_{1}+K^{o}_{l}},e^{\beta_{1}+r^{o}_{l}})$.

Instead of producing a fitness plot in terms of $MDR\times MDP$, it can also be useful to analyse carrying capacity $K$ and growth rate $r$ fitness plots as\hl{, in the JHM,} evidence for genetic interaction \hl{comes from both of} these parameters \hl{simultaneously}, see Figures~\ref{fig:JHM_K}~and~\ref{fig:JHM_r}.
Fitness plots in terms of logistic growth parameters are useful for identifying some unusual characteristics of $\emph{orf}\Delta$s.
For example, an $\emph{orf}\Delta$ may be defined as a suppressor in terms of $K$ but an enhancer in terms of $r$.
\hl{To enable direct} comparison with the \cite{QFA1} analyses we generated a $MDR\times MDP$ fitness plot, Figure~\ref{fig:JHM}D.
An extract from the list of top interactions identified by the JHM is included in Table~\ref{app:JHM_interactions}.
\FloatBarrier

\section{\label{Application3}Comparison with previous analysis}
\subsection{\label{individual_interactions}Significant genetic interactions}
Of the genes identified as interacting with \emph{cdc13-1} (1038, see Table~\ref{tab:overlap}A) some are identified consistently across all four approaches (215 out of 1038, see Table~\ref{tab:overlap}A).  Of the hits identified by the JHM (939), the majority (639) are common with those in the previously published \citet{QFA1} approach.  However, 231 of 939 are uniquely identified by the JHM and could be subtle interactions which are the result of previously unknown biological processes.

To examine the evidence for some interactions uniquely identified by the JHM in more detail we compared the growth curves for three examples from the group of interactions identified only by the JHM.  These examples (\emph{chz1}$\Delta$, \emph{pre9}$\Delta$ and \emph{pex6}$\Delta$) are genetic interactions which can be identified in terms of carrying capacity $K$, but not in terms of growth rate $r$ (see Figure~\ref{fig:JHM_only}). 
By observing the difference between the fitted growth curve (red) and the expected growth curve, given no interaction (green) in Figure~\ref{fig:JHM_only}A, \ref{fig:JHM_only}B and \ref{fig:JHM_only}C we test for genetic interaction.  Since the expected growth curves in the absence of genetic interaction are not representative of either the data or the fitted curves on the repeat and \emph{orf}$\Delta$ level, there is evidence for genetic interaction.

\begin{table}
\caption[Overlap between methods for genes interacting with \emph{cdc13-1} at $\boldsymbol{{27}^{\circ}}$C and gene ontology terms over-represented in lists of interactions]{\label{tab:overlap}Genes interacting with \emph{cdc13-1} at $\boldsymbol{{27}^{\circ}}$C and GO terms over-represented in the list of interactions according to each approach A) Number of genes identified for each approach (Add \cite{QFA1}, REM, IHM and JHM) and the overlap between the approaches. 4135 genes from the \emph{S. cerevisiae} single deletion library tested overall.
B) Number of GO terms identified for each approach (Add \cite{QFA1}, REM, IHM and JHM) and the overlap between the approaches.  6107 \emph{S. cerevisiae} GO Terms available.}
\centering
\resizebox{\columnwidth}{!}{%
\begin{tabular}{*{6}{c}}
\multicolumn{1}{l}{\bf{A.}}&  & \multicolumn{2}{c}{\emph{REM:0}} & \multicolumn{2}{c}{\emph{REM:1}}\\
\cline{3-6}
& &\emph{Add:0} &\emph{Add:1} &\emph{Add:0} &\emph{Add:1}\\\hline
\multirow{2}{*}{\emph{IHM:0}} &\emph{JHM:0} &3097&54&31&10\\
& \emph{JHM:1} &231&78&29&29\\\hline
\multirow{2}{*}{\emph{IHM:1}} &\emph{JHM:0} &1&2&1&0\\
&\emph{JHM:1} &30&327&0&215\\\hline
\end{tabular}
\qquad
\begin{tabular}{*{6}{c}}
\multicolumn{1}{l}{\bf{B.}}&  & \multicolumn{2}{c}{\emph{REM:0}} & \multicolumn{2}{c}{\emph{REM:1}}\\
\cline{3-6}
& &\emph{Add:0} &\emph{Add:1} &\emph{Add:0} &\emph{Add:1}\\\hline
\multirow{2}{*}{\emph{IHM:0}} &\emph{JHM:0} &5813&21&58&7\\
& \emph{JHM:1} &46&8&6&10\\\hline
\multirow{2}{*}{\emph{IHM:1}} &\emph{JHM:0} &20&15&3&12\\
&\emph{JHM:1} &13&54&2&147\\\hline
\end{tabular}
}
\end{table}

\begin{figure}[h!]
  \centering
\includegraphics[width=14cm]{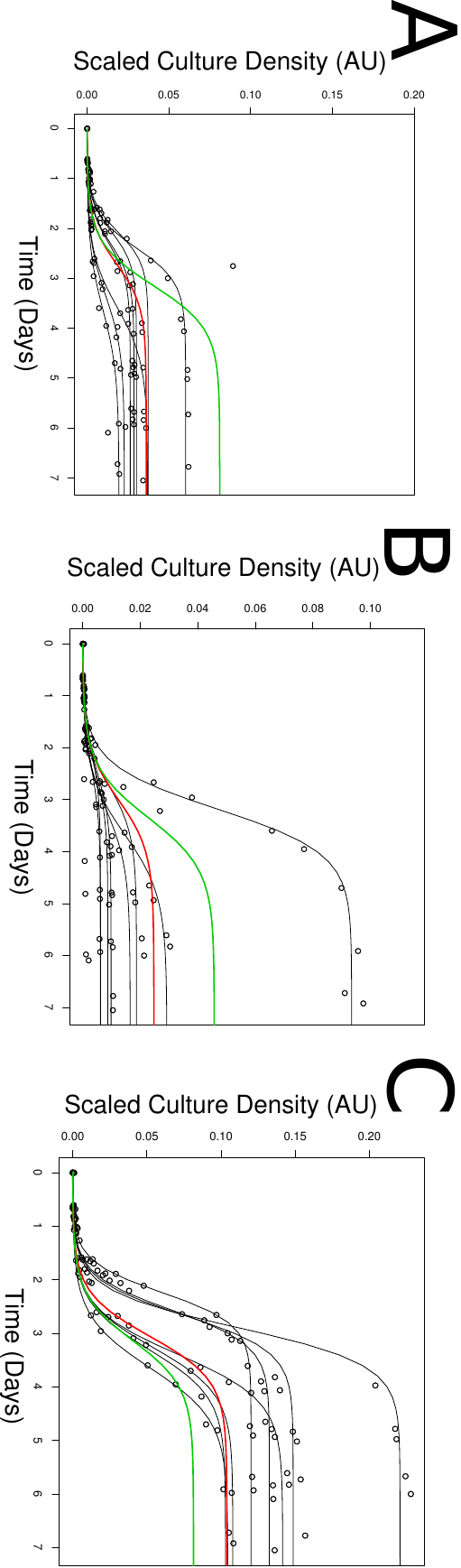}
 \caption[Joint hierarchical model logistic growth curve fits]{
Joint hierarchical model (JHM) logistic growth curve fitting. JHM data for $\emph{orf}\Delta$ repeats have been plotted in A, B and C, with fitted curves overlaid in black for repeat level parameters, red for the  $\emph{orf}\Delta$ level query parameter fit and green for the expected $\emph{orf}\Delta$ level query parameter fit with no genetic interaction. 
A) JHM scatter plot for 8 \emph{chz1}$\Delta$ \emph{cdc13-1} repeats. 
B) JHM scatter plot for 8 \emph{pre9}$\Delta$ \emph{cdc13-1} repeats. 
C) JHM scatter plot for 8 \emph{pex6}$\Delta$ \emph{cdc13-1} repeats. 
}
\label{fig:JHM_only}
\end{figure} 

We chose a prior for the probability $p$ of a gene interacting with the background mutation as 0.05.  We therefore expected to find 215 genes interacting. The Bayesian models, for which a prior is applicable (IHM and JHM), find more genes than expected (576 and 939 interactions respectively, Table~\ref{tab:sup_enh}), demonstrating that information in this dataset can overcome prior expectations. The JHM identifies the highest proportion of genes as hits out of all methods considered, particularly identifying suppressors of \emph{cdc13-1} (Table~\ref{tab:sup_enh}). In fact, the JHM identifies more hits than the \citet{QFA1} approach, even when constrained to using only half of the available data.
An important advantage to our new Bayesian approach is
that we no longer have the difficulty of choosing a q-value threshold.
For the \citet{QFA1} approach to have similar numbers of interactions to the JHM, a less stringent q-value threshold would have to be justified \emph{a posteriori} by the experimenter.

\subsection{\label{Application5}Previously known genetic interactions}
In order to compare the quality of our new, Bayesian hierarchical models with existing, frequentist alternatives, we examined the lists of genetic interactions identified by all the methods discussed and presented here.
Comparing results with expected or \hl{previously} known lists of interactions from the relevant literature, we find that genes coding for the MRX complex (\emph{MRE11}, \emph{XRS2} \& \emph{RAD50}), which are known to interact with \emph{cdc13-1} \citep{MRX}, are identified by all four approaches considered and can be seen in a similar position in all four fitness plots (Figure~\ref{fig:old}A, \ref{fig:REM}B, \ref{fig:IHM}C and \ref{fig:JHM}D).

By observing the genes labelled in Figure~\ref{fig:old}A~and~\ref{fig:REM}B we can see that the frequentist approaches are unable to identify many of the interesting genes identified by the JHM as these methods are unable to detect interactions for genes close to the genetic independence line.
The JHM has extracted more information from deletion strain fitnesses observed with high variability than the \cite{QFA1} approach by sharing more information between levels, consequently improving our ability to identify interactions for genes close to the line of genetic independence (subtle interactions).  \emph{CTI6}, \emph{RTC6} and  \emph{TGS1} are three examples of subtle interactors identified only by the JHM (interaction in terms of $r$ but not $K$) which all have previously known telomere-related functions \citep{TGS1,CTI6,RTC6}.

We tested the biological relevance of results from the various approaches by carrying out unbiased Gene Ontology (GO) term enrichment analyses on the hits (lists of genes classified as having a significant interaction with \emph{cdc13-1}) using the {bioconductoR package GOstats \citep{GOstats}}. 
For the GO term enrichment analysis R code used, see Section~\ref{app:GOstats} of the Appendix.

All methods identify a large proportion of the genes in the yeast genome annotated with the GO terms ``telomere maintenance'' and ``response to DNA damage stimulus'' (see Table~\ref{tab:sup_enh}), which were the targets of the original screen, demonstrating that they all correctly identify previously known hits of biological relevance.  Interestingly, the JHM identifies many more genes annotated with the ``ageing'' GO term, which we also expect to be related to telomere biology (though the role of telomeres in ageing remains controversial) suggesting that the JHM is identifying novel, relevant interactions not previously identified by the \citet{QFA1} screen (see Table~\ref{tab:sup_enh}).  Similarly, the JHM identifies a much larger proportion of the PEX ``peroxisomal'' complex (included in GO term: ``peroxisome organisation'') as interacting with \emph{cdc13-1} (see Table~\ref{tab:sup_enh}) including all of those identified in \citet{QFA1}. Many of the PEX genes show large variation in both $K$ and $r$, an example can be seen in Figure~\ref{fig:JHM_only}C for \emph{pex6$\Delta$}. Members of the PEX complex cluster tightly, above the fitted line in the fitness plot Figure~\ref{fig:JHM}D {(fitness plots with highlighted genes for GO terms in Table~\ref{tab:sup_enh} are given in Section~\ref{app:GO_fit} of the Appendix)}, demonstrating that although these functionally related genes are not strong interactors, they do behave consistently with each other, suggesting that the interactions are real.  The results of tests for significant over-representation of all GO terms are given in a spreadsheet document, freely available online at \sloppy\url{http://research.ncl.ac.uk/qfa/HeydariQFABayes/}.\sloppy

Overall, within the genes interacting with \emph{cdc13-1} identified by the \citet{QFA1}, REM, IHM and JHM approaches, 274, 245, 266 and 286 GO terms were significantly over-represented respectively (out of 6235 possible GO terms, see Table \ref{tab:overlap}B).  147 were common to all approaches and examples from the group of GO terms over-represented in the JHM analysis and not in the \citet{QFA1} analysis seem internally consistent (e.g. ``peroxisome organisation'' GO term) and consistent with the biological target of the screen, telomere biology (significant GO terms for genes identified only by the JHM are also included in the spreadsheet document).

{Extracts from the list of top interactions identified by both the IHM and JHM are provided in Section~\ref{app:interactions}.
Files including the full lists of genetic interactions for the IHM and JHM are freely available online at \sloppy\url{http://research.ncl.ac.uk/qfa/HeydariQFABayes/}.}\sloppy
Alternative fitness plots to Figure~\ref{fig:old}A, B, C \& D with gene labels for those showing significant evidence of genetic interaction are provided in Figure~\ref{fig:old_first} and Section~\ref{app:alt_fitness}.
As suppressors and enhancers in the JHM may be in terms of both $K$ and $r$, fitness plots in terms of $K$ and $r$ with gene labels for those showing significant evidence of genetic interaction are given in Figure~\ref{fig:JHM_K_full} and Figure~\ref{fig:JHM_r_full} respectively.   

To further compare the similarity of the Bayesian hierarchical models and frequentist analysis, a table of Spearman's rank correlation coefficients \citep{spearman} between genetic strengths and a $MDR\times MDP$ correlation plot of the JHM versus the \citet{QFA1} are given in Section~\ref{app:corr} of the Appendix. 

\subsection{\label{Application6}Hierarchy and model parameters}
\hl{The hierarchical structure and model choices included in the Bayesian JHM and IHM are derived from the known experimental structure of QFA.
Different levels of variation for different $\emph{orf}\Delta$s are expected and can be observed \hl{by comparing} distributions of frequentist estimates or by visual inspection of yeast culture images.} 
\hl{The direct relationship between experimental and model structure, together with the richness of detail and number of replicates included in QFA experimental design, reassures us that overfitting is not an issue in this analysis.} 
For the \emph{ura3$\Delta$}~${{27}^{\circ}}$C and \mbox{\emph{cdc13-1}}~${27}^{{\circ}}$C experiment with 4294 $\emph{orf}\Delta$s there are ~1.25 times the number of parameters in the JHM ($\sim$200,000) compared to the two stage REM approach ($\sim$160,000) but when compared to the large number of pairs of data points ($\sim$830,000) there are sufficient degrees of freedom to justify our proposed Bayesian models.

\subsection{\label{Application7}Computing requirements}
Our Bayesian hierarchical models require significant computational time. 
{As expected, the mixing of chains in our models is weakest at population level parameters such as $K_p$ and $\alpha_c$.}
For the \emph{ura3$\Delta$} ${{27}^{\circ}}$C and \mbox{\emph{cdc13-1}} ${27}^{{\circ}}$C \hl{dataset}, the JHM takes ${\sim}2$ weeks to converge and produce a sufficiently large sample.  The two stage Bayesian approach takes one week (with the IHM part taking ${\sim}1$ day), whereas the REM takes ${\sim}3$ days and the \cite{QFA1} approach takes ${\sim}3$ hours. 
\hl{A QFA experiment can take over a month from start to finish and so analysis time is acceptable in comparison to the time taken for the creation of the data set but still a notable inconvenience.}
{We expect that with further research effort, computational time can be decreased by using an improved inference scheme and that inference for the JHM could be completed in less than a week without parallelisation.}
MCMC algorithms are inherently sequential so, parallelisation is not completely trivial and may be considered for future development.
Parallelisation may reduce computational time by partitioning the state space into segments that can be updated in parallel \citep{parallel}.
For the JHM it may be possible to partition by QFA screens to reduce computational time.
Further, parallelisation may be possible across $\emph{orf}\Delta$s for even further reduction to computational time.
\FloatBarrier

\subsection{\label{conv_diag}Convergence diagnostics}
Evidence of convergence for our Bayesian models in Section~\ref{sub:two_sta_bay_app}~and~\ref{sub:one_sta_app} can be shown by observing posterior samples from the MCMC samplers used.
Figures~\ref{fig:SHMdiag}, \ref{fig:IHMdiag} and \ref{fig:JHMdiag} show evidence of convergence for a subset of population level parameters from the SHM, IHM and JHM respectively.
Posterior samples of 1000 particles are obtained after a burn-in period of 800k and a thinning of every 100 observations for the SHM, IHM and JHM.

Population level parameters are found to have the worst mixing in our models due to the large number of lower level parameters that population level parameter sampling distributions are conditioned upon. 
We demonstrate how our population parameters have converged with Trace plots, ACF and density plots in Figures~\ref{fig:SHMdiag}, \ref{fig:IHMdiag} and \ref{fig:JHMdiag}.
Trace plots show that the posterior samples are bound between a fixed range of values, indicating convergence.
Auto-correlation functions do not have any large peaks above the dashed blue line for significant evidence of dependence, showing that each sequential sample value from the posterior distributions are largely uncorrelated with previous values and ensuring that the effective sample size is similar to the actual sample size.
ACF plots in Figures~\ref{fig:IHMdiag}~and~\ref{fig:JHMdiag} do show some dependence within our posterior samples but as the ACF decays rapidly before a lag of 5, there is only a small amount that will not be a problem for inference.
Density plots show that that there is enough information within the models to give sufficiently peaked single modes, converging around a fixed region of plausible values.

Table~\ref{tab:modelconv} gives diagnostic statistics for the population parameters considered in Figures~\ref{fig:SHMdiag}, \ref{fig:IHMdiag} and \ref{fig:JHMdiag}.
We can see in Table~\ref{tab:modelconv} that the lowest effective sample size of our model parameters is $324$, for the JHM $P$ parameter, followed by $378$ for the SHM $P$ parameter. 
Of all our model parameters, $P$ was found to have the lowest effective sample size, but we are still able to find a large enough sample for our inference. 
Heidelberg and Welch P-values do not show evidence against the stationary of our chains, using a cut-off of $0.10$.
The above statistics are calculated for all model parameters and are used to identify where mixing is poor and if our model has reached convergence.
All chains are accepted for parameter posterior samples in Section~\ref{sub:two_sta_bay_app}~and~\ref{sub:one_sta_app} as effective sample sizes are found to be greater than $300$ and Heidelberg and Welch P-values greater than $0.10$ for every chain.

\begin{table}
\caption[Bayesian model convergence statistics]{Bayesian model convergence statistics for the two-stage approach in Section~\ref{sub:two_sta_bay_app} and one-stage approach in Section~\ref{sub:one_sta_app}. Heidelberg and Welch P-values and the effective sample size have been calculated for a subset of population level parameters. \label{tab:modelconv}}
\centering
\resizebox{\columnwidth}{!}{%
\npdecimalsign{.}
\nprounddigits{2}
\begin{tabular}{c c c c}
\hline
\emph{Model} & \emph{Parameter} & \emph{Effective sample size}  & \emph{Heidelberg and Welch P-value} \\ 
\hline
SHM & $K_p$ & 521 & 0.49 \\ 
     & $r_p$ & 441 & 0.11\\ 
     & $P$ & 378 & 0.56\\
		 & $\nu_p$ & 1000 & 0.17 \\
IHM & $Z_p$ & 677 & 0.35\\
     & $\sigma_z$ & 430 & 0.14 \\ 
     & $\nu_p$ & 1000 & 0.46\\
		     & $\alpha_c$ & 914 & 0.59\\
JHM & $K_p$ & 473 & 0.72\\ 
     & $r_p$ & 566 & 0.12\\ 
     & $P$ & 324 & 0.12\\
		     & $\nu_p$ & 1000 & 0.13 \\
				     & $\alpha$ & 407 & 0.36\\
		     & $\beta$ & 808 & 0.67\\
\hline
\end{tabular}
\npnoround%
}
\end{table}

\begin{figure}[h!]
  \centering
	\resizebox{\columnwidth}{!}{%
\includegraphics[width=14cm]{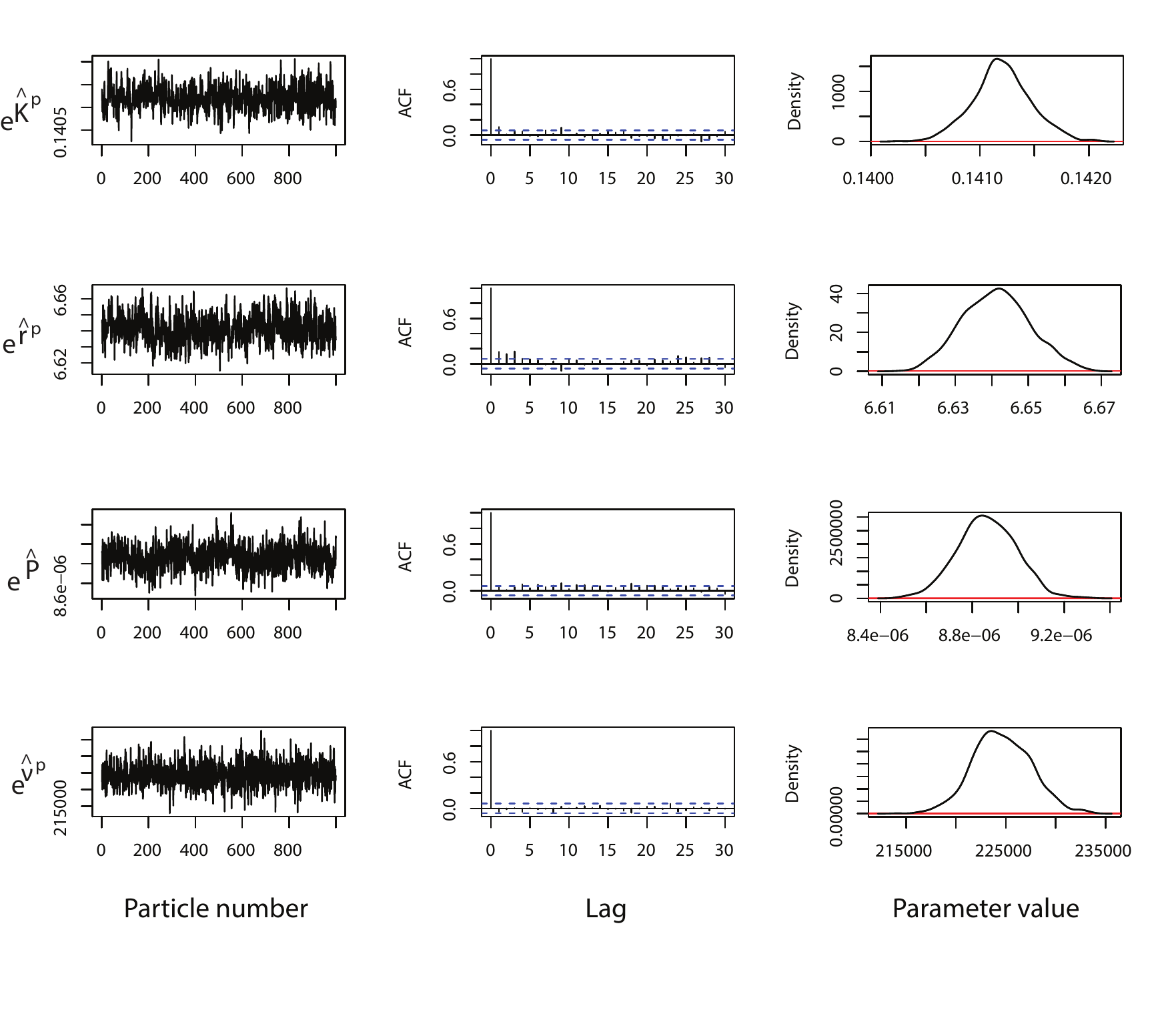}
}
\caption[Convergence diagnostics for the separate hierarchical model]{Convergence diagnostics for the separate hierarchical model (SHM). Trace, auto-correlation and density plots for the SHM parameter posteriors (sample size = 1000, thinning interval = 100 and burn-in = 800000), see Section~\ref{sub:two_sta_bay_app}. Posterior (black) and prior (red) densities are shown in the right hand column. \label{fig:SHMdiag}
}
\end{figure}

\begin{figure}[h!]
  \centering
	\resizebox{\columnwidth}{!}{%
\includegraphics[width=14cm]{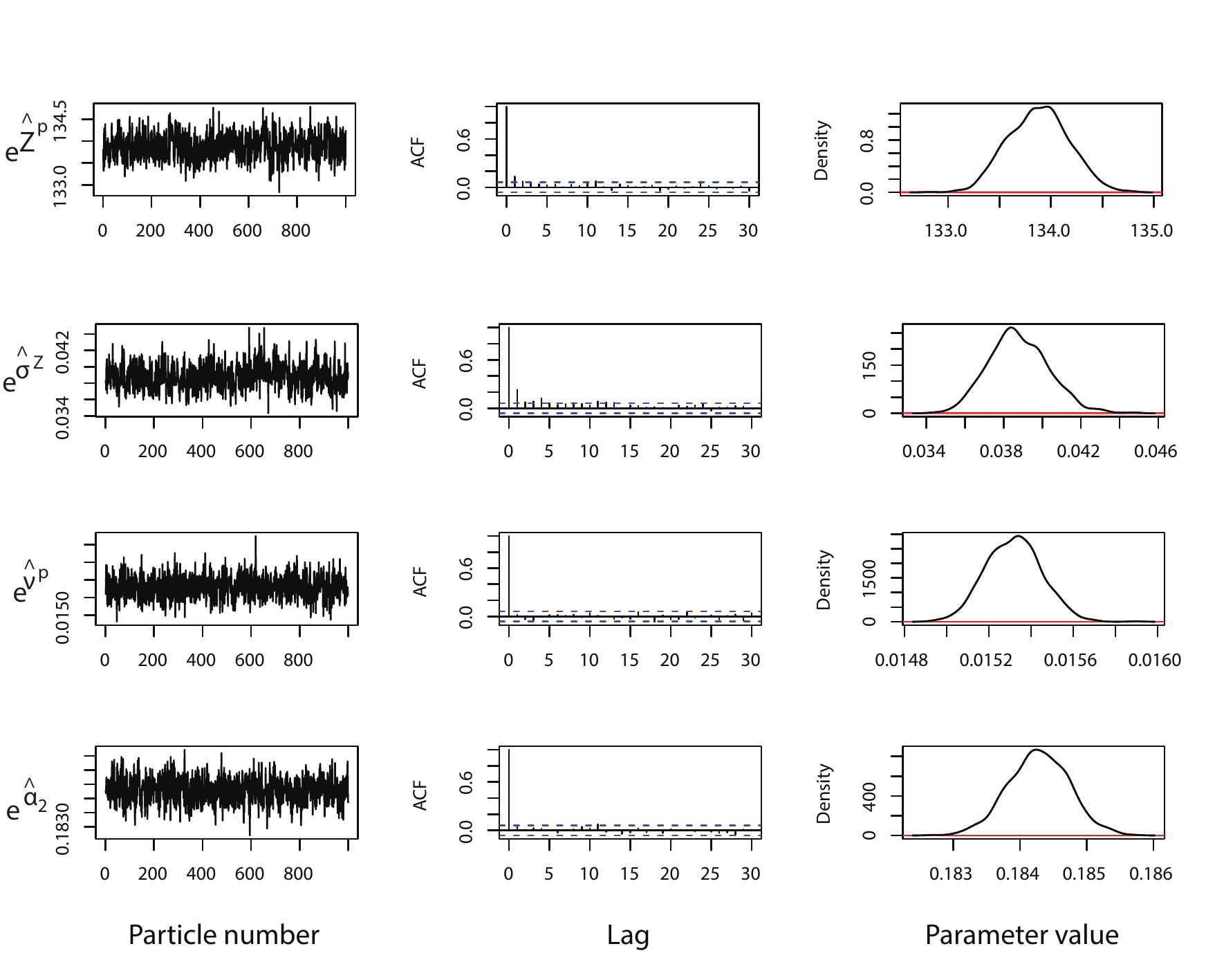}
}
\caption[Convergence diagnostics for the interaction hierarchical model]{Convergence diagnostics for the interaction hierarchical model (IHM). Trace, auto-correlation and density plots for the IHM parameter posteriors (sample size = 1000, thinning interval = 100 and burn-in = 800000), see Section~\ref{sub:two_sta_bay_app}. Posterior (black) and prior (red) densities are shown in the right hand column.\label{fig:IHMdiag}
}
\end{figure}

\begin{figure}[h!]
  \centerline{
	\resizebox{\columnwidth}{!}{%
\includegraphics[width=14cm]{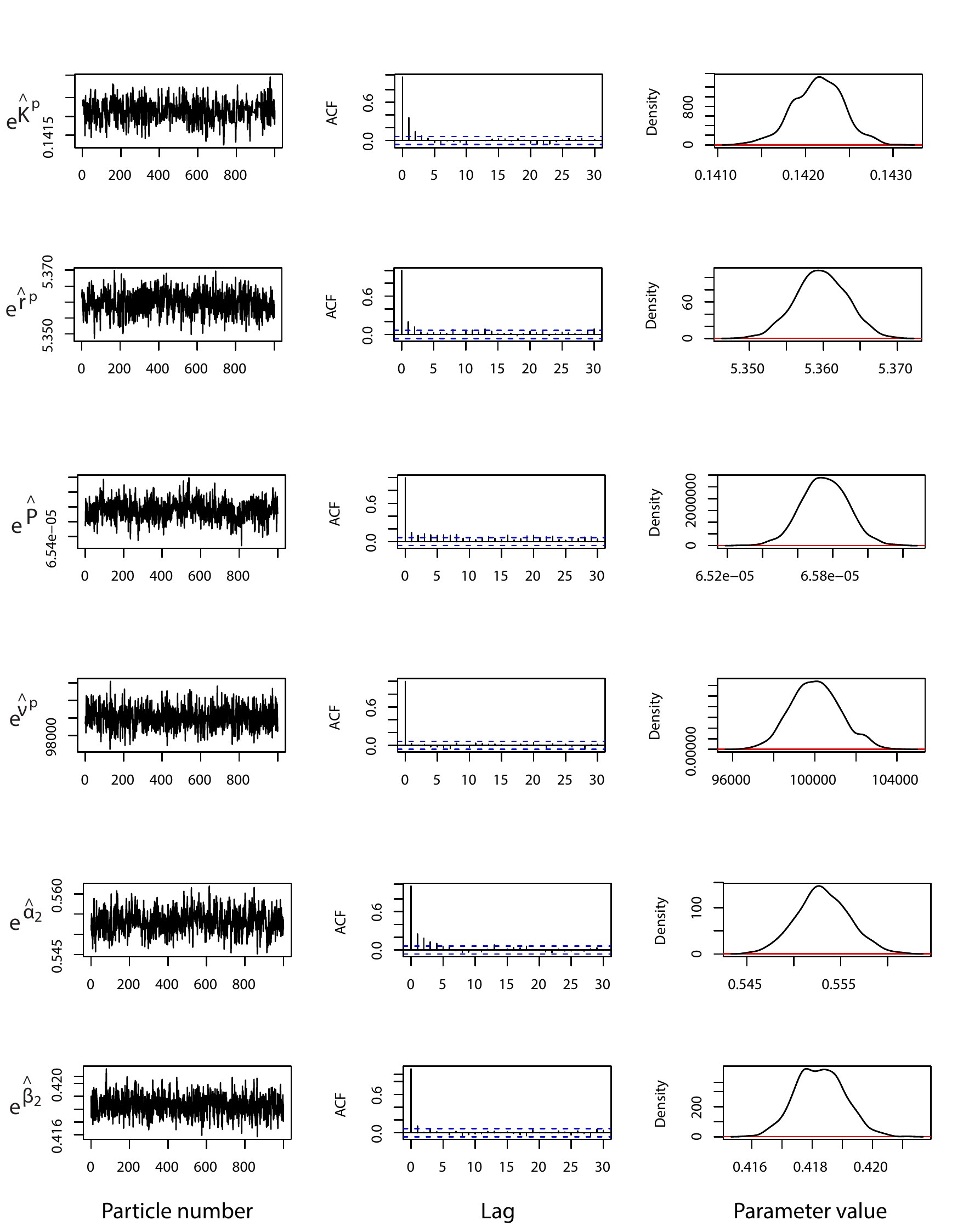}
}
}
\caption[Convergence diagnostics for the joint hierarchical model]{Convergence diagnostics for the joint hierarchical model (JHM). Trace, auto-correlation and density plots for the JHM parameter posteriors (sample size = 1000, thinning interval = 100 and burn-in = 800000), see Section~\ref{sub:one_sta_app}. Posterior (black) and prior (red) densities are shown in the right hand column.\label{fig:JHMdiag}
}
\end{figure}
\FloatBarrier

\subsection{\label{sim_study}Simulation study}
A simulation study was carried out to compare the performance of the different approaches considered for a simulated QFA screen comparison from the JHM.
We believe that the JHM closely models a QFA screen comparison and so by simulating a QFA screen comparison data set from the JHM we will obtain a data set for which we know the full set of true genetic interactions.
Simulated JHM data will include important features of QFA screen comparison data, such as a hierarchical structure and genetic interaction in terms of both $K$ and $r$.

Two simulated QFA screens where generated, a control and query screen with some condition effect in the query.
Each screen consists of 4300 \emph{orf}$\Delta$s and 8 logistic growth time-course repeats for each \emph{orf}$\Delta$. 
Each time-course consists of 10 measurements, evenly distributed across 6 days.
430 genes were set as genetic interactors in the query screen. 
The true Population level parameters are chosen from frequentist estimates of 10 historic data sets, \emph{orf}$\Delta$ and repeat level parameters are then generated from the JHM structure in Table~\ref{tab:JHM} and growth time-course data simulated.

Table~\ref{tab:simstudy} shows the number of true genetic interactions identified, suppressors and enhancers, as well as false positives (FPs) and false negatives (FN) for each of the approaches considered.
As expected, the JHM identifies the largest number of true genetic interactions.
The number of suppressors identified by the JHM is higher than the \citet{QFA1}, REM and IHM but for enhancers, all methods perform very similarly.
Performance of the different methods can be observed through the FP and FN rates.
From Table~\ref{tab:simstudy} we can calculate FP and FN rates, where FP rate$=1-$``sensitivity'' and FN rate$=1-$``specificity''.
FP rates for the \cite{QFA1}, REM, IHM and JHM are $0.078$, $0.042$, $0.006$ and $0.002$ respectively.
The JHM has the lowest FP rate when compared to the other approaches available.
Frequentist approaches \citet{QFA1} and REM have large FP rates when compared to the two Bayesian approaches.
The \citet{QFA1} approach has more false positives than true genetic interactions.
FN rates for the \cite{QFA1}, REM, IHM and JHM are $0.488$, $0.570$, $0.593$ and $0.270$ respectively.
Two-stage approaches \citet{QFA1}, REM and IHM have large FP rates when compared to the JHM.
The \cite{QFA1}, REM and IHM  have ${\sim}200$ false negatives, approximately double the number identified by the JHM (${\sim}100$).
Observing the genes that have been missed by the two-stage approaches, we find that they often fail to identify genetic interactions when evidence is weak in only $K$ or $r$, even if there is sufficient evidence in the other parameter such that the JHM can identify the genetic interaction.

From our simulation study we have been able to show that the two-stage frequentist approaches have high false positives and false negatives. 
From the number of false positives identified for each method, we can see that the non-hierarchical \citet{QFA1} approach has the worst performance, followed by the hierarchical two-stage approaches.
As expected, the JHM is the best approach when we consider a simulated hierarchical data set with genetic interaction in terms of $K$ and $r$, as the two-stage approaches fail to capture more subtle genetic interactions.

\begin{table}[h!]
\caption[Simulation study with a joint hierarchical model simulated dataset.]{Simulation study with a joint hierarchical model (JHM) simulated dataset.
A QFA screen comparison was generated from the JHM and  430 genes are set as genetic interactors, see Section~\ref{sim_study}. Applications of the \citep{QFA1}, REM, two-stage Bayesian (IHM) and one-stage Bayesian (JHM) approaches are made to the JHM simulated dataset and performance compared. Suppressors and enhancers are defined in terms of $MDR{\times}MDP$.\label{tab:simstudy}}
\centering
\resizebox{\columnwidth}{!}{%
\npdecimalsign{.}
\nprounddigits{2}
\begin{tabular}{c c c c c c c c}
\hline
\emph{Model} & \emph{True interactions} & \emph{True Suppressors} & \emph{True Enhancers } & \emph{False Positives}  & \emph{False Negatives} & \emph{Sensitivity} & \emph{Specificity} \\
  & \emph{identified (N=430)} & \emph{(N=274)} & \emph{(N=156)} &  &  & &\\
\hline
\cite{QFA1} & 220 & 158 & 62 & 303 & 210 &0.922 &0.512\\
REM & 185 & 100 & 85 & 163 & 245 &0.958 &0.430\\
IHM & 175 & 130 & 45 & 23 & 255 &0.994 &0.407\\ 
JHM & 314 & 256 & 58 & 8 & 116 &0.998	 & 0.730\\ 
\hline
\end{tabular}
\npnoround%
}
\end{table}




\FloatBarrier
\section{\label{sec:candjags}Bayesian inference code comparison}
Inference for the Bayesian hierarchical models in this thesis is carried out using code written in the C programming language.
To see how our code compares to commonly used software available for carrying out inference for Bayesian models, we have tested posterior samples for our C code and equivalent code using Just Another Gibbs Sampler (JAGS) software (written in C++) \citep{Plummer2003} .
We carry out our JAGS analysis within the R package ``rjags'' \citep{rjags} which provides a more familiar framework for an R user implementing the JAGS software.
The BUGS (Bayesian inference Using Gibbs Sampling) language \citep{BUGS} is used to describe models in JAGS.
The SHM, IHM and JHM have each been described with the BUGS language in Section~\ref{app:jags_code} of the Appendix.

For the following comparison we use a subset from the \emph{cdc13-1}~$\boldsymbol{{27}^{\circ}}$C~vs~\emph{ura3}$\Delta$~$\boldsymbol{{27}^{\circ}}$C suppressor/enhancer data set described in Section~\ref{sec:ura3_cdc13-1_27_27}.
A subset of 50 \emph{orf}$\Delta$s (for both the control and query) are chosen, each with 8 time-course repeats.
With a smaller data set we are able to collect large posterior sample sizes, sufficient to carry out a comparison between posterior samples.
Density plots are used to visually compare the similarity of the posterior samples from the C and JAGS code.
The Kolmogorov–Smirnov test \citep{hubergoodness} and unpaired two-sample Student's t-test \citep{degrootprobability} are used to test for significant difference between posterior samples from our C and JAGS code.

A comparison of posterior samples for our most sophisticated model, the JHM, is given below.
Posterior samples of 100k particles are obtained after a burn-in period of 1000k and a thinning of every 100 observations for both the C and JAGS code.
Computational time for the C and JAGS code is $\sim{30}$ hours and $\sim{400}$ hours respectively.
The minimum effective sample size per second (ESS\textsubscript{min}/sec) for the C and JAGS code is $\sim${1} and $\sim${0.1} respectively, demonstrating that the C code is $\sim{10}\times$ faster.

\begin{table}
\caption[Unpaired t-test and Kolmagorov-Smirnov p-values comparing posterior samples from the joint hierarchical model using both C programming language and Just Another Gibbs Sampler software]{Unpaired t-test and Kolmagorov-Smirnov p-values comparing posterior samples from the joint hierarchical model (JHM) using both C and Just Another Gibbs Sampler (JAGS) software. An extract of JHM parameters are given for both the C programming language and JAGS software. Posterior means are also included for both approaches. t-tests are carried out on the log posterior samples i.e. $\hat{K_p}$ in place of $e^{\hat{K_p}}$ to assume normality. \label{tab:jagscompare}}
\centering 
\resizebox{14cm}{!}{%
\npdecimalsign{.}
\nprounddigits{3}
  \begin{tabular}{c n{1}{3} n{1}{3} n{1}{3} n{1}{3}}
	\hline
\emph{Parameter}&\multicolumn{1}{c}{\emph{C Code posterior mean}}&\multicolumn{1}{c}{\emph{JAGS posterior mean}}&\multicolumn{1}{c}{\emph{t-test (with log posterior samples)}}&\multicolumn{1}{c}{\emph{Kolmagorov-Smirnov test}} \\ 
\hline
$e^{\hat{K_p}}$&0.1432044&0.1431957&0.4522&0.4005\\
$e^{\hat{r_p}}$&4.639228&4.640523&0.4236&0.4815\\
$e^{\hat{P}}$&2.53738e04&2.516994e04&0.1366&0.1162\\
$e^{\hat{\nu_p}}$&7.401522e04&7.416073e04&0.2497&0.1901\\
$e^{\hat{\alpha_c}}$&0.3038192&0.3037929&0.2034&0.1401\\
$e^{\hat{\beta_c}}$&0.384091&0.3841624&0.1563&0.1462\\
\hline
  		 \end{tabular}
			}
\end{table}

Figure~\ref{tab:jagscompareplot} gives density plots for an extract of JHM parameters for the C and JAGS software.
Visually there is no significant difference between the posterior sample density plots in Figure~\ref{tab:jagscompareplot}. Of the parameters shown, the weakest effective sample size ($\sim{80000}$ESS) is for the initial inoculum parameter $P$, but this is sufficiently large enough ESS to test if posterior samples show a significant difference.
Table~\ref{tab:jagscompare} demonstrates further that there is no significant difference found between the parameters shown.
The unpaired t-test for log posterior samples (for normality assumption) and Kolmogorov-Smirnov test p-values are all greater than 0.10 for the parameters given, including the inoculum density parameter $P$.
Overall we find no significant evidence against the C code and JAGS code sampling from the same posterior distributions.
\\
\\
As carrying out inference using C is $\sim${10} times faster than the JAGS equivalent code we prefer the C code for our Bayesian hierarchical models.
Obtaining sufficiently sized independent posterior samples of our posterior distributions for a larger data set of $\sim${4000} \emph{orf}$\Delta$s, we estimate our C code to be at least more than $\sim${50}$\times$ faster than the equivalent JAGS as we find the JAGS code to have exponential computational costs as we introduce larger data sets.
JAGS is very useful for model exploration as it is fast and simple to describe complex models.
The JAGS software is so prohibitively slow for the JHM, that an experimenter is likely to not carry out such inference and use a more simple or faster method, justifying the use of the C programming language to carry out inference.
Further improvements such as the introduction of parallelisation may lead to more favourable computational times in the future.
\begin{figure}
  \centering
\includegraphics[width=9.4cm]{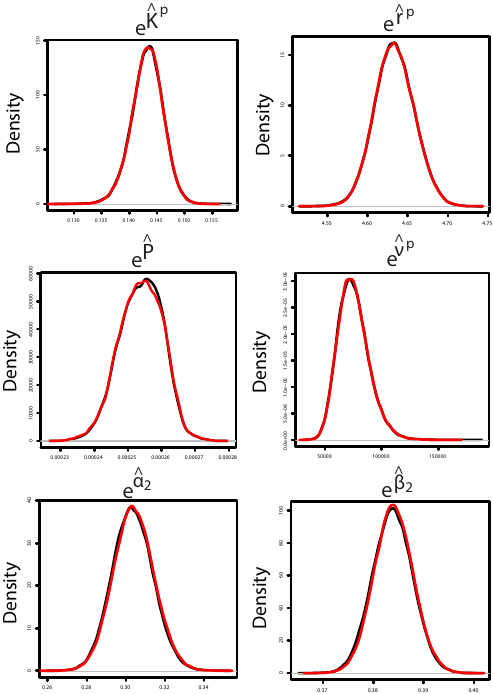}
\caption[Density plots for posterior samples from the joint hierarchical model using the C programming language and Just Another Gibbs Sampler software]{Density plots for posterior samples from the joint hierarchical model (JHM) using the C programming language (red) and Just Another Gibbs Sampler (black) software. Density plots for the JHM parameter posteriors (sample size = 100000, thinning interval = 100 and burn-in = 1000000). 
\label{tab:jagscompareplot}
}
\end{figure}
\clearpage
\section{\label{sec:fur_case_stu}Further case studies}
In this section we briefly introduce different data sets that may be considered for further investigation and research.
We can also see how the JHM performs for different experimental conditions by applying the JHM to different QFA screen comparisons, see $MDR{\times}MDP$ fitness plots in Figures~\ref{JHM_CDC13-1_CDC13-1EXO1_27_27}-\ref{JHM_URA_URA_20_37_reversefix}.
The data sets used in Figures~\ref{JHM_CDC13-1_CDC13-1EXO1_27_27}-\ref{JHM_URA_URA_20_37_reversefix} are currently unpublished from the Lydall lab.
For each of the data sets, the JHM in Table~\ref{tab:JHM} is applied with the prior hyper-parameters in Table~\ref{tab:SHM_priors}.
Posterior samples of 1000 particles are obtained after a burn-in period of 800k, and a thinning of every 100 observations.
Similarly to Section~\ref{conv_diag}, chains from our MCMC sampler are accepted where the effective sample sizes are greater than $300$ and Heidelberg and Welch P-values are greater than $0.10$ for every chain.
As in the \cite{QFA1} analysis, each experiment has a list of 159 genes stripped from our final list of genes for biological and experimental reasons. 
Results for the \emph{cdc13-1}\emph{exo1}$\Delta$~$\boldsymbol{{27}^{\circ}}$C vs \emph{cdc13-1}~$\boldsymbol{{27}^{\circ}}$C and \emph{cdc13-1}\emph{rad9}$\Delta$~$\boldsymbol{{27}^{\circ}}$C vs \emph{cdc13-1}~$\boldsymbol{{27}^{\circ}}$C experiments have further genes removed for biological and experimental reasons, 23 and 13 genes respectively (a total of 182 and 172 genes respectively).

Figure~\ref{JHM_CDC13-1_CDC13-1EXO1_27_27} is a \emph{cdc13-1}\emph{exo1}$\Delta$~$\boldsymbol{{27}^{\circ}}$C~vs~\emph{cdc13-1}~$\boldsymbol{{27}^{\circ}}$C suppressor/enhancer analysis for finding genes that interact with $\emph{exo1}$ in a telomere maintenance defective background (\emph{cdc13-1} at $\boldsymbol{{27}^{\circ}}$C).
Similarly, Figure~\ref{JHM_CDC13-1RAD_CDC13-1_27_27} is a \emph{cdc13-1}\emph{rad9}$\Delta$~$\boldsymbol{{27}^{\circ}}$C~vs~\emph{cdc13-1}~$\boldsymbol{{27}^{\circ}}$C suppressor/enhancer analysis for finding genes that interact with $\emph{rad9}$ in a telomere maintenance defective background.
Figure~\ref{JHM_URA_YKU70_37_37} is a \emph{yku70}$\Delta$~$\boldsymbol{{37}^{\circ}}$C~vs~\emph{ura3}$\Delta$~$\boldsymbol{{37}^{\circ}}$C suppressor/enhancer analysis for finding genes that interact with \emph{yku70} at high temperature. 
Figure~\ref{JHM_URA_URA_20_37_reversefix} is an example of a temperature sensitivity experiment, for finding genes that interact with the high temperature of $\boldsymbol{{37}^{\circ}}$C.
Figures~\ref{JHM_CDC13-1_CDC13-1EXO1_27_27}-\ref{JHM_URA_URA_20_37_reversefix} demonstrate that the JHM can capture different linear relationships that are above or below the 1-1 line.
Curvature of the data in Figures~\ref{JHM_CDC13-1_CDC13-1EXO1_27_27}-\ref{JHM_URA_URA_20_37_reversefix} suggests that the linear relationships modelled by the JHM may be improved through linearising transformations of the data. Extending the JHM to account for the curvature in the data may improve our model fit and allow to better determine genes which significantly interact.

Table~\ref{tab:JHM_hits} compares the number of suppressors and enhancers estimated for each of the experiments considered. 
The experiments in Table~\ref{tab:JHM_hits} have similar numbers of genetic interactions, ranging from 358 to 511, but much lower than the \emph{cdc13-1}$\boldsymbol{{27}^{\circ}}$C vs \emph{ura3}$\Delta$~$\boldsymbol{{27}^{\circ}}$C experiment which has $939$.
The experiments introduced in this section also differ from the \emph{cdc13-1}~$\boldsymbol{{27}^{\circ}}$C vs \emph{ura3}$\Delta$~$\boldsymbol{{27}^{\circ}}$C experiment as they have more enhancers than suppressors, further demonstrating the JHM's ability to model different experimental situations and the non-restrictive choice of priors (Table~\ref{tab:SHM_priors}). 

\begin{table}[h!]
\caption[Number of interactions identified for further case studies and applications of the joint hierarchical model extensions]{Number of joint hierarchical model (JHM) interactions for QFA datasets given in Section~\ref{sec:fur_case_stu}. Interactions for each dataset is split into suppressors and enhancers. \label{tab:JHM_hits} The number of interactions found with the extensions to the joint hierarchical model (see Section~\ref{sec:batch_eff}) are also given. Each QFA screen comparison consists of 4294 \emph{orf}$\Delta$s. Results for all experiments have a list of 159 genes removed from the final list of interactions for biological and experimental reasons.
Results for the \emph{cdc13-1}\emph{exo1}$\Delta$~$\boldsymbol{{27}^{\circ}}$C vs \emph{cdc13-1}~$\boldsymbol{{27}^{\circ}}$C and \emph{cdc13-1}\emph{rad9}$\Delta$~$\boldsymbol{{27}^{\circ}}$C vs \emph{cdc13-1}~$\boldsymbol{{27}^{\circ}}$C experiments have further genes removed for biological and experimental reasons, 23 and 13 genes respectively (a total of 182 and 172 genes respectively).}
\centering
\resizebox{\columnwidth}{!}{%
\npdecimalsign{.}
\nprounddigits{2}
\begin{tabular}{c c c c c c}
\hline
\emph{Query screen} & \emph{Control screen} & \emph{Interactions}  & \emph{Suppressors} & \emph{Enhancers} \\ 
\hline
\noalign{\vskip 0.2mm} 
\emph{cdc13-1}\emph{exo1}$\Delta$~$\boldsymbol{{27}^{\circ}}$C&\emph{cdc13-1}~$\boldsymbol{{27}^{\circ}}$C
  & 388 & 81 & 307 \\ 
\emph{cdc13-1}\emph{rad9}$\Delta$~$\boldsymbol{{27}^{\circ}}$C&\emph{cdc13-1}~$\boldsymbol{{27}^{\circ}}$C
 & 358 & 73 & 285\\ 
\emph{yku70}$\Delta$~$\boldsymbol{{37}^{\circ}}$C&\emph{ura3}$\Delta$~$\boldsymbol{{37}^{\circ}}$C  & 511 & 104 & 407\\ 
\emph{ura3}$\Delta$~$\boldsymbol{{37}^{\circ}}$C&\emph{ura3}$\Delta$~$\boldsymbol{{20}^{\circ}}$C
 & 460 & 138 & 322 \\ 
\\
\hline
\\
\multicolumn{2}{c}{Model for \emph{cdc13-1}~$\boldsymbol{{27}^{\circ}}$C~vs} & \emph{Interactions}  & \emph{Suppressors} & \emph{Enhancers} \\ 
\multicolumn{2}{c}{\emph{ura3}$\Delta$~$\boldsymbol{{27}^{\circ}}$C experiment}&&&&\\
\hline
\noalign{\vskip 0.2mm} 
\multicolumn{2}{c}{JHM}  & 939 & 665 & 274 \\ 
\multicolumn{2}{c}{JHM-Batch}  & 553 & 378 & 174 \\ 
\multicolumn{2}{c}{JHM-Transformation}  & 901 & 658 & 243\\
\hline
\end{tabular}
\npnoround%
}
\end{table}

Table~\ref{tab:overlap_further}A shows the overlap in genes with significant evidence of genetic interactions between the different QFA comparisons considered.
The largest number of overlapping genetic interactions are found with the \mbox{\emph{cdc13-1}}$\Delta$~$\boldsymbol{{27}^{\circ}}$C vs \emph{ura}$\Delta$~$\boldsymbol{{27}^{\circ}}$C experiment, overlapping with  301 and 263 genes from the \mbox{\emph{cdc13-1}}\emph{exo1}$\Delta$~$\boldsymbol{{27}^{\circ}}$C vs \mbox{\emph{cdc13-1}}~$\boldsymbol{{27}^{\circ}}$C and \mbox{\emph{cdc13-1}}\emph{rad9}$\Delta$~$\boldsymbol{{27}^{\circ}}$C vs \mbox{\emph{cdc13-1}}~$\boldsymbol{{27}^{\circ}}$C experiment respectively.
The \mbox{\emph{cdc13-1}}$\Delta$~$\boldsymbol{{27}^{\circ}}$C vs \emph{ura}$\Delta$~$\boldsymbol{{27}^{\circ}}$C, \mbox{\emph{cdc13-1}}\emph{exo1}$\Delta$~$\boldsymbol{{27}^{\circ}}$C vs \mbox{\emph{cdc13-1}}~$\boldsymbol{{27}^{\circ}}$C and \mbox{\emph{cdc13-1}}\emph{rad9}$\Delta$~$\boldsymbol{{27}^{\circ}}$C vs \mbox{\emph{cdc13-1}} $\boldsymbol{{27}^{\circ}}$C experiments are expected to overlap most as they are designed to find genes interacting in a \mbox{\emph{cdc13-1}} background.
The smallest number of overlapping genetic interactions are found with the \emph{ura3}$\Delta$~$\boldsymbol{{37}^{\circ}}$C vs \emph{ura3}$\Delta$~$\boldsymbol{{20}^{\circ}}$C and \emph{yku70}$\Delta$~$\boldsymbol{{37}^{\circ}}$C vs \emph{ura3}$\Delta$~$\boldsymbol{{37}^{\circ}}$C experiment.
The \emph{ura3}$\Delta$~$\boldsymbol{{37}^{\circ}}$C vs \emph{ura3}$\Delta$~$\boldsymbol{{20}^{\circ}}$C and \emph{yku70}$\Delta$~$\boldsymbol{{37}^{\circ}}$C vs \emph{ura3}$\Delta$~$\boldsymbol{{37}^{\circ}}$C experiments are expected to have the least overlap as they are not designed to find genes interacting in a \mbox{\emph{cdc13-1}} background.
The \emph{yku70}$\Delta$~$\boldsymbol{{37}^{\circ}}$C vs \emph{ura3}$\Delta$~$\boldsymbol{{37}^{\circ}}$C experiment is designed to look at telomeres, but instead of disrupting the telomere capping protein Cdc13 using \mbox{\emph{cdc13-1}}, a \emph{yku70}$\Delta$ mutation is made such that the protein Yku70 (a telomere binding protein which guides the enzyme telomerase to the telomere \citep{QFA1}) is no longer produced by the cell.
Further \emph{ura3}$\Delta$~$\boldsymbol{{37}^{\circ}}$C vs \emph{ura3}$\Delta$~$\boldsymbol{{20}^{\circ}}$C is designed to investigate temperature sensitivity only.

Table~\ref{tab:overlap_further}B shows the overlap in significant GO terms between the different QFA comparisons considered.
The largest number of overlapping significant GO terms are found with the \mbox{\emph{cdc13-1}}$\Delta$~$\boldsymbol{{27}^{\circ}}$C experiment, overlapping with $\sim${150} GO terms for each experiment.
The smallest overlap with \mbox{\emph{cdc13-1}}$\Delta$~$\boldsymbol{{27}^{\circ}}$C vs \emph{ura}$\Delta$~$\boldsymbol{{27}^{\circ}}$C experiment is 110 GO terms with the \emph{ura3}$\Delta$~$\boldsymbol{{37}^{\circ}}$C vs \emph{ura3}$\Delta$~$\boldsymbol{{20}^{\circ}}$C experiment.
The smallest number of overlapping genetic interactions are for the \emph{ura3}$\Delta$~$\boldsymbol{{37}^{\circ}}$C vs \emph{ura3}$\Delta$~$\boldsymbol{{20}^{\circ}}$C experiment, followed by \emph{yku70}$\Delta$~$\boldsymbol{{37}^{\circ}}$C vs \emph{ura3}$\Delta$~$\boldsymbol{{37}^{\circ}}$C, with $\sim${110} and $\sim${120} GO terms overlapping with the other experiments respectively.
Similarly to the overlap of genes with significant evidence of genetic interaction, the overlap of significant GO terms shows that our \mbox{\emph{cdc13-1}} background experiments share the most GO terms and that the temperature sensitivity experiment \emph{ura3}$\Delta$~$\boldsymbol{{37}^{\circ}}$C vs \emph{ura3}$\Delta$~$\boldsymbol{{20}^{\circ}}$C has the least overlap.

We have shown that the JHM can successfully model different experimental data sets, Figures~\ref{JHM_CDC13-1_CDC13-1EXO1_27_27}-\ref{JHM_URA_URA_20_37_reversefix} are included as a reference for further research.
Of the different experiments we can see that \mbox{\emph{cdc13-1}}~$\boldsymbol{{27}^{\circ}}$C vs \emph{ura3}$\Delta$~$\boldsymbol{{27}^{\circ}}$C is the most dissimilar to the other experiments due to the large number of genetic interactions, 939 in total (see Table~\ref{tab:JHM_hits}).
The next largest number of genetic interactions is 511 with the \emph{yku70}$\Delta$~$\boldsymbol{{37}^{\circ}}$C vs\ emph{ura3}$\Delta$~$\boldsymbol{{37}^{\circ}}$C experiment, which is approximately half the genes found for the \mbox{\emph{cdc13-1}}~$\boldsymbol{{27}^{\circ}}$C vs \emph{ura3}$\Delta$~$\boldsymbol{{27}^{\circ}}$C experiment.
Tables~\ref{tab:overlap_further}A and \ref{tab:overlap_further}B show that the overlap between QFA comparisons is as expected using the JHM, with the closer related experiments sharing the most overlap.
To account for the curvature of the data observed in Figures~\ref{JHM_CDC13-1_CDC13-1EXO1_27_27}-\ref{JHM_URA_URA_20_37_reversefix} we introduce a JHM with linearising transformations in the next section.
Further research may include developing models that can incorporate multiple QFA comparisons to find evidence of genetic interactions between query screens and incorporate more information within our models.


\begin{table}
\caption[Overlap between different QFA comparisons for genes interacting and gene ontology terms over-represented in lists of interactions]{\label{tab:overlap_further}Overlap between different QFA comparisons for genes interacting and gene ontology terms over-represented in lists of interactions.
For a fair comparison, any genes removed from the results of a QFA comparison for biological and experimental reasons are removed for all experiments,
therefore results for all experiments have a list of 195 genes (159+23+13, see Table~\ref{tab:JHM_hits}) removed from the final list of interactions for biological and experimental reasons.
A) Number of genes identified for each QFA comparison and the overlap between QFA comparisons. 4099 genes from the \emph{S. cerevisiae} single deletion library are considered.
B) Number of GO terms identified for each approach and the overlap between QFA comparisons. 6094 \emph{S. cerevisiae} GO Terms available.}
\centering
\resizebox{\columnwidth}{!}{%
\begin{tabular}{*{6}{c}}
\multicolumn{1}{l}{\bf{A.}}&\emph{cdc13-1}$\Delta$~$\boldsymbol{{27}^{\circ}}$C &\emph{cdc13-1}\emph{exo1}$\Delta$~$\boldsymbol{{27}^{\circ}}$C&\emph{cdc13-1}\emph{rad9}$\Delta$~$\boldsymbol{{27}^{\circ}}$C&\emph{yku70}$\Delta$~$\boldsymbol{{37}^{\circ}}$C&\emph{ura3}$\Delta$~$\boldsymbol{{37}^{\circ}}$C\\
&vs \emph{ura}$\Delta$~$\boldsymbol{{27}^{\circ}}$C&vs \emph{cdc13-1}~$\boldsymbol{{27}^{\circ}}$C&vs \emph{cdc13-1}~$\boldsymbol{{27}^{\circ}}$C&vs \emph{ura3}$\Delta$~$\boldsymbol{{37}^{\circ}}$C &vs \emph{ura3}$\Delta$~$\boldsymbol{{20}^{\circ}}$C\\\hline
\emph{cdc13-1}$\Delta$~$\boldsymbol{{27}^{\circ}}$C vs \emph{ura}$\Delta$~$\boldsymbol{{27}^{\circ}}$C&926&N/A&N/A&N/A&N/A\\
\emph{cdc13-1}\emph{exo1}$\Delta$~$\boldsymbol{{27}^{\circ}}$C vs \emph{cdc13-1}~$\boldsymbol{{27}^{\circ}}$C&301&386&N/A&N/A&N/A\\
\emph{cdc13-1}\emph{rad9}$\Delta$~$\boldsymbol{{27}^{\circ}}$C vs \emph{cdc13-1}~$\boldsymbol{{27}^{\circ}}$C&263&245&355&N/A&N/A\\
\emph{yku70}$\Delta$~$\boldsymbol{{37}^{\circ}}$C vs \emph{ura3}$\Delta$~$\boldsymbol{{37}^{\circ}}$C&252&155&146&506&N/A\\
\emph{ura3}$\Delta$~$\boldsymbol{{37}^{\circ}}$C vs \emph{ura3}$\Delta$~$\boldsymbol{{20}^{\circ}}$C&223&152&149&164&455\\\hline
\end{tabular}
}
\\ \qquad
\\ \qquad
\\
\resizebox{\columnwidth}{!}{%
\begin{tabular}{*{6}{c}}
\multicolumn{1}{l}{\bf{B.}}&\emph{cdc13-1}$\Delta$~$\boldsymbol{{27}^{\circ}}$C &\emph{cdc13-1}\emph{exo1}$\Delta$~$\boldsymbol{{27}^{\circ}}$C&\emph{cdc13-1}\emph{rad9}$\Delta$~$\boldsymbol{{27}^{\circ}}$C&\emph{yku70}$\Delta$~$\boldsymbol{{37}^{\circ}}$C&\emph{ura3}$\Delta$~$\boldsymbol{{37}^{\circ}}$C\\
&vs \emph{ura}$\Delta$~$\boldsymbol{{27}^{\circ}}$C&vs \emph{cdc13-1}~$\boldsymbol{{27}^{\circ}}$C&vs \emph{cdc13-1}~$\boldsymbol{{27}^{\circ}}$C&vs \emph{ura3}$\Delta$~$\boldsymbol{{37}^{\circ}}$C &vs \emph{ura3}$\Delta$~$\boldsymbol{{20}^{\circ}}$C\\\hline
\emph{cdc13-1}$\Delta$~$\boldsymbol{{27}^{\circ}}$C vs \emph{ura}$\Delta$~$\boldsymbol{{27}^{\circ}}$C&282&N/A&N/A&N/A&N/A\\
\emph{cdc13-1}\emph{exo1}$\Delta$~$\boldsymbol{{27}^{\circ}}$C vs \emph{cdc13-1}~$\boldsymbol{{27}^{\circ}}$C&142&188&N/A&N/A&N/A\\
\emph{cdc13-1}\emph{rad9}$\Delta$~$\boldsymbol{{27}^{\circ}}$C vs \emph{cdc13-1}~$\boldsymbol{{27}^{\circ}}$C&151&130&212&N/A&N/A\\
\emph{yku70}$\Delta$~$\boldsymbol{{37}^{\circ}}$C vs \emph{ura3}$\Delta$~$\boldsymbol{{37}^{\circ}}$C&150&119&125&245&N/A\\
\emph{ura3}$\Delta$~$\boldsymbol{{37}^{\circ}}$C vs \emph{ura3}$\Delta$~$\boldsymbol{{20}^{\circ}}$C&110&100&112&119&195\\\hline
\end{tabular}
}
\end{table}

\begin{figure}[h!]
  \centering
\includegraphics[width=14cm]{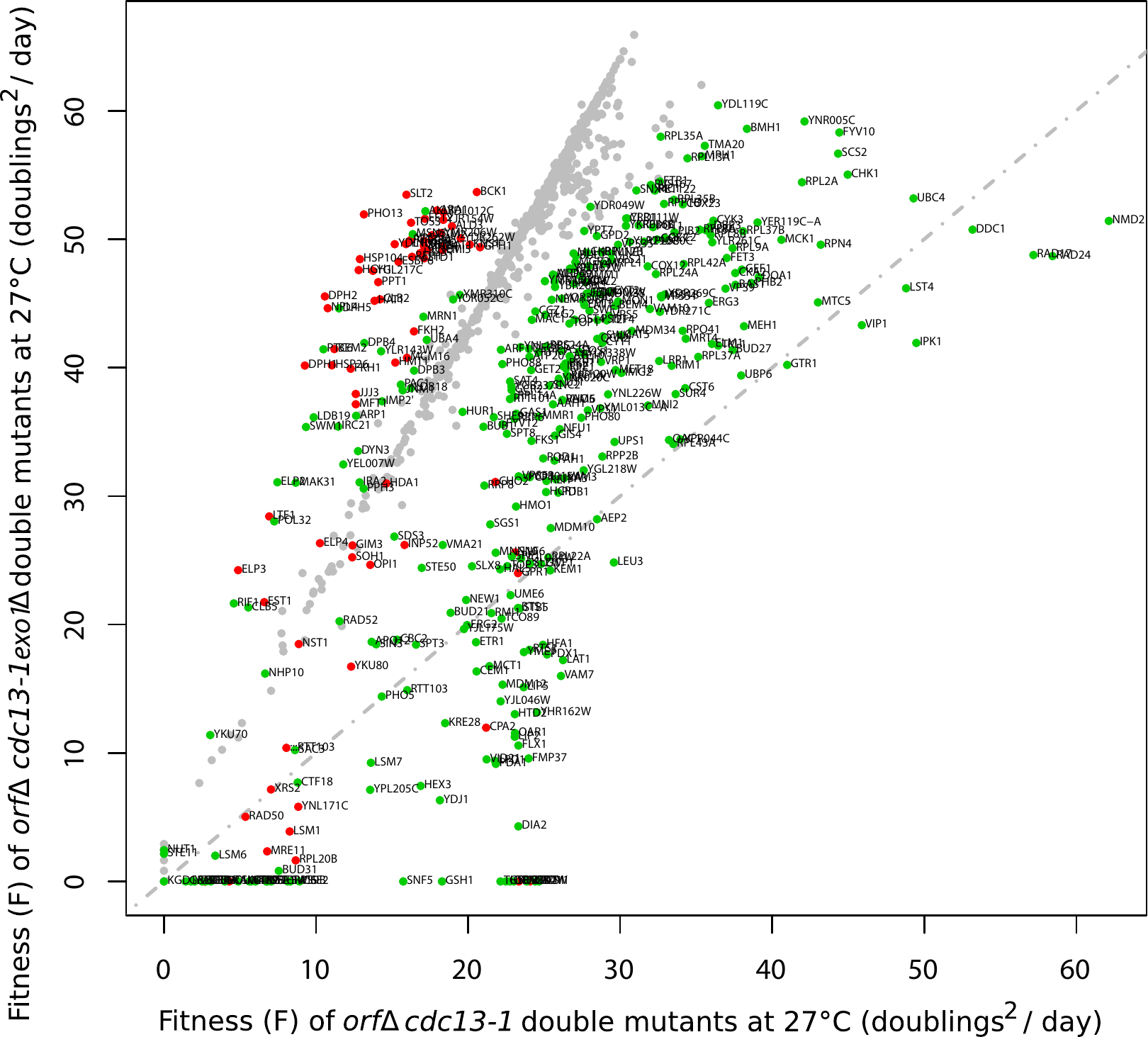}
\caption[\emph{cdc13-1}\emph{exo1}$\Delta$~$\boldsymbol{{27}^{\circ}}$C~vs~\emph{cdc13-1}~$\boldsymbol{{27}^{\circ}}$C joint hierarchical model fitness plot]{\emph{cdc13-1}\emph{exo1}$\Delta$~$\boldsymbol{{27}^{\circ}}$C~vs~\emph{cdc13-1}~$\boldsymbol{{27}^{\circ}}$C joint hierarchical model (JHM) fitness plot with $\emph{orf}\Delta$ posterior mean fitnesses.
The JHM does not does not make use of a fitness measure such as $MDR\times{MDP}$ but the fitness plot is given in terms of $MDR\times{MDP}$ for comparison with other approaches which do. 
$\emph{orf}\Delta$ strains are classified as being a suppressor or enhancer based on one of the two parameters used to classify genetic interaction, growth parameter $r$, this means occasionally strains can be more fit in the query experiment in terms of $MDR\times MDP$ but be classified as enhancers (green).
Further fitness plot explanation and notation is given in Figure~\ref{fig:JHM}.\label{JHM_CDC13-1_CDC13-1EXO1_27_27}
}
\end{figure}
\clearpage


\begin{figure}[h!]
  \centering
\includegraphics[width=14cm]{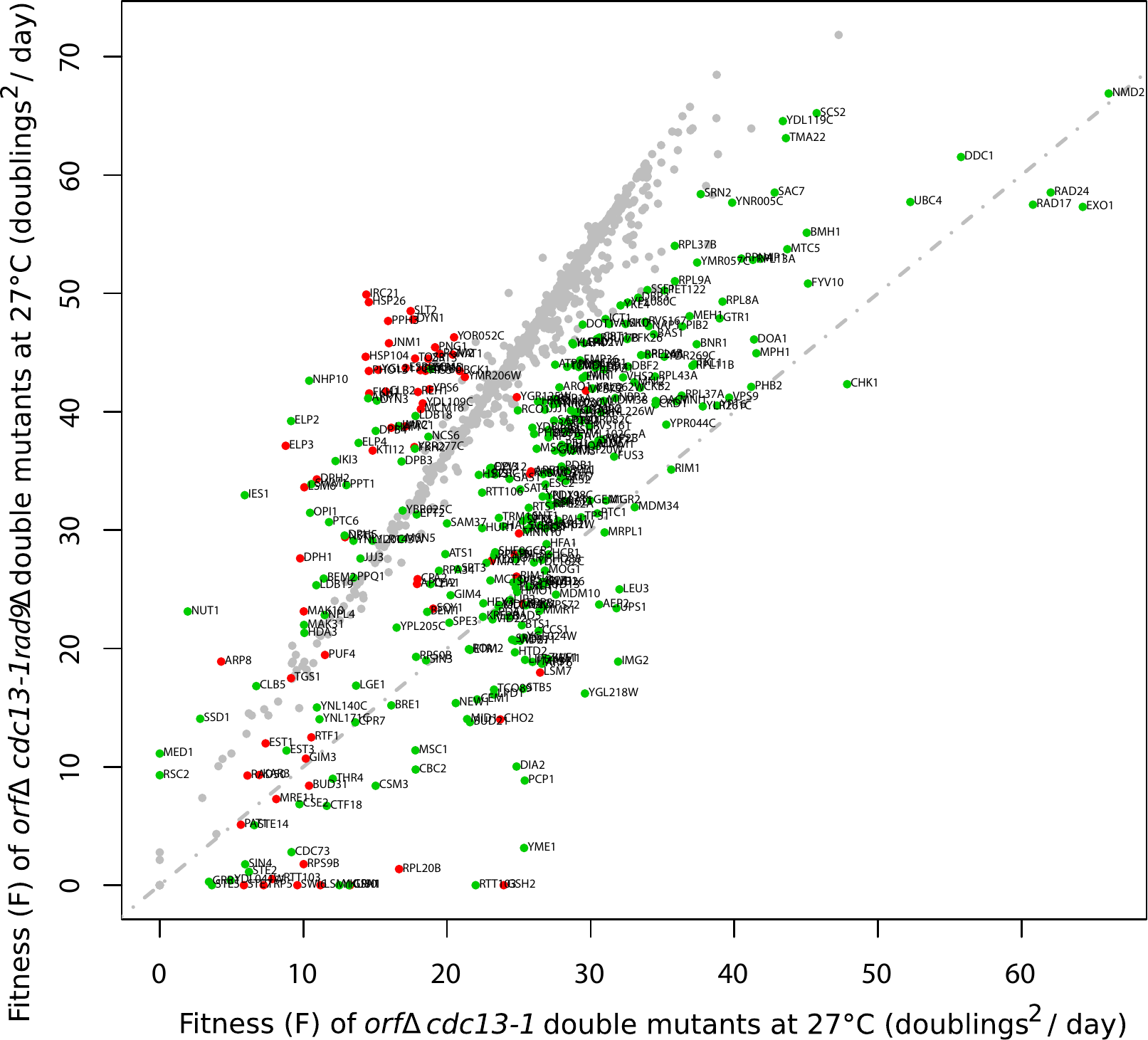}
\caption[\emph{cdc13-1}\emph{rad9}$\Delta$~$\boldsymbol{{27}^{\circ}}$C~vs~\emph{cdc13-1}~$\boldsymbol{{27}^{\circ}}$C joint hierarchical model fitness plot]{\emph{cdc13-1}\emph{rad9}$\Delta$~$\boldsymbol{{27}^{\circ}}$C~vs~\emph{cdc13-1}~$\boldsymbol{{27}^{\circ}}$C joint hierarchical model (JHM) fitness plot with $\emph{orf}\Delta$ posterior mean fitnesses.
The JHM does not does not make use of a fitness measure such as $MDR\times{MDP}$ but the fitness plot is given in terms of $MDR\times{MDP}$ for comparison with other approaches which do. 
$\emph{orf}\Delta$ strains are classified as being a suppressor or enhancer based on one of the two parameters used to classify genetic interaction, growth parameter $r$, this means occasionally strains can be more fit in the query experiment in terms of $MDR\times MDP$ but be classified as enhancers (green).
Further fitness plot explanation and notation is given in Figure~\ref{fig:JHM}.\label{JHM_CDC13-1RAD_CDC13-1_27_27}
}
\end{figure}
\clearpage


\begin{figure}[h!]
  \centering
\includegraphics[width=14cm]{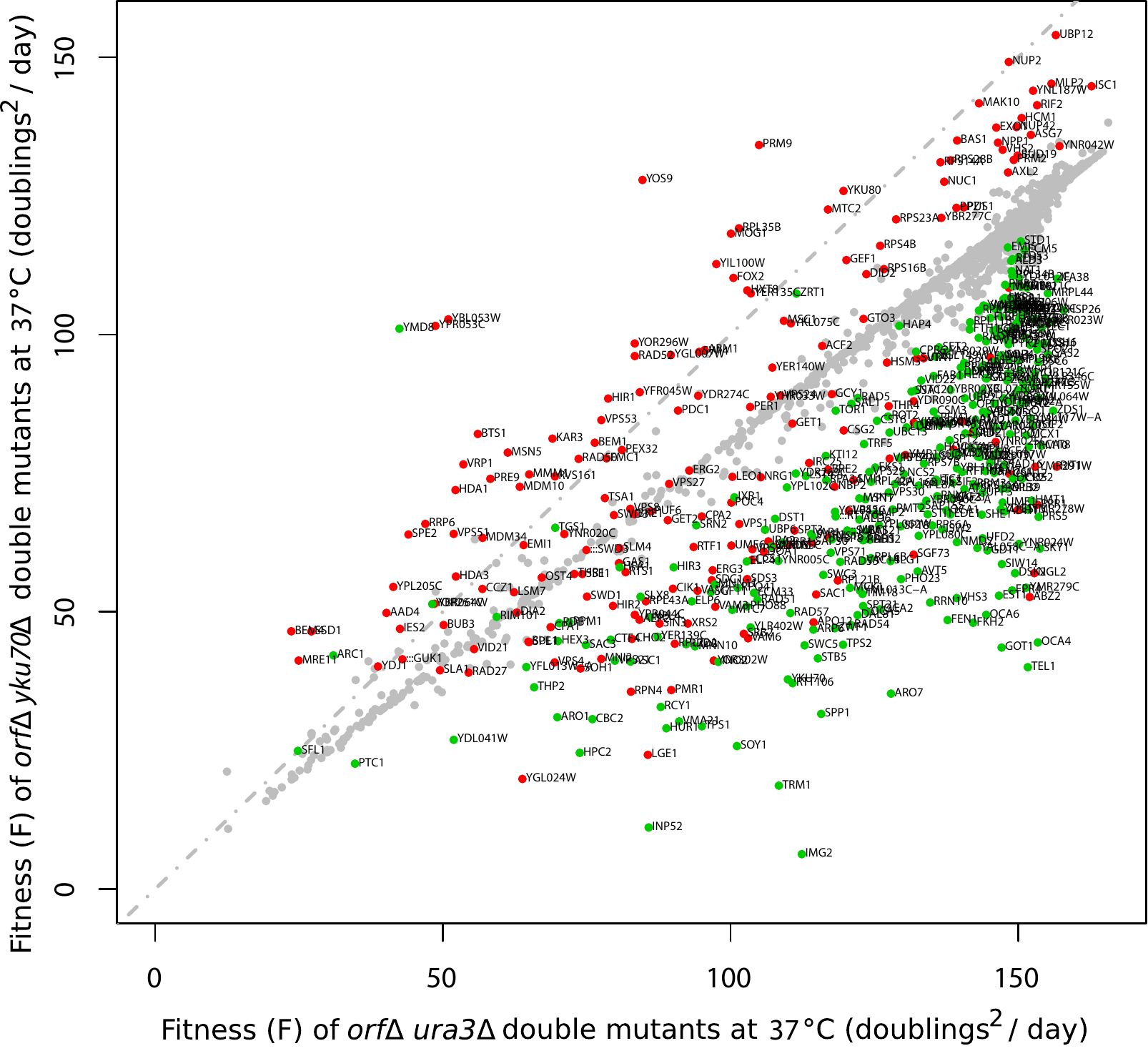}
\caption[\emph{yku70}$\Delta$~$\boldsymbol{{37}^{\circ}}$C~vs~\emph{ura3}$\Delta$~$\boldsymbol{{37}^{\circ}}$C joint hierarchical model fitness plot]{\emph{yku70}$\Delta$~$\boldsymbol{{37}^{\circ}}$C~vs~\emph{ura3}$\Delta$~$\boldsymbol{{37}^{\circ}}$C joint hierarchical model (JHM) fitness plot with $\emph{orf}\Delta$ posterior mean fitnesses.
The JHM does not does not make use of a fitness measure such as $MDR\times{MDP}$ but the fitness plot is given in terms of $MDR\times{MDP}$ for comparison with other approaches which do. 
$\emph{orf}\Delta$ strains are classified as being a suppressor or enhancer based on one of the two parameters used to classify genetic interaction, growth parameter $r$, this means occasionally strains can be more fit in the query experiment in terms of $MDR\times MDP$ but be classified as enhancers (green).
Further fitness plot explanation and notation is given in Figure~\ref{fig:JHM}.\label{JHM_URA_YKU70_37_37}
}
\end{figure}
\clearpage

\begin{figure}[h!]
  \centering
\includegraphics[width=14cm]{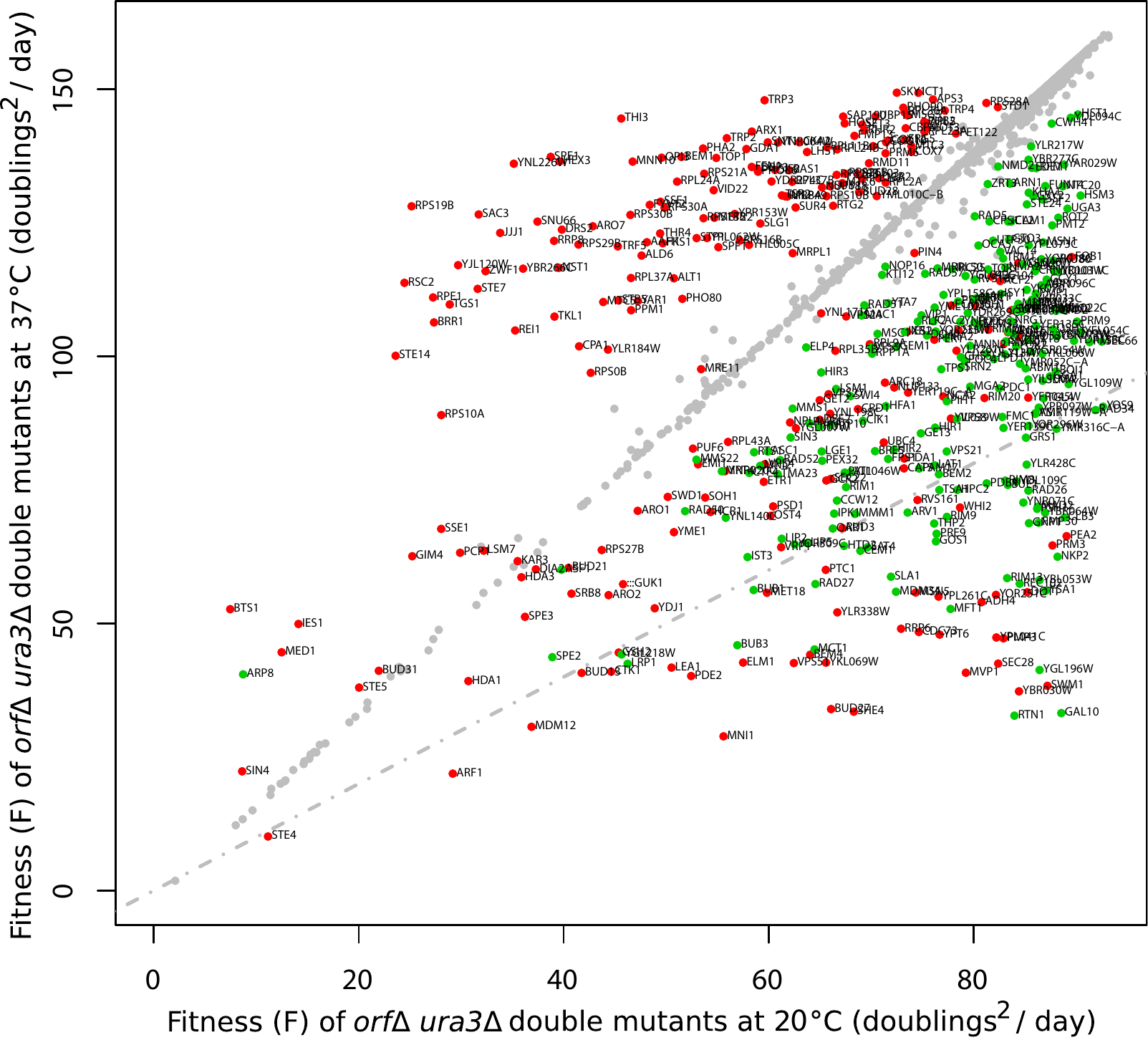}
\caption[\emph{ura3}$\Delta$~$\boldsymbol{{37}^{\circ}}$C~vs~\emph{ura3}$\Delta$~$\boldsymbol{{20}^{\circ}}$C joint hierarchical model fitness plot]{\emph{ura3}$\Delta$~$\boldsymbol{{37}^{\circ}}$C~vs~\emph{ura3}$\Delta$~$\boldsymbol{{20}^{\circ}}$C joint hierarchical model (JHM) fitness plot with $\emph{orf}\Delta$ posterior mean fitnesses.
The JHM does not does not make use of a fitness measure such as $MDR\times{MDP}$ but the fitness plot is given in terms of $MDR\times{MDP}$ for comparison with other approaches which do. 
$\emph{orf}\Delta$ strains are classified as being a suppressor or enhancer based on one of the two parameters used to classify genetic interaction, growth parameter $r$, this means occasionally strains can be more fit in the query experiment in terms of $MDR\times MDP$ but be classified as enhancers (green).
Further fitness plot explanation and notation is given in Figure~\ref{fig:JHM}.\label{JHM_URA_URA_20_37_reversefix}
}
\end{figure}
\clearpage

\section{\label{sec:batch_eff}Extensions of the joint hierarchical model}
In this section we briefly introduce two new extensions of the JHM for further investigation and research.
An extension to the JHM, given in Table~\ref{tab:JHM}, is to consider a batch effect.
Batch effects are technical sources of variation from the handling of experimental cultures \citep{batch1,batch2}. 
Batch effects can be confounded with the biology of interest, leading to misleading results and conclusions.

A QFA screen comparison is carried out between two QFA screens.
Each QFA screen consists of multiple 384 plates grown over time (see Figure~\ref{fig:int:spot}), typically with each \emph{orf}$\Delta$ repeat on a different 384 plate.
For the \emph{cdc13-1}~$\boldsymbol{{27}^{\circ}}$C~vs~\emph{ura3}$\Delta$~$\boldsymbol{{27}^{\circ}}$C experiment, each QFA screen is built of 120 384 spot plates (240 total unique plates).
Each 384 plate is created sequentially and may be created by a different experimenter.
The 384 plates may therefore differ due to factors that the experimenters do their best to control such as the amount of nutrition in a plate, temperature, or other environmental effects. 
Where \emph{orf}$\Delta$ repeats are carried out across multiple plates, differences in plates can therefore be captured by introducing a batch effect into the model. 

Through careful planning and improved experimental design, batch effects can be reduced or removed.
When we are unable to improve our experimental design any further we may be interested in accounting for a batch effect within our model.
Introducing parameters to model batch effects in our experiment we can account for any differences between the 240 384 spot plates.
A JHM with batch effects (JHM-B), described in Table~\ref{tab:BATCH}, will be able to improve inference by including more of the experimental structure.
The model in Table~\ref{tab:BATCH} introduces a batch effect $\kappa_b$ and $\lambda_b$, for a plate $b$, to capture any batch effect in carrying capacity $K$ and growth rate $r$ respectively.
A batch effect will be estimated within the model and consequently any confounding with \emph{orf}$\Delta$ level carrying capacity $K$ and growth rate $r$ parameters will be removed.
Using frequentist estimates of the batch effects in the QFA screens, a normal prior was chosen to describe batch effect parameters, allowing either a positive or negative effect to be incorporated for each \emph{orf}$\Delta$ repeat in terms of $K$ and $r$.
\\
\\
Another extension of the JHM is to consider a transformation to linearise the relationship describing genetic independence in the JHM.
When carrying out linear regression we may be interested in linearising the data to improve the linear relationship \citep{trans1}.
There are many different transformations used for linearising data, the most common are log and power transformations. 
Power transformations are families of power functions that are typically used to stabilise variance and make our data more Normal distribution-like.
For a variable $x$, a power function is of the form $f:x\mapsto cx^r$, for $c,r\in\mathbb{R}$, where $c$ and $r$ are constant real numbers.
The Box-Cox transformation \citep{trans2} is a particular case of power transformation that is typically used to transform data and linearise a relationship within a data set.

Without linearising our data, we may not be describing genetic independence within our model correctly, leading to misleading results and conclusions.
A JHM with transformations (JHM-T), described in Table~\ref{tab:TRANS}, will be able to improve inference by ensuring a more linear relationship is made between the control and query screen.
Genetic independence within the JHM is described as a linear relationship (see Sections~\ref{int:defining_epi}~and~\ref{joi:JHM}) for both carrying capacity $K$ and growth rate $r$. 
We may not believe there to be a perfectly linear relationship between the control and query for both $K$ and $r$.
Introducing a power transformation for the model of genetic independence in terms of $K$ and $r$ can allow us to linearise the relationship and better model genetic independence.
The model in Table~\ref{tab:TRANS} introduces the transformation parameters $\phi$ and $\chi$ at an \emph{orf}$\Delta$ level for both the carrying capacity $K$ and growth rate $r$ respectively, where $\phi>0$ and $\chi>0$. 
The ``vanilla'' JHM assumes an additive model of epistasis with $(\alpha_{c}+K^{o}_{l}+\delta_{l}\gamma_{cl},\beta_{c}+r^{o}_{l}+\delta_{l}\omega_{cl})$, where $\alpha_{c}$ and $\beta_{c}$ are the scale parameters, as we are considering log \emph{orf}$\Delta$ parameters. 
The ``vanilla'' JHM effectively assuming a multiplicative model on the original scale of the data i.e. $(e^{\alpha_{c}}e^{K^{o}_{l}+\delta_{l}\gamma_{cl}},e^{\beta_{c}}e^{r^{o}_{l}+\delta_{l}\omega_{cl}})$.
By introducing new parameters $\phi$ and $\chi$ to scale the control and query data $\left(\frac{\alpha_{c}+K^{o}_{l}+\delta_{l}\gamma_{cl}}{\phi},\frac{\beta_{c}+r^{o}_{l}+\delta_{l}\omega_{cl}}{\chi}\right)$ we can expect to have a power transformation with the control and query on the original scale of the data $\left[\left(e^{\alpha_{c}}e^{K^{o}_{l}+\delta_{l}\gamma_{cl}}\right)^{\frac{1}{\phi}},\left(e^{\beta_{c}}e^{r^{o}_{l}+\delta_{l}\omega_{cl}}\right)^{\frac{1}{\chi}}\right]$.
The transformation parameters give the same transformation to both the control and query screens.
Our model will learn about $\phi$ and $\chi$, adjusting the relationship of genetic independence and consequently those identified as genetic interaction.
Choosing to include a multiplicative transformation parameter where the model describes genetic independence (as an additive model) will give the model the flexibility to adjust the linear relationship between the control and query screens. 
Prior hyper-parameter choice for the transformation effect must be strictly positive and centred at $1$ (no transformation effect) and so a gamma distribution with a mean of $1$ is chosen for both $\chi$ and $\phi$. 
\\
\\
Figures~\ref{fig:BATCH}~and~\ref{fig:TRANS}~show JHM-B and JHM-T $MDR{\times}MDP$ fitness plots respectively, for the \emph{cdc13-1}~$\boldsymbol{{27}^{\circ}}$C~vs~\emph{ura3}$\Delta$~$\boldsymbol{{27}^{\circ}}$C experiment.
Prior hyper-parameter choices for the models are given Table~\ref{tab:SHM_priors}.
Bayesian inference and MCMC methods for the JHM in Table~\ref{tab:JHM} is carried out similarly for both the JHM-B and JHM-T.
Posterior samples of 1000 particles are obtained after a burn-in period of 800k, and a thinning of every 100 observations.
Similarly to Section~\ref{conv_diag}, chains from our MCMC sampler are accepted where the effective sample sizes are greater than $300$ and Heidelberg and Welch P-values are greater than $0.10$ for every chain.
Similarly to the other previous modelling approaches considered (including the ``vanilla'' JHM), a list of 159 are stripped from our final list of genes for biological and experimental reasons.
 
The JHM-B fit in Figure~\ref{fig:BATCH} has many less interactions on the plot than the ``vanilla'' JHM fitness plot, this may be evidence of a plate effect existing.
The JHM-T fit in Figure~\ref{fig:TRANS} is largely the same as the ``vanilla'' JHM fitness plot.
It is worth noting that the JHM-T model fit in Figure~\ref{fig:TRANS} has posterior mean estimates of $\hat{\phi}=0.96$ and $\hat{\chi}=0.87$, 2dp, suggesting that a transformation may only exist in terms of $r$.

Table~\ref{tab:JHM_hits} compares the number of suppressors and enhancers estimated for the two extensions of the JHM. 
The JHM-B reduces the number of genetic interactions from the ``vanilla'' JHM from $939$ to $553$, and similarly reduces the number of suppressors and enhancers.
Therefore from the ``vanilla'' JHM to the JHM-B, there is approximately a $41\%$ reduction of genes identified as showing significant evidence of genetic interaction, strong evidence for the presence of a batch effect. 
The JHM-T is more similar to the JHM with $901$ interactions, reducing both suppressors and enhancers by a small amount. 
Therefore from the ``vanilla'' JHM to the JHM-T, there is approximately a $4\%$ reduction of genes identified as showing significant evidence of genetic interaction, a much smaller reduction from the JHM than that observed with the JHM-B. 

Table~\ref{tab:overlap_ext}A shows that the number of genes that overlap with the genes identified by the ``vanilla'' JHM is 531 and 886 for the JHM-B and JHM-T respectively. 
Therefore the number of genes identified as interacting by the ``vanilla'' JHM and now no longer identified is $408$ and $53$ for the JHM-B and JHM-T respectively.
This further demonstrates the large reduction in genetic interactions when using the JHM-B, suggesting that a batch effect is present within the data.
The number of genes newly identified as showing significant evidence of genetic interaction by the JHM-B and JHM-T is $22$ and $15$ respectively.
These numbers are small relative to the number of genes that are no longer identified, indicating that the biggest change from the ``vanilla'' JHM is that the JHM-B and JHM-T are more stringent for determining significant genetic interactions.
Table~\ref{tab:overlap_ext}A shows that the ``vanilla'' JHM and JHM-T have similar overlap with the \citet{QFA1}, REM and IHM approaches. The JHM-B has much less overlap with the \citet{QFA1} approach than the ``vanilla'' JHM does, reducing the overlap from $649$ to $498$, indicating that the changes lead to an approach that is even more dissimilar from the \citet{QFA1} approach.

Table~\ref{tab:overlap_ext}B shows that the overlap in significant GO terms for the JHM-T and JHM-B with the JHM is 204 and 267 respectively. 
There are 286 (see Table~\ref{tab:overlap_ext}B) significant GO terms found with the ``vanilla'' JHM, meaning there is a reduction of approximately $29\%$ and $7\%$ with the JHM-B and JHM-T respectively, demonstrating the difference of our new approaches from ``vanilla'' JHM.
Table~\ref{tab:overlap_ext}B also shows that the ``vanilla'' JHM, JHM-B and JHM-T all have a similar number of overlap in significant GO terms with the \citet{QFA1}, REM and IHM approaches. 

We have introduced two potential ways of further extending the JHM to better model a QFA screen comparison, Figures~\ref{fig:BATCH}~and~\ref{fig:TRANS} are included as a reference for further research.
The JHM-B has made large changes to our results by reducing the number of hits, see Table~\ref{tab:JHM_hits}. 
Further research may involve investigating the behaviour of an alternative JHM-B with tighter priors for the batch effect parameters so we can see how the additional parameters affect the model fit in more detail.
Further research for the JHM-T would involve developing an alternative JHM-T where different transformations are made for the control and query screens. 
We find that the largest difference with the JHM-B and JHM-T is that they are more stringent for determining genetic interactions than the ``vanilla'' JHM.
Currently we prefer the ``vanilla'' JHM until further model exploration and analysis such as simulation studies are carried out to further investigate how the JHM-B and JHM-T affect our results. 
\begin{table}
\caption[Overlap with joint hierarchical model extensions for genes interacting with \emph{cdc13-1} at $\boldsymbol{{27}^{\circ}}$C and gene ontology terms over-represented in lists of interactions]{\label{tab:overlap_ext}Genes interacting with \emph{cdc13-1} at $\boldsymbol{{27}^{\circ}}$C and GO terms over-represented in the list of interactions according to each approach A) Number of genes identified for each approach (Add \cite{QFA1}, REM, IHM, JHM, JHM-B and JHM-T) and the overlap between the approaches. 4135 genes from the \emph{S. cerevisiae} single deletion library are considered.
B) Number of GO terms identified for each approach (Add \cite{QFA1}, REM, IHM, JHM, JHM-B and JHM-T) and the overlap between the approaches. 6107 \emph{S. cerevisiae} GO Terms available. See Tables~\ref{tab:overlap}A and~\ref{tab:overlap}B for further details on the overlap between the ``vanilla'' models (Add \cite{QFA1}, REM, IHM, JHM).}
\centering
\resizebox{\columnwidth}{!}{%
\begin{tabular}{*{7}{c}}
\multicolumn{1}{l}{\bf{A.}}&\emph{Add} &\emph{REM} &\emph{IHM} &\emph{JHM}&\emph{JHM-B}&\emph{JHM-T}\\\hline
\emph{JHM}&649&273&572&939&N/A&N/A\\
\emph{JHM-B}&498&239&468&531&553&N/A\\
\emph{JHM-T}&628&276&572&886&535&901\\\hline
\end{tabular}
\qquad
\begin{tabular}{*{7}{c}}
\multicolumn{1}{l}{\bf{B.}}&\emph{Add} &\emph{REM} &\emph{IHM} &\emph{JHM}&\emph{JHM-B}&\emph{JHM-T}\\\hline
\emph{JHM}&219&165&216&286&N/A&N/A\\
\emph{JHM-B}&223&170&217&204&265&N/A\\
\emph{JHM-T}&215&160&219&267&206&293\\\hline
\end{tabular}
}
\end{table}

\FloatBarrier
				\begin{table}
  \caption[Description of the joint hierarchical model with batch effects]{Description of the joint hierarchical model with batch effects. $b$ identifies the batch which an \emph{orf}$\Delta$ repeat belongs to. Further model notation is defined in Table~\ref{tab:JHM}\label{tab:BATCH}}
\begin{align*}
c&=0,1   &&\; \text{Condition level}\\
l&=1,...,L_{c}   &&\; \text{$\emph{orf}\Delta$ level}\\
m&=1,...,M_{cl}       &&\; \text{Repeat level}\\
n&=1,...,N_{clm}      &&\; \text{Time point level}\\
b&=1,...,B      &&\; \text{Batch}\\
\shortintertext{Time point level}
y_{clmn} &\sim \operatorname{N}(\hat{y}_{clmn},({\nu_{cl}})^{-1} )\;
&\hat{y}_{clmn} &= x(t_{clmn};{ K_{clm} } ,{ r_{clm} } , { P })\\
\shortintertext{Repeat level}
\resizebox{!}{2.8mm}{$\log~K_{clm}$} &\resizebox{!}{3.2mm}{$\:\sim \operatorname{N}(\alpha_{c}+\kappa_{b}+K_{l}^o+\delta_{l}\gamma_{cl},({ \tau_{cl}^K })^{-1})I_{(-\infty,0]}$}
 & \resizebox{!}{2.8mm}{$\log~\tau_{cl}^K$} &\resizebox{!}{3.2mm}{$\:\sim \operatorname{N}(\tau^{K,p}_{c}, ({\sigma^{\tau,K}_{c}})^{-1} )I_{[0,\infty)}$}\\
\resizebox{!}{2.8mm}{$\log~r_{clm}$} &\resizebox{!}{3.2mm}{$\:\sim \operatorname{N}(\beta_{c}+\lambda_{b}+r_{l}^o+\delta_{l}\omega_{cl},({ \tau_{cl}^r })^{-1})I_{(-\infty,3.5]}$}
 &\resizebox{!}{2.8mm}{$\log~\tau_{cl}^r$} &\resizebox{!}{3.2mm}{$\:\sim \operatorname{N}(\tau^{r,p}_{c}, ({\sigma^{\tau,r}_{c}})^{-1} )$}\\
\shortintertext{$\emph{orf}\Delta$ level}
e^{K_{l}^o} &\sim t(K^p, ({ \sigma^{K,o} })^{-1},3 )I_{[0,\infty)}\qquad &\log~\sigma^{K,o} &\sim \operatorname{N}(\eta^{K,o}, (\psi^{K,o})^{-1} )\\
e^{r_{l}^o} &\sim t(r^p, ({ \sigma^{r,o} })^{-1},3 )I_{[0,\infty)}\qquad &\log~\sigma^{r,o} &\sim \operatorname{N}(\eta^{r,o}, (\psi^{r,o})^{-1} )\\
\log~\nu_{cl} &\sim \operatorname{N}(\nu^p,({ \sigma^{\nu} })^{-1})\qquad& \log~\sigma^{\nu} &\sim \operatorname{N}(\eta^{\nu}, (\psi^{\nu})^{-1} )\\
\delta_{l} &\sim Bern(p)\\
e^{\gamma_{cl}}&=\begin{cases}
1  & \text{if } c=0;\\
t(1,{({\sigma^{\gamma}})}^{-1},3)I_{[0,\infty)} & \text{if } c=1.
\end{cases}
\qquad
&\log~\sigma^{\gamma}&\sim
\operatorname{N}(\eta^{\gamma},\psi^{\gamma})  
\\
e^{\omega_{cl}}&=\begin{cases}
1  & \text{if } c=0;\\
t(1,({{\sigma^{\omega}})}^{-1},3)I_{[0,\infty)} & \text{if } c=1.
\end{cases}
\qquad
&\log~\sigma^{\omega}&\sim
\operatorname{N}(\eta^{\omega},\psi^{\omega})
\\
\shortintertext{Condition level}
\alpha_{c}&=\begin{cases}
0  & \text{if } c=0;\\
\operatorname{N}(\alpha^{\mu},\eta^{\alpha})  & \text{if } c=1.
\end{cases}
&\beta_{c}&=\begin{cases}
0  & \text{if } c=0;\\
\operatorname{N}(\beta^{\mu},\eta^{\beta}) & \text{if } c=1.
\end{cases}
\\
\tau^{K,p}_{c} &\sim \operatorname{N}(\tau^{K,\mu}, ({\eta^{\tau,K,p}})^{-1} )
& \log~\sigma^{\tau,K}_{c} &\sim \operatorname{N}(\eta^{\tau,K}, (\psi^{\tau,K})^{-1} )\\
\tau^{r,p}_{c} &\sim \operatorname{N}(\tau^{r,\mu}, ({\eta^{\tau,r,p}})^{-1} )
& \log~\sigma^{\tau,r}_{c} &\sim \operatorname{N}(\eta^{\tau,r}, (\psi^{\tau,r})^{-1} )\\
\shortintertext{Population level}
\log~K^p &\sim \operatorname{N}(K^\mu, ({\eta^{K,p}})^{-1} )
& \log~r^p &\sim \operatorname{N}(r^\mu, ({\eta^{r,p}})^{-1} )\\
\nu^p &\sim \operatorname{N}(\nu^\mu, ({\eta}^{\nu,p})^{-1} )
& \log~P &\sim \operatorname{N}(P^\mu, ({\eta^{P}})^{-1} )\\
\shortintertext{Batch}
Log~\kappa_{b} &\sim \operatorname{N}(\kappa^p,(\eta^\kappa)^{-1})
& Log~\lambda_{b} &\sim \operatorname{N}(\lambda^p,(\eta^\lambda)^{-1})\\
\end{align*}
\end{table}
				\begin{table}
  \caption[Description of the joint hierarchical model with transformations]{Description of the joint hierarchical model with transformations. Model notation is defined in Table~\ref{tab:JHM}\label{tab:TRANS}}
	\begin{align*}
c&=0,1   &&\; \text{Condition level}\\
l&=1,...,L_{c}   &&\; \text{$\emph{orf}\Delta$ level}\\
m&=1,...,M_{cl}       &&\; \text{Repeat level}\\
n&=1,...,N_{clm}      &&\; \text{Time point level}\\
\shortintertext{Time point level}
y_{clmn} &\sim \operatorname{N}(\hat{y}_{clmn},({\nu_{cl}})^{-1} )\;
&\hat{y}_{clmn} &= x(t_{clmn};{ K_{clm} } ,{ r_{clm} } , { P })\\
\shortintertext{Repeat level}
\log~K_{clm} &\sim \operatorname{N}(\frac{\alpha_{c}+K_{l}^o+\delta_{l}\gamma_{cl}}{\phi},({ \tau_{cl}^K })^{-1})I_{(-\infty,0]}
\; & \log~\tau_{cl}^K &\sim \operatorname{N}(\tau^{K,p}_{c}, ({\sigma^{\tau,K}_{c}})^{-1} )I_{[0,\infty)}\\
\log~r_{clm} &\sim \operatorname{N}(\frac{\beta_{c}+r_{l}^o+\delta_{l}\omega_{cl}}{\chi},({ \tau_{cl}^r })^{-1})I_{(-\infty,3.5]}
\; &\log~\tau_{cl}^r &\sim \operatorname{N}(\tau^{r,p}_{c}, ({\sigma^{\tau,r}_{c}})^{-1} )\\
\shortintertext{$\emph{orf}\Delta$ level}
e^{K_{l}^o} &\sim t(K^p, ({ \sigma^{K,o} })^{-1},3 )I_{[0,\infty)}\qquad &\log~\sigma^{K,o} &\sim \operatorname{N}(\eta^{K,o}, (\psi^{K,o})^{-1} )\\
e^{r_{l}^o} &\sim t(r^p, ({ \sigma^{r,o} })^{-1},3 )I_{[0,\infty)}\qquad &\log~\sigma^{r,o} &\sim \operatorname{N}(\eta^{r,o}, (\psi^{r,o})^{-1} )\\
\log~\nu_{cl} &\sim \operatorname{N}(\nu^p,({ \sigma^{\nu} })^{-1})\qquad& \log~\sigma^{\nu} &\sim \operatorname{N}(\eta^{\nu}, (\psi^{\nu})^{-1} )\\
\delta_{l} &\sim Bern(p)\\
e^{\gamma_{cl}}&=\begin{cases}
1  & \text{if } c=0;\\
t(1,{({\sigma^{\gamma}})}^{-1},3)I_{[0,\infty)} & \text{if } c=1.
\end{cases}
\qquad
&\log~\sigma^{\gamma}&\sim
\operatorname{N}(\eta^{\gamma},\psi^{\gamma})  
\\
e^{\omega_{cl}}&=\begin{cases}
1  & \text{if } c=0;\\
t(1,({{\sigma^{\omega}})}^{-1},3)I_{[0,\infty)} & \text{if } c=1.
\end{cases}
\qquad
&\log~\sigma^{\omega}&\sim
\operatorname{N}(\eta^{\omega},\psi^{\omega})
\\
\shortintertext{Condition level}
\alpha_{c}&=\begin{cases}
0  & \text{if } c=0;\\
\operatorname{N}(\alpha^{\mu},\eta^{\alpha})  & \text{if } c=1.
\end{cases}
&\beta_{c}&=\begin{cases}
0  & \text{if } c=0;\\
\operatorname{N}(\beta^{\mu},\eta^{\beta}) & \text{if } c=1.
\end{cases}
\\
\tau^{K,p}_{c} &\sim \operatorname{N}(\tau^{K,\mu}, ({\eta^{\tau,K,p}})^{-1} )
& \log~\sigma^{\tau,K}_{c} &\sim \operatorname{N}(\eta^{\tau,K}, (\psi^{\tau,K})^{-1} )\\
\tau^{r,p}_{c} &\sim \operatorname{N}(\tau^{r,\mu}, ({\eta^{\tau,r,p}})^{-1} )
& \log~\sigma^{\tau,r}_{c} &\sim \operatorname{N}(\eta^{\tau,r}, (\psi^{\tau,r})^{-1} )\\
\shortintertext{Population level}
\log~K^p &\sim \operatorname{N}(K^\mu, ({\eta^{K,p}})^{-1} )
& \log~r^p &\sim \operatorname{N}(r^\mu, ({\eta^{r,p}})^{-1} )\\
\nu^p &\sim \operatorname{N}(\nu^\mu, ({\eta}^{\nu,p})^{-1} )
& \log~P &\sim \operatorname{N}(P^\mu, ({\eta^{P}})^{-1} )\\
\phi &\sim \Gamma(\phi^{shape},\phi^{scale})
& \chi &\sim \Gamma(\chi^{shape},\chi^{scale})\\
\end{align*}
\end{table}

\begin{figure}[h!]
  \centering
\includegraphics[width=14cm]{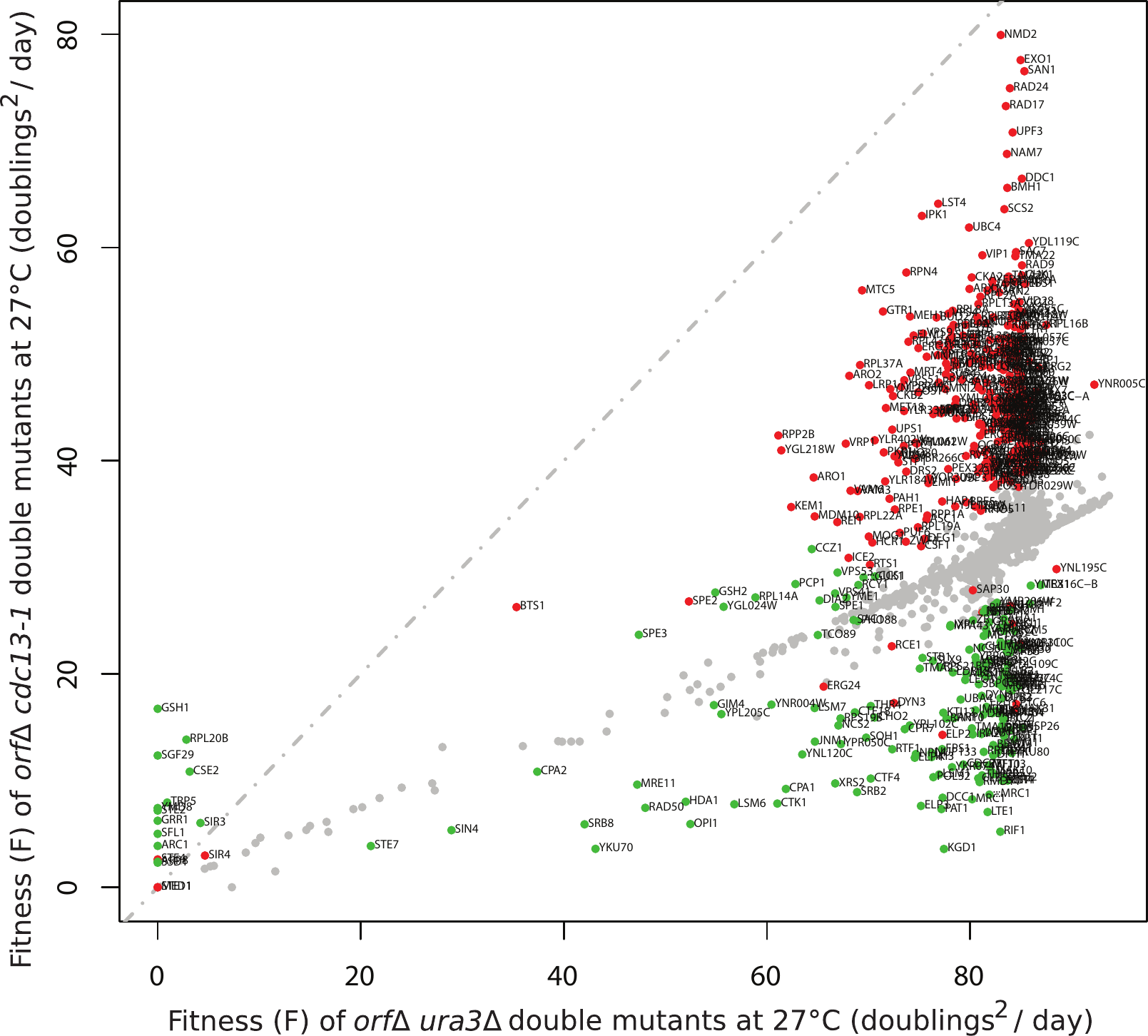}
\caption[\emph{cdc13-1}~$\boldsymbol{{27}^{\circ}}$C~vs~\emph{ura3}$\Delta$~$\boldsymbol{{27}^{\circ}}$C joint hierarchical model with batch effects fitness plot]{\emph{cdc13-1}~$\boldsymbol{{27}^{\circ}}$C~vs~\emph{ura3}$\Delta$~$\boldsymbol{{27}^{\circ}}$C joint hierarchical model with Batch effect (JHM-B) fitness plot with $\emph{orf}\Delta$ posterior mean fitnesses.
The JHM does not does not make use of a fitness measure such as $MDR\times{MDP}$ but the fitness plot is given in terms of $MDR\times{MDP}$ for comparison with other approaches which do. 
$\emph{orf}\Delta$ strains are classified as being a suppressor or enhancer based on one of the two parameters used to classify genetic interaction, growth parameter $r$, this means occasionally strains can be more fit in the query experiment in terms of $MDR\times MDP$ but be classified as enhancers (green).
Further fitness plot explanation and notation is given in Figure~\ref{fig:JHM}.\label{fig:BATCH}
}
\end{figure}

\begin{figure}[h!]
  \centering
\includegraphics[width=14cm]{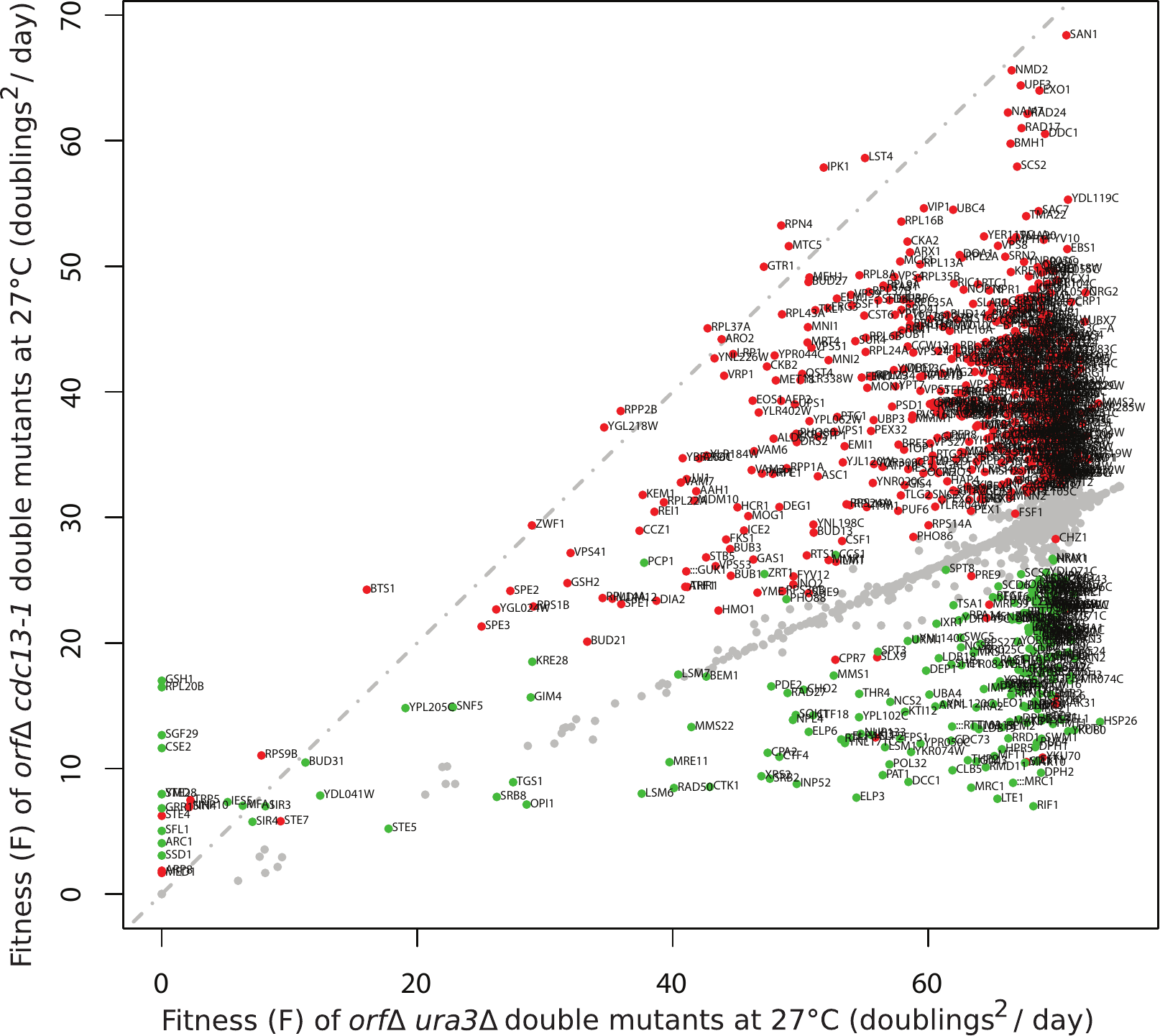}
\caption[\emph{cdc13-1}~$\boldsymbol{{27}^{\circ}}$C~vs~\emph{ura3}$\Delta$~$\boldsymbol{{27}^{\circ}}$C joint hierarchical model with transformations fitness plot]{\emph{cdc13-1}~$\boldsymbol{{27}^{\circ}}$C~vs~\emph{ura3}$\Delta$~$\boldsymbol{{27}^{\circ}}$C joint hierarchical model with transformations (JHM-T) fitness plot with $\emph{orf}\Delta$ posterior mean fitnesses.
The JHM does not does not make use of a fitness measure such as $MDR\times{MDP}$ but the fitness plot is given in terms of $MDR\times{MDP}$ for comparison with other approaches which do. 
$\emph{orf}\Delta$ strains are classified as being a suppressor or enhancer based on one of the two parameters used to classify genetic interaction, growth parameter $r$, this means occasionally strains can be more fit in the query experiment in terms of $MDR\times MDP$ but be classified as enhancers (green).
Further fitness plot explanation and notation is given in Figure~\ref{fig:JHM}.\label{fig:TRANS}
}
\end{figure}

\end{chapter}
 \begin{chapter}{\label{cha:stochastic_app}Fast Bayesian parameter estimation for stochastic logistic growth models}
\section{\label{sto:intro}Introduction}
In this Chapter, fast approximations to the stochastic logistic growth model (SLGM) \citep{capo_slgm} (see Section~\ref{int:stochastic_logistic_gro}) are presented. 
The SLGM is given by the following diffusion equation:
\begin{align}
\label{eq_det_sde}
dX_t&=rX_t\left(1-\frac{X_t}{K}\right)dt+\sigma X_t dW_t, 
\end{align}
where $X_{t_0}=P$ and is independent of $W_t$, $t\geq t_0$.

A deterministic logistic growth model (see Section\ref{int:logistic_gro}) is unable to describe intrinsic error within stochastic logistic growth time course data.
Consequently a deterministic model may lead to less accurate estimates of logistic growth parameters than a SDE, which can describe intrinsic noise.
So that random fluctuations present within observed yeast QFA data (\ref{int:QFA}) can accounted for as intrinsic noise instead of being confounded within our measurement error we are interested in using the SLGM in (\ref{eq_det_sde}), instead of its deterministic counterpart (\ref{eq_det}).
Alternative stochastic logistic growth equations exist (see Section~\ref{int:stochastic_logistic_gro}) but we find (\ref{eq_det_sde}) to be the most appropriate as intrinsic noise does not tend to zero with larger population sizes.

The SLGM (\ref{eq_det_sde}) is analytically intractable and therefore inference requires relatively slow numerical simulation.
Where fast inference is of importance such as real-time analysis or big data problems, we can use model approximations which do have analytically tractable densities, enabling fast inference.
For large hierarchical Bayesian models (see Chapter~\ref{cha:modelling_den_int}), computational time for inference is typically long, ranging from one to two weeks using a deterministic logistic growth model.
Inference for large hierarchical Bayesian models using the SLGM would increase computational time considerably (computational time is roughly proportional to the number of time points longer) with relatively slow numerical simulation approaches, therefore we may be interested in using approximate models that will allow us to carry out fast inference.

First an approximate model developed by \citet{roman} is introduced.
Two new approximate models are then presented using the linear noise approximation (LNA) \citep{LNA,komorowski} of the SLGM.
The model proposed by \citet{roman} is found to be a zero-order noise approximation.

The approximate models considered are compared against each other for both simulated and observed logistic growth data.
Finally, the approximate models are compared to ``exact'' approaches.

\section{\label{sec:roman}The \cite{roman} diffusion process}

\cite{roman} present a logistic growth diffusion process (RRTR) which has a transition density that can be written explicitly, allowing inference for model parameter values from discrete sampling trajectories.
 \\
The RRTR is derived from the following ODE:
\begin{align}
\label{eq_ode}
\frac{dx_t}{dt}&=\frac{Qr}{e^{rt}+Q}x_t,
\end{align}
where  $Q=\left(\frac{K}{P}-1\right)e^{rt_0}$, $P=x_{t_0}$ and $t\geq t_0$.
The solution to (\ref{eq_ode}) is given in (\ref{eq:logistic}) (it has the same solution as (\ref{eq_det})).

\cite{roman} see (\ref{eq_ode}) as a generalisation of the Malthusian growth model with a deterministic, time-dependent fertility $h(t)=\frac{Qr}{e^{rt}+Q}$, and replace this with $\frac{Qr}{e^{rt}+Q}+\sigma W_t$ to obtain the following approximation to the SLGM:
\begin{align}
\label{eq_sde}
dX_t&=\frac{Qr}{e^{r{t}}+Q}X_td{t}+{\sigma}X_tdW_t,
\end{align}
where  $Q=\left(\frac{K}{P}-1\right)e^{rt_0}$, $P=X_{t_0}$ and is independent of $W_t$, $t\geq t_0$.
The process described in (\ref{eq_sde}) is a particular case of the Log-normal process with exogenous factors, therefore an exact transition density is available \citep{gutierrez}.
The transition density for $Y_t$, where $Y_t=\log(X_t)$, can be written:
\begin{align}
\begin{split}\label{eq:RRTR_tran}
(Y_{t_i}|Y_{t_{i-1}} &= y_{t_{i-1}}) \sim\operatorname{N}\left(
\mu_{t_i}
,
\Xi_{t_i}
\right),\\
\text{where } a &= r,\quad b=\frac{r}{K}, \\
\mu_{t_i}=& \log(y_{t_{i-1}})
+\log\left(\frac{1+be^{-at_i}}{1+be^{-at_{i-1}}}\right)
-\frac{\sigma^2}{2}(t_i-t_{i-1}) \text{ and}\\
\Xi_{t_i} &= \sigma^2(t_i-t_{i-1}).
\end{split}
\end{align}

\section{\label{sec:LNAM}Linear noise approximation with multiplicative noise}
We now take a different approach to approximating the SLGM (\ref{eq_det_sde}), which will turn out to be closer to the exact solution of the SLGM than the RRTR (\ref{eq_sde}). 
Starting from the original model (\ref{eq_det_sde}),
we apply It\^{o}'s lemma \citep{ito,sdebook}:
\begin{equation}\label{eq_itolemma}
df(t,X_t)=\frac{df}{dt}dt+\mu\frac{df}{dx}dt+\frac{1}{2}\sigma^2\frac{d^2f}{dx^2}dt+\sigma\frac{df}{dx}dW_t,
\end{equation}
with the transformation $f(t,X_t)\equiv Y_t=\log X_t$.
After deriving the following partial derivatives:
\begin{equation*}
\frac{df}{dt}=0,\qquad\frac{df}{dx}=\frac{1}{X_t}\quad\text{and}\quad\frac{d^2f}{dx^2}=-\frac{1}{X_t^2},
\end{equation*}
we can obtain the following It\^{o} drift-diffusion process:
\begin{align}
\label{eq:SDE2}
dY_t=\left(r-\frac{1}{2}\sigma^2-\frac{r}{K}e^{Y_t}\right)dt+\sigma dW_t.
\end{align}
The log transformation from multiplicative to additive noise, gives a constant diffusion term, so that the LNA will give a good approximation to (\ref{eq_det_sde}).
The LNA reduces a non-linear SDE to a linear SDE with additive noise. 
The LNA can be viewed as a first order Taylor expansion of an approximating SDE about a deterministic solution.
We now separate the process $Y_t$ into a deterministic part $v_t$ and a stochastic part $Z_t$ so that $Y_t=v_t+Z_t$ and consequently $dY_t=dv_t+dZ_t$.
We choose $v_t$ to be the solution of the deterministic part of (\ref{eq:SDE2}):
\begin{equation}\label{eq:SDEV}
dv_t=\left(r-\frac{1}{2}\sigma^2-\frac{r}{K}e^{v_t}\right)dt.
\end{equation}
We now redefine our notation as follows: $a=r-\frac{\sigma^2}{2}$ and $b=\frac{r}{K}$. 
Equation~\ref{eq:SDEV} is then solved for $v_t$:
\begin{equation}\label{eq:LNAM_det_sol}
v_t=\log\left(\frac{aPe^{aT}}{bP(e^{aT}-1)+a}\right),
\end{equation}
where $T=t-t_0$. We now write down an expression for $dZ_t$, where $dZ_t=dY_t-dv_t$:
\begin{equation*} 
dZ_t=\left(a-be^{Y_t}\right)dt+\sigma dW_t-\left(a-be^{v_t}\right)dt
\end{equation*}
We then substitute in $Y_t=v_t+Z_t$ and simplify the expression to give
\begin{equation} \label{eq:zero_start}
dZ_t=b\left(e^{v_t}-e^{v_t+Z_t}\right)dt+\sigma dW_t.
\end{equation} 
As $dZ_t$ is a non-linear SDE it cannot be solved explicitly, we use the LNA (see Section~\ref{lit:LNA}) to obtain a linear SDE that we can solve explicitly.
We apply the LNA by making a first-order approximation of $e^{Z_t}\approx 1+Z_t$ and then simplify to give 
\begin{equation}\label{eq:LNAM_dz}
dZ_t=-be^{v_t}Z_tdt+\sigma dW_t.
\end{equation}
This process is a particular case of the time-varying Ornstein-Uhlenbeck process, which can be solved explicitly.  
The transition density for $Y_t$ (derivation in Appendix~\ref{app:LNAM_sol}) is then:
\begin{align}
\begin{split}\label{eq:LNAM_tran}
(Y_{t_i}|Y_{t_{i-1}}&=y_{t_{i-1}})\sim\operatorname{N}\left(\mu_{t_i},\Xi_{t_i}\right),\\
\text{redefine } y_{t_{i-1}}&=v_{t_{i-1}}+z_{t_{i-1}}, Q=\left(\frac{\frac{a}{b}}{P}-1\right)e^{at_{0}},\\
\mu_{t_i}&=y_{t_{i-1}}+\log\left(\frac{1+Qe^{-at_{i-1}}}{1+Qe^{-at_i}}\right)+e^{-a(t_i-t_{i-1})}\frac{1+Qe^{-at_{i-1}}}{1+Qe^{-at_i}}z_{t_{i-1}} \text{ and}\\
\Xi_{t_i}&=\sigma^2\left[\frac{4Q(e^{at_i}-e^{at_{i-1}})+e^{2at_i}-e^{2at_{i-1}}+2aQ^2(t_i-t_{i-1})}{2a(Q+e^{at_i})^2}\right].
\end{split}
\end{align}
The LNA of the SLGM with multiplicative intrinsic noise (LNAM) can then be written as
\begin{align*}
d\log X_t=\left[dv_t+be^{v_t}v_t-be^{v_t}\log X_t\right]dt+\sigma dW_t,
\end{align*}
where $P=X_{t_0}$ and is independent of $W_t$, $t\geq t_0$.
\\
Note that the RRTR given in (\ref{eq_sde}) can be similarly derived using a zero-order noise approximation ($e^{Z_t}\approx 1$) instead of the LNA.
 
\section{\label{sec:LNAA}Linear noise approximation with additive noise}
As in Section~\ref{sec:LNAM}, we start from the SLGM, given in (\ref{eq_det_sde}).
Without first log transforming the process, the LNA will lead to a worse approximation to the diffusion term of the SLGM, but we will see in the coming sections that there are nevertheless advantages.
We separate the process $X_t$ into a deterministic part $v_t$ and a stochastic part $Z_t$ so that $X_t=v_t+Z_t$ and consequently $dX_t=dv_t+dZ_t$.
We chose $v_t$ to be the solution of the deterministic part of (\ref{eq_det_sde}):
\begin{equation}
dv_t=\left(rv_t-\frac{r}{K}v_t^2\right)dt.\label{eq:SDEV2}
\end{equation}
We now redefine our previous notation as follows: $a=r$ and $b=\frac{r}{K}$.
Equation~\ref{eq:SDEV2} is then solved for $v_t$:
\begin{equation}\label{eq:LNAA_det_sol}
v_t=\frac{aPe^{aT}}{bP(e^{aT}-1)+a}.
\end{equation}
We now write down an expression for $dZ_t$, where $dZ_t=dX_t-dv_t$:
\begin{equation*}
dZ_t=\left(aX_t-bX_t^2\right)dt+\sigma X_t dW_t-\left(av_t-bv_t^2\right)dt.
\end{equation*}
We then substitute in $X_t=v_t+Z_t$ and simplify the expression to give
\begin{equation*}
dZ_t=(a-2bv_t)Z_t-bZ_t^2dt+\left( \sigma v_t +\sigma Z_t\right) dW_t.
\end{equation*}
As $dZ_t$ is a non-linear SDE it cannot be solved explicitly, we use the LNA (see Section~\ref{lit:LNA}) to obtain a linear SDE that we can solve explicitly.
We now apply the LNA, by setting second-order term $-bZ_t^2dt=0$ and $\sigma Z_t dW_t=0$ to obtain
\begin{equation}\label{eq:LNAA_dz}
dZ_t=(a-2bv_t)Z_tdt+\sigma v_t  dW_t.
\end{equation}
This process is a particular case of the Ornstein-Uhlenbeck process, which can be solved. 
The transition density for $X_t$ (derivation in Appendix~\ref{app:LNAA_sol}) is then
\begin{align}
\begin{split}\label{eq:LNAA_tran}
(X_{t_i}|X_{t_{i-1}}&=x_{t_{i-1}})\sim N(\mu_{t_i},\Xi_{t_i}),\\
\text{where } x_{t_{i-1}}&=v_{t_{i-1}}+z_{t_{i-1}},\\
\mu_{t_i}&=x_{t_{i-1}}+\left(\frac{aPe^{aT_i}}{bP(e^{aT_i}-1)+a}\right)-\left(\frac{aPe^{aT_{i-1}}}{bP(e^{aT_{i-1}}-1)+a}\right)\\
&+e^{a(t_i-t_{i-1})}\left(\frac{bP(e^{aT_{i-1}}-1)+a}{bP(e^{aT_i}-1)+a}\right)^2Z_{t_{i-1}}\text{ and}\\
 \Xi_t&=\frac{1}{2}\sigma^2aP^2e^{2aT_i}\left(\frac{1}{bP(e^{aT_i}-1)+a}\right)^4\\
&\times[
b^2P^2(e^{2aT_i}-e^{2aT_{i-1}})
+4bP(a-bP)(e^{aT_i}-e^{aT_{i-1}})\\
&\;\:\:\:\:+2a(t_i-t_{i-1})(a-bP)^2
].
\end{split}
\end{align}
The LNA of the SLGM, with additive intrinsic noise (LNAA) can then be written as
\begin{align*}
dX_t=\left[b{v_t}^2+\left(a-2bv_t\right)X_t\right]dt+\sigma v_t dW_t,
\end{align*}
where $P=X_{t_0}$ and is independent of $W_t$, $t\geq t_0$.
\section{\label{sec:SDE_application}Simulation and Bayesian inference for the stochastic logistic growth model and approximations}
To compare the accuracies of each of the three approximate models in representing the SLGM, we first compare simulated forward trajectories from the RRTR, LNAM and LNAA with simulated forward trajectories from the SLGM (Figure~\ref{4nonu}).
We use the Euler-Maruyama method \citep{embook} (see Section~\ref{lit:em}) with very fine discretisation to give arbitrarily exact simulated trajectories from each SDE.

The LNAA and LNAM trajectories are visually indistinguishable from the SLGM (Figures~\ref{4nonu} A, C \& D).
On the other hand, population sizes simulated with the RRTR display large deviations from the mean as the population approaches its stationary phase (Figures~\ref{4nonu}A \& B). 
Figure~\ref{4nonu}E further highlights the increases in variation as the population approaches stationary phase for simulated trajectories of the RRTR, in contrast to the SLGM and LNA models.
\begin{figure}[h!]
  \centering
\includegraphics[width=14cm]{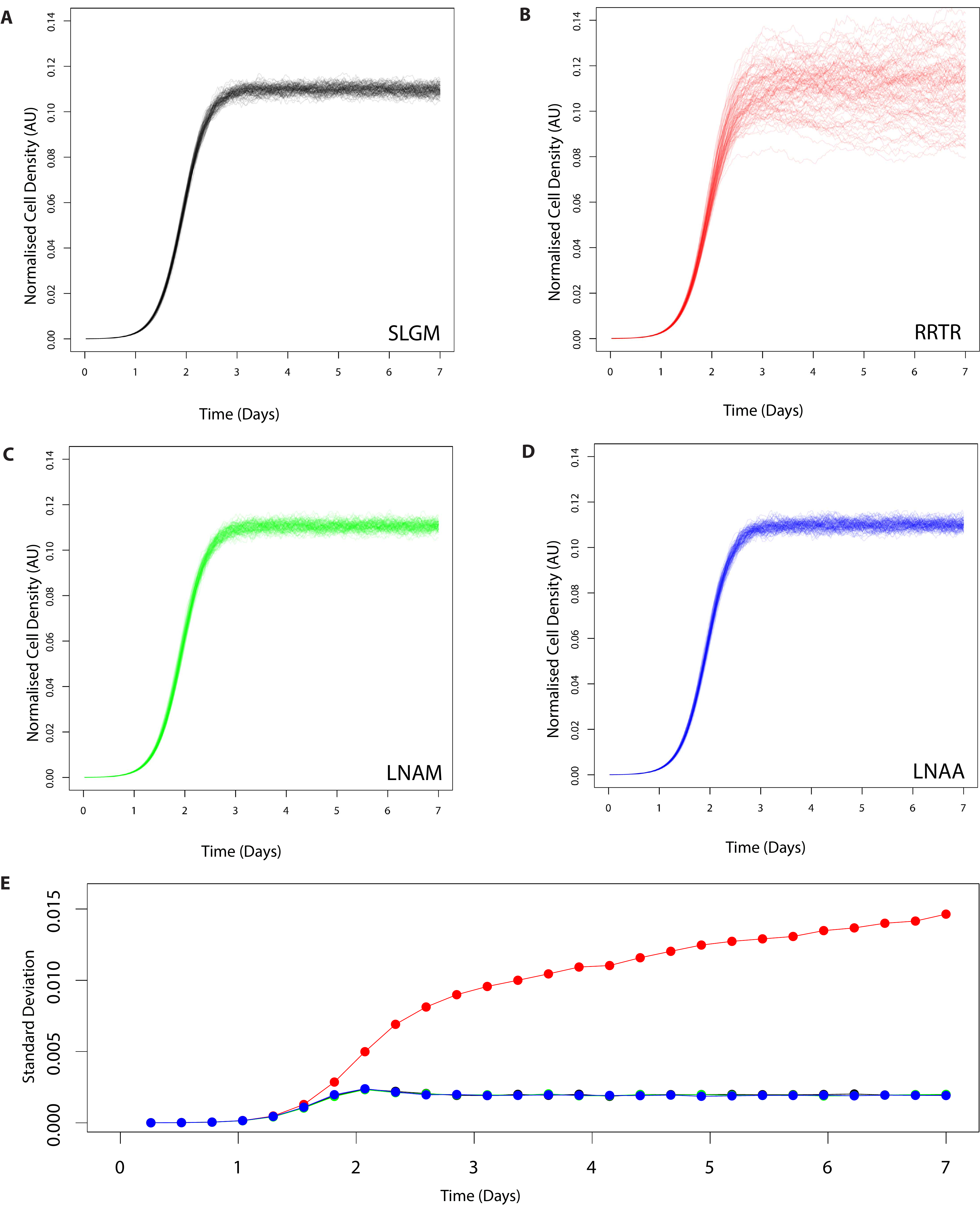}
\caption[Forward trajectories for the stochastic logistic growth model and approximations]{\label{4nonu} 
Forward trajectories (No. of simulations=100) for the stochastic logistic growth model and approximations.
See Table~\ref{app:sde_val_fur} for parameter values. 
A) The stochastic logistic growth model (SLGM). 
B) The \cite{roman} (RRTR) approximation. 
C) The linear noise approximation with multiplicative intrinsic noise (LNAM).
D) The linear noise approximation with additive intrinsic noise (LNAA).
E) Standard deviations of simulated trajectories over time for the SLGM (black), RRTR (red), LNAM (green) and LNAA (blue).
}
\end{figure}			
\clearpage
\subsection{\label{sec:simulation_stu}Bayesian parameter inference with approximate models}
To compare the quality of parameter inference using each of these approximations we simulated synthetic time-course data from the SLGM and combined this with either Log-normal or Normal measurement error. 
Carrying out Bayesian inference with broad priors (see (\ref{app:LNAM_sta_spa_mod}) and (\ref{app:LNAA_sta_spa_mod})) we compared the parameters recovered using each approximation with those used to generate the synthetic dataset. 
The synthetic time-course datasets consist of 27 time points generated using the Euler-Maruyama method with very fine intervals \citep{embook}.

We formulate our inference problem as a dynamic linear state space model \citep{dynamicmodels}. 
The advantage of a state space formulation is that we are then able to build a Kalman filter to carry out fast parameter inference. 
We can take advantage of a linear Gaussian structure and construct a Kalman filter recursion for marginal likelihood computation (Appendix~\ref{app:kalman_fil}).
By choosing to match the measurement error structure to the intrinsic error of our models we can build a linear Gaussian structure.
We therefore assume Log-normal (multiplicative) error for the RRTR and LNAM, and for the LNAA we assume Normal (additive) measurement error.
Dependent variable $y_{t_i}$ and independent variable $\{t_{i},i=1,...,N\}$ are data input to the model (where $t_i$ is the time at point $i$ and $N$ is the number of time points). 
$X_t$ is the state process, describing the population size.

The state space model for the RRTR and LNAM is as follows:
\begin{align}
\log(y_{t_i}) &\sim \operatorname{N}(X_{t_i},{\nu}^{2} ),\notag\\
(X_{t_i}|X_{t_{i-1}}=x_{t_{i-1}})&\sim\operatorname{N}\left(\mu_{t_i},\Xi_{t_i}\right), \text{ where } x_{t_{i}}=v_{t_{i}}+z_{t_{i}},\label{app:LNAM_sta_spa_mod}
\end{align}
$\mu_{t_i}$ and $\Xi_{t_i}$ are given by (\ref{eq:RRTR_tran}) and (\ref{eq:LNAM_tran}) for the RRTR and LNAM respectively. Priors are as follows:
\begin{align*}
\log X_0 \equiv \log~P &\sim \operatorname{N}({\mu}_P,{\tau_P}^{-1}),& \quad
\log~K &\sim \operatorname{N}({\mu}_K,{\tau_K}^{-1}),& \quad
\log~r &\sim \operatorname{N}({\mu}_r,{\tau_r}^{-1}),\\
\log~\nu^{-2} &\sim \operatorname{N}({\mu}_\nu,{\tau_\nu}^{-1}),&
\log~{\sigma^{-2}} &\sim \operatorname{N}({\mu}_\sigma,{\tau_\sigma}^{-1})I_{[1,\infty]}.&
\end{align*}

Bayesian inference is carried out with broad priors such that estimated parameter values are not heavily influenced by our choice. 
See Table~\ref{table:SDE_priors} for prior hyper-parameter values.
Log-normal prior distributions are chosen to ensure positive logistic growth parameters and precision parameters are strictly positive.
Our prior for $\log~{\sigma^{-2}}$ is truncated below 1 to avoid unnecessary exploration of extremely low probability regions, which could be caused by problems identifying $\nu$, for example when $\log~\nu^{-2}$ takes large values, and to ensure that intrinsic noise does not dominate the process.
Our choice of 1 for the truncation threshold is made by observing forward simulations from our processes and choosing a value for $\log~{\sigma^{-2}}$ where intrinsic noise is so large that the deterministic part of the process is masked, consequently making the LNA a bad approximation.
We also find that truncating $\log~{\sigma^{-2}}$ is more preferable to truncating $\log~\nu^{-2}$ as truncating $\log~\nu^{-2}$ does not alleviate the identifiability problem without being very restrictive for the measurement error structures.

The state space model for the LNAA is as follows: 
\begin{align}
y_{t_i} &\sim \operatorname{N}(X_{t_i},{\nu}^{2} ),\notag\\
(X_{t_i}|X_{t_{i-1}}=x_{t_{i-1}})&\sim\operatorname{N}\left(\mu_{t_i},\Xi_{t_i}\right)
, \text{ where } x_{t_{i}}=v_{t_{i}}+z_{t_{i}},\label{app:LNAA_sta_spa_mod}
\end{align}
$\mu_{t_i}$ and $\Xi_{t_i}$ are given by (\ref{eq:LNAA_tran}).
Priors are as in (\ref{app:LNAM_sta_spa_mod}).
Measurement error for the observed values is Normal so that we have a linear Gaussian structure.
The state space models in (\ref{app:LNAM_sta_spa_mod}) and (\ref{app:LNAA_sta_spa_mod}) have different measurement error structures.
So that a fair comparison can be made between (\ref{app:LNAM_sta_spa_mod}) and (\ref{app:LNAA_sta_spa_mod}), we choose our priors so that the marginal moments for the measurement error of our models is not too dissimilar, particularly at the earliest stage where most growth is observed.

To see how the inference from our approximate models compares with slower ``exact'' models, we consider Euler-Maruyama approximations \citep{euler_maruyama} of (\ref{eq_det_sde}) and of the log transformed process, using fine intervals.
We use the approach of \citep{darren2005} to carry out inference of our ``exact'' models. 
A single site update algorithm is used to update model parameters and the Euler-Maruyama approximation of the latent process in turn.
Given these approximations we can construct a state space model for an ``exact'' SLGM with Log-normal measurement error (SLGM+L) and similarly for the SLGM with Normal measurement error (SLGM+N), priors are as in (\ref{app:LNAM_sta_spa_mod}).

Our inference makes use of a Kalman filter to integrate out the state process. 
The Kalman filer allows for fast inference compared to slow numerical simulation approaches that impute all states.
The algorithm for our approximate models is the Metropolis-within-Gibbs sampler with a symmetric proposal \citep{gamerman}. Full-conditionals are sampled in turn to give samples from the joint posterior distribution: 
\begin{equation*}
\pi{(K,r,P,\sigma,\nu,X_{t_{1:N}},y_{t_{1:N}})},
\end{equation*}
where $X_{t_{1:N}}$ is the latent process and $y_{t_{1:N}}$ is the observed data, for $N$ observed data points.
The Metropolis-within-Gibbs sampler algorithm is as follows:\\\\
1) Initialise counter $i=1$ and parameters $K_{(0)},r_{(0)},\sigma_{(0)},P_{(0)},\nu_{(0)}$\\
\\
2) Simulate $K_{(i)}$ from $K\sim\pi{(K|\nu_{(i-1)},r_{(i-1)},\sigma_{(i-1)},P_{(i-1)},y_{t_{1:N}})}$\\
\\
3) Simulate $r_{(i)}$ from $r\sim\pi{(r|\nu_{(i-1)},K_{(i)},\sigma_{(i-1)},P_{(i-1)},y_{t_{1:N}}}$\\
\\
4) Simulate $\sigma_{(i)}$ from $\sigma\sim\pi{(\sigma|\nu_{(i-1)},K_{(i)},r_{(i)},P_{(i-1)},y_{t_{1:N}})}$\\
\\
5) Simulate $P_{(i)}$ from $\nu\sim\pi{(P|\nu_{(i-1)},K_{(i)},r_{(i)},\sigma_{(i)},y_{t_{1:N}})}$\\
\\
6) Simulate $\nu_{(i)}$ from $\nu\sim\pi{(\nu|K_{(i)},r_{(i)},\sigma_{(i)},P_{(i)},y_{t_{1:N}})}$\\
\\
7) Repeat steps 2-6 until the sample size required is obtained.\\ 

We find the mixing for our algorithm is improved when we have intermediate steps between sampling from the $\sigma_{(i)}$ and $\nu_{(i)}$ full conditionals.
Each update in our algorithm is accomplished by a Metropolis-Hastings step using a Kalman filter.
Acceptance ratios are calculated for each update during a burn-in period. 
To improve the computational speed of our inference, further research may involve using an algorithm where we jointly update our parameters.
Posterior means are used to obtain point estimates and standard deviations for describing variation of inferred parameters. 
The Heidelberger and Welch convergence diagnostic \citep{Heidelberger} is used to determine whether convergence has been achieved for all parameters. 

Computational times for convergence of our MCMC schemes (code is available at \url{https://github.com/jhncl/LNA.git}) can be compared using estimates for the minimum effective sample size per second (ESS\textsubscript{min}/sec) \citep{coda}. 
The average ESS\textsubscript{min}/sec of our approximate model (coded in C) is $\sim$100 and ``exact'' model $\sim$1 (coded in JAGS \citep{rjags} with 15 imputed states between time points, chosen to maximise ESS\textsubscript{min}/sec).
We find that our C code is typically twice as fast as the simple MCMC scheme used by JAGS, indicating that our inference is ${\sim}50\times$  faster than an ``exact'' approach.
A more efficient ``exact'' approach could speed up further, say by another factor of 5, but our approximate approach will at least be an order of magnitude faster.
We use a burn-in of 600,000 and a thinning of 4,000 to obtain a final posterior sample size of 1,000 for MCMC convergence of all our models.


To compare the approximate models ability to recover parameters from the SLGM with simulated Log-normal measurement error, we simulate data and carry out Bayesian inference. Figure~\ref{simlog} shows that all three approximate models can capture the synthetic time-course well, but that the RRTR model is the least representative with the largest amount of drift occurring at the saturation stage, a property not found in the SLGM or the two new LNA models.
Comparing forwards trajectories with measurement error (Figure~\ref{simlog}), the ``exact'' model is visually similar to all our approximate models, but least similar to the RRTR.
Further, Table~\ref{app:sde_val_fur} demonstrates that parameter posterior means are close to the true values and that standard deviations are small for all models and each parameter set.
By comparing posterior means and standard deviations to the true values, Table~\ref{app:sde_val_fur} shows that all our models are able to recover the three different parameter sets considered. 
	\begin{figure}[h!]
  \centering
\includegraphics[width=14cm]{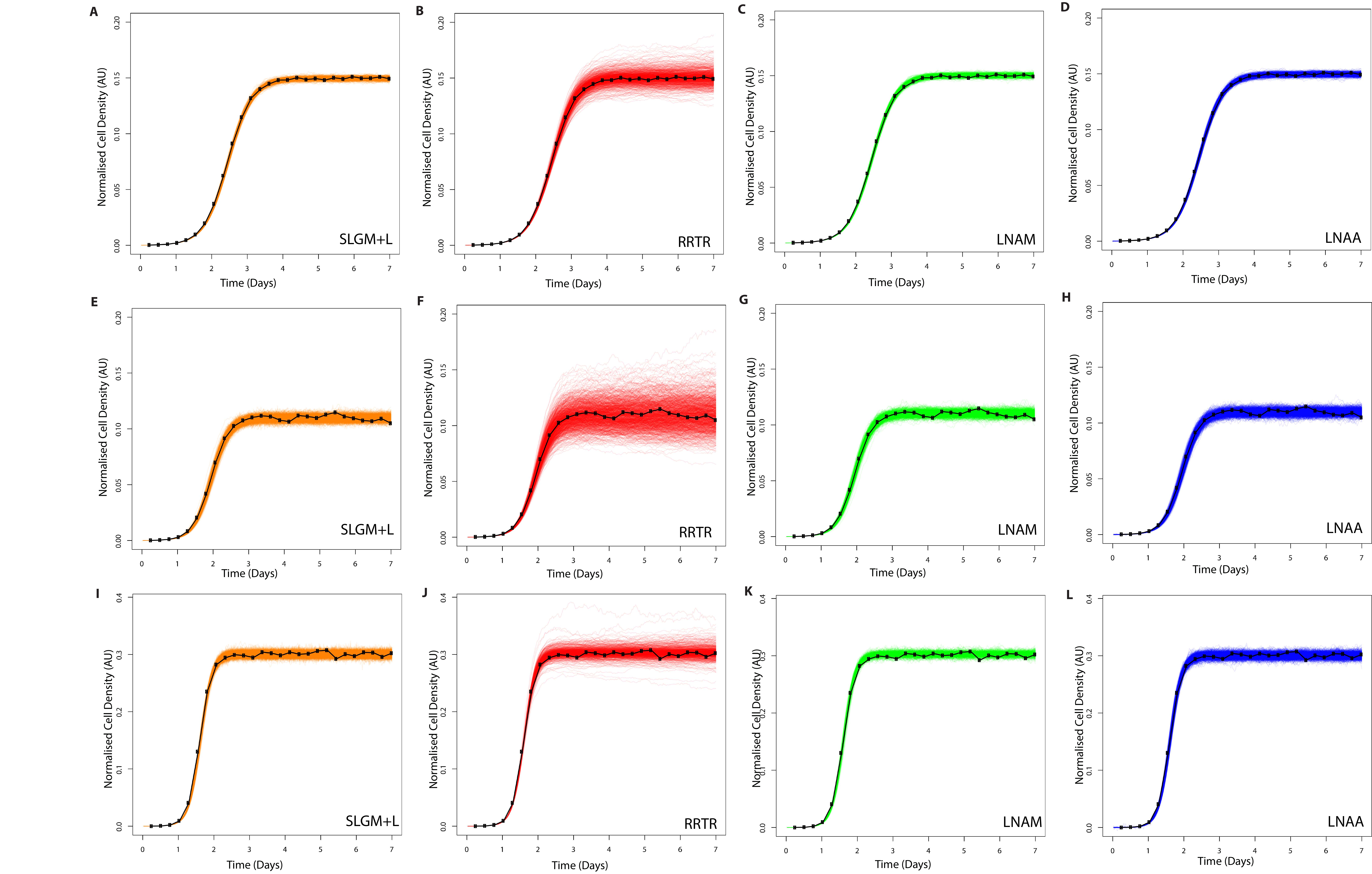}
\caption[Forward trajectories of logistic growth models and stochastic logistic data with Log-normal measurement error]{\label{simlog}
Forward trajectories with measurement error for the stochastic logistic growth model and approximations, simulated from parameter posterior samples (sample size=1000). 
Model fitting is carried out on SLGM forward trajectories with Log-normal measurement error (black), for three different sets of parameters (see Table~\ref{app:sde_val_fur}).  See (\ref{app:LNAM_sta_spa_mod}) or (\ref{app:LNAA_sta_spa_mod}) for model and Table~\ref{table:SDE_priors} for prior hyper-parameter values.
Each row of figures corresponds to a different time course data set, simulated from a different set of parameter values, see Table~\ref{app:sde_val_fur}.  
Each column of figures corresponds to a different model fit:
A), E) \& I) SLGM+L (orange).
B), F) \& J) RRTR model with lognormal error (red).
C), G) \& K) LNAM model with lognormal error (green).
D), H) \& L) LNAA model with normal error (blue).
See Table~\ref{app:sde_val_fur} for parameter posterior means and true values.
}
\end{figure}


To compare the approximations to the SLGM with simulated Normal measurement error, we simulate data and carry out Bayesian inference. 
Figure~\ref{sim} shows that of our approximate models, only the LNAA model can appropriately represent the simulated time-course as both our models with Log-normal measurement error, the RRTR and LNAM do not closely bound the data. 
Comparing forwards trajectories with measurement error (Figure~\ref{sim}), the ``exact'' model is most visually similar to the LNAA, which shares the same measurement error structure.
Further, Table~\ref{app:sde_val_fur} demonstrates that only our models with Normal measurement error have posterior means close to the true values and that standard deviations are larger in the models with Log-normal measurement error. 
Observing the posterior means for $K$ for each parameter set (Table~\ref{app:sde_val_fur}), we can see that the RRTR has the largest standard deviations and that, of the approximate models, its posterior means are furthest from both the true values and the ``exact'' model posterior means. 
Comparing LNA models to the ``exact'' models with matching measurement error, we can see in Table~\ref{app:sde_val_fur} that they share similar posterior means and only slightly larger standard deviations.
Example posterior diagnostics given in Figure~\ref{app:diag_sim_n}, demonstrate that posteriors are distributed tightly around true values for our LNAA and data from the SLGM with Normal measurement error.

\begin{figure}[h!]
  \centering
\includegraphics[width=13cm]{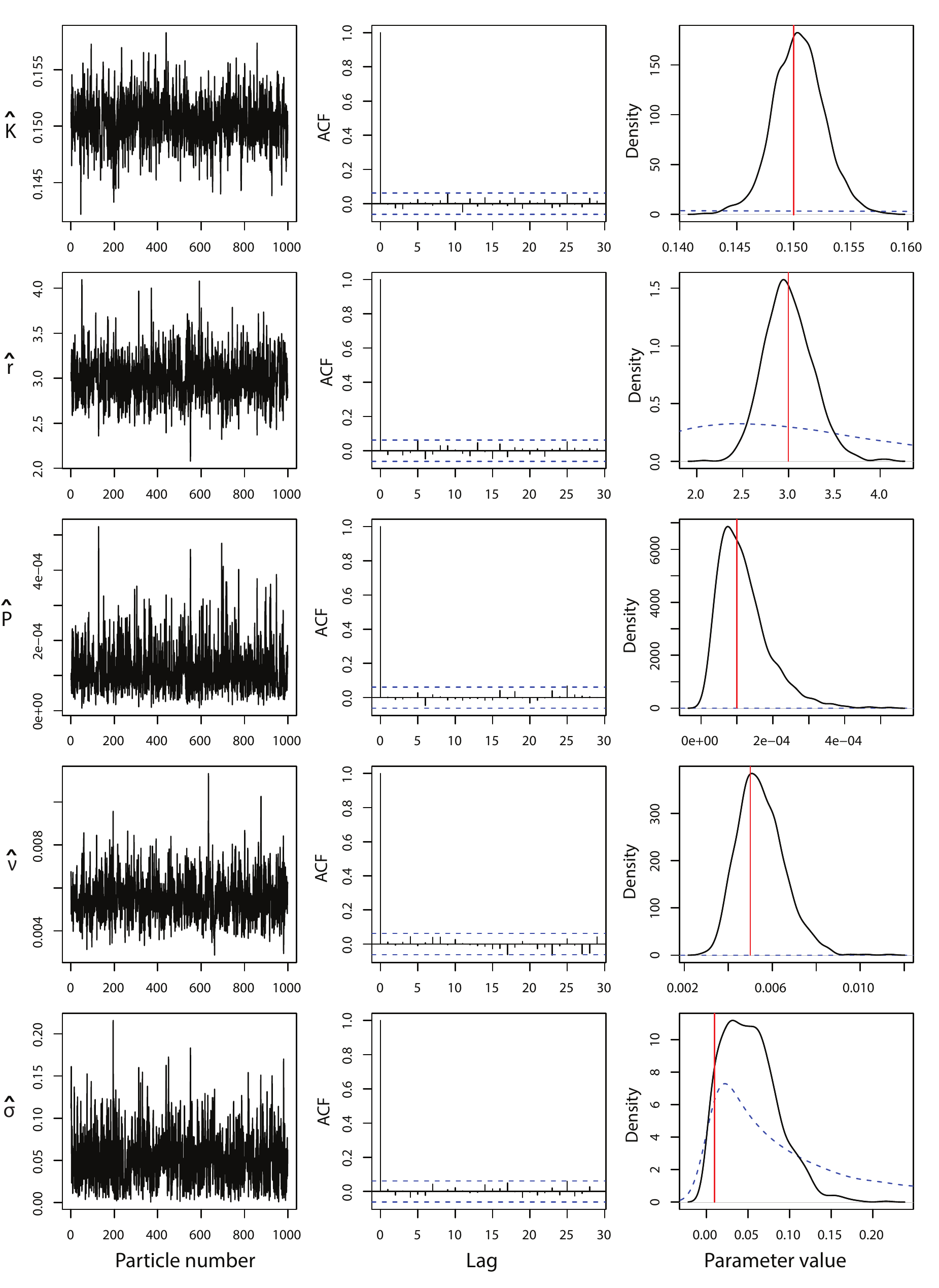}
\caption[Convergence diagnostics for the linear noise approximation of the stochastic logistic growth model with additive intrinsic noise]{
Convergence diagnostics for the linear noise approximation of the stochastic logistic growth model with additive intrinsic noise (LNAA) fit to simulated stochastic logistic growth data with Normal measurement error, see Figure~\ref{sim}D.
Trace, auto-correlation and density plots for the (LNAA) parameter posteriors (sample size = 1000, thinning interval = 4000).
Posterior density (black), prior density (dashed blue) and true parameter values (red) are shown in the right hand column.\label{app:diag_sim_n}
}
\end{figure}	
	
	\begin{figure}[h!]
  \centering
\includegraphics[width=14cm]{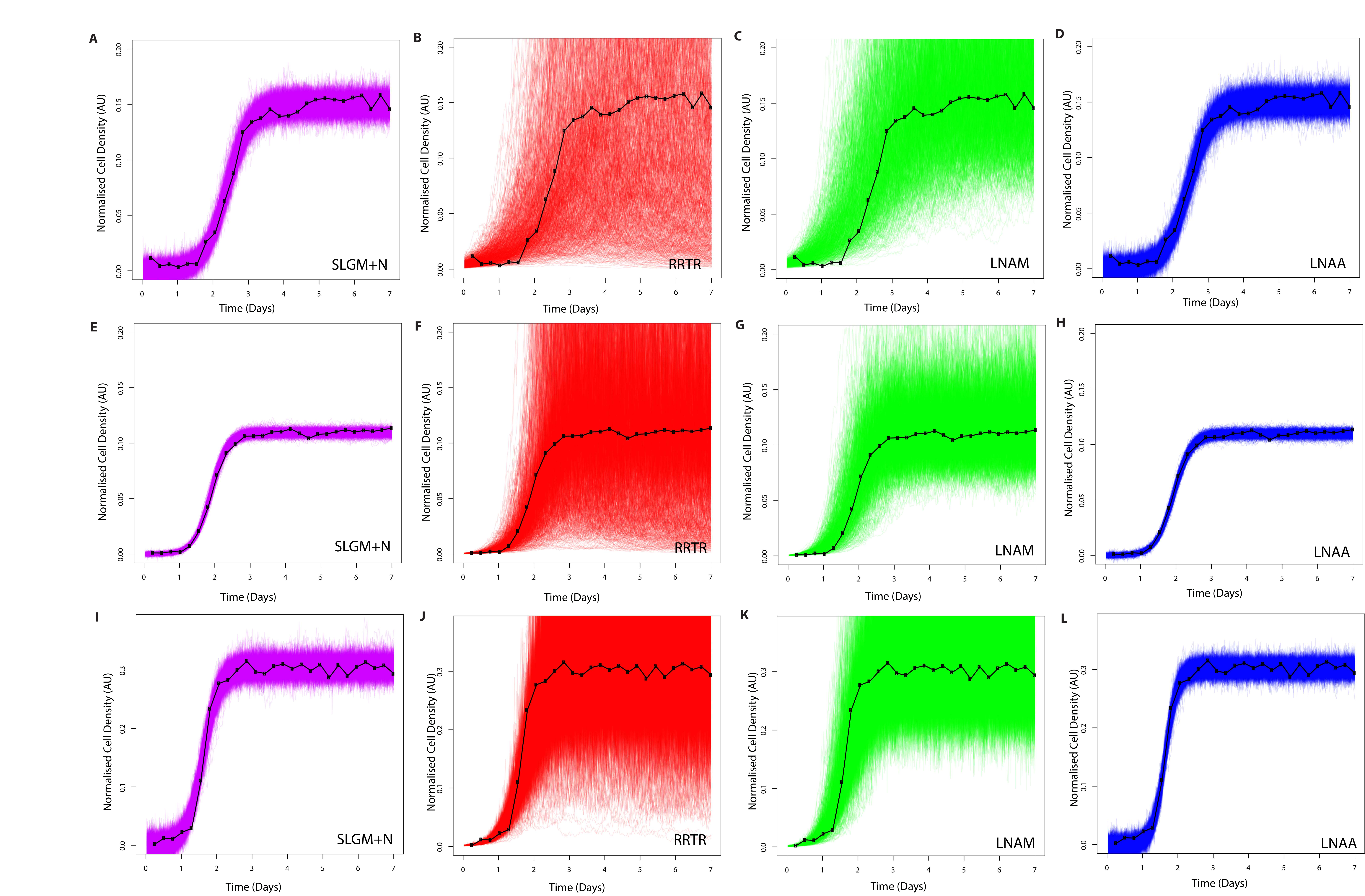}
\caption[Forward trajectories of logistic growth models and stochastic logistic data with Normal measurement error]{\label{sim}
Forward trajectories with measurement error, simulated from inferred parameter posterior samples (sample size=1000).  
Model fitting is carried out on SLGM forward trajectories with Normal measurement error (black), for three different sets of parameters (see Table~\ref{app:sde_val_fur}). 
See (\ref{app:LNAM_sta_spa_mod}) or (\ref{app:LNAA_sta_spa_mod}) for model and Table~\ref{table:SDE_priors} for prior hyper-parameter values.
Each row of figures corresponds to a different time course data set, simulated from a different set of parameter values, see Table~\ref{app:sde_val_fur}.  
Each column of figures corresponds to a different model fit:
A), E) \& I) SLGM+N (pink).
B), F) \& J) RRTR model with lognormal error (red).
C), G) \& K) LNAM model with lognormal error (green).
D), H) \& L) LNAA model with normal error (blue).
See Table~\ref{app:sde_val_fur} for parameter posterior means and true values.
}
\end{figure}
\begin{table}[h!]
\caption[Bayesian state space model parameter posterior means, standard deviations and true values]{\label{app:sde_val_fur}Bayesian state space model parameter posterior means, standard deviations and true values for Figures \ref{simlog}, \ref{sim} and \ref{real}. True values for the simulated data used for Figures \ref{4nonu}, \ref{simlog} and \ref{sim} are also given.}
\centering
\resizebox{\columnwidth}{!}{%
\npdecimalsign{.}
\nprounddigits{3}
\begin{tabular}{c c n{2}{5} n{1}{3} n{1}{3} n{1}{3} n{1}{3} n{1}{3} n{1}{3} n{1}{3} n{1}{3} n{1}{3}}
\hline
\emph{Panel} & \emph{Model} & \multicolumn{2}{c}{\emph{$\hat{K}$}}  & \multicolumn{2}{c}{\emph{$\hat{r}$}}  & \multicolumn{2}{c}{\emph{$\hat{P}$}} & \multicolumn{2}{c}{\emph{$\hat{\nu}$}} & \multicolumn{2}{c}{\emph{$\hat{\sigma}$}}  \\ 
\hline
\noalign{\vskip 0.4mm} 
\multicolumn{4}{l}{\emph{Figure \ref{simlog}, SLGM with lognormal error}}&&&&&&&&\\
\noalign{\vskip 0.2mm} 
A & SGLM+L & 0.1495369660 &( 0.001)&
2.9822427445 &(0.01350133)&
1.001567e-04 &(1.111598e-06)&
3.8597809e-03 &(2.126679e-03)&
0.0173228498 &(0.005037563)\\ 
B & RRTR & 0.149999335 &(0.00341)&
2.99015285 &(0.01092)&
9.93119e-05 &(1.06863e-06)&
5.6843212256e-03 &(2.36e-03)&
0.011547265557 &(0.00618)\\ 
C & LNAM & 0.149614787 &(0.001)&
2.98773783 &(0.01318)&
9.97960e-05 &(1.124219e-06)&
4.1396496436e-03 &(2.179554e-03)&
0.016364920984 &(0.00517)\\
D & LNAA & 0.14953488 &(0.0005)&
3.00514737 &(0.01995)&
9.64713e-05 &(2.945503e-06)&
3.098832e-05 &(2.53443e-05)&
0.0194997429 &(0.00344)\\ 
E & SGLM+L & 0.1096583 &(0.0007301528)&
3.974696 &(0.04724553)&
5.054189e-05 &(1.568196e-06)&
6.158735e-03 &(5.527275e-03)&
0.05052818 &(0.01393793)\\ 
F  & RRTR & 0.1086907   &(0.006533453)&
3.983695  &(0.03462259)&
5.046032e-05  &(1.137058e-06)&
5.928172e-03  &(4.596061e-03)&
0.03704644 &(0.00901516)\\ 
G  & LNAM & 0.1097371  &(0.0007490187)&
 3.985208  &(0.04606133)&
 5.043313e-05&(1.580017e-06)&
 6.187716e-03 &(5.190529e-03)&
  0.05156039 &(0.0126579)\\
H  & LNAA  &  0.1097912  &(0.0007867848)&
 3.959346  &(0.06704208)&
 5.207434e-05  &(4.309876e-06)&
 4.539877e-05   &(4.394698e-05)&
 0.05912248 &(0.00978808)\\ 
I
& SGLM+L & 0.3003784 &(0.0009748994)&
  5.996904  &(0.02851697)&
 1.962434e-05  &(4.0414e-07)&
 9.543053e-03 &(4.03492e-03)&
0.02406215 &(0.01543261)\\ 
J & RRTR & 0.3005997   &(0.0044928)&
6.015265  &(0.01697153)&
 1.942819e-05 &(2.835413e-07)&
 1.240917e-02 &(2.306528e-03)&
 0.007745952  &(0.006357571)\\ 
K & LNAM &  0.3004755 &(0.0009411441)&
  6.014510 &(0.03065974)&
1.953061e-05  &(4.202024e-07)&
8.942722e-03  &(4.251644e-03)&
0.02690633 &(0.01575695)\\
L & LNAA  &  0.3004586  &(0.001146215)&
 6.037006   &( 0.06747846)&
 1.895225e-05 &(1.502188e-06)&
  8.122229e-05  &(1.595971e-04)&
 0.04733552  &(0.007512705)\\ 
\hline\noalign{\vskip 1mm} 
\multicolumn{4}{l}{\emph{Figure \ref{sim}, SLGM with normal error}}&&&&&&&&\\
\noalign{\vskip 0.2mm} 
A
& SLGM+N & 0.1502746 &(0.002206958)&
3.099127 &(0.08534504)&
9.298825e-05 &(7.304899e-06)&
5.326174e-03 &(1.008844e-03)&
0.05948689 &(0.02986999)\\ 
B & RRTR & 0.2127014398 &(0.12323)&
1.367943557 &(0.26339)&
4.55242737e-03 &(2.1184933e-03)&
2.5392518626e-01 &(1.0969e-01)&
0.41885683478 &(0.12943)\\ 
C  & LNAM & 0.1714519199 &(0.03280)&
1.579950397 &(0.27107)&
5.2409643e-03 &(2.0479954e-03)&
2.05421122203e-01 &(7.805e-02)&
0.4730700594  &(0.05100)\\ 
D & LNAA & 0.150452393 &(0.00222)&
2.9899791 &(0.26239)&
1.189e-04 &(7.098716e-05)&
5.49023564e-03 &(1.06e-03)&
0.052829431599  &(0.03326)\\
E
& SLGM+N &  0.1094272  &(0.0007586429)&
4.183277 &(0.07356718)&
 4.389757e-05 &(4.128667e-06)&
9.679088e-04  &(2.806267e-04)&
0.05662970 &(0.01165919)\\ 
F & RRTR & 0.1574587528 &(0.08749706)&
2.6313040010 &(0.3367076)&
4.398197e-04 &(1.677808e-04)&
1.040098786e-01 &(1.008875e-01)&
0.3735663206 &(0.1616813)\\ 
G  & LNAM & 0.1159608439 &(0.009422523)&
3.0185447100 &(0.373545)&
 4.967272e-04 &(1.396969e-04)&
3.34648648e-02 &(4.308955e-02)&
0.4753192595 &(0.04357369)\\ 
H & LNAA &  0.1095810  &(0.0007913119)&
4.009893 &( 0.1581109)&
5.011655e-05 &( 1.442577e-05)&
1.093103e-03 &(3.638245e-04)&
0.05275491 &(0.01322628)\\ 
I
& SLGM+N & 0.3052235554 &(0.00282216)&
5.2670916442 &(0.1249732)&
3.263387e-04 &(3.407446e-05)&
1.11938047e-02 &(1.974384e-03)&
0.0446024422 &(0.0310588)\\ 
J & RRTR & 0.313643513 &(0.05677519)&
3.029806490 &(0.2325609)&
1.307166e-03 &(2.897121e-04)&
2.22843445e-01 &(3.707941e-02)&
0.074919987 &(0.08628192)\\ 
K  & LNAM  & 0.312629722 &(0.02045301)&
3.391999960 &(0.4300475)&
1.118432e-03 &(3.2685e-04)&
1.17641652e-01 &(8.435387e-02)&
0.360004398 &(0.1647072)\\ 
L & LNAA &  0.3022532 &(0.002361439)&
 5.862218 &(0.5228873)&
2.889783e-05 &(2.598836e-05)&
8.773735e-03 &(1.466338e-03)&
0.04084204  &(0.02844386)\\
\hline\noalign{\vskip 1mm} 
\multicolumn{4}{l}{\emph{ Figure \ref{real}, observed yeast data}}&&&&&&&&\\
\noalign{\vskip 0.2mm} 
A & SLGM+L &  0.1096916 &(0.007456264)&
4.098380 &(0.2994734)&
7.603410e-06 &(3.205816e-06)&
3.456518e-01 &(5.318567e-02)&
0.1128021 &(0.109331)\\ 
B & SLGM+N &  0.1098586 &(0.003273399)&
3.905401 &(0.1725181)&
1.043704e-05 &(3.085561e-06)&
1.851678e-04 &( 7.460388e-05)&
0.1666784 &(0.02814871)\\ 
C & RRTR &  0.1143094 &(0.02602507)&
3.763728 &(0.2006525)&
1.079002e-05 &(3.154911e-06)&
3.378525e-01 &(4.839998e-02)&
0.07771955 &(0.07732242)\\ 
D & LNAM & 0.1103702 &(0.01113706)&
3.776766 &(0.216272)&
1.077129e-05 &(3.277218e-06)&
3.36179e-01 &(5.137415e-02)&
0.1036413 &(0.1079923)\\
E & LNAA & 0.109176 &(0.003306315)&
3.832318 &(0.1984635)&
1.068951e-05 &(3.680477e-06)&
1.768693e-04 &(6.606697e-05)&
0.1637506 &(0.03259627)\\
\\ 
\hline
\multicolumn{2}{c}{\emph{True values}} & \multicolumn{2}{c}{\emph{K}}  & \multicolumn{2}{c}{\emph{r}}  & \multicolumn{2}{c}{\emph{P}} & \multicolumn{2}{c}{\emph{$\nu$}} & \multicolumn{2}{c}{\emph{$\sigma$}}   \\
\hline
\multicolumn{2}{c}{Figures~\ref{4nonu}, panels A, B, C and D} & \multicolumn{2}{c}{0.11} & \multicolumn{2}{c}{4} & \multicolumn{2}{c}{0.00005} & \multicolumn{2}{c}{N/A} & \multicolumn{2}{c}{0.05}  \\
\multicolumn{2}{c}{Figures~\ref{simlog} and \ref{sim}, panels A, B, C \& D}  & \multicolumn{2}{c}{0.15} & \multicolumn{2}{c}{3} & \multicolumn{2}{c}{0.0001} & \multicolumn{2}{c}{0.005} & \multicolumn{2}{c}{0.01} \\ 
\multicolumn{2}{c}{Figures~\ref{simlog} and \ref{sim}, panels E, F, G and H} & 
\multicolumn{2}{c}{0.11} & \multicolumn{2}{c}{4} & \multicolumn{2}{c}{0.00005} & \multicolumn{2}{c}{0.001} & \multicolumn{2}{c}{0.05}  \\
\multicolumn{2}{c}{Figures~\ref{simlog} and \ref{sim}, panels I, J, K and L} &
 \multicolumn{2}{c}{0.3} & \multicolumn{2}{c}{6} & \multicolumn{2}{c}{0.0002} & \multicolumn{2}{c}{0.01} & \multicolumn{2}{c}{0.02}  \\
\hline
\end{tabular}
\npnoround%
}
\end{table}

\subsection{\label{sec:application_obs}Application to observed yeast data}
We now consider which diffusion equation model can best represent observed microbial population growth curves taken from a Quantitative Fitness Analysis (QFA) experiment (Section~\ref{int:QFA}) \citep{QFA1,jove}, see Figure~\ref{real}. 
The data consists of scaled cell density estimates over time for budding yeast \emph{Saccharomyces cerevisiae}. 
Independent replicate cultures are inoculated on plates and photographed over a period of 5 days.
The images captured are then converted into estimates of integrated optical density (IOD, which we assume are proportional to cell population size), by the software package Colonyzer \citep{Colonyzer}.
The dataset chosen for our model fitting is a representative set of 10 time-courses, each with 27 time points.
Once we have chosen the most appropriate stochastic model we can then look to apply our chosen model to logistic growth data from the QFA screens used throughout Chapter~\ref{cha:case_stu} in the future.

As in Figure~\ref{sim}, we see that the LNAA model is the only approximation that can appropriately represent the time-course and that both the RRTR and LNAM fail to bound the data as tightly as the LNAA (Figure~\ref{real}).
Our two ``exact'' models are visually similar to our approximate models with the same measurement error, with the SLGM+N most similar to the LNAA and the SLGM+L to the RRTR and LNAM. This is as  expected due to matching measurement error structures.
Table~\ref{app:sde_val_fur} summarises parameter estimates for the observed yeast data using each model. The variation in the LNAA model parameter posteriors is much smaller than the RRTR and LNAM, indicating a more appropriate model fit. 
Comparing the LNA models and ``exact'' models with matching measurement error, we can see in Table~\ref{app:sde_val_fur} that they share similar posterior means and standard deviations for all parameters and in particular, they are very similar for both $K$ and $r$, which are important phenotypes for calculating fitness \citep{QFA1}.

	\begin{figure}[h!]
  \centering
\includegraphics[width=14cm]{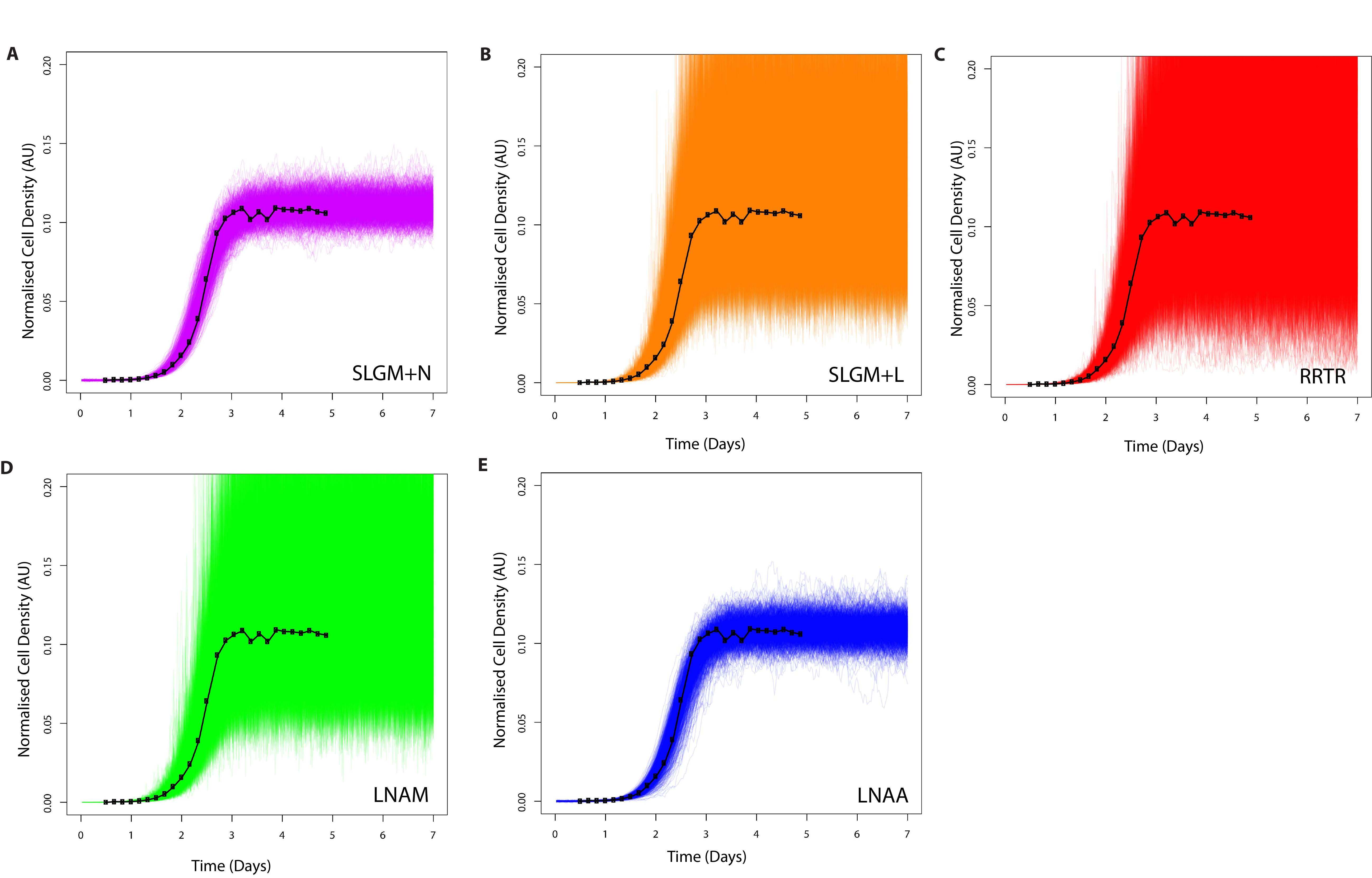}
\caption[Forward trajectories of logistic growth models and observed yeast data]{\label{real}
Forward trajectories with measurement error, simulated from inferred parameter posterior samples (sample size=1000).  
Model fitting is carried out on observed yeast time-course data (black). 
See (\ref{app:LNAM_sta_spa_mod}) or (\ref{app:LNAA_sta_spa_mod}) and Table~\ref{table:SDE_priors} for prior hyper-parameter values.
See Table~\ref{app:sde_val_fur} for parameter posterior means.
A) SLGM+N (pink).
B) SLGM+L (orange).
A) RRTR model with Log-normal error (red).
B) LNAM model with log-normal error (green).
C) LNAA model with Normal error (blue).
}
\end{figure}		
\begin{table}[h!]
\caption[Total mean squared error for 10 observed yeast growth time courses]{\label{tab:MSE} Total mean squared error (MSE) for 10 observed yeast growth time courses, each with 1000 forward simulated time-courses with measurement error. Parameter values are taken from posterior samples. Standard Deviations give the variation between the sub-total MSEs for each yeast time course fit (n=10). }
\centering
\npdecimalsign{.}
\nprounddigits{3}
\begin{tabular}{c n{1}{3} n{1}{3} n{1}{3} n{1}{3} n{1}{3}}
\hline
\emph{Model} & \multicolumn{1}{c}{{SLGM+N}}  & \multicolumn{1}{c}{{SLGM+L}}  &  \multicolumn{1}{c}{{RRTR}} & \multicolumn{1}{c}{{LNAM}} & \multicolumn{1}{c}{{LNAA}} \\ 
\hline
\noalign{\vskip 0.4mm} 
Total MSE & 29.84698 & 100.165 & 600.6007 & 99.39689 &  30.95872 \\
Standard Deviation & 1.688649 & 8.390729 &55.72009 &9.263094& 2.029686\\
\noalign{\vskip 0.4mm} 
\hline
\end{tabular}
\npnoround%
\end{table}

In Table~\ref{tab:MSE}, to compare quality of parameter inference for 10 observed yeast time-courses with each approximate model. Mean squared error (MSE) for 1000 posterior sample forward simulations are calculated for each yeast time course and summed to give a Total MSE for each model.
It is clear that the RRTR is the worst overall representation of the 10 yeast time courses, with the highest total MSE and a much larger total MSE than the ``exact'' SLGM+L. It is interesting to see there is a very similar total MSE for the SLGM+L and LNAM, and similarly for the SLGM+N and LNAA, demonstrating that our approximations perform well.

Once the most appropriate approximate stochastic model is chosen, we can incorporate the SDE within our Bayesian hierarchical models described in Section~\ref{cha:modelling_den_int}.
Currently the Bayesian hierarchical models described in Section~\ref{cha:modelling_den_int} have long computational times, $\sim$2 weeks for the joint hierarchical model (JHM) ($\sim$1 week with further optimisations) and so extending these models using slow numerical methods would lead to prohibitively slow computational times that we estimate to take $\sim$3-6 months (with 4294 \emph{orf}$\Delta$s, $\sim$8 repeats and $\sim$27 time points).
Inference using the Kalman filter will allow the Bayesian hierarchical models to carry out stochastic modelling at a greatly reduced computational time ($\sim$10$\times$ faster) compared to an arbitrarily exact approach.

\end{chapter}
  

\begin{chapter}{\label{cha:conclusion}Conclusions and future work}
We have joined a hierarchical model of microbial growth with a model for genetic interaction in order to learn about strain fitnesses, evidence for genetic interaction and interaction strengths simultaneously. By introducing Bayesian methodology to QFA we have been able to model the hierarchical nature of the experiment and expand the multiplicative model for genetic interaction to incorporate many sources of variation that previously had to be ignored.

\hl{We proposed} two new Bayesian hierarchical model approaches to replace the current statistical analysis for identifying genetic interactions within a QFA screen comparison. 
Both the new two-stage and one-stage approaches give similar results but have different interpretations. 
The two-stage approach fits the SHM followed by the IHM, \hl{with univariate point estimate fitness} definitions \hl{generated} as an intermediate step. 
The two-stage approach can therefore be regarded as a Bayesian hierarchical version of the \cite{QFA1} approach.
\hl{In contrast,} the one-stage approach fits the JHM, which does not require \hl{a univariate definition of} fitness, recognising that fitness is a multi-faceted concept, allowing interaction to be identified \hl{by either growth rate (logistic parameter $r$) or final biomass (logistic parameter $K$) achievable by a given genotype}. 
Our one-stage approach is a new method of detecting genetic interaction that further develops the interpretation of epistasis within QFA screens.

Hierarchical methods are able to account for the many sources of variation that exist within QFA data \hl{by accurately reflecting QFA experimental design, which is known}. 
A hierarchical, frequentist approach using random effects, namely the REM is \hl{presented} in order to improve on the \cite{QFA1} approach. Due to the lack of flexibility with modelling assumptions in the standard frequentist modelling paradigm, the REM is \hl{unsuitable for modelling} the distribution of \emph{orf}$\Delta$ level variation on a log scale or for simultaneously modelling genetic interaction and logistic growth curves.

The data from which logistic parameter estimates are derived during QFA are the result of a technically challenging, high-throughput experimental procedure with a diverse range of possible technical errors.
Our Bayesian, hierarchical models allow us the flexibility to make distributional assumptions that more closely match the data.
This allows us to switch between modelling parameter uncertainty with Normal, Log-Normal and Student's t distribution where appropriate.

QFA experimental design is intrinsically multilevel and is therefore more closely modelled in our hierarchical scheme.
Consequently the JHM and IHM capture sources of variation not considered by  \cite{QFA1}.
By sharing information across levels in the hierarchy, our models have allowed us to learn more about $\emph{orf}\Delta$s with weaker genetic interaction.  \hl{Our more flexible model of variance also avoids misclassification of individual genotypes with high variance as having significant interactions}.
Without fully accounting for the variation described in the Bayesian hierarchical models, the previous \cite{QFA1} approach may have relatively poor power to detect subtle interactions, obscuring potential novel observations.

Many subtle, interesting genetic interactions may remain to be investigated for the example dataset we present: QFA to understand telomere capping using \emph{cdc13-1}.
\hl{The JHM is better able to identify subtle interactions  (see Figure~\ref{fig:JHM_only}).
In our two-stage approaches, univariate fitness measures such as $MDR\times MDP$ are used in the intermediate steps, occasionally causing interaction in terms of one parameter to be masked by the other.
For example, strains with little evidence for interaction with a background mutation in terms of growth rate but with strong evidence of interaction in terms of carrying capacity are sometimes classified as interactors using the JHM (see Figure~\ref{fig:JHM_only}).}
The JHM has identified genes that have not been identified as showing genetic interaction in the \cite{QFA1} or two-stage Bayesian analysis, for example \emph{CHZ1}, which is thought to be related to telomere activity \citep{chz1}.

\hl{As expected, many genes previously unidentified by \cite{QFA1} have been identified as showing evidence of interaction using both of our Bayesian hierarchical modelling approaches.}
Some genes which have been identified only by the JHM (see Figure~\ref{fig:old}D), such as those showing interaction only in terms of $r$, are found to be related to telomere biology in the literature.
Currently there is not sufficient information available to identify the proportion of identified interactions that are true hits and so we use unbiased GO term enrichment analyses to confirm that the lists of genetic interactions closely reflect the true underlying biology.
GO term annotations relevant to telomere biology are available for well-studied genes in the current literature. Unsurprisingly all of the approaches considered closely reflect the most well-known GO terms (see Table~\ref{tab:sup_enh}).


Computational time for the new Bayesian approach ranges from one to two weeks for one of the datasets presented in \cite{QFA1}. This compares favourably with the time taken to design and execute the experimental component of QFA (approximately six weeks). 
Time and resources used to follow up the results of a QFA screen comparison can be saved with the Bayesian approaches suggested, allowing genes to be chosen for further investigation with increased confidence.
With an improved analysis it may be possible to detect more genetic interactions with the same sample size, \hl{allowing us to systematically detect and rank interactions genome-wide}. 
Overall \hl{we recommend} a JHM or ``Bayesian QFA'' for analysis of current and future QFA data sets as it accounts for more sources of variation than the \cite{QFA1} QFA methodology.
With the JHM we have outlined new genes with significant evidence of interaction in the \emph{ura3}$\Delta$~${{27}^{\circ}}$C~and~\mbox{\emph{cdc13-1}}~${27}^{{\circ}}$C experiment.
The full lists of genetic interactions for both the two and one stage Bayesian hierarchical approaches as well as lists of significant GO terms are freely available online at \sloppy\url{http://research.ncl.ac.uk/qfa/HeydariQFABayes/}.\sloppy
The new Bayesian hierarchical models we present here will also be suitable for identifying new genes showing evidence of genetic interaction in backgrounds other than telomere activity.
\hl{We hope that further, reductionist} lab work by experimental biologists will give additional insight into the mechanisms by which the new genes we have uncovered interact with the telomere.
\\
\\
In this thesis we have also presented two new diffusion processes for modelling logistic growth data where fast inference is required: the linear noise approximation (LNA) of the stochastic logistic growth model (SLGM) with multiplicative noise and the LNA of the SLGM with additive intrinsic noise (labelled as the LNAM and LNAA respectively). 
Both the LNAM and LNAA are derived from the linear noise approximation of the stochastic logistic growth model (SLGM).
The new diffusion processes approximate the SLGM more closely than an alternative approximation (RRTR) proposed by \cite{roman}.
The RRTR lacks a mean reverting property that is found in the SLGM, LNAM and LNAA, resulting in increasing variance during the stationary phase of population growth (see Figure~\ref{4nonu}).

We compared the ability of each of the three approximate models and the SLGM to recover parameter values from simulated datasets using standard MCMC techniques.  
When modelling stochastic logistic growth with Log-normal measurement error we find that our approximate models are able to represent data simulated from the original process and that the RRTR is least representative, with large variation over the stationary phase (see Figure~\ref{simlog}).
When modelling stochastic logistic growth with Normal measurement error we find that only our models with Normal measurement error can appropriately bound data simulated from the original process (see Figure~\ref{sim}).
We also compared parameter posterior distribution summaries with parameter values used to generate simulated data after inference using both approximate and ``exact'' models (see Table~\ref{app:sde_val_fur}).  We find that, when using the RRTR model, posterior distributions for the carrying capacity parameter $K$ are less precise than for the LNAM and LNAA approximations.  We also note that it is not possible to model additive measurement error while maintaining a linear Gaussian structure (which allows fast inference with the Kalman filter) when carrying out inference with the RRTR.  We conclude that when measurement error is additive, the LNAA model is the most appropriate approximate model.  

To test model performance during inference with real population data, we fitted our approximate models and the ``exact'' SLGM to microbial population growth curves generated by quantitative fitness analysis (QFA) (see Figure~\ref{real}).
We found that the LNAA model was the most appropriate for modelling experimental data.  It seems likely that this is because a Normal error structure best describes this particular dataset, placing the LNAM and RRTR models at a disadvantage.  We demonstrate that arbitrarily exact methods and our fast approximations perform similarly during inference for 10 diverse, experimentally observed, microbial population growth curves (see Table~\ref{tab:MSE}) which shows that, in practise, our fast approximations are as good as ``exact'' methods.
We conclude that our LNA models are preferable to the RRTR for modelling QFA data. 

It is interesting to note that, although the LNAA is not a better approximation of the original SGLM process than the LNAM, it is still quite reasonable. Figures~\ref{4nonu}A~and~\ref{4nonu}D show that the SLGM and LNAA processes are visually similar. Figure~\ref{4nonu}E demonstrates that forward trajectories of the LNAA also share similar levels of variation over time with the SLGM and LNAM.  

Fast inference with the LNAA gives us the potential to develop large hierarchical Bayesian models for genome-wide QFA datasets, using a diffusion equation and realistic computational resources

Here, we have concentrated on a biological model of population growth.  However, we expect that the approach we have demonstrated: generating linear noise approximations of stochastic processes to allow fast Bayesian inference with Kalman filtering for marginal likelihood computation, will be useful in a wide range of other applications where simulation is prohibitively slow.
\\
\\	
Further work involves extending the Bayesian hierarchical models in Chapter~\ref{cha:modelling_den_int} with the approximate stochastic logistic growth models and methods for carrying out inference described in Chapter~\ref{cha:stochastic_app}. 
By accounting for the random fluctuations within the logistic growth data we will be able to improve our logistic growth parameter estimates.

We have demonstrated how to incorporate a batch effect or a transformation effect to the joint hierarchical model in Section~\ref{sec:fur_case_stu}.
Introducing a batch or transformation effect into our models will allow us to capture even further experimental variation.
Fitness plots for further case studies given in Section~\ref{sec:fur_case_stu} and extensions of the joint hierarchical model given in Section~\ref{sec:batch_eff} are included for experimental biologists to investigate further.

A related experiment to the QFA screen comparison analysed within this thesis is the ``all-by-all'' QFA experiment (in early development at the time of writing). 
The ``all-by-all'' QFA experiment begins with a control plate consisting of $N$ \emph{orf}$\Delta$s. For each of the $N$ \emph{orf}$\Delta$s a new query plate is created, each query plate consists of the control plate and an additional background mutation related to one of the $N$ \emph{orf}$\Delta$s. 
In total there will be $N+1$ unique plates (including the control plate).
Where a standard QFA comparison looks for genes that interact with a single query mutation (or condition), the ``all-by-all'' QFA experiment aims to find genetic interactions for multiple query mutations ($N$) simultaneously.
The ``all-by-all'' experiment therefore incorporates more information and investigates more potential genetic interactions than a standard QFA comparison.
We expect that the Bayesian hierarchical modelling and genetic interaction modelling developed in this thesis will be used to create models for describing the ``all-by-all'' QFA experiment as well as many other similar experiments in the future.

By improving our software we may be able to reduce computational time for inference.
Currently the code for implementing the Bayesian models described in this thesis is written in the C programming language which can be run as standalone software or through an R package ``qfaBayes'', available at \sloppy\url{https://r-forge.r-project.org/projects/qfa}.\sloppy
The computational speed of our C code used for inference could be improved by parallel implementation, taking advantage of a multi-core processor to carry out tasks simultaneously.
With faster computational times we expect to reduce the time for a typical QFA comparison with the JHM from $\sim$2 weeks to less than a week.

Currently the information available on true genetic interactions and biological processes in yeast is limited and so we rely on objective analyses such as simulation studies to give unbiased comparisons between the approaches considered.
The biological processes of many genes in the yeast genome are yet to be identified so we are unable to use GO term enrichment analysis as a ``gold standard'' for comparing the results of our approaches.
Information used to build a gene ontology is typically well known and taken from well understood experiments, we expect that subtle genetic interactions which we are interested in finding will have little information available.
QFA screen comparisons are designed to learn biology which is not already fully understood and so a biological comparison between the different approaches considered is difficult.
Simulation studies (see Section~\ref{sim_study}) give us the ability to compare the different approaches and the effects of modelling more experimental structure.

A typical QFA comparison is a large and complex data set corresponding to around 400,000 time series, posing considerable computational, as well as statistical challenges. 
With a Bayesian approach we are able to evaluate complex hierarchical models to better reflect the structure or design of genome-wide QFA experiments. 
Bayesian variable selection methods embedded within a large hierarchical model allow us to describe genetic interaction and use prior information to incorporate physical and biological constraints within our models.
We have shown that Bayesian hierarchical modelling of large and complex data gives us the advantage of increased modelling flexibility compared to a frequentist approach, allowing us to better describe the experimental structure or design.
For the reasons above, a QFA screen comparison or any other highly structured experimental dataset is better modelled using a Bayesian hierarchical modelling approach when compared to an alternative frequentist approach.
\\
\\
Overall this thesis presents improved modelling approaches to the current non-hierarchical frequentist approach for a QFA screen comparison.
The research contained in this thesis illustrates how Bayesian inference gives us further modelling flexibility, allowing us to better describe the known experimental structure. 
Further, our modelling approaches and assumptions are transferable outside QFA screen experiments where we wish to capture as much experimental structure as possible.
The results from our temperature sensitive \emph{cdc13-1} QFA experiment results will give further insight to the telomere and consequentially aging and cancer in yeast and potentially the human genome \citep{yeastorg}.
\end{chapter}
\clearpage

  \appendix

\titleformat{\chapter}{\centering\fontsize{16pt}{1em}\normalfont\bfseries}{Appendix \thechapter.}{0.5em}{}

\chapter{QFA data set sample, solving the logistic growth model and random effects model R code\label{cha:appendix_A}}

\section{\label{app:QFA_set_sam}\emph{cdc13-1} Quantitative Fitness Analysis data set sample}

\begin{figure}[h!]
  \centering
\includegraphics[width=14cm]{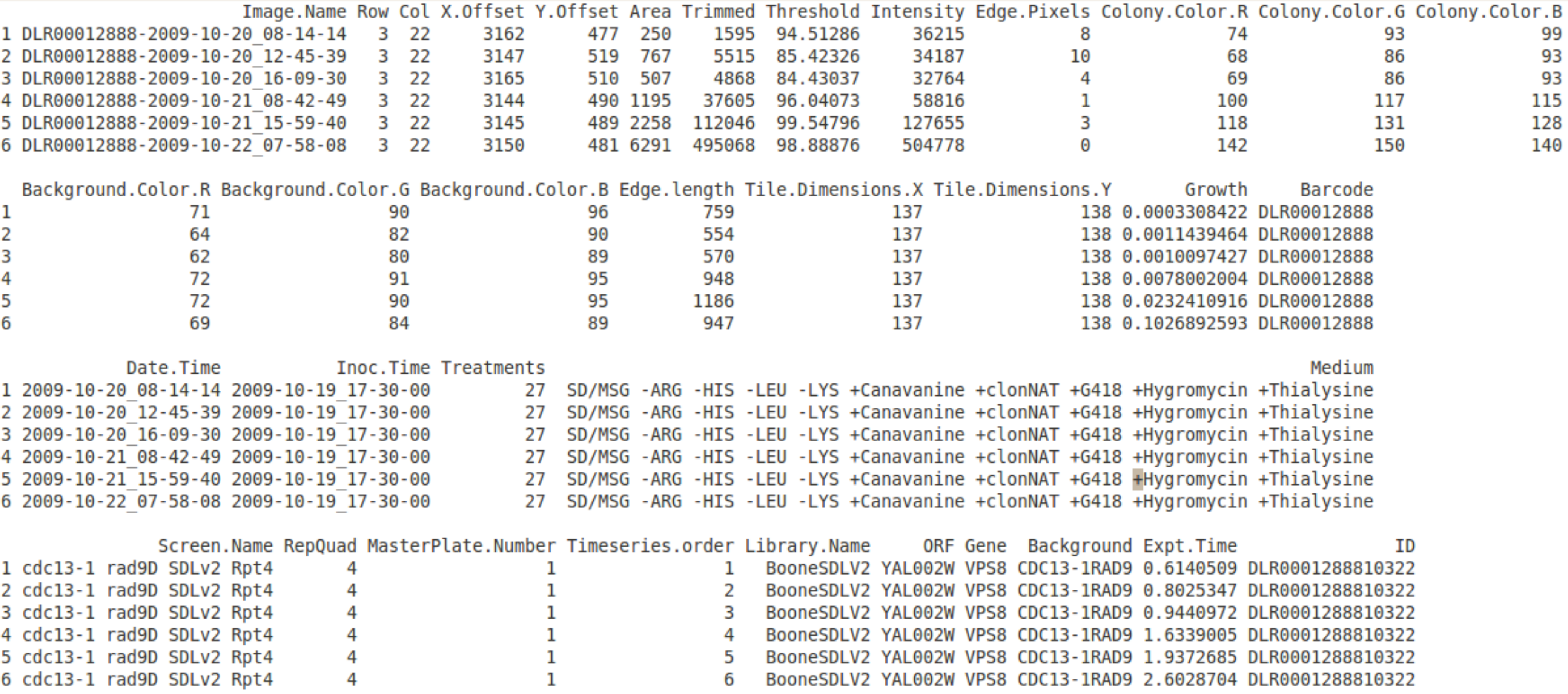}
\caption[\emph{cdc13-1} QFA data set sample]{\emph{cdc13-1} Quantitative Fitness Analysis data set sample. 
Notable columns include ``ORF", ``Expt.Time" and ``Growth". ``ORF" indicates which $\emph{orf}\Delta$ strain the row corresponds to. ``Expt.Time" indicates the time in days from the $\emph{orf}\Delta$ strain being spotted \citep{QFA1}. 
``Growth" gives an adjusted measure of cell culture density from the image analysis for a given $\emph{orf}\Delta$ strain and time point.
Generated from Colonyzer output files with the qfa R package, freely available at \url{http://qfa.r-forge.r-project.org/}. \label{app:appendix_label}
}
\end{figure}
\clearpage

\section{\label{app:solving_log_gro}Solving the logistic growth model}  

The solution to the logistic growth ODE (\ref{eq_det}) can be obtained as follows.
First we factor the right side of (\ref{eq_det}) and rearrange to give:
			\begin{equation*}
	\frac{d{x(t)}}{{x(t)}\left(1 - \frac{{x(t)}}{K}\right)}=rdt.
				\end{equation*}
We now rearrange further using a partial fractions expansion and integrate over both sides of the equation:			
\begin{equation}\label{eq:log_int}
	\int\frac{d{x(t)}}{{x(t)}}+\frac{\frac{1}{K}d{x(t)}}{\left(1 - \frac{{x(t)}}{K}\right)}=\int rdt.
\end{equation}
Integrating the first component on the left side of (\ref{eq:log_int}) we obtain the following, where $c_1$ is an unknown constant:
				\begin{equation*}
	\int\frac{d{x(t)}}{{x(t)}}=\log({x(t)})+c_1.
	\end{equation*}
	Integrating the second component on the left side of (\ref{eq:log_int}) we obtain the following, where $c_2$ is an unknown constant:
						\begin{equation*}
	\frac{1}{K}\int\frac{d{x(t)}}{1-\frac{{x(t)}}{K}}=-\log(1-\frac{{x(t)}}{K})+c_2.
		\end{equation*}
Integrating the right side of (\ref{eq:log_int}) we obtain the following, where $c_3$ is an unknown constant:
						\begin{equation*}
\int rdt=rt+c_3.
		\end{equation*}
Solving the integrals in (\ref{eq:log_int}) we obtain the following, where $c_4=c_3-c_1-c_2$ is an unknown constant:
						\begin{equation*}
\log\left(\frac{{x(t)}}{1-\frac{{x(t)}}{K}}\right)=rt+c_4.
		\end{equation*}
					Rearranging our equation, we obtain the following:
											\begin{equation*}
  \frac{{x(t)}}{1-\frac{{x(t)}}{K}}=e^{rt+c_4}.
			\end{equation*}
We now apply initial conditions, $P={x_{0}}$ and rearrange to obtain an expression for $c_4$:
											\begin{equation*}
	  c_4=\log\left(\frac{P}{1-\frac{P}{K}}\right). 
					\end{equation*}
		We now substitute in our expression for $c_4$ to give:
													\begin{equation*}
\log\left(\frac{{x(t)}}{1-\frac{{x(t)}}{K}}\right)=rt+\log\left(\frac{P}{1-\frac{P}{K}}\right)
					\end{equation*}
			Finally, we rearrange to give (\ref{eq:logistic}).
		\clearpage	
\section{Random effects model R code\label{app:remcode}}
{\fontsize{7.6}{7.6}\selectfont
\begin{verbatim}
library(lme4) #http://cran.r-project.org/web/packages/lme4/index.html
#http://research.ncl.ac.uk/colonyzer/AddinallQFA/Logistic.zip and extract zip file
#alternatively http://research.ncl.ac.uk/colonyzer/AddinallQFA/ 
#"Table S8 Logistic Output Files - 36MB .zip file"
aa<-read.delim("cSGA_v2_r1_Logistic.txt",header=T,skip=1,sep="\t")
#...
bb<-read.delim("Adam_cdc13-1_SDLV2_REP1_Logistic.txt",header=T,skip=0,sep="\t")
#...
aa<-aa[aa$Treatments==27,]
bb<-bb[bb$Treatments==27,]
aa<-aa[!aa$Row==1,]
aa<-aa[!aa$Row==16,]
aa<-aa[!aa$Col==1,]
aa<-aa[!aa$Col==24,]
bb<-bb[!bb$Row==1,]
bb<-bb[!bb$Row==16,]
bb<-bb[!bb$Col==1,]
bb<-bb[!bb$Col==24,]

ORFuni=ORFuni_a=unique(aa$ORF)
ORFuni_b=unique(bb$ORF)
L=length(ORFuni_a)
NoORF_a=NoORF_b=aaa=bbb=numeric()
for (i in 1:L){
 NoORF_a[i]=nrow(aa[aa$ORF==ORFuni[i],])
 NoORF_b[i]=nrow(bb[bb$ORF==ORFuni[i],])
 aaa<-rbind(aaa,aa[aa$ORF==ORFuni[i],])
 bbb<-rbind(bbb,bb[bb$ORF==ORFuni[i],])
}
a=b=numeric(0)
K_lm=aaa$Trimmed.K
P_a=43
r_lm=aaa$Trimmed.r
for (i in 1:length(r_lm)){
 if(K_lm[i]<=2*P_a){K_lm[i]=2*P_a+0.01;r_lm[i]=0;}
 a[i]=(r_lm[i]/log(2*max(0,K_lm[i]-P_a)/max(0,K_lm[i]-2*P_a)))*(log(K_lm[i]/P_a)/log(2));
}
K_lmb=bbb$Trimmed.K
P_b=43
r_lmb=bbb$Trimmed.r
for (i in 1:length(r_lmb)){
 if(K_lmb[i]<=2*P_b){K_lmb[i]=2*P_b+0.01;r_lmb[i]=0;}
 b[i]=(r_lmb[i]/log(2*max(0,K_lmb[i]-P_b)/max(0,K_lmb[i]-2*P_b)))*(log(K_lmb[i]/P_b)/log(2));
}

condition<-factor(c(rep("a",length(a)),rep("b",length(b))))
subject=numeric()
for (i in 1:L){
 subject=c(subject,rep(i,NoORF_a[i]))
}
for (i in 1:L){
 subject=c(subject,rep(i,NoORF_b[i]))
}
subcon=subject
subcon[1:length(a)]=0
subcon<-factor(subcon)
subject<-factor(subject)
f=c(a,b)
data=data.frame(f,subject,condition,subcon)
data$lf=log(data$f+1)
data$subcon<-C(data$subcon,sum)
bk<-contrasts(data$subcon)
contrasts(data$subcon)=bk[c(nrow(contrasts(data$subcon)),1:(nrow(contrasts(data$subcon))-1)),]
model1<-lmer(lf~subcon+(1|subject),data=(data),REML=F)
\end{verbatim}
}
\clearpage
			
\chapter{Bayesian hierarchical modelling\label{cha:appendix}}
\section{Hyper-parameter values for Bayesian hierarchical modelling}

\begin{table}[h!]
\caption[Hyper-parameter values for Bayesian hierarchical modelling of quantitative fitness analysis data]{
Hyper-parameter values for Bayesian hierarchical modelling of quantitative fitness analysis data. Hyper-parameter values for the separate hierarchical model (SHM), interaction hierarchical model (IHM) and joint hierarchical model (JHM) are provided. \label{tab:SHM_priors}}
\centering 
\resizebox{!}{0.9in}{%
\npdecimalsign{.}
\nprounddigits{2}
\centering 
\begin{tabular}{c n{2}{2} c n{2}{2} c n{2}{2} c n{2}{2} c n{2}{2}}
    \hline
\noalign{\vskip 0.4mm} 
		\multicolumn{2}{c}{SHM \& JHM} &\multicolumn{2}{c}{SHM \& JHM}& \multicolumn{2}{c}{JHM} & \multicolumn{2}{c}{IHM} & \multicolumn{2}{c}{JHM-B \& JHM-T}\\
Parameter Name  & \multicolumn{1}{c}{Value} & Parameter Name  & \multicolumn{1}{c}{Value} & Parameter Name  & \multicolumn{1}{c}{Value}& Parameter Name  & \multicolumn{1}{c}{Value} &Parameter Name  & \multicolumn{1}{c}{Value}\\ \hline
$\tau^{K,\mu}$ & 2.20064039227566   & $\eta^{r,p}$ & 0.133208648543871 &$\alpha^{\mu}$ & 0 & $Z_{\mu}$ & 3.65544229414228 &$\kappa^p$ & 0\\ 
$\eta^{\tau,K,p}$ & 0.0239817523340161  &       $\nu^{\mu}$ & 19.8220570630669 &  $\eta^{\alpha}$ & 0.25 & $\eta^{Z,p}$ & 0.697331530063874 & $\eta^\kappa$& 1.166666666666\\  
 $\eta^{K,o}$ & -0.79421175992029 & $\eta^{\nu,p}$ & 0.0174869367984725 &$\beta^{\mu}$ & 0 &  $\eta^{Z}$ & 0.104929506383255 &$\lambda^p$ & 0\\ 
  $\psi^{K,o}$ & 0.610871036009521 &  $P^{\mu}$ & -9.03928728018792 &  $\eta^{\beta}$ & 0.25  &   $\psi^{Z}$ & 0.417096744759774 & $\eta^\lambda$& 1.166666666666\\ 
     $\tau^{r,\mu}$ &  3.64993037268256 &  $\eta^{P}$ & 0.469209463148874 &  $p$ & 0.05 & $\eta^{\nu}$ & 0.101545024587153 & $\phi^{shape}$ & 100\\   
 $\eta^{\tau,r,p}$ &  0.0188443648965434 &&&$\eta^{\gamma}$ & -0.79421175992029 &  $\psi^{\nu}$ & 2.45077729037385 & $\phi^{scale}$ & 0.01\\ 
$\eta^{r,o}$ & 0.468382435659566  &&& $\psi^{\gamma}$ & 0.610871036009521 &    $\nu^{\mu}$ & 2.60267545154548 & $\chi^{shape}$ & 100\\  
 $\psi^{r,o}$ & 0.0985295312016232  &&&  $\eta^{\omega}$ & 0.468382435659566 &    $\eta^{\nu,p}$ &0.0503202367841729 & $\chi^{scale}$ & 0.01\\ 
  $\eta^{\nu}$ & -0.834166609695065  &&& $\psi^{\omega}$ & 0.0985295312016232 &    $\alpha^{\mu}$ & 0 &&\\ 
 $\psi^{\nu}$ & 0.855886535578262  &&& $\eta^{\tau,K}$ &  2.20064039227566&    $\eta^{\alpha}$ & 0.309096075088720 &&\\   
  $K^{\mu}$ & -2.01259579112252  &&&   $\psi^{\tau,K}$ & 0.0239817523340161 &    $p$ & 0.05 &&\\ 
 $\eta^{K,p}$ & 0.032182397822033  &&& $\eta^{\tau,r}$ & 3.64993037268256 & $\eta^{\gamma}$ & 0.104929506383255&& \\ 
    $r^{\mu}$ & 0.97398228941848  &&&   $\psi^{\tau,r}$ & 0.0188443648965434 &  $\psi^{\gamma}$ & 0.417096744759774&& \\  
    \hline
    \end{tabular}
    \npnoround
		}
\end{table}

\clearpage

\section{\label{app:GO_fit}\emph{cdc13-1}~$\boldsymbol{{27}^{\circ}}$C~vs~\emph{ura3}$\Delta$~$\boldsymbol{{27}^{\circ}}$C fitness plots with gene ontology terms highlighted}
\begin{figure}[h!]
  \centering
\includegraphics[width=14cm]{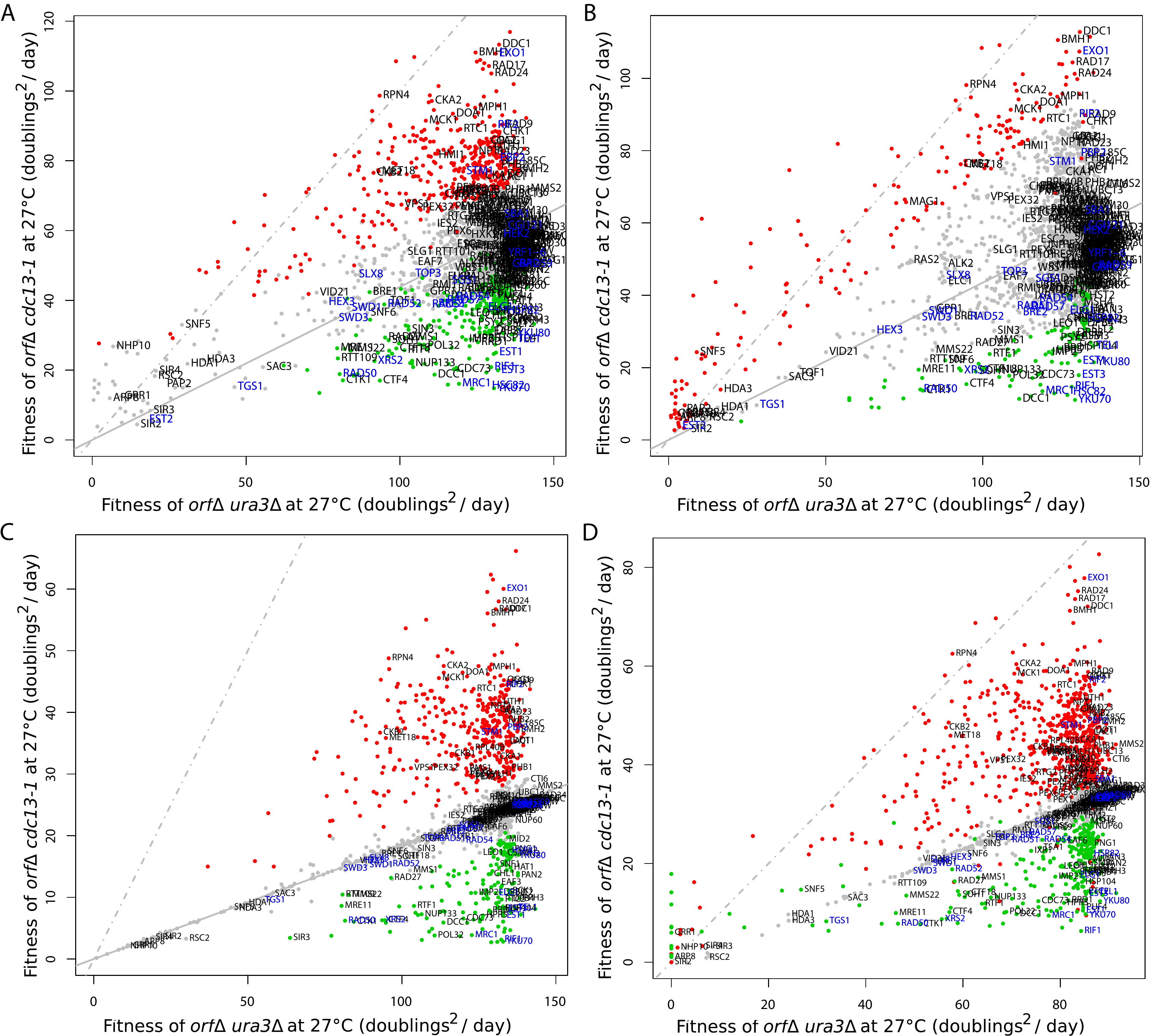}
\caption[Alternative fitness plots with \emph{orf}$\Delta$ posterior mean fitnesses and labels for the ``telomere maintenance'' gene ontology term]{Alternative fitness plots with \emph{orf}$\Delta$ posterior mean fitnesses. Labels for the ``telomere maintenance'' gene ontology term are highlighted in blue.
A) Non-Bayesian, non-hierarchical fitness plot, based on Table~S6 from Addinall et al. (2011) $(F=MDR\times MDP)$.
B) Non-Bayesian, hierarchical fitness plot, \hl{from fitting REM to data} in Table~S6 from Addinall et al. (2011) $(F=MDR\times MDP)$.
C) IHM fitness plot with $\emph{orf}\Delta$ posterior mean fitness $(F=MDR\times MDP)$.
D) JHM fitness plot with $\emph{orf}\Delta$ posterior mean fitnesses.
$\emph{orf}\Delta$ strains are classified as being a suppressor or enhancer based on analysis of growth parameter $r$.
Further fitness plot explanation and notation is given in Figure~\ref{fig:old}.
}
\end{figure}
\begin{figure}[h!]
  \centering
\includegraphics[width=14cm]{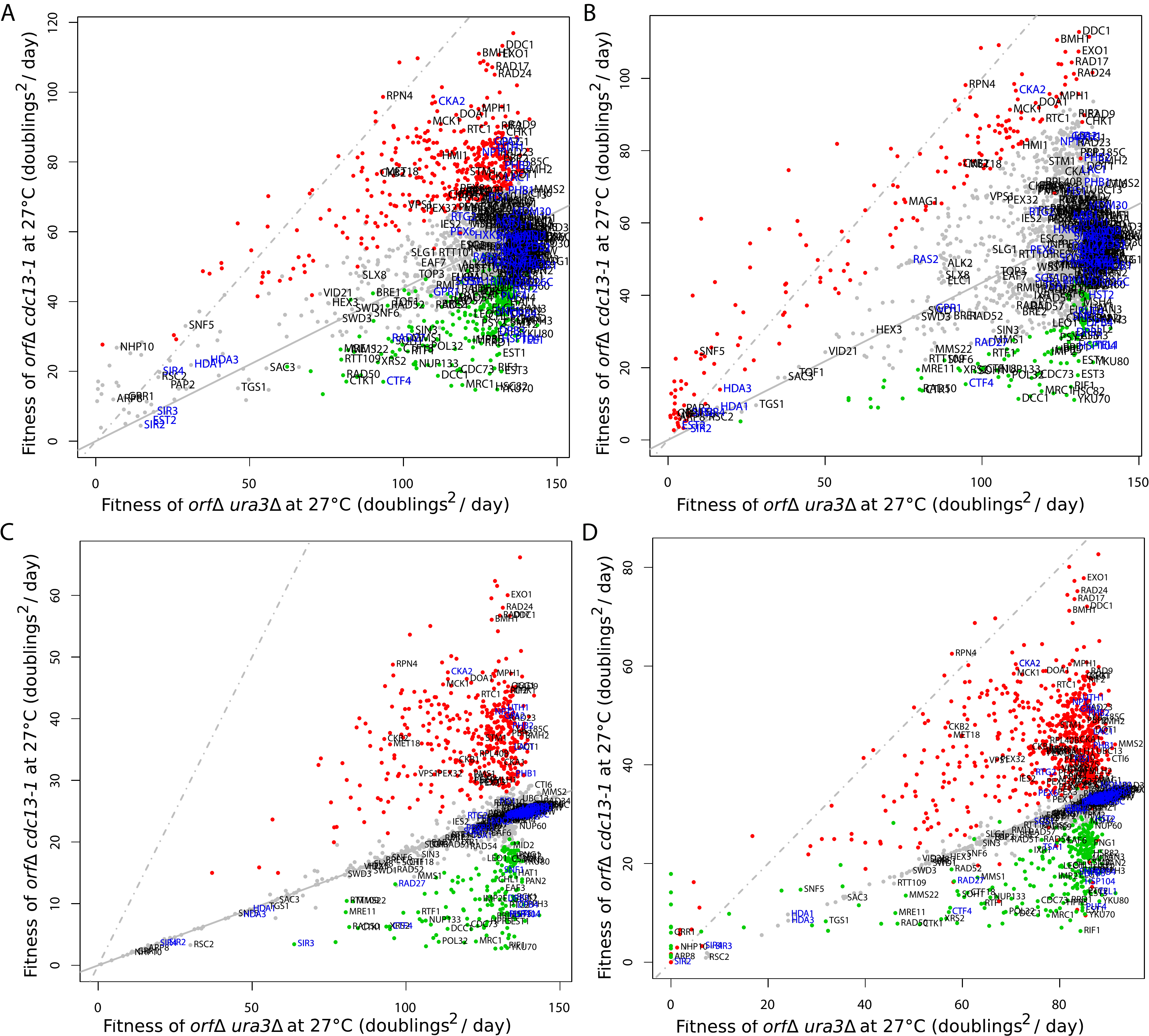}
\caption[Alternative fitness plots with \emph{orf}$\Delta$ posterior mean fitnesses and labels for the ``ageing'' gene ontology term]{Alternative fitness plots with $\emph{orf}\Delta$ posterior mean fitnesses. Labels for the ``ageing'' gene ontology term are highlighted in blue.
A) Non-Bayesian, non-hierarchical fitness plot, based on Table~S6 from Addinall et al. (2011) $(F=MDR\times MDP)$.
B) Non-Bayesian, hierarchical fitness plot, \hl{from fitting REM to data} in Table~S6 from Addinall et al. (2011) $(F=MDR\times MDP)$.
C) IHM fitness plot with $\emph{orf}\Delta$ posterior mean fitness $(F=MDR\times MDP)$.
D) JHM fitness plot with $\emph{orf}\Delta$ posterior mean fitnesses.
$\emph{orf}\Delta$ strains are classified as being a suppressor or enhancer based on analysis of growth parameter $r$.
Further fitness plot explanation and notation is given in Figure~\ref{fig:old}.
}
\end{figure}
\begin{figure}[h!]
  \centering
\includegraphics[width=14cm]{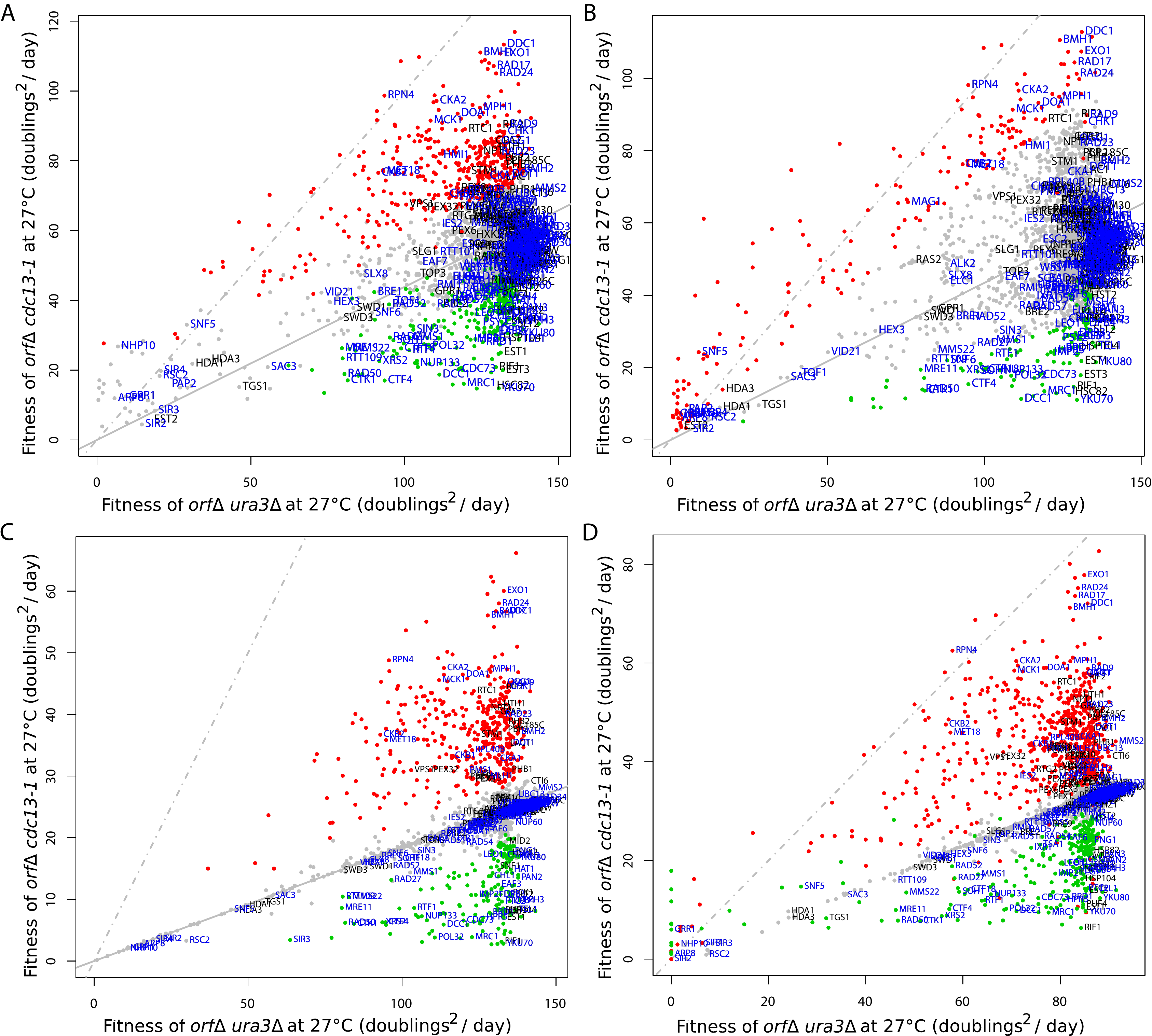}
\caption[Alternative fitness plots with \emph{orf}$\Delta$ posterior mean fitnesses and labels for the ``response to DNA damage'' gene ontology term]{Alternative fitness plots with $\emph{orf}\Delta$ posterior mean fitnesses. Labels for the ``response to DNA damage'' gene ontology term are highlighted in blue.
A) Non-Bayesian, non-hierarchical fitness plot, based on Table~S6 from Addinall et al. (2011) $(F=MDR\times MDP)$.
B) Non-Bayesian, hierarchical fitness plot, \hl{from fitting REM to data} in Table~S6 from Addinall et al. (2011) $(F=MDR\times MDP)$.
C) IHM fitness plot with $\emph{orf}\Delta$ posterior mean fitness $(F=MDR\times MDP)$.
D) JHM fitness plot with $\emph{orf}\Delta$ posterior mean fitnesses.
$\emph{orf}\Delta$ strains are classified as being a suppressor or enhancer based on analysis of growth parameter $r$.
Further fitness plot explanation and notation is given in Figure~\ref{fig:old}.
}
\end{figure}
\begin{figure}[h!]
  \centering
\includegraphics[width=14cm]{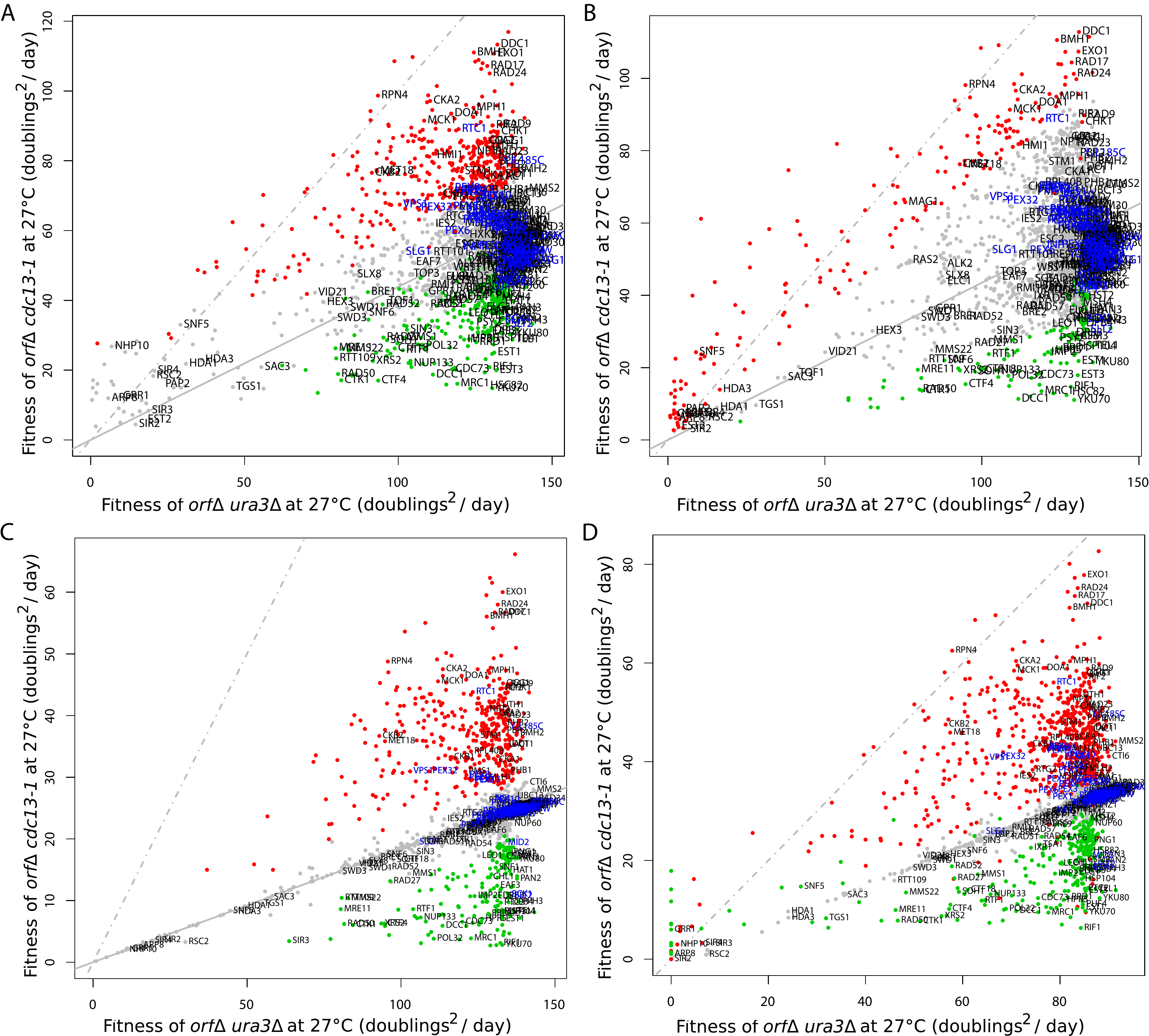}
\caption[Alternative fitness plots with \emph{orf}$\Delta$ posterior mean fitnesses and labels for the ``peroxisomal organisation'' gene ontology term]{Alternative fitness plots with $\emph{orf}\Delta$ posterior mean fitnesses. Labels for the ``peroxisomal organisation'' gene ontology term are highlighted in blue.
A) Non-Bayesian, non-hierarchical fitness plot, based on Table~S6 from Addinall et al. (2011) $(F=MDR\times MDP)$.
B) Non-Bayesian, hierarchical fitness plot, \hl{from fitting REM to data} in Table~S6 from Addinall et al. (2011) $(F=MDR\times MDP)$.
C) IHM fitness plot with $\emph{orf}\Delta$ posterior mean fitness $(F=MDR\times MDP)$.
D) JHM fitness plot with $\emph{orf}\Delta$ posterior mean fitnesses.
$\emph{orf}\Delta$ strains are classified as being a suppressor or enhancer based on analysis of growth parameter $r$.
Further fitness plot explanation and notation is given in Figure~\ref{fig:old}.
}
\end{figure}

\clearpage
\section{\label{app:interactions}Lists of top genetic interactions for the two-stage and one-stage Bayesian approaches}
\begin{table}[h!]
\caption{Sample of interaction hierarchical model top genetic interactions with \emph{cdc13-1} at $\boldsymbol{{27}^{\circ}}$C\label{app:IHM_interactions}}
\centering
\resizebox{14cm}{!}{%
\npdecimalsign{.}
\nprounddigits{2}
\begin{tabular}{ c c n{2}{2} n{2}{2} c}
\hline
\emph{Type of}  & \emph{Gene} & \multicolumn{1}{c}{\emph{Probability of}} & \multicolumn{1}{c}{\emph{Strength of}}& \emph{Position in} \\ \emph{Interaction} & \emph{Name}  & \multicolumn{1}{c}{\emph{Interaction} ${\delta}_{l}$} & \multicolumn{1}{c}{\emph{Interaction} \emph{$e^{({\delta}_{l}{\gamma}_{l})}$}} & \emph{Addinall (2011)} \\ 
\hline
Suppressor & IPK1 & 1.00 & 2.87 & 10 \\ 
   & LST4 & 1.00 & 2.77 & 13 \\ 
   & RPN4 & 1.00 & 2.76 & 17 \\ 
   & MTC5 & 1.00 & 2.66 & 20 \\ 
   & GTR1 & 1.00 & 2.64 & 38 \\ 
   & NMD2 & 1.00 & 2.62 & 3 \\ 
   & SAN1 & 1.00 & 2.62 & 16 \\ 
   & UPF3 & 1.00 & 2.58 & 21 \\ 
   & RPL37A & 1.00 & 2.56 & 121 \\ 
   & NAM7 & 1.00 & 2.53 & 22 \\ 
   & RPP2B & 1.00 & 2.52 & 120 \\ 
   & YNL226W & 0.99 & 2.49 & 126 \\ 
   & YGL218W & 1.00 & 2.46 & 250 \\ 
   & MEH1 & 1.00 & 2.45 & 45 \\ 
   & ARO2 & 1.00 & 2.45 & 68 \\ 
   & EXO1 & 1.00 & 2.45 & 1 \\ 
   & BUD27 & 1.00 & 2.43 & 46 \\ 
   & RAD24 & 1.00 & 2.39 & 4 \\ 
   & RPL16B & 1.00 & 2.39 & 33 \\ 
   & RPL43A & 1.00 & 2.39 & 150 \\ 
\hline\noalign{\vskip 0.5mm} 
Enhancer  & :::MRC1 & 1.00 & 0.11 & 35 \\ 
   & YKU70 & 1.00 & 0.11 & 31 \\ 
   & STI1 & 1.00 & 0.11 & 42 \\ 
   & RIF1 & 1.00 & 0.13 & 36 \\ 
   & ELP3 & 1.00 & 0.16 & 82 \\ 
   & CLB5 & 1.00 & 0.17 & 58 \\ 
   & MRC1 & 1.00 & 0.17 & 63 \\ 
   & DPH2 & 1.00 & 0.18 & 24 \\ 
   & POL32 & 1.00 & 0.19 & 113 \\ 
   & MAK31 & 1.00 & 0.19 & 37 \\ 
   & SWM1 & 1.00 & 0.20 & 25 \\ 
   & LTE1 & 1.00 & 0.21 & 48 \\ 
   & MAK10 & 1.00 & 0.22 & 44 \\ 
   & ELP2 & 1.00 & 0.22 & 77 \\ 
   & PAT1 & 1.00 & 0.24 & 144 \\ 
   & DPH1 & 1.00 & 0.25 & 55 \\ 
   & SRB2 & 0.99 & 0.25 & 174 \\ 
   & THP2 & 1.00 & 0.26 & 67 \\ 
   & MFT1 & 1.00 & 0.26 & 52 \\ 
   & LSM6 & 0.97 & 0.26 & 389 \\ 
\hline
\multicolumn{5}{c}{See \url{http://research.ncl.ac.uk/qfa/HeydariQFABayes/IHM_strip.txt} for the full list.}
\end{tabular}
\npnoround%
}
\end{table}

\begin{table}
\caption{Sample of joint hierarchical model top genetic interactions with \emph{cdc13-1} at $\boldsymbol{{27}^{\circ}}$C\label{app:JHM_interactions}}
\centering
\resizebox{\columnwidth}{!}{%
\npdecimalsign{.}
\nprounddigits{2}
\begin{tabular}{ c c n{2}{2} n{2}{2} n{2}{2} n{2}{2} c}
\hline
\emph{Type of}  & \emph{Gene} & \multicolumn{1}{c}{\emph{Probability of}} & \multicolumn{1}{c}{\emph{Strength of}} & \multicolumn{1}{c}{\emph{Strength of}} & \emph{Strength of} & \multicolumn{1}{c}{\emph{Position in}} \\ 
 \emph{Interaction} & \emph{Name} & \multicolumn{1}{c}{\emph{Interaction}} & \multicolumn{1}{c}{\emph{Interaction}} & \multicolumn{1}{c}{\emph{Interaction}} & \emph{Interaction} & \emph{Addinall (2011)} \\ 
 & &\multicolumn{1}{c}{${\delta}_l$} & \multicolumn{1}{c}{ \emph{$e^{({\delta}_{l}{\gamma}_l)}$}} & \multicolumn{1}{c}{\emph{$e^{({\delta}_{l}{\omega}_{l})}$}} & \emph{$MDR \times MDP$} & \\
\hline
Suppressor & CSE2 & 1.00 & 490.51 & 0.48 & 11.71 & 838 \\ 
in K   & SGF29 & 1.00 & 273.69 & 0.68 & 14.16 & 580 \\ 
   & GSH1 & 1.00 & 78.79 & 0.92 & 17.89 & 281 \\ 
   & YMD8 & 1.00 & 59.31 & 0.65 & 7.05 & 2022 \\ 
   & YGL024W & 1.00 & 28.13 & 1.18 & 13.33 & 151 \\ 
   & RPS9B & 1.00 & 24.67 & 1.12 & 10.24 & 801 \\ 
   & GRR1 & 1.00 & 22.51 & 0.67 & 5.99 & 1992 \\ 
\hline\noalign{\vskip 0.5mm} 
Suppressor   & BTS1 & 1.00 & 19.27 & 2.29 & 19.65 & 201 \\ 
in r   & IPK1 & 1.00 & 5.56 & 2.26 & 44.81 & 10 \\ 
   & NMD2 & 1.00 & 2.96 & 2.19 & 48.51 & 3 \\ 
   & SAN1 & 1.00 & 2.37 & 2.17 & 48.70 & 16 \\ 
   & LST4 & 1.00 & 5.79 & 2.14 & 44.14 & 13 \\ 
   & RPN4 & 1.00 & 8.00 & 2.12 & 40.46 & 17 \\ 
   & UPF3 & 1.00 & 3.16 & 2.07 & 45.25 & 21 \\ 
\hline\noalign{\vskip 0.5mm} 
Suppressor in   & SAN1 & 1.00 & 2.37 & 2.17 & 48.70 & 16 \\ 
 $MDR\times MDP$  & NMD2 & 1.00 & 2.96 & 2.19 & 48.51 & 3 \\ 
   & UPF3 & 1.00 & 3.16 & 2.07 & 45.25 & 21 \\ 
   & EXO1 & 1.00 & 2.89 & 2.06 & 45.04 & 1 \\ 
   & IPK1 & 1.00 & 5.56 & 2.26 & 44.81 & 10 \\ 
   & LST4 & 1.00 & 5.79 & 2.14 & 44.14 & 13 \\ 
   & NAM7 & 1.00 & 3.02 & 2.04 & 43.00 & 22 \\
\hline\noalign{\vskip 0.5mm} 
Enhancer   & YKU70 & 1.00 & 0.01 & 1.09 & -23.44 & 31 \\ 
 in K  & STI1 & 1.00 & 0.01 & 1.20 & -21.60 & 42 \\ 
   & RIF1 & 1.00 & 0.01 & 0.63 & -26.17 & 36 \\ 
   & :::MRC1 & 1.00 & 0.01 & 0.83 & -23.15 & 35 \\ 
   & MAK31 & 1.00 & 0.02 & 1.18 & -18.19 & 37 \\ 
   & CLB5 & 1.00 & 0.02 & 0.87 & -19.54 & 58 \\ 
   & MRC1 & 1.00 & 0.02 & 0.81 & -20.40 & 63 \\ 
\hline\noalign{\vskip 0.5mm} 
Enhancer   & PAT1 & 1.00 & 1.71 & 0.28 & -18.30 & 144 \\ 
 in r  & PUF4 & 1.00 & 2.00 & 0.31 & -21.61 & 34 \\ 
   & YKU80 & 1.00 & 2.15 & 0.33 & -21.68 & 32 \\ 
   & RTT103 & 1.00 & 2.54 & 0.34 & -17.87 & 153 \\ 
   & LSM1 & 0.99 & 2.13 & 0.34 & -16.20 & 101 \\ 
   & GIM3 & 0.99 & 0.93 & 0.35 & -19.70 & 132 \\ 
   & INP52 & 0.96 & 0.86 & 0.36 & -14.50 & 345 \\

\hline\noalign{\vskip 0.5mm} 
Enhancer in   & RIF1 & 1.00 & 0.01 & 0.63 & -26.17 & 36 \\ 
  $MDR\times MDP$  & LTE1 & 1.00 & 0.06 & 0.40 & -23.96 & 48 \\ 
   & YKU70 & 1.00 & 0.01 & 1.09 & -23.44 & 31 \\ 
   & :::MRC1 & 1.00 & 0.01 & 0.83 & -23.15 & 35 \\ 
   & DPH2 & 1.00 & 0.04 & 0.56 & -23.11 & 24 \\ 
   & EST1 & 1.00 & 0.12 & 0.46 & -22.20 & 5 \\ 
   & MAK10 & 1.00 & 0.04 & 0.59 & -21.92 & 44 \\ 

\hline
\multicolumn{7}{c}{See \url{http://research.ncl.ac.uk/qfa/HeydariQFABayes/JHM_strip.txt} for the full list.}
\end{tabular}
\npnoround%
}
\end{table}
 
\clearpage

\section{\label{app:alternative_JHM}\emph{cdc13-1}~$\boldsymbol{{27}^{\circ}}$C~vs~\emph{ura3}$\Delta$~$\boldsymbol{{27}^{\circ}}$C fitness plots for the joint hierarchical model in terms of carrying capacity and growth rate parameters}
\begin{figure}[h!]
  \centering
\includegraphics[width=14cm]{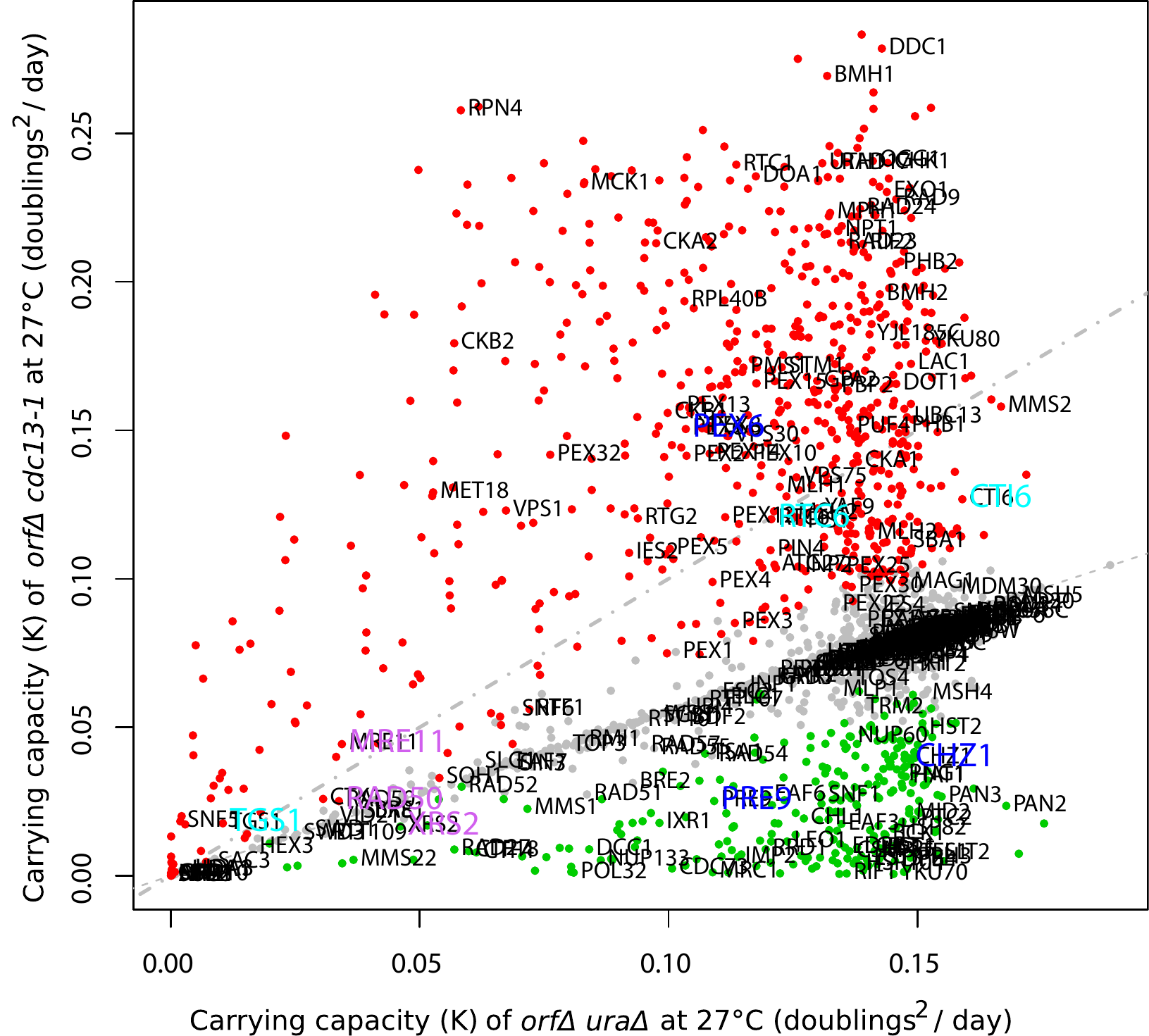}
\caption[Joint hierarchical model carrying capacity fitness plot]{Joint hierarchical model (JHM) carrying capacity fitness plot with $\emph{orf}\Delta$ posterior mean fitnesses.
$\emph{orf}\Delta$ strains are classified as being a suppressor or enhancer based on carrying capacity parameter $K$.
Significant interactors have posterior probability $\Delta>0.5$.
To compare fitness plots, labelled genes are those belonging to the following gene ontology terms in Table~\ref{tab:sup_enh}: ``telomere maintenance'', ``ageing'', ``response to DNA damage stimulus'' or ``peroxisomal organization'', as well as the genes identified as interactions only in $K$ with the JHM (see Figure~\ref{fig:JHM_only}) (blue), genes interacting only in $r$ with the JHM (cyan) and the MRX complex genes (pink).
\label{fig:JHM_K}
}
\end{figure}

\begin{figure}[h!]
  \centering
\includegraphics[width=14cm]{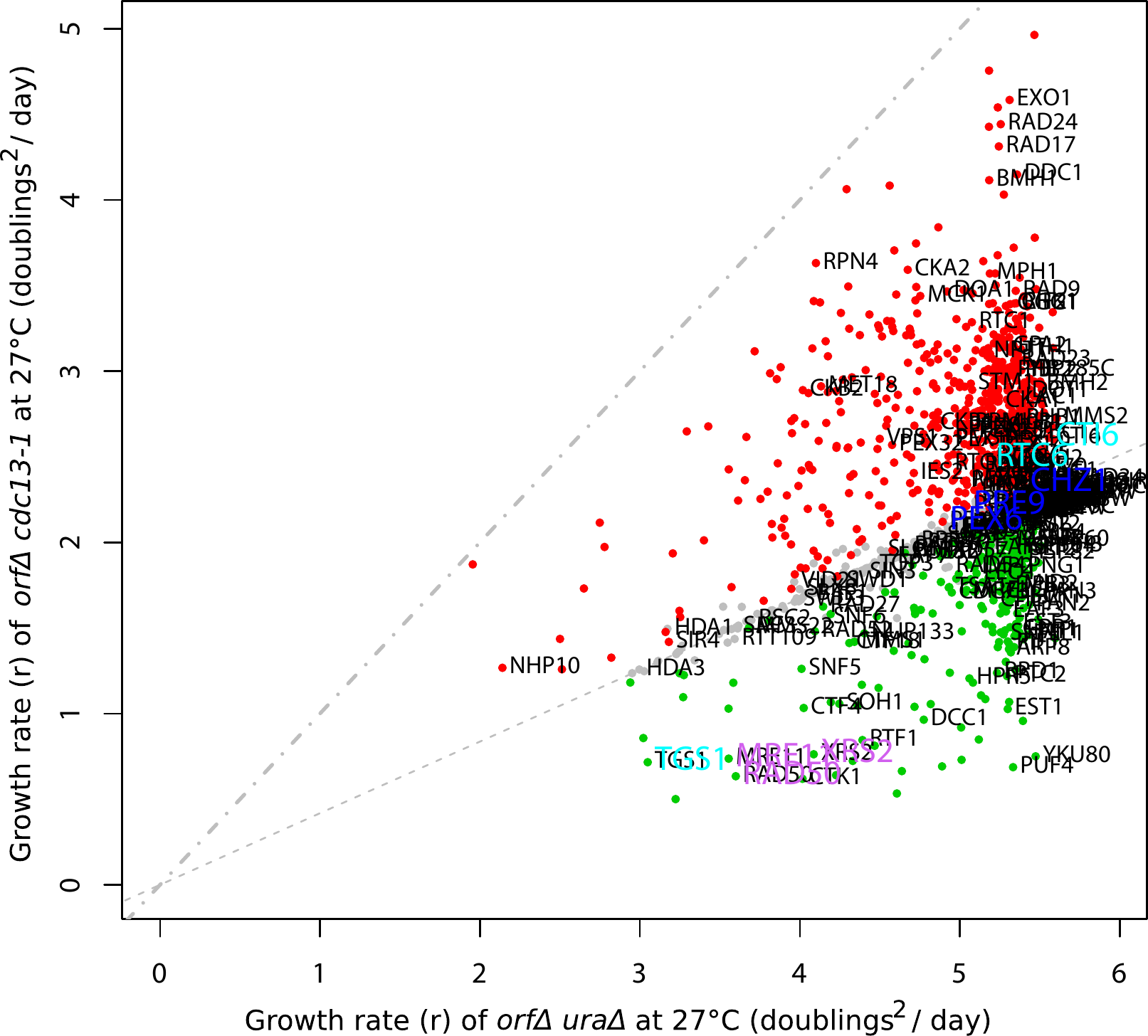}
\caption[Joint hierarchical model growth rate fitness plot]{Joint hierarchical model (JHM) growth rate fitness plot with $\emph{orf}\Delta$ posterior mean fitnesses.
$\emph{orf}\Delta$ strains are classified as being a suppressor or enhancer based on growth parameter $r$.
Significant interactors have posterior probability $\Delta>0.5$.
To compare fitness plots, labelled genes are those belonging to the following gene ontology terms in Table~\ref{tab:sup_enh}: ``telomere maintenance'', ``ageing'', ``response to DNA damage stimulus'' or ``peroxisomal organization'', as well as the genes identified as interactions only in $K$ with the JHM (see Figure~\ref{fig:JHM_only}) (blue), genes interacting only in $r$ with the JHM (cyan) and the MRX complex genes (pink).
\label{fig:JHM_r}
}
\end{figure}
\FloatBarrier
\clearpage
\section{Gene ontology term enrichment analysis in R\label{app:GOstats}}
{\fontsize{9}{9}\selectfont
\begin{verbatim}
source("http://bioconductor.org/biocLite.R")
biocLite("GOstats")
biocLite("org.Sc.sgd.db")
###################
library(GOstats) # GO testing tool package
library(org.Sc.sgd.db) # yeast gene annotation package
genes=read.table("JHM_strip.txt", header=T)
UNIVSTRIP=genes[,2]
genes<-as.vector(genes[genes[,3]>0.5,2])
genes<-unique(genes)
ensemblIDs=as.list(org.Sc.sgdPMID2ORF)
univ=unlist(ensemblIDs)
univ=univ[!is.na(univ)]
length(univ)
length(unique(univ))
univ=unique(univ)
all=as.vector(univ)
all=all[all%in%UNIVSTRIP]
length(all)
ontology=c("BP")
vec<-genes%in%univ
genes<-genes[vec]
params_temp=new("GOHyperGParams", geneIds=genes, 
universeGeneIds=all,
 annotation="org.Sc.sgd.db", categoryName="GO",
 ontology=ontology, pvalueCutoff=1, 
 testDirection = "over")
results=hyperGTest(params_temp)
results=summary(results)
results$qvalue<-p.adjust(results$Pvalue,method="BH")
\end{verbatim}
}
\clearpage
\section{Code for Just Another Gibbs Sampler software\label{app:jags_code}}
\subsection{Separate hierarchical model code}
{\fontsize{7.4}{7.4}\selectfont
\begin{verbatim}
model {
for (l in 1:N){
 for (m in 1:NoORF[l]){
  for (n in 1:NoTime[(NoSum[l]+m)]){
   y[m,n,l] ~ dnorm(y.hat[m,n,l], exp(nu_l[l]))
   y.hat[m,n,l] <- (K_lm[(NoSum[l]+m)]
    *P*exp(r_lm[(NoSum[l]+m)]*x[m,n,l]))
    /(K_lm[(NoSum[l]+m)]+P*(exp(r_lm[(NoSum[l]+m)]*x[m,n,l])-1))
  }
  K_lm[(NoSum[l]+m)]<- exp(K_lm_L[(NoSum[l]+m)])
  K_lm_L[(NoSum[l]+m)] ~ dnorm(K_o_l_L[l],exp(tau_K_l[l]))T(,0)
  r_lm[(NoSum[l]+m)]<- exp(r_lm_L[(NoSum[l]+m)])
  r_lm_L[(NoSum[l]+m)] ~ dnorm(r_o_l_L[l],exp(tau_r_l[l]))T(,3.5)
 }
 K_o_l_L[l]<- log(K_o_l[l])
 K_o_l[l] ~ dt( exp(K_p), exp(sigma_K_o),3)T(0,)
 r_o_l_L[l]<- log(r_o_l[l])
 r_o_l[l] ~ dt( exp(r_p), exp(sigma_r_o),3)T(0,)
 nu_l[l] ~ dnorm(nu_p,  exp(sigma_nu) )
 tau_K_l[l]~dnorm(tau_K_p,exp(sigma_tau_K))T(0,)
 tau_r_l[l]~dnorm(tau_r_p,exp(sigma_tau_r))
}
K_p ~ dnorm(K_mu,eta_K_p)
r_p ~ dnorm(r_mu,eta_r_p)
nu_p ~ dnorm(nu_mu,eta_nu_p)
P<-exp(P_L)
P_L ~ dnorm(P_mu,eta_P)
tau_K_p ~ dnorm(tau_K_mu,eta_tau_K_p)
sigma_tau_K ~ dnorm(eta_tau_K,psi_tau_K)
tau_r_p ~ dnorm(tau_r_mu,psi_tau_r)
sigma_tau_r ~ dnorm(eta_tau_r,psi_tau_r)
sigma_nu~dnorm(eta_nu,psi_nu)
sigma_K_o ~ dnorm(eta_K_o,psi_K_o)
sigma_r_o ~ dnorm(eta_r_o,psi_r_o)
}
\end{verbatim}
}
\subsection{Interaction hierarchical model code}
{\fontsize{7.4}{7.4}\selectfont
\begin{verbatim}
model {
for (l in 1:N){
 for (c in 1:2){
  for (m in 1:NoORF[l,c]){
   y[m,c,l]~ dnorm(exp(alpha_c[c]
    +delta_l[l,c]*gamma_cl_L[l,c])*Z_l[l],exp(nu_cl[l+(c-1)*N]))
  }
  nu_cl[l+(c-1)*N]~dnorm(nu_p,exp(sigma_nu))
 }
 Z_l[l]~dt(exp(Z_p),exp(sigma_Z),3)T(0,)
 delta_l[l,1]<-0
 delta_l[l,2]~dbern(p)
 gamma_cl_L[l,1]<-0
 gamma_cl_L[l,2]<-log(gamma_l[l])
 gamma_l[l]~dt(1,exp(sigma_gamma),3)T(0,)
}
alpha_c[1]<-0
alpha_c[2]~dnorm(alpha_mu,eta_alpha)
Z_p~dnorm(Z_mu,eta_Z_p)
nu_p~dnorm(nu_mu,eta_nu_p)
sigma_Z~dnorm(eta_Z,psi_Z)
sigma_nu~dnorm(eta_nu,psi_nu_p)
sigma_gamma~dnorm(eta_gamma,psi_gamma)
}
\end{verbatim}
}
\clearpage
\subsection{Joint hierarchical model code}
{\fontsize{7.6}{7.6}\selectfont
\begin{verbatim}
model {
for (l in 1:N){
 for (c in 1:2){
  for (m in 1:NoORF[l,c]){
   for (n in 1:NoTime[NoSum[l,c]+m,c]){
    y[m,n,l,c] ~ dnorm(y.hat[m,n,l,c],exp(nu_cl[l+(c-1)*N]))
    y.hat[m,n,l,c] <- (K_clm[(SHIFT[c]+NoSum[l,c]+m)]
     *P*exp(r_clm[(SHIFT[c]+NoSum[l,c]+m)]*x[m,n,l,c]))
     /(K_clm[(SHIFT[c]+NoSum[l,c]+m)]+P*(exp(r_clm[(SHIFT[c]+NoSum[l,c]+m)]
     *x[m,n,l,c])-1))
   }
   K_clm[(SHIFT[c]+NoSum[l,c]+m)]<-exp(K_clm_L[(SHIFT[c]+NoSum[l,c]+m)])
   K_clm_L[(SHIFT[c]+NoSum[l,c]+m)] ~ dnorm(alpha_c[c]+K_o_l_L[l]
   +(delta_l[l,c]*gamma_cl_L[l,c]),exp(tau_K_cl[l+(c-1)*N]))T(,0)
   r_clm[(SHIFT[c]+NoSum[l,c]+m)]<-exp(r_clm_L[(SHIFT[c]+NoSum[l,c]+m)])
   r_clm_L[(SHIFT[c]+NoSum[l,c]+m)] ~ dnorm(beta_c[c]+r_o_l_L[l]
   +(delta_l[l,c]*omega_cl_L[l,c]),exp(tau_r_cl[l+(c-1)*N]))T(,3.5)
  }
  tau_K_cl[l+(c-1)*N]~dnorm(tau_K_p_c[c],exp(sigma_tau_K_c[c]))T(0,)
  tau_r_cl[l+(c-1)*N]~dnorm(tau_r_p_c[c],exp(sigma_tau_r_c[c]))
	nu_cl[l+(c-1)*N]~dnorm(nu_p,exp(sigma_nu))
 }
 K_o_l_L[l]<- log(K_o_l[l])
 K_o_l[l] ~ dt(exp(K_p),exp(sigma_K_o),3)T(0,)
 r_o_l_L[l]<- log(r_o_l[l])
 r_o_l[l] ~ dt(exp(r_p),exp(sigma_r_o),3)T(0,)
 delta_l[l,1]<-0
 delta_l[l,2]~dbern(p)
 gamma_cl_L[l,1]<-0
 gamma_cl_L[l,2]<-log(gamma_l[l])
 gamma_l[l]~dt(1,exp(sigma_gamma),3)T(0,)
 omega_cl_L[l,1]<-0
 omega_cl_L[l,2]<-log(omega_l[l])
 omega_l[l]~dt(1,exp(sigma_omega),3)T(0,)
}
alpha_c[1]<-0
alpha_c[2]~dnorm(alpha_mu,eta_alpha)
beta_c[1]<-0
beta_c[2]~dnorm(beta_mu,eta_beta)
K_p~dnorm(K_mu,eta_K_p)
r_p~dnorm(r_mu,eta_r_p)
nu_p~dnorm(nu_mu,eta_nu_p)
P <- exp(P_L)
P_L ~dnorm(P_mu,eta_P)
sigma_K_o~dnorm(eta_K_o,psi_K_o)
sigma_r_o~dnorm(eta_r_o,psi_r_o)
tau_K_p_c[1]~dnorm(tau_K_mu,eta_tau_K_p)
tau_K_p_c[2]~dnorm(tau_K_mu,eta_tau_K_p)
tau_r_p_c[1]~dnorm(tau_r_mu,eta_tau_r_p)
tau_r_p_c[2]~dnorm(tau_r_mu,eta_tau_r_p)
sigma_tau_K_c[1]~dnorm(eta_tau_K,psi_tau_K)
sigma_tau_K_c[2]~dnorm(eta_tau_K,psi_tau_K)
sigma_tau_r_c[1]~dnorm(eta_tau_r,psi_tau_r)
sigma_tau_r_c[2]~dnorm(eta_tau_r,psi_tau_r)
sigma_nu~dnorm(eta_nu,psi_nu)
sigma_gamma~dnorm(eta_gamma,psi_gamma)
sigma_omega~dnorm(eta_omega,psi_omega)
}
}
\end{verbatim}
}
\clearpage
\section{Additional \emph{cdc13-1}~$\boldsymbol{{27}^{\circ}}$C~vs~\emph{ura3}$\Delta$~$\boldsymbol{{27}^{\circ}}$C fitness plots\label{app:alt_fitness}}

\begin{figure}[h!]
  \centering
\includegraphics[width=14cm]{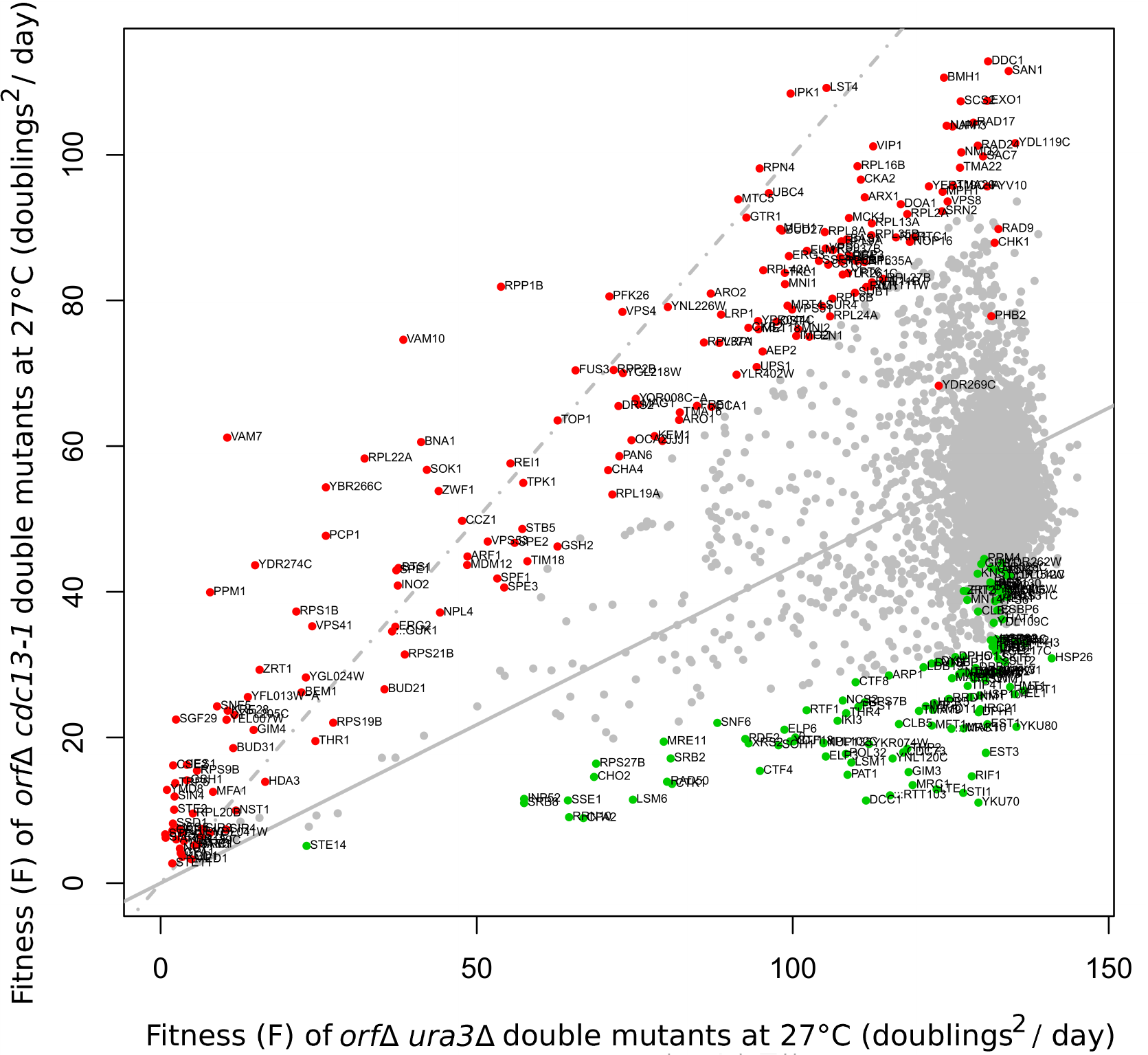}
\caption[Alternative non-Bayesian, hierarchical random effects model fitness plot]{Alternative non-Bayesian, hierarchical fitness plot, \hl{from fitting the random effects model (REM) to data} in Table~S6 from \cite{QFA1} $(F=MDR\times MDP)$.
$\emph{orf}\Delta$s with significant evidence of interaction are highlighted in red and green for suppressors and enhancers respectively.
$\emph{orf}\Delta$s without significant evidence of interaction are in grey and have no \emph{orf} name label.
Significant interactors are classified as those with FDR corrected p-values $<0.05$.
}
\label{fig:REM_app}
\end{figure}

\begin{figure}[h!]
  \centering
\includegraphics[width=14cm]{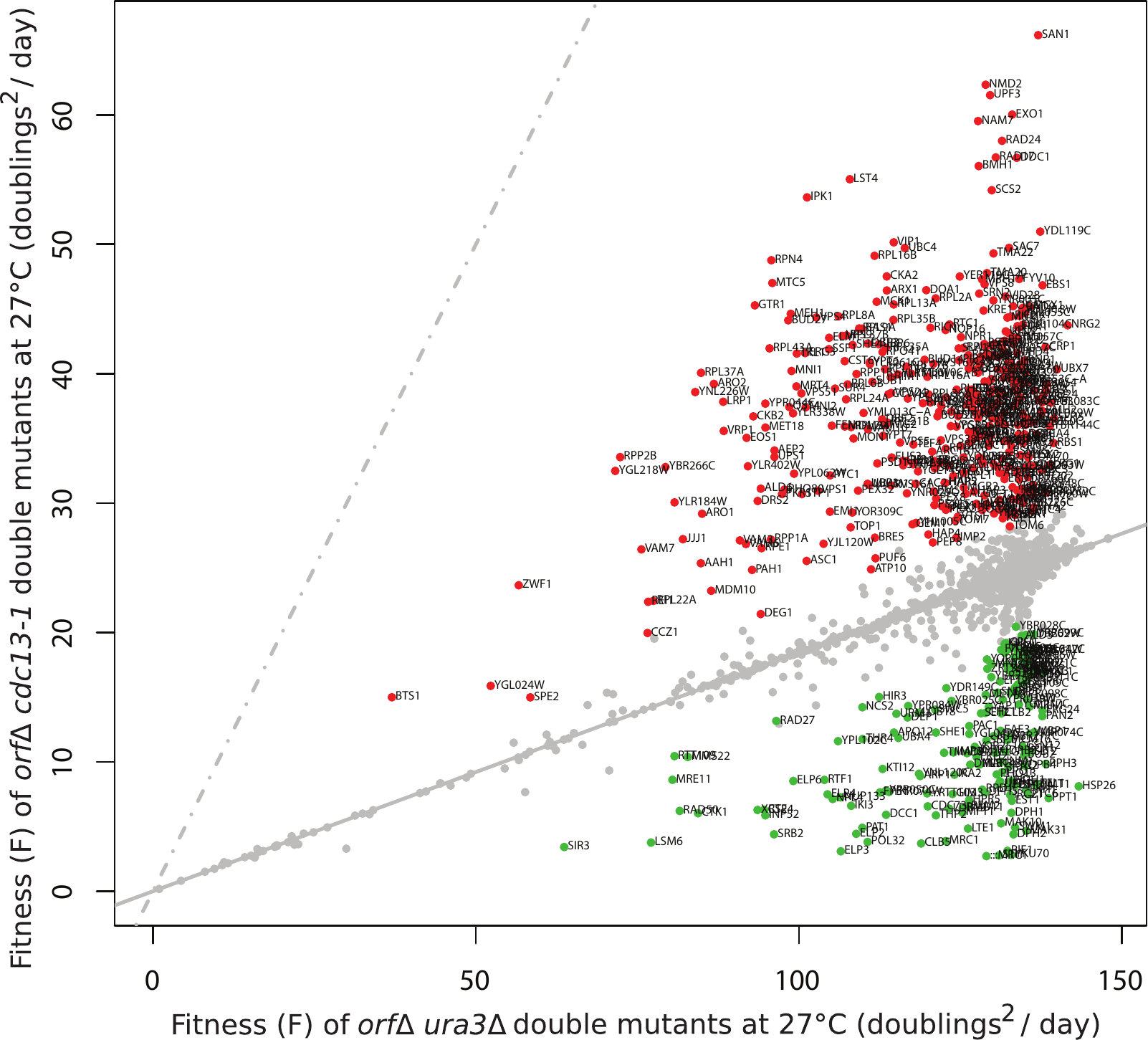}
\caption[Alternative interaction hierarchical model fitness plot]{Alternative interaction hierarchical model (IHM) fitness plot with $\emph{orf}\Delta$ posterior mean fitness. 
$\emph{orf}\Delta$s with significant evidence of interaction are highlighted on the plot as red and green for suppressors and enhancers respectively $(F=MDR\times MDP)$. 
Solid and dashed grey fitted lines are for the IHM linear model fit.
$\emph{orf}\Delta$s with significant evidence of interaction are highlighted in red and green for suppressors and enhancers respectively.
$\emph{orf}\Delta$s without significant evidence of interaction are in grey and have no \emph{orf} name label.
Significant interactors have posterior probability $\Delta>0.5$.
\label{fig:IHM_app}
}
\end{figure}

\begin{figure}[h!]
  \centering
\includegraphics[width=14cm]{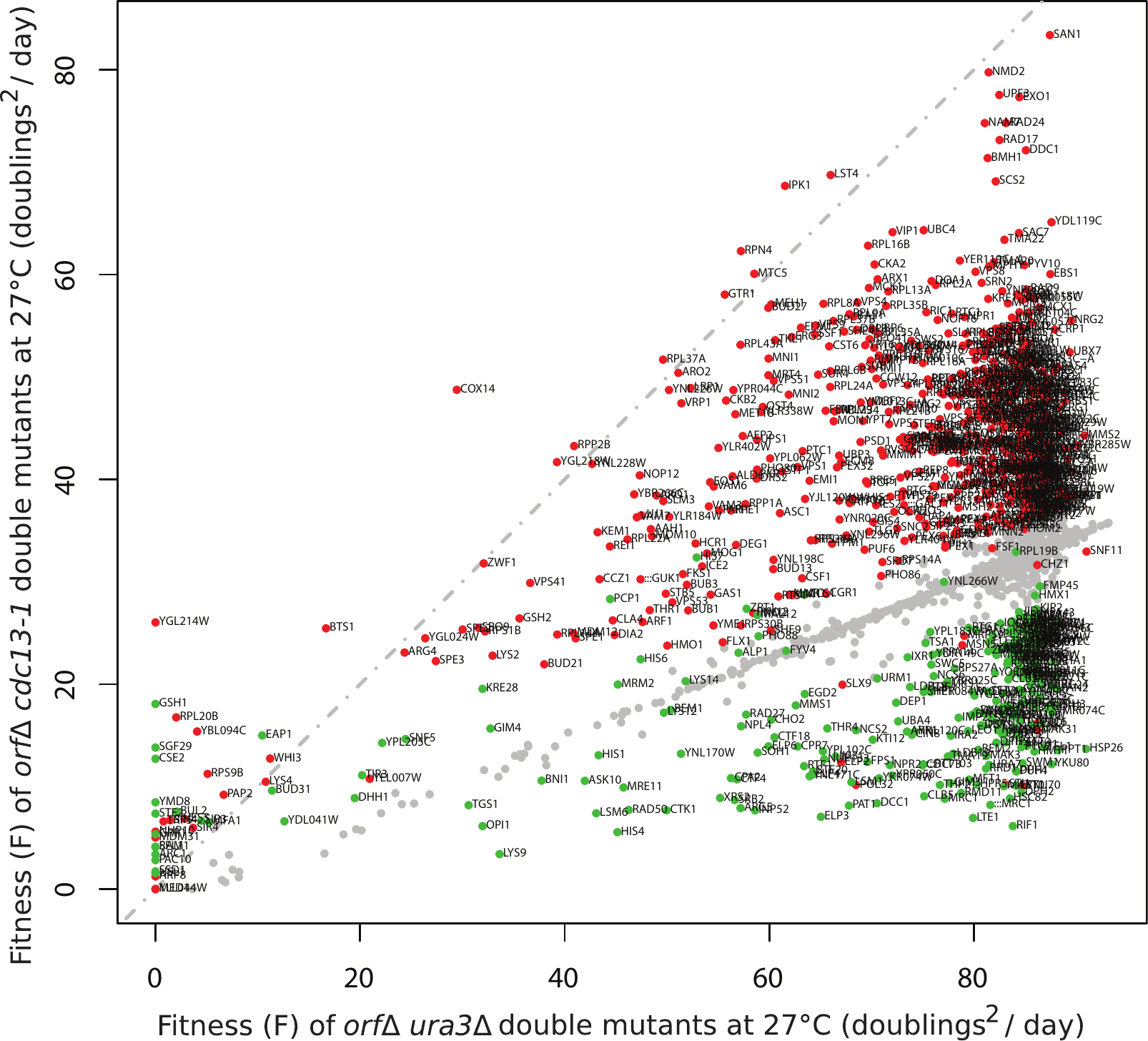}
\caption[Alternative joint hierarchical model fitness plot]{Alternative joint hierarchical model (JHM) fitness plot with $\emph{orf}\Delta$ posterior mean fitnesses.
The JHM does not does not make use of a fitness measure such as $MDR\times{MDP}$ but the fitness plot is given in terms of $MDR\times{MDP}$ for comparison with other approaches which do. 
$\emph{orf}\Delta$ strains are classified as being a suppressor or enhancer based on one of the two parameters used to classify genetic interaction, growth parameter $r$, this means occasionally strains can be more fit in the query experiment in terms of $MDR\times MDP$ but be classified as enhancers (green).
$\emph{orf}\Delta$s with significant evidence of interaction are highlighted in red and green for suppressors and enhancers respectively.
$\emph{orf}\Delta$s without significant evidence of interaction are in grey and have no \emph{orf} name label.
Significant interactors have posterior probability $\Delta>0.5$.
\label{fig:JHM_app}
}
\end{figure}

\begin{figure}[h!]
  \centering
\includegraphics[width=14cm]{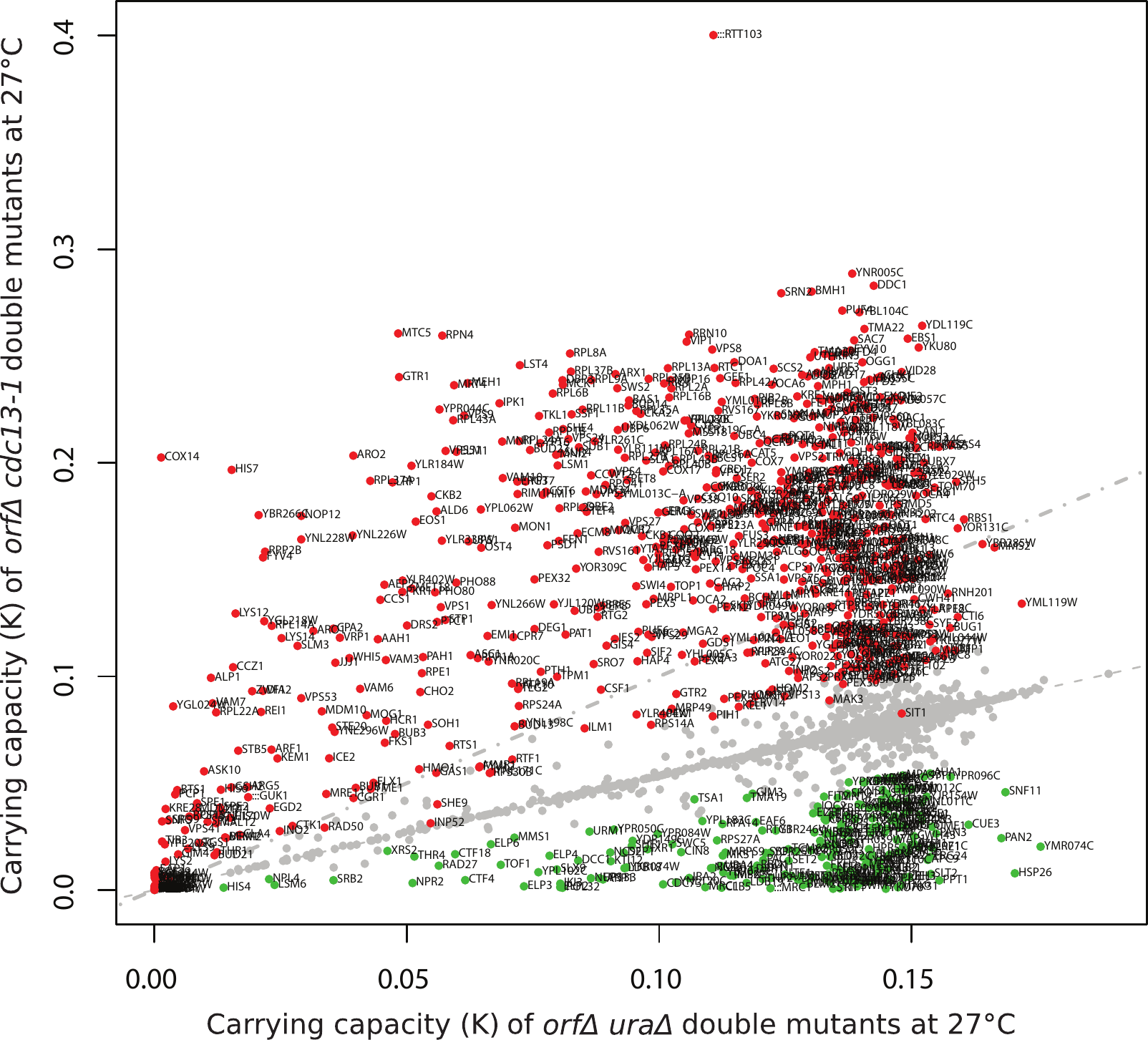}
\caption[Alternative joint hierarchical model carrying capacity fitness plot]{Joint hierarchical model (JHM) carrying capacity fitness plot with $\emph{orf}\Delta$ posterior mean fitnesses.
$\emph{orf}\Delta$ strains are classified as being a suppressor or enhancer based on carrying capacity parameter $K$.
$\emph{orf}\Delta$s with significant evidence of interaction are highlighted in red and green for suppressors and enhancers respectively.
$\emph{orf}\Delta$s without significant evidence of interaction are in grey and have no \emph{orf} name label.
Significant interactors have posterior probability $\Delta>0.5$.
\label{fig:JHM_K_full}
}
\end{figure}

\begin{figure}[h!]
  \centering
\includegraphics[width=14cm]{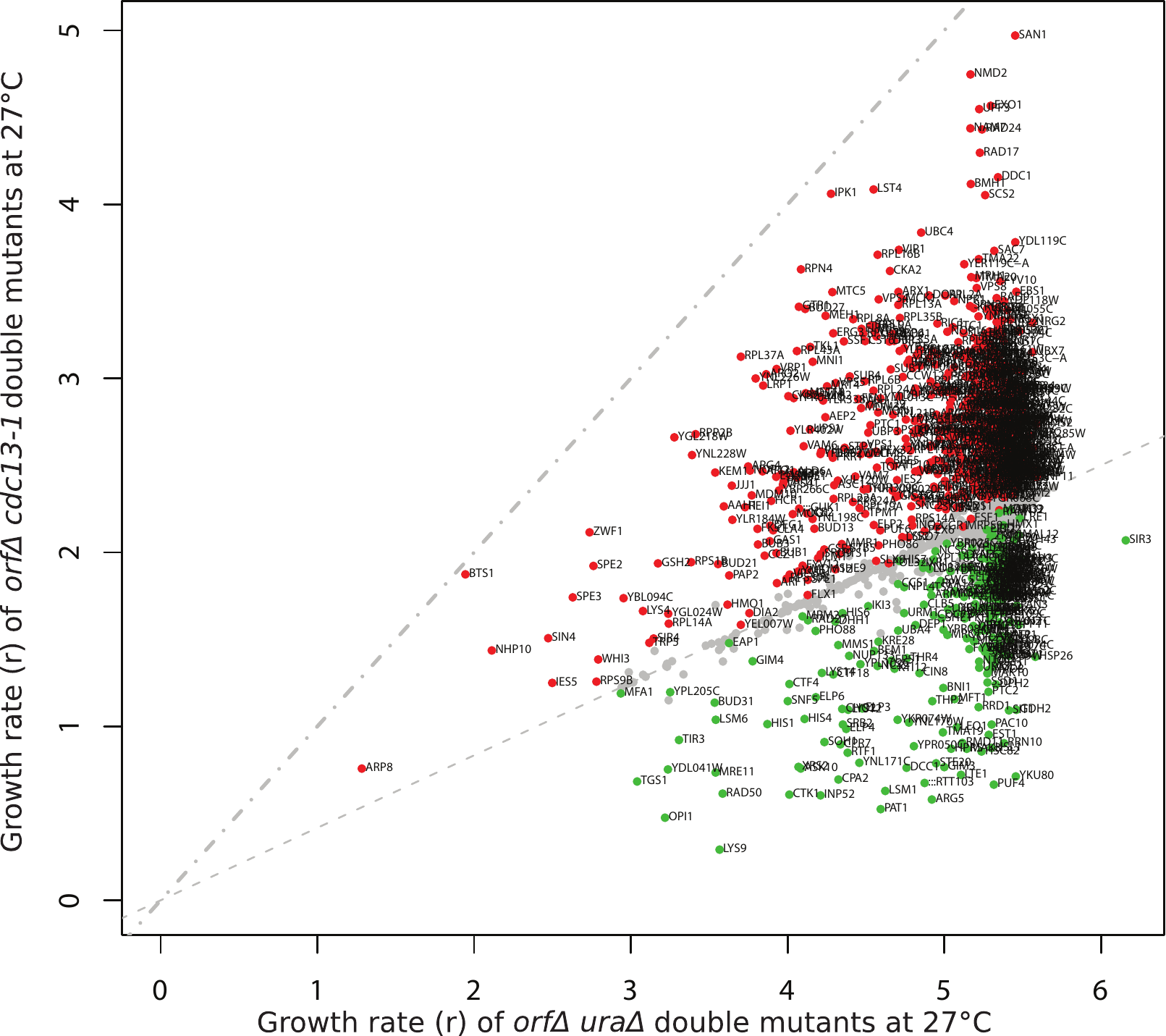}
\caption[Alternative joint hierarchical model growth rate fitness plot]{Joint hierarchical model (JHM) growth rate fitness plot with $\emph{orf}\Delta$ posterior mean fitnesses.
$\emph{orf}\Delta$ strains are classified as being a suppressor or enhancer based on growth parameter $r$.
$\emph{orf}\Delta$s with significant evidence of interaction are highlighted in red and green for suppressors and enhancers respectively.
$\emph{orf}\Delta$s without significant evidence of interaction are in grey and have no \emph{orf} name label.
Significant interactors have posterior probability $\Delta>0.5$.
\label{fig:JHM_r_full}
}
\end{figure}
\clearpage
\FloatBarrier

\section{Correlation between methods\label{app:corr}}
The Addinall et al. (2011) approach has its highest correlation with the IHM, followed by the JHM and then the REM. 
The REM correlates least well with the JHM while showing the same correlation with both the Addinall et al. (2011) approach and the IHM.
The correlation between the IHM and the JHM is the largest observed between any of the methods, demonstrating the similarity of our Bayesian hierarchical methods. 
\begin{table}[h!]
\caption[Spearman's rank correlation coefficients for magnitudes from genetic independence, between approaches]{Spearman's rank correlation coefficients for magnitudes from genetic independence, between Addinall et al. (2011), random effects approach (REM), interaction hierarchical model (IHM) and joint hierarchical model (JHM) approaches \label{tab:spearman}}
\centering 
\resizebox{\columnwidth}{!}{%
  \begin{tabular}{*{5}{c}}
    \\
  	   \hline
  \\
\emph{Method} &\multicolumn{4}{c}{\emph{Method}}\\ 
&&& \\ \cline{2-5}
&&&\\
   		  & \emph{Addinall et al. (2011)} & \emph{REM} & \emph{IHM} & \emph{JHM QFA} \\
			   		  & \emph{QFA} & \emph{QFA} & \emph{QFA} & \emph{($MDR\times MDP$)} \\	
   		  \\
   		   \hline
\\   		  
Addinall et al. (2011) QFA,   	&1	& 0.77 & 0.89 & 0.88 \\ 	

REM QFA,    					& 	&1	 & 0.77	& 0.75 \\

IHM QFA,   						&	  &    &1	& 0.95 \\

JHM QFA ($MDR\times MDP$),    			& 	&	   &  & 1 \\

\\		
\hline
  		 \end{tabular}
			}
\end{table}

The $MDR\times MDP$ correlation plot of the JHM versus the Addinall et al. (2011) approach demonstrates the similarity (Pearson correlation=0.90) and differences between the two approaches in terms of $MDR\times MDP$.
{We can see how the results differ between the JHM and Addinall et al. (2011), with a kink at the origin due to the JHM allowing shrinkage of non-interacting genes towards the fitted line.}

\clearpage
\begin{figure}[h!]
  \centering
\includegraphics[width=14cm]{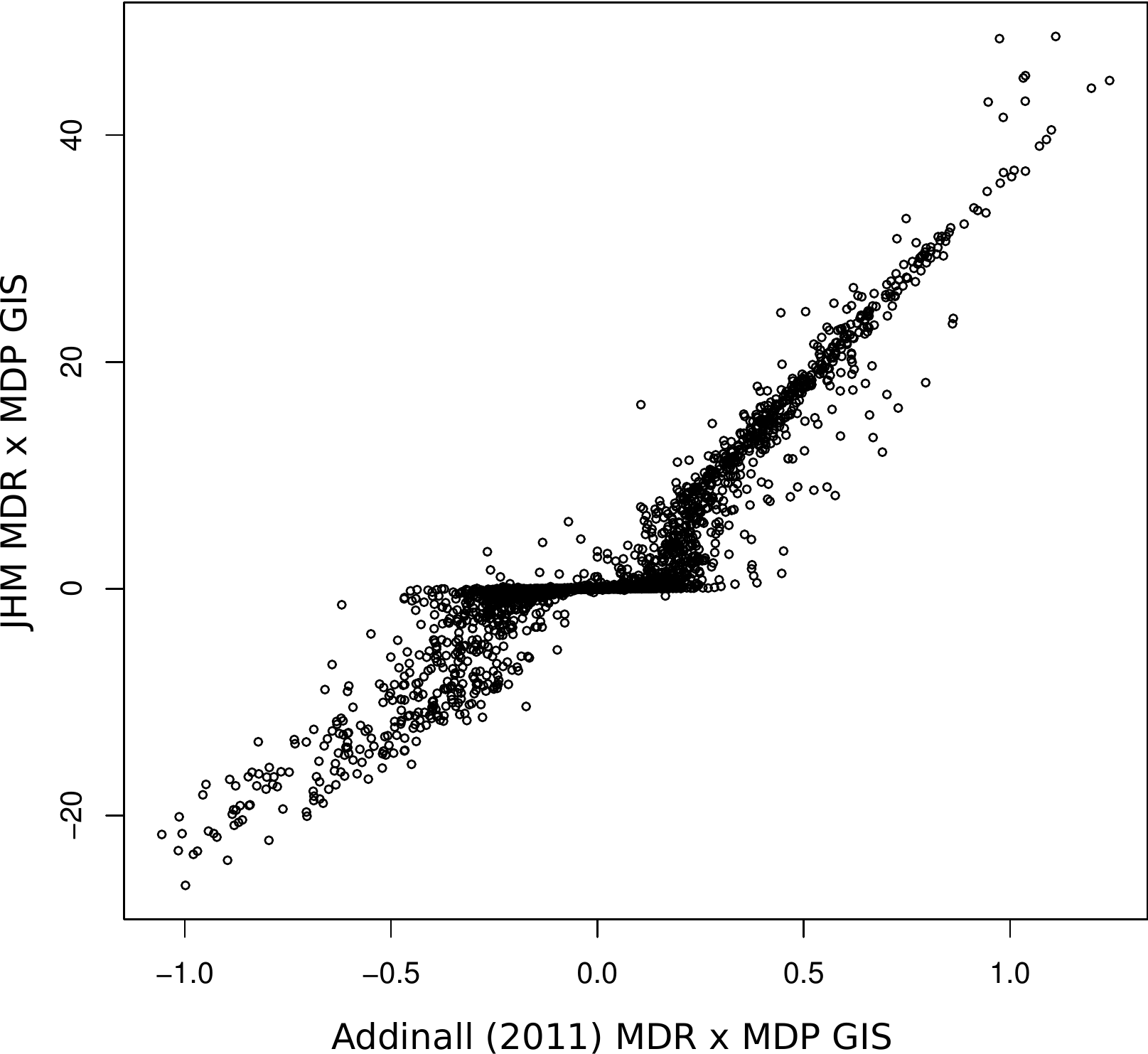}
\caption[$MDR\times MDP$ genetic interaction correlation plot of the joint hierarchcial model versus Addinall et al. (2011)]{ $MDR\times MDP$ genetic interaction correlation plot of JHM versus Addinall et al. (2011) (Pearson correlation=0.90).
}
\label{app:correlation_JHM_ADD}
\end{figure}
\clearpage
\FloatBarrier

\chapter{Stochastic logistic growth modelling}
\section{\label{app:LNAM_sol}Linear noise approximation of the stochastic logistic growth model with multiplicative intrinsic noise solution}
First we look to solve $dZ_t$, given in equation (\ref{eq:LNAM_dz}).
We define $f(t)=-be^{v_t}=-\frac{baPe^{aT}}{bP(e^{aT}-1)+a}$ to obtain the following,
\begin{equation*}
dZ_t=f(t)Z_tdt+\sigma dW_t.
\end{equation*}
In order to match our initial conditions correctly, $Z_0=0$.  
Define a new process $U_t=e^{-\int^t_{t_0}f(s)ds}Z_t$ and solve the integral,
\begin{equation*}
\int^t_{t_0}f(s)ds=\int^t_{t_0}-\frac{baPe^{aS}}{bP(e^{aS}-1)+a}ds=\log\left(\frac{a}{bP(e^{aT}-1)+a}\right),
\end{equation*}
where, $S=s-{t_0}$ and $T=t-{t_0}$.
Apply the chain rule to $U_t$,
\begin{equation*}
dU_t=e^{-\int^t_{t_0}f(s)ds}dZ_t-f(t)e^{-\int^t_{t_0}f(s)ds}Z_tdt.
\end{equation*}
Now substitute in $dZ_t=f(t)Z_tdt+\sigma dW_t$ and simplify to give
\begin{equation*}
dU_t= e^{-\int^t_{t_0}f(s)ds}\sigma dW_t.
\end{equation*}
Apply the following notation
$\phi(t)=e^{\int^t_{t_0}f(s)ds}=\frac{a}{bP(e^{aT}-1)+a}$ and $\psi(t)=\sigma$ to give
\begin{equation*}
dU_t=\phi(t)^{-1}\psi(t) dW_t.
\end{equation*}
$U_t$, has the following solution,
\begin{equation*}
U_t=U_0+\int^t_{t_0} \phi(s)^{-1} \psi(s)dW_s.
\end{equation*}
As $U_t=\phi(t)^{-1}Z_t$, $Z_t$ then has the following solution \citep{arnold2013stochastic},
\begin{equation*}
Z_t=\phi(t)\left[Z_0+\int^t_{t_0}\phi(s)^{-1}\psi(s) dW_s\right].
\end{equation*}
Finally, the distribution at time t is $Z_t|Z_0\sim N(M_t,E_t)$ \citep{arnold2013stochastic}, where \\
$M_t=\phi(t)Z_0$ and 
$E_t=\phi(t)^2\int^t_{t_0}\left[{\phi(s)}^{-1}\psi(s)\right]^2ds$.
\\
Further, $M_t=\frac{a}{bP(e^{aT}-1)+a}Z_0 $ and 
$E_t=\sigma^2\left[\frac{a}{bP(e^{aT}-1)+a}\right]^2
\int^t_{t_0}\left[
\frac{a}{bP(e^{aS}-1)+a}
\right]^{-2}$ds.\\
As 
$\int^t_{t_0}\left[
\frac{a}{bP(e^{aS}-1)+a}
\right]^{-2}ds=\frac{
b^2P^2(e^{2aT}-1)
+4bP(a-bP)(e^{aT}-1)
+2aT(a-bP)^2
}{2a^3}
$,
{\fontsize{11.5}{11.5}\selectfont
\begin{align*}
E_t=&\sigma^2\left[\frac{a}{bP(e^{aT}-1)+a}\right]^2
\left[
\frac{
b^2P^2(e^{2aT}-1)
+4bP(a-bP)(e^{aT}-1)
+2aT(a-bP)^2
}{2a^3}
\right]\\
=&\sigma^2\left[\frac{
b^2P^2(e^{2aT}-1)
+4bP(a-bP)(e^{aT}-1)
+2aT(a-bP)^2
}
{2a\left(bP(e^{aT}-1)+a\right)^2}\right].
\end{align*}
}
Taking our solutions for $v_t$ (\ref{eq:LNAM_det_sol}) and $Z_t$, we can now write our solution for the LNA to the log of the logistic growth process (\ref{eq:SDE2}).\\
As $Y_t=v_t+Z_t$, 
\begin{equation*}
Y_t|Y_0\sim \mathcal{N}\left(\log\left[\frac{aPe^{aT}}{bP(e^{aT}-1)+a}\right]+M_t,E_t\right).
\end{equation*}
Note: $\frac{aPe^{aT}}{bP(e^{aT}-1)+a}$ has the same functional form as the solution to the deterministic part of the logistic growth process (\ref{eq_det_sde}) and is equivalent when $\sigma=0$ (such that $a=r-\frac{\sigma^2}{2}=r$).\\
\\
Further, as $Y_t$ is normally distributed, we know $X_t=e^{Y_t}$ will be log normally distributed and
\begin{equation*}
X_t|X_0\sim \log\:\mathcal{N}(\log\left(\frac{aPe^{aT}}{bP(e^{aT}-1)+a}\right)+M_t,E_t).
\end{equation*}
Alternatively set $Q=\left(\frac{\frac{a}{b}}{P}-1\right)e^{at_{0}}$,
\begin{equation*}
X_t|X_{0}\sim \log\:\mathcal{N}(\log\left(\frac{\frac{a}{b}}{1+Qe^{-at}}\right)+M_t,E_t).
\end{equation*}

\noindent From our solution to the log process we can obtain the following transition density
\begin{align*}
\begin{split}
(Y_{t_i}|Y_{t_{i-1}}&=y_{t_{i-1}})\sim\operatorname{N}\left(\mu_{t_i},\Xi_{t_i}\right),\\
\text{where } y_{t_{i-1}}&=v_{t_{i-1}}+z_{t_{i-1}}, Q=\left(\frac{\frac{a}{b}}{P}-1\right)e^{at_{0}},\\
\mu_{t_i}&=y_{t_{i-1}}+\log\left(\frac{1+Qe^{-at_{i-1}}}{1+Qe^{-at_i}}\right)+e^{-a(t_i-t_{i-1})}\frac{1+Qe^{-at_{i-1}}}{1+Qe^{-at_i}}z_{t_{i-1}} \text{ and}\\
\Xi_{t_i}&=\sigma^2\left[\frac{4Q(e^{at_i}-e^{at_{i-1}})+e^{2at_i}-e^{2at_{i-1}}+2aQ^2(t_i-t_{i-1})}{2a(Q+e^{at_i})^2}\right].
\end{split}
\end{align*}

\clearpage

\section{\label{app:zero_ord}Zero-order noise approximation of the stochastic logistic growth model}
After obtaining (\ref{eq:SDEV}) in Section~\ref{sec:LNAM}, we can derive the RRTR logistic growth diffusion process as follows.
First our expression for $dv_t$, given in (\ref{eq:SDEV}), is approximated by setting~$\sigma^2=0$,
\begin{equation*}
dv_t=\left(r-\frac{1}{2}\sigma^2-\frac{r}{K}e^{v_t}\right)dt=\left(r-\frac{r}{K}e^{v_t}\right)dt.
\end{equation*}
We now write down an expression for $dZ_t$, where $dY_t$ is given in (\ref{eq:SDE2}) and $dZ_t=dY_t-dv_t$,
\begin{equation*}
dZ_t=
\left(r-\frac{1}{2}\sigma^2-\frac{r}{K}e^{Y_t}\right)dt+\sigma dW_t-\left(r-\frac{r}{K}e^{v_t}\right)dt.
\end{equation*}
We can then rearrange and simplify to give the following,
\begin{equation*}
dZ_t=\left(\frac{r}{K}\left[e^{v_t}-e^{Y_t}\right]-\frac{1}{2}\sigma^2\right)dt+\sigma dW_t.
\end{equation*}
We now substitute in $Y_t=v_t+Z_t$, 
\begin{equation*}
dZ_t=\left(\frac{r}{K}\left[e^{v_t}-e^{v_t+Z_t}\right]-\frac{1}{2}\sigma^2\right)dt+\sigma dW_t.
\end{equation*}
We now apply a zero order LNA by setting $e^{Z_t}=1$ to obtain,
\begin{equation*}
dZ_t=\left(\frac{r}{K}\left[e^{v_t}-e^{v_t}\right]-\frac{1}{2}\sigma^2\right)dt+\sigma dW_t.
\end{equation*}
We can then simplify to give the following,
\begin{equation}\label{eq:RRTRa}
dZ_t=-\frac{1}{2}\sigma^2 dt+\sigma dW_t.
\end{equation}
Differentiating $v_t$, given in (\ref{eq:LNAM_det_sol}), with respect to t we can obtain an alternative expression for $dv_t$,
\begin{equation}\label{eq:RRTRb}
dv_t=\frac{a(a-bP)}{bP(e^{aT}-1)+a}dt=\frac{r(K-P)}{K+P(e^{rT}-1)}dt,
\end{equation}
where $T=t-t_0$. We now write down our new expression for $Y_t$, where $dY_t=dv_t+dZ_t$, given (\ref{eq:RRTRb}) and (\ref{eq:RRTRa}),
\begin{equation*}
dY_t=\left(\frac{r(K-P)}{K+P(e^{aT}-1)}-\frac{1}{2}\sigma^2\right)dt+\sigma dW_t
\end{equation*}
or alternatively by setting $Q=\left(\frac{K}{P}-1\right)e^{at_{0}}$,
\begin{equation*}
dY_t=\left(\frac{Qr}{e^{rt}+Q}-\frac{1}{2}\sigma^2\right)dt+\sigma dW_t.
\end{equation*}
We can then apply It\^{o}'s lemma (\ref{eq_itolemma}) \citep{ito} with the transformation ${f(t,Y_t)\equiv X_t=e^{Y_t}}$.
After deriving the following partial derivatives:
\begin{equation*}
\frac{df}{dt}=0,\qquad\frac{df}{dx}=e^{Y_t}\quad\text{and}\quad\frac{d^2f}{dx^2}=e^{Y_t},
\end{equation*}
we can obtain the following It\^{o} drift-diffusion process:
\begin{equation*}
dX_t=\frac{Qr}{e^{rt}+Q}X_tdt+\sigma dW_t,
\end{equation*}
which is exactly the RRTR logistic diffusion process presented by \cite{roman}.
\clearpage

\section{\label{app:LNAA_sol}Linear noise approximation of the stochastic logistic growth model with additive intrinsic noise solution}
First we look to solve $dZ_t$, given in (\ref{eq:LNAA_dz}).
We define $f(t)=a-2bv_t$ to obtain the following,
\begin{equation*}
dZ_t=f(t)Z_tdt+\sigma v_t dW_t.
\end{equation*}
In order to match our initial conditions correctly, $Z_0=0$.  
Define a new process $U_t=e^{-\int^t_{t_0}f(s)ds}Z_t$ and solve the integral,
\begin{equation*}
\int^t_{t_0}f(s)ds=\int^t_{t_0}(a-2bV_s)ds=aT-2\log\left(\frac{bP(e^{aT}-1)+a}{a}\right),
\end{equation*}
as $\int^t_{t_0}V_sds=\frac{1}{b}\log \left(\frac{bP(e^{aT}-1)+a}{a}\right)$, where $S=s-{t_0}$ and $T=t-{t_0}$.
Apply the chain rule to $U_t$,
\begin{equation*}
dU_t=e^{-\int^t_{t_0}f(s)ds}dZ_t-f(t)e^{-\int^t_{t_0}f(s)ds}Z_tdt.
\end{equation*}
Now substitute in $dZ_t=f(t)Z_tdt+\sigma v_t dW_t$ and simplify to give,
\begin{equation*}
dU_t=e^{-\int^t_{t_0}f(s)ds}\sigma v_t dW_t.
\end{equation*}
Apply the following notation $\phi(t)=e^{\int^t_{t_0}f(s)ds}=e^{aT}\left(\frac{a}{bP(e^{aT}-1)+a}\right)^2$ and $\psi(t)=\sigma v_t$ to give, 
\begin{equation*}
dU_t=\phi(t)^{-1}\psi(t) dW_t.
\end{equation*}
$U_t$ has the following solution,
\begin{equation*}
U_t=U_0+\int^t_{t_0} \phi(s)^{-1} \psi(s)dW_s.
\end{equation*}
As $U_t=\phi(t)^{-1}Z_t$, $Z_t$ has the following solution \citep{arnold2013stochastic},
\begin{equation*}
Z_t=\phi(t)\left[Z_0+\int^t_{t_0}\phi(s)^{-1}\psi(s) dW_s\right].
\end{equation*}
Finally the distribution at time t is $Z_t|Z_0\sim N(M_t,E_t)$ \citep{arnold2013stochastic}, where \\
{\fontsize{11}{11}\selectfont
$M_t=\phi(t)Z_0$ and 
$E_t=\phi(t)^2\int^t_{t_0}\left[{\phi(s)}^{-1}\psi(s)\right]^2ds$.
}
\begin{equation*}
M_t=e^{aT}\left(\frac{a}{bP(e^{aT}-1)+a}\right)^2Z_0
\end{equation*}
 and 
\begin{align*}
E_t=&\left(e^{aT}\left(\frac{a}{bP(e^{aT}-1)+a}\right)^2\right)^2
\int^t_{t_0}\left[e^{aS}\left(\frac{a}{bP(e^{aS}-1)+a}\right)^2\right]^{-2} \sigma^2 V_s^2 ds
\\
=&\sigma^2\left(e^{aT}\left(\frac{a}{bP(e^{aT}-1)+a}\right)^2\right)^2
\\
&\times\int^t_{t_0}\left[
e^{aS}\left(\frac{a}{bP(e^{aS}-1)+a}\right)^2
\right]^{-2} \left[
\frac{aPe^{aS}}{bP(e^{aS}-1)+a}
\right]^{2} ds
\\
=&\sigma^2\left(e^{aT}\left(\frac{a}{bP(e^{aT}-1)+a}\right)^2\right)^2
\\
&\times\int^t_{t_0}\left[
e^{-2aS}\left(\frac{a}{bP(e^{aS}-1)+a}\right)^{-4}
\right] \left[
\frac{aPe^{aS}}{bP(e^{aS}-1)+a}
\right]^{2} ds
\\
=&\sigma^2\left(e^{aT}\left(\frac{1}{bP(e^{aT}-1)+a}\right)^2\right)^2
\int^t_{t_0}\left[a^2P^2
\left(\frac{1}{bP(e^{aS}-1)+a}\right)^{-2}
\right] ds,
\end{align*}
as $\int^{t}_{t_{0}}\left(\frac{1}{bP(e^{aS}-1)+a}\right)^{-2}ds
=\frac{
b^2P^2(e^{2aT}-1)
+4bP(a-bP)(e^{aT}-1)
+2aT(a-bP)^2
}{2a}$,
\begin{align*}
E_t=&\frac{1}{2}\sigma^2aP^2e^{2aT}\left(\frac{1}{bP(e^{aT}-1)+a}\right)^4\\
&\times
\left[
b^2P^2(e^{2aT}-1)
+4bP(a-bP)(e^{aT}-1)
+2aT(a-bP)^2
\right].
\end{align*}
\noindent Taking our solutions for $v_t$ (\ref{eq:LNAA_det_sol}) and $Z_t$, we can obtain the following transition density
\begin{align*}
\begin{split}
(X_{t_i}|X_{t_{i-1}}=&x_{t_{i-1}})\sim N(\mu_{t_i},\Xi_{t_i}),\\
\text{where }x_{t_{i-1}}=&v_{t_{i-1}}+z_{t_{i-1}},\\
\mu_{t_i}=&x_{t_{i-1}}+\left(\frac{aPe^{aT_i}}{bP(e^{aT_i}-1)+a}\right)-\left(\frac{aPe^{aT_{i-1}}}{bP(e^{aT_{i-1}}-1)+a}\right)\\
&+e^{a(t_i-t_{i-1})}\left(\frac{bP(e^{aT_{i-1}}-1)+a}{bP(e^{aT_i}-1)+a}\right)^2Z_{t_{i-1}}\text{ and}\\
 \Xi_{t_i}=&\frac{1}{2}\sigma^2aP^2e^{2aT_i}\left(\frac{1}{bP(e^{aT_i}-1)+a}\right)^4\\
&\times[
b^2P^2(e^{2aT_i}-e^{2aT_{i-1}})
+4bP(a-bP)(e^{aT_i}-e^{aT_{i-1}})\\
&\;\:\:\:\:+2a(t_i-t_{i-1})(a-bP)^2
].
\end{split}
\end{align*}
\clearpage

\section{\label{app:prior_hyp}Prior hyper-parameters for Bayesian state space models}
\begin{table}[h!]
\caption[Prior hyper-parameters for Bayesian sate space models]{Prior hyper-parameters for Bayesian sate space models, Log-normal with mean ($\mu$) and precision ($\tau$) \label{table:SDE_priors}}
\centering     
\begin{tabular}{c c}
    \hline
		\noalign{\vskip 0.4mm} 
    Parameter Name  & \multicolumn{1}{c}{Value} \\ \hline
 ${\mu}_K$ & $\log(0.1)$   \\ 
 $\tau_{K}$ & 2 \\ 
 ${\mu}_r$ & $\log(3)$  \\ 
 $\tau_{r}$ & 5\\ 
 ${\mu}_P$ &  $\log(0.0001)$\\ 
 $\tau_{P}$ & 0.1 \\ 
 ${\mu}_\sigma$ & $\log(100)$  \\ 
 $\tau_{\sigma}$ & 0.1 \\ 
 ${\mu}_\nu$ & $\log(10000)$  \\ 
 $\tau_{\nu}$ & 0.1 \\ 
    \hline
    \end{tabular}
    \npnoround
\end{table}
\clearpage

	\clearpage

\section{\label{app:kalman_fil}Kalman filter for the linear noise approximation of the stochastic logistic growth model with additive intrinsic noise and Normal measurement error}
To find $\pi(y_{t_{1:N}})$ for the LNAA with Normal measurement error we can use the following Kalman Filter algorithm. First we assume the following:
\begin{align*}
\theta_{{t_{i}}}|y_{1:{{t_{i}}}}&\sim \operatorname{N}(m_{t_{i}},C_{t_{i}}),\\
m_{t_{i}}&=a_{t_{i}}+R_{t_{i}}F(F^{T}R_{{t_{i}}}F+U)^{-1}[y_{t_{i}}-F^{T}a_{t_{i}}],\\
C_{t_{i}}&=R_{t_{i}}-R_{t_{i}}F(F^TR_{t_{i}}F+U)^{-1}F^{T}R_{t_{i}}
\end{align*}
and initialize with $m_0=P$ and $C_0=0$. Now suppose that,
\begin{align*}
\theta_{t_{i}}|y_{1:{t_{i-1}}}&\sim \operatorname{N}(a_{t_{i}},R_{t_{i}}),\\
a_{t_{i}}&=G_{{t_{i}}}m_{{t_{i-1}}}\\  
\text{and }R_{t_{i}}&=G_{{t_{i}}}C_{{t_{i-1}}}G_{t_{i}}^T+W_{t_{i}}.
\end{align*}
The transition density distribution, see (\ref{eq:LNAA_tran}) is as follows:
\begin{align*}
\theta_{{t_{i}}}|\theta_{{t_{i-1}}}&\sim\operatorname{N}(G_{{t_{i}}}\theta_{{t_{i-1}}},W_{t_{i}})\\
\text{or equivalently }(X_{t_i}|X_{t_{i-1}}=x_{t_{i-1}})&\sim\operatorname{N}\left(\mu_{t_i},\Xi_{t_i}\right),\text{ where }x_{t_{i-1}}=v_{t_{i-1}}+z_{t_{i-1}},\\
\theta_{t}
 &=
\begin{pmatrix}
  1 \\
	X_{t_{i}}
 \end{pmatrix}
 =
 \begin{pmatrix}
  1 & 0 \\
  H_{\alpha,t_{i}} & H_{\beta,t_{i}}
 \end{pmatrix}
 \begin{pmatrix}
  1 \\
	X_{t_{i-1}}
 \end{pmatrix}
\\&=
 G_{t_{i}}\theta_{{t_{i-1}}},\\
 G_{t_{i}}&= \begin{pmatrix}
 1 & 0\\
 H_{\alpha,t_{i}} & H_{\beta,t_{i}}
 \end{pmatrix}, \quad
  W_{t_{i}}= \begin{pmatrix}
  0 & 0 \\
  0 & \Xi_{t_i}
	\end{pmatrix} \\
\text{where }H_{\alpha,t_{i}}=H_\alpha({t_{i}},{t_{i-1}})=&v_t-V_{t-1}e^{a(t_i-t_{i-1})}\left(\frac{bP(e^{aT_{i-1}}-1)+a}{bP(e^{aT_i}-1)+a}\right)^2\\
\text{and }H_{\beta,t_{i}}=&H_\beta({t_{i}},{t_{i-1}})=e^{a(t_i-t_{i-1})}\left(\frac{bP(e^{aT_{i-1}}-1)+a}{bP(e^{aT_i}-1)+a}\right)^2.
\end{align*}
The measurement error distribution is as follows:
\begin{align*}
y_{t_{i}}|\theta_{{t_{i}}}{\sim}&\operatorname{N}(F^T\theta_{{t_{i}}},U)\\
\text{or equivalently }y_{t_{i}}|\theta_{{t_{i}}}{\sim}&\operatorname{N}(X_{{t_{i}}},\sigma_{\nu}^2),\\
 \text{where }
 F=& \begin{pmatrix}
  0 \\
	1 
 \end{pmatrix}\text{ and }
  U=  \sigma_{\nu}^2.
  \end{align*}  
  Matrix Algebra:
\begin{align*}
a_{t_{i}}=&G_{{t_{i}}}m_{{t_{i-1}}}\\
=&\begin{pmatrix}
 1 & 0\\
H_{\alpha,t_{i}} & H_{\beta,t_{i}}
 \end{pmatrix}
 \begin{pmatrix}
  1\\
	m_{{t_{i-1}}} 
 \end{pmatrix}
=\begin{pmatrix}
  1\\
	H_{\alpha,t_{i}}+H_{\beta,t_{i}}m_{{t_{i-1}}}
 \end{pmatrix}
  \end{align*}  
  \begin{align*}
R_{t_{i}}&=G_{{t_{i}}}C_{{t_{i-1}}}G_{t_{i}}^T+W_{t_{i}}
\\
 &=
  \begin{pmatrix}
  0 & 0\\
  0 & {H_{\beta,t_{i}}}^2c_{{t_{i-1}}}^2
 \end{pmatrix}
 +
  \begin{pmatrix}
  0 & 0\\
  0 & \Xi_{t_i}
 \end{pmatrix}
   =\begin{pmatrix}
  0 & 0\\
  0 & {H_{\beta,t_{i}}}^2 c_{{t_{i-1}}}^2+\Xi_{t_i}
 \end{pmatrix}
\end{align*}
  \begin{align*}
  C_{t_{i-1}}&=
  \begin{pmatrix}
	0 & 0\\
  0 & c_{{t_{i-1}}}^2
 \end{pmatrix}
    \end{align*}
 \begin{align*}
  R_{t_{i}}F(F^{T}R_{{t_{i}}}F+U)^{-1}=&
  \begin{pmatrix}
	0 & 0\\
  0 & {H_{\beta,t_{i}}}^2c_{{t_{i-1}}}^2+\Xi_{t_i}
 \end{pmatrix}
 \begin{pmatrix}
  0 \\
	1
 \end{pmatrix}
\\
&\times
 \left[
 \begin{pmatrix}
 0 & 1\\
 \end{pmatrix}
   \begin{pmatrix}
	0 & 0\\
  0 & {H_{\beta,t_{i}}}^2c_{{t_{i-1}}}^2+\Xi_{t_i}
 \end{pmatrix}
  \begin{pmatrix}
  0 \\
	1
 \end{pmatrix}
 +\sigma_{\nu}^2
 \right]^{-1}
 \\
  =&\left[
    \begin{pmatrix}
{H_{\beta,t_{i}}}^2c_{{t_{i-1}}}^2+\Xi_{t_i} +\sigma_{\nu}^2
 \end{pmatrix}
 \right]^{-1}
 \begin{pmatrix}
	0 \\
  {H_{\beta,t_{i}}}^2c_{{t_{i-1}}}^2+\Xi_{t_i}
 \end{pmatrix}
\end{align*}
\begin{align*}
m_{t_{i}}=&a_{t_{i}}+R_{t_{i}}F(F^{T}R_{{t_{i}}}F+U)^{-1}[y_{t_{i}}-F^{T}a_{t_{i}}]\\
=&\begin{pmatrix}
  1\\
	H_{\alpha,t_{i}}+H_{\beta,t_{i}}m_{{t_{i-1}}}
 \end{pmatrix}\\
 &+
 \left[
    \begin{pmatrix}
{H_{\beta,t_{i}}}^2c_{{t_{i-1}}}^2+\Xi_{t_i} +\sigma_{\nu}^2
 \end{pmatrix}
 \right]^{-1}\\
&\times
 \begin{pmatrix}
	0 \\
  {H_{\beta,t_{i}}}^2c_{{t_{i-1}}}^2+\Xi_{t_i}
 \end{pmatrix}
 \left[
 y_{t_{i}}-
 \begin{pmatrix}
  0 & 1 \\
 \end{pmatrix}
 \begin{pmatrix}
  1\\
	H_{\alpha,t_{i}}+H_{\beta,t_{i}}m_{{t_{i-1}}}
 \end{pmatrix}
 \right]\\
   =&\begin{pmatrix}
	0\\
   H_{\alpha,t_{i}}+H_{\beta,t_{i}}m_{{t_{i-1}}}+\frac{{H_{\beta,t_{i}}}^2c_{{t_{i-1}}}^2+\Xi_{t_i}}{ {H_{\beta,t_{i}}}^2c_{{t_{i-1}}}^2+\Xi_{t_i}+\sigma_{\nu}^2}\left[y_{t_{i}}-H_{\alpha,t_{i}}-H_{\beta,t_{i}}m_{{t_{i-1}}}\right]
 \end{pmatrix}
 \end{align*}
   \begin{align*}
   C_{t_{i}}=&R_{t_{i}}-R_{t_{i}}F(F^TR_{t_{i}}F+U)^{-1}F^{T}R_{t_{i}}\\
 =&\begin{pmatrix}
	0 & 0\\
  0 & {H_{\beta,t_{i}}}^2c_{{t_{i-1}}}^2+\Xi_{t_i}
 \end{pmatrix}\\
&-
 \left[
    \begin{pmatrix}
{H_{\beta,t_{i}}}^2c_{{t_{i-1}}}^2+\Xi_{t_i} +\sigma_{\nu}^2
 \end{pmatrix}
 \right]^{-1}\\
&\times
 \begin{pmatrix}
	0\\
  {H_{\beta,t_{i}}}^2c_{{t_{i-1}}}^2+\Xi_{t_i}
 \end{pmatrix}
 \left[ 
  \begin{pmatrix}
 0 & 1
 \end{pmatrix}
  \begin{pmatrix}
  0 & 0\\
  0 & {H{\beta,t_{i}}}^2c_{{t_{i-1}}}^2+\Xi_{t_i}
 \end{pmatrix}
 \right]
 \\
     =&
       \begin{pmatrix}
  0 & 0 \\
  0 & {H_{\beta,t_{i}}}^2c_{{t_{i-1}}}^2+\Xi_{t_i}-\frac{\left({H_{\beta,t_{i}}}^2c_{{t_{i-1}}}^2+\Xi_{t_i}\right)^2}{{H_{\beta,t_{i}}}^2c_{{t_{i-1}}}^2+\Xi_{t_i}+\sigma_{\nu}^2}
 \end{pmatrix}
      \end{align*}
\\
With $m_{t_{i}}$ and $C_{t_{i}}$ for $i=1:N$, we can evaluate $a_{t_{i}}$, $R_{t_{i}}$ and $\pi(x_{t_{i}}|y_{t_{1:(i-1)}})$ for $i=1:N$.
We are interested in $\pi(y_{t_{1:i}})=\prod^N_{i=1}\pi(y_{t_{i}}|y_{t_{1:(i-1)}})$, where
$\pi(y_{t_{i}}|y_{t_{1:(i-1)}})=\int_{x}\pi(y_{t_{i}}|x_{t_{i}})\pi(x_{t_{i}}|y_{t_{1:(i-1)}})dx_{t_{i}}$ gives a tractable Gaussian integral.
Finally, 
\begin{align*}
\log\pi(y_{{t_{1:N}}})&=\sum^N_{i=1}\log\pi(y_{t_{i}}|y_{{t_{1:(i-1)}}})\\
&=\sum^N_{i=1}\left[-\log\left({\sqrt{2\pi(\sigma_{f}^2+\sigma_{g}^2)}}\right){-\frac{(\mu_f-\mu_g)^2}{2(\sigma_{f}^2+\sigma_{g}^2)}}\right],
\end{align*}
\begin{align*}
\text{where }\mu_f-\mu_g=&y_{t_{i}}-a_{t_{i}}=y_{t_{i}}-H_{\alpha,t_{i}}-H_{\beta,t_{i}}m_{t_{i-1}}\\
\text{and }\sigma_{f}^2+\sigma_{g}^2=&\sigma_{\nu}^2+R_{t_{i}}=\sigma_{\nu}^2+{H_{\beta,t_{i}}}^2c_{{t_{i-1}}}^2+\Xi_{t_i}.
\end{align*}
  \clearpage
\underline{Procedure}\\
\\
1. Set $i=1$. Initialize $m_0=P$ and $C_0=0$.\\
\\
2. Evaluate and store the following log likelihood term: 
\begin{align*}
\log\pi(y_{t_{i}}|y_{t_{1:(i-1)}})=&\left[-\log\left({\sqrt{2\pi(\sigma_{f}^2+\sigma_{g}^2)}}\right){-\frac{(\mu_f-\mu_g)^2}{2(\sigma_{f}^2+\sigma_{g}^2)}}\right],\\
\text{where }\mu_f-\mu_g=&y_{t_{i}}-H_{\alpha,t_{i}}-H_{\beta,t_{i}}m_{{t_{i-1}}}
\text{ and }\sigma_{f}^2+\sigma_{g}^2=\sigma_{\nu}^2+{H_{\beta,t_{i}}}^2c_{{t_{i-1}}}^2+\Xi_{t_i}.
\end{align*}
3. Create and store both $m_{t_i}$, and $C_{t_{i}}$,
\begin{align*}
 \text{where }m_{t_{i}}=&H_{\alpha,t_{i}}+H_{\beta,t_{i}}m_{t_{i-1}}+\frac{{H_{\beta,t_{i}}}^2c_{{t_{i-1}}}^2+\Xi_{t_i}}{ {H_{\beta,t_{i}}}^2c_{{t_{i-1}}}^2+\Xi_{t_i}+\sigma_{\nu}^2}\left[y_{t_{i}}-H_{\alpha,t_{i}}-H_{\beta,t_{i}}m_{{t_{i-1}}}\right]\\
\text{and }c_{{t_{i}}}^2=&{H_{\beta,t_{i}}}^2c_{{t_{i-1}}}^2+\Xi_{t_i}-\frac{\left({H_{\beta,t_{i}}}^2c_{{t_{i-1}}}^2+\Xi_{t_i}\right)^2}{{H_{\beta,t_{i}}}^2c_{{t_{i-1}}}^2+\Xi_{t_i}+\sigma_{\nu}^2}.
 \end{align*}
\\
4. Increment $i$, $i$=$(i+1)$ and repeat steps 2-3 till $\log\pi(y_{t_{N}}|y_{t_{1:(N-1)}})$ is evaluated.\\
\\
5. Calculate the sum:
\begin{equation*}
\log\pi(y_{{t_{1:N}}})=\sum^N_{i=1}\log\pi(y_{t_{i}}|y_{{t_{1:(i-1)}}}).
\end{equation*}


  \bibliography{references}             
 \bibliographystyle{jfm2}

\end{document}